\documentclass[12pt]{unhthesis}

\usepackage{amsfonts}
\usepackage{textcomp}
\usepackage{amsmath,amssymb}
\usepackage{amsthm}
\usepackage{graphicx}
\usepackage{subcaption}

\usepackage{rotating}
\usepackage{natbib}

\usepackage[T1]{fontenc}
\usepackage{times}
\usepackage{textcomp,wasysym,comment,titlesec,color,booktabs} 
\usepackage[notintoc]{nomencl}
\usepackage{appendix}
\usepackage{multirow}

\usepackage{caption}								
\usepackage{indentfirst}                  

\usepackage{relsize,units}
\usepackage[multiple]{footmisc}
\usepackage{ulem}
\usepackage{longtable}
\usepackage{url}
\usepackage{float}
\usepackage{pdfpages}
\usepackage{color,soul}

\usepackage[bookmarks=true, colorlinks=true, citecolor=black,
urlcolor=black, linkcolor=black]{hyperref}

\graphicspath{{thesisimages/}}

\definecolor{myGrey}{rgb}{0.85,0.85,0.85}

\includecomment{figs}

\titlespacing*{\section}{0in}{0.1in}{0.1in}
\titlespacing*{\subsection}{0in}{0.05in}{0.05in}
\titlespacing*{\subsubsection}{0in}{0.025in}{0.025in}

\makenomenclature
\RequirePackage{ifthen}
\renewcommand{\nomgroup}[1]{%
\ifthenelse{\equal{#1}{A}}{\item[\textbf{Roman Symbols}]}{%
\ifthenelse{\equal{#1}{G}}{\vspace{0.5cm}\item[\textbf{Greek Symbols}]}{%
\ifthenelse{\equal{#1}{Z}}{\vspace{0.5cm}\item[\textbf{Abbreviations}]}{%
\ifthenelse{\equal{#1}{S}}{\vspace{0.5cm}\item[\textbf{Subscripts/Superscripts}]}
{}
}
}
}
}

\begin{document}

\title{DATA ANALYSIS FOR A BALLOON BORNE COMPTON POLARIMETER}

\author{Sambid Kumar Wasti}

\prevdegrees
{
    BA, Hiram College, Hiram, Ohio, 2012
 }

\major{Physics}

\degree{Doctor of Philosophy}

\degreemonth{December	}

\degreeyear{2020}

\thesisdate{\today}

\frontmatter

\maketitle

\copyrightyear{2020}
\makecopyright

\newpage
\vspace*{0.7 in}

\begin{singlespace}

\noindent \small{This dissertation has been examined and approved in partial fulfillment of the requirements for the degree of Doctor of Philosophy in Physics by:}
\vspace{0.9in}

\hfill                                      
\parbox{4in} {                               
\textbf{Dissertation Director, Mark McConnell,}\\ \small{Professor of Physics \& Astronomy}
\vspace{0.8in}

 {\textbf{Joe Dwyer,}\\ \small{Professor of Physics \& Astronomy}}
\vspace{0.8in}

 {\textbf{Karsten Pohl,}\\   \small{Professor of Physics \& Astronomy}}
\vspace{0.8in}

 {\textbf{Fabian Kislat,}\\  \small {Assistant Professor of Physics \& Astronomy}}
\vspace{0.8in}

{\textbf{Lynn Kistler,}\\   \small{Professor of Physics \& Astronomy}}

\vspace{0.4in}
         on 11/25/2020.}

\vspace{0.5 in}
\noindent \small{Original approval signatures are on file with the University of New Hampshire Graduate School.}


\end{singlespace}
\vspace*{\fill}

\begin{Dedication}
\vfill
\begin{flushright}
\textit
{
    This dissertation is dedicated to my parents.
}
\end{flushright}
\end{Dedication}


\begin{Acknowledgments}
\vspace{0.3in}
\setlength{\baselineskip}{2\baselineskip}
{
    First and foremost, I would like to thank my advisor, Dr. Mark McConnell, for his guidance through out my research years at UNH.  
    This would not have been possible without his guidance, advice and patience. 
    I would also like to acknowledge and thank Camden Ertley, Peter Bloser, James Ryan and Jason Legere for their help with the GRAPE and the balloon campaign.  
    I would also like to thank Taylor Connor, Chris Bancroft, Amanda Madden, and rest of the GRAPE collaboration.
     I would also like to thank my committee members, Dr. Joseph Dwyer, Dr. Karsten Pohl, Dr. Lynn Kister and Dr. Fabian Kislat for their insightful remarks and guidance to make this thesis a reality.

	Words cannot express how grateful I am to my family. 
	My mom Sushama, who has been more worried about my thesis completion than a graduate student like myself.
	And my dad Sharat, whose wisdom has guided me during difficult times, throughout my life.
	My brother Shashanka and my cousins, Reshmi, Asphota, Ayush, Ashish, Abhijan, Prapti and Prafulla who have continuously reminded me to be grounded and be myself.
	I would also like to thank my friends Jason Shuster, Cristian Ferradas, Kate Luksha, Tejaswita Sharma, Karla O\~nate, Nick Lubinsky, Pranav Shakya, and Aaditya Raj Bhandari. The board game nights and the philosophical discussions were an escape to life outside research and classes.  	
	All of this would not have been possible without the support of my friends and family. 

	Finally, I would like to acknowledge the funding that made my work possible: NASA NNX13AB96G, NASA NNX13AB96G, and NASA 80NSSC19K0624. I would also like thank CUA for supporting me during the last few months of completion of this thesis.
}
\end{Acknowledgments}

\clearpage
\tableofcontents
\listofnomenclature
\listoftables
\listoffigures
\clearpage

\doublespace
\begin{Abstractpage}
\doublespacenormalsize
The Gamma RAy Polarimeter Experiment (GRAPE), a balloon borne Compton polarimeter for 50-500 keV gamma rays, was successfully flown for the second time in 2014.
GRAPE uses events in which a photon scatters between 2 detector elements to measure the polarization of the measured photons. 
GRAPE consists of 24 collimated polarimeter modules.
Each module is made up of 64 rectangular scintillators elements in a grid of 8 $\times$ 8 (36 plastic scintillators surrounded by 28 CsI scintillators). 
GRAPE was flown from Fort Sumner, New Mexico on the morning of September 26$^{\text{th}}$ 2014. 
The experiment was at float altitude for 14.4 hours. 
The Crab (the pulsar and the nebula) was the primary target of the experiment.
A polarization measurement of the Crab would not only validate the instrument design (for future missions observing GRBs), but also improve our understanding of the emission mechanisms of the Crab emission.
The flight plan included 8 hours of Crab observation but the flight was terminated before all data could be collected.
The Crab was observed for only 1.8 hours. 
Background dominates the observations at flight altitudes, therefore a good background estimation is crucial for the analysis. 
The background depends on many flight and instrument parameters (altitude, instrument pointing, temperatures, etc).
Estimating the Crab background using these parameters were a challenging task.
A technique based on the Principle Component Analysis (PCA) was implemented which used the varying parameters to estimate the background for the Crab. 
 Our instrument design allows for an optical crosstalk (a known issue) that arises from light exiting the scintillator element and spilling to neighboring anodes.  
A model was developed to represent the crosstalk, which was subsequently incorporated in our simulations and the instrument response which further improved our analysis.
The analysis was focused on the phase-integrated data (due to limited statistics of the shortened observation period).
A power-law spectrum with  photon index of 1.70$\pm$0.24 and a normalization of 1.01$\pm$1.35 ph/keV/s/cm$^2$ was measured. 
A polarization fraction of 0.43$\pm$0.4 and a polarization angle of 56$^\circ \pm$30$^\circ$ was measured for the phase integrated Crab observation in the 70-200 keV energy range.  
This result was not sufficiently significant enough to further our understanding of the emission mechanism but was statistically consistent with other experiments. 

\end{Abstractpage}

\mainmatter
\chapter{Introduction}
Astronomical objects are studied using the radiation that they produce. 
 There are four properties of this electromagnetic radiation that can be measured. 
 These four properties are energy, intensity, direction, and polarization \citep{Weisskopf2009}. 
  The spectral characteristics and time variability of the electromagnetic radiation observed is usually used to understand the emission mechanisms of these radiations. 
However, this analysis often allows two or more very different models to successfully explain the observations. 
 The polarization measurements adds two more parameters (polarization fraction and polarization angle) to the observational parameters and helps to narrow down the emission models \citep{Lei1997}.
 For high energies (X-rays and gamma-rays) the first three properties have been extensively studied.
 However, the fourth property (polarization) hasn't been studied extensively until recently. 
 This lack of study can be accounted for by several reasons. 
 First, the efficiency of the polarimetry measurements are relatively lower than that of spectral and flux measurements. 
 The major problem with polarimetry is that the measurements are subject to subtle systematic effects which both cover up and mimic a real polarization signal.
The low signal rate (low efficiency in a high background environment) is the second obstacle \citep{Weisskopf2009}.

 The polarization measurements consist of measuring the polarization faction (polarization degree) and the polarization angle. 
 The polarization vector is defined by the electric field vector of the electromagnetic radiation. 
  For a beam of photons, the polarization fraction is defined by the fraction of the photons that share the same electric field vector and the angle of this electric field vector defines the polarization angle. 
  The techniques used to measure the polarization depends on the energy of the incoming photons. 
  The photoelectric effect is used at lower energies ($\sim$1 to $\sim$30 keV). Pair production can be used at higher energies ($\sim$20 MeV). At intermediate energies ($\sim$30  keV to 20 MeV) polarization measurements are done using Compton scattering \citep{Abdel-Rahman2010}.
  The Gamma RAy Polarimeter Experiment (GRAPE) is optimized for the energy range 50-500 keV using Compton polarimetry.

GRAPE's primary objective is to measure the gamma-ray polarization of the Crab Nebula. 
The Crab Nebula is the remnant of a supernova explosion first observed in 1954 by Chinese astronomers.
The Crab pulsar resides in the center of the Crab Nebula, whose luminosity is powered by the pulsar.
The Crab is one of the most observed object in the sky as the Crab emits photons from radio to gamma energies.
At optical energies and below, the Crab has been studied extensively.
At X-rays and higher, there are still some debate about the origin these high energy photons. 
The polarization measurements of these photons are expected to provide additional information that would help identify the emission mechanism and the emission regions \citep{Hester2008,Lei1997}.

There are only handful of polarization measurements done of the Crab in the gamma ray energy range. 
One of the early polarization measurements of the Crab Nebula, in the hard X-ray and gamma energy, was done by \citet{Weisskopf1978} using the Eighth Orbiting Solar Observatory (OSO-8). 
The polarization measurements were done at two energies. 
At 2.6 keV, they measured a polarization fraction of 19.2 $\pm$ 1.0\% and at 5.2 keV, they measured a polarization of 19.5 $\pm$ 2.8\% for the off-pulse phase that corresponds to the Crab Nebula. 
More than 20 years, the first measurements at higher energies were made using the instruments on board INTErnational Gamma-Ray Astrophysics Laboratory (INTEGRAL). 
 
The two instruments that have been used to measure the polarization (even though they are not primary a polarimeters) are the Spectrometer on INTEGRAL (SPI) and the Imager on-board INTEGRAL (IBIS). 
These instruments and the polarimeter is described in Chapter \ref{sec:polarimetry}. 
SPI measured the polarization fraction of 46 $\pm$ 10\% and the polarization angle of 124.0$^\circ$ $\pm$ 0.1$^\circ$ for the the Crab Nebula (the off-pulse phase observations) in the 0.1 MeV to 1MeV \citep{Dean2008,Chauvin2013}.
IBIS measured the polarization fraction of $>72\%$ and the polarization angle of 120.0$^\circ$ $\pm$ 7.7$^\circ$ for the the Crab Nebula in the200-800 keV \citep{Forot2008}.

In addition to these orbital instruments, The Polarized Gamma-ray Observer (PoGO Lite)  (a balloon borne experiment) have also made polarization measurements of the Crab Nebula. 
The PoGO+ (a newer version of PoGO Lite) flew in 2016 and did the polarization measurement in the energy interval of 20-160 keV of the Crab Nebula (Off-pulse phase of the observation).
The PoGO Lite operated between the energy interval of 20-120 keV and was flown on 2013. 
\citet{Chauvin2017,Chauvin2016} summarizes the polarization measurements that are outlined in Chapter \ref{sec:polarimetry} and 	\ref{sec:results_discussion}.
The PoGO missions filled a gap between the $\sim$10 keV and $\sim$100 keV region for the polarization measurements. 

Gamma Ray Polarimeter Experiment (GRAPE) is a balloon borne polarimeter which was first launched in 2011 to  measure the polarization of the Crab as a demonstration of the instrument design.
GRAPE was optimized to operate in the energy interval of 50-500 keV.
The analysis of the data from the 2011 flight did not result in a significant polarization measurement of the Crab. 
The GRAPE was launched again in 2014 with improvements.
The 2014 GRAPE was also optimized to operate between the energy range of 50-500 keV and the primary goal was still to measure the polarization of the Crab. 
The results from this experiment would fill in the energy gaps that are left out by other missions and also would compare the results with the PoGO mission. 
This dissertation discusses the 2014 GRAPE experiment, the campaign and the results. 

\clearpage
\chapter{Science}

This chapter provides an overview of the science relevant to Gamma RAy Polarimeter Experiment (GRAPE). 
It introduces the emission mechanisms that have the potential of creating polarized photons that are relevant to the 50-500 keV energy range of GRAPE.
The main astrophysical object of interest was the Crab pulsar (and the associated nebula), whose polarized emission was the focus of the 2014 GRAPE balloon flight.

	\section{Emission mechanism}		
	We focus this section on emission mechanisms that are important for the 50-500 keV energy range of GRAPE. 
	This includes Bremsstrahlung, magneto-Bremsstrahlung, curvature radiation and inver Compton. 
	These are summarized here from \citet{Lei1997} and \citet{Longair2011}.
			\subsection{Bremsstrahlung radiation}
\begin{figure}[!t]
\centering 
\includegraphics[width=0.6\textwidth]{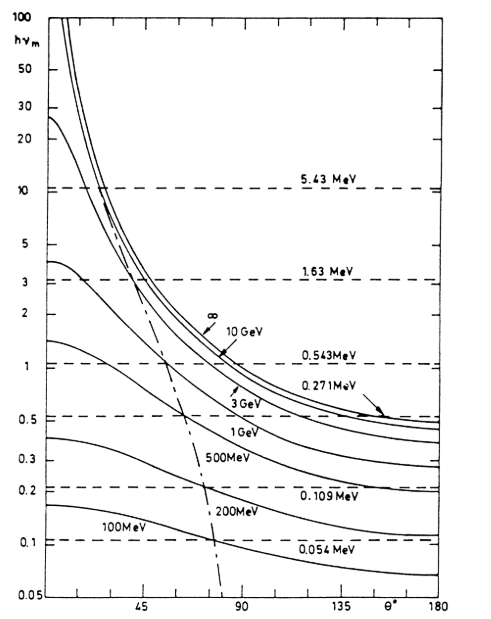}
\caption{Variation of maximum energy (in units of $m_ec^2$) of emitted photons from electron-proton (dashed lines) and proton-electron (solid lines) Bremmstrahlung with the direction of emission for various incident particle energies \citep{Heristchi1986}. The lines correspond to equivalent electron and proton velocities . The electron-proton is isotropic but the proton-electron is beamed in a focused direction \citep{Lei1997}. }
\label{fig:sci_brem}       
\end{figure}

			This radiation was called 'braking radiation' or, in German, \textit{bremsstrahlung}.
			This radiation is mainly associated with the acceleration of a charged particle in an electrostatic field (e.g., that of an ion or an atomic nucleus). 
			 The maximum energy of a photon emitted from bremsstrahlung radiation for any two charged particles is given by \citep{Heristchi1986}
			\begin{equation}
					E_{max} = \frac{m_2\ E_1}{m_2 + m_1 + E_1 - M \cos \theta}
					\label{eqn:max_brem}
			\end{equation}
			where $E_1$, $M$, and $m_1$ are the kinetic energy, momentum and rest mass of the incident particle respectively, $m_2$ is the rest mass of the target particle and $\theta$ is the emission angle measured from incident particle's direction. 
			The bremsstrahlung radiation from an electron-proton interaction (the electron is incident and accelerated by proton) and proton-electron (where the proton is the incident) with maximum emitted energy with the angle of emission  is shown in Figure \ref{fig:sci_brem}.

			The degree of linear polarization for electron-proton can be expressed using 
			\begin{equation}
					\Pi \equiv \frac{d \sigma_\perp - d\sigma_\parallel}{d \sigma_\perp + d\sigma_\parallel}
			\end{equation}						
			where $\sigma_\perp$ is the differential cross section of the emitted photons polarized perpendicular to the plane of emission and  $\sigma_\parallel$ is the differential cross section of the emitted photons polarized parallel to the plane of emission \citep{Lei1997}. \citet{gluckstern1953} derived these parallel and perpendicular component and derived that the degree of linear polarization for the electron-proton bremsstrahlung radiation is given by  	
			\begin{equation}
						\Pi = \frac{m_ec^2  E \Delta_0 - 2m_e^3c^6}{E^2\Delta_0 - m_ec^2E\Delta_0 +2m_e^3c^6}
			\end{equation}
			where E is the energy of the emitted photon and $\Delta_0 = E_0 - M\, \cos\theta$, and $E_e$ is the initial energy of electron. The polarization vector tends to be parallel to the direction of acceleration and the photons tend to be emitted perpendicular to the electron's plane of motion. The degree of linear polarization can reach high levels of the order of 80\% and reaches a maximum for a scattering angle which is dependent upon the incident electron's energy \citep{Longair2011,Lei1997}.
			
			\subsection{Magneto Bremsstrahlung}
			Magneto bremsstrahlung radiation represents the radiation due to acceleration of a particle in presence of a magnetic field.
			For a charged particle moving at a velocity, \textbf{v}, in an uniform magnetic field, \textbf{B}, in the absence of a static electric field, the external force on the particle is given by
			\begin{equation}
					\textbf{F} = \frac{d}{dt}(\gamma m_0\textbf{v})=\gamma m_0\frac{d\textbf{v}}{dt} =\frac{z e}{c} (\textbf{v} \times \textbf{B})
			\label{eqn:force_equation}
			\end{equation}
			where z is the atomic number, $m_0$ is mass of the charged particle, and e is the elementary charge, c is the speed of light and $\gamma$ is the Lorentz factor defined by $\gamma = (1-v^2/c^2)^{-1/2}$. 
			The force on the particle is perpendicular to both \textbf{v} and \textbf{B}. 
			The angle between \textbf{v} and \textbf{B} is defined as the pitch angle, $\alpha$. If the \textbf{v} is perpendicular to \text{B} ($\alpha$ = $90^\circ$) then the particle's path is circular about the magnetic field direction.  
			If the pitch angle is not $90^\circ$ then the particle will follow a helical path in the direction of the magnetic field with relativistic gyrofrequency, $\nu_g$ which is defined by 
			\begin{equation}
					\nu_g = \frac{z e|\textbf{B}|}{2 \pi \gamma m_0} 
			\label{eqn:gyrofrequency}
			\end{equation}
			As the magnetic field is uniform, the radius of the circular motion of the particle corresponding to the above gyrofrequency is given by 
			\begin{equation}
					r = \frac{ \gamma m_0 |\textbf{v}|  \sin\alpha}{ze |\textbf{B}|} = \bigg(\frac{pc}{ze}\bigg) \frac{\sin \alpha}{|\textbf{B}|c}						
			\label{eqn:radius_gyro}
			\end{equation}
			The radiation loss rate for a charge q, with an acceleration $a_\perp$ and $a_\parallel$ in the laboratory frame of reference can be written as 
			\begin{equation}
				-\frac{dE}{dt} = \frac{q^2 \gamma^4}{6 \pi \epsilon_0 c^3} (|a_\perp|^2 + \gamma^2 |a_\parallel|^2)
				\label{eqn:general_radiation_rate_a}
			\end{equation}
			Since the acceleration is always perpendicular to the velocity vector of the particle, $a_\perp = qvB\, \sin \alpha$ and $a_\parallel =0$ \citep{Longair2011}. For an electron, the total radiation loss rate is
			 \begin{equation}
				-\frac{dE}{dt} = \frac{e^4 B^2}{6\pi \epsilon_0 c m_e^2} \bigg( v/c \bigg)^2 \gamma^2 B^2 \sin^2 \alpha
			\label{eqn:general_radiation_rate_elec}
			\end{equation}
			This equation can be rewritten as 			      
			\begin{equation}
				-\frac{dE}{dt} = 2 \, \sigma_T\, c\, U_{mag} \bigg( \frac{v}{c} \bigg)^2 \gamma^2 \sin^2 \alpha
			\label{eqn:general_radiation_rate}
			\end{equation}
			where where $\sigma_T = 8\pi(Ze)^4 / 3m_c^2 c^2$ is the Thompson cross section, $U_{mag} =  B^2/2\mu_0$ is the energy density of the magnetic field, and $c^2 = (\mu_0 \epsilon_0)^{-1}$. The pitch angle ($\alpha$) distribution is often isotropic as it is likely to be randomized either by irregularities in magnetic field distribution or by streaming instabilities.
			 Averaging over the pitch angles, $p(\alpha)\,d\alpha = 1/2 \,\sin\alpha \,d\alpha$,  the average energy loss rate can be defined as \citep{Longair2011}
			\begin{equation}
				-\bigg(\frac{dE}{dt}\bigg)_{Avg} = 2 \, \sigma_T\, c\, U_{mag} \bigg( \frac{v}{c} \bigg)^2\gamma^2 \int_{0}^{\pi} \frac{1}{2} \sin^3 \,\alpha \, d\alpha = \frac{4}{3}\sigma_T\, c\, U_{mag} \bigg( \frac{v}{c} \bigg)^2  \,\gamma^2
			\label{eqn:general_radiation_rate_avg}
			\end{equation}

			There are two special cases of this radiation - the non-relativistic cyclotron and the relativistic synchrotron.			
			\subsubsection{Cyclotron Radiation}
\begin{figure}[hbtp]
 \centering
\begin{subfigure}[b]{0.85\textwidth}
 		 \includegraphics[width=1\linewidth]{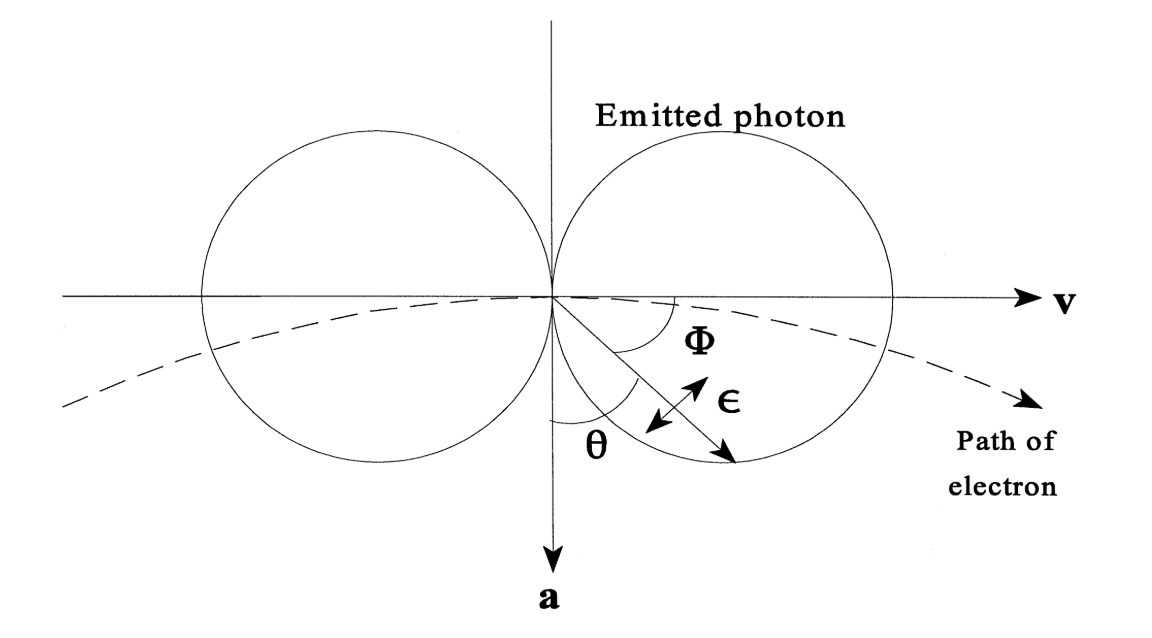}
		 \caption{}
\end{subfigure}    

 \begin{subfigure}[b]{0.60\textwidth}
 		 \includegraphics[width=1\linewidth]{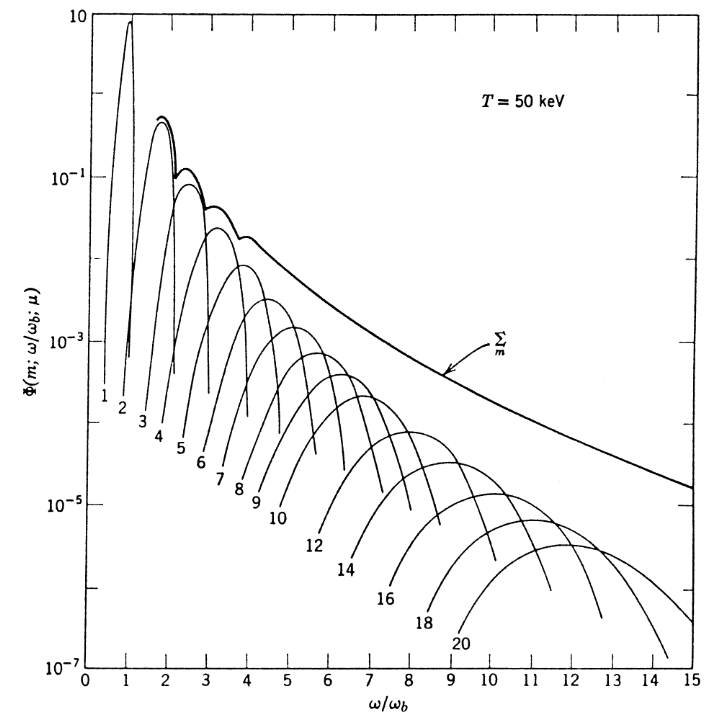}
		 \caption{}
\end{subfigure} 
  \caption{a) Polar diagram showing the dipole radiation emitted by an accelerated electron \citep{Lei1997}. b) The spectrum of emission of the first 20 harmonics of mildly relativistic cyclotron radiation for electrons with $v = 0.04c$ \citep{Bekefi1966}}
\label{fig:sci_cyclotron}
\end{figure}

			Cyclotron radiation is the non-relativistic case of the magneto-Bremsstrahlung radiation. In the non-relativistic limit, where $v \ll c, \gamma =1$, Equation \ref{eqn:general_radiation_rate} for the energy loss rate becomes
			\begin{equation}
				-\frac{dE}{dt} = 2\, \sigma_T\, c\, U_{mag} \bigg( \frac{v}{c} \bigg)^2 \sin^2 \alpha
			\end{equation}
The emitted radiation is dipolar, as shown in Figure \ref{fig:sci_cyclotron}a.
The gyrofrequency defined in Equation \ref{eqn:gyrofrequency} for the cyclotron radiation simplifies to 
			\begin{equation}
				\nu_c = \frac{eB}{2\pi \,m_e c}
				\label{eqn:gyro_cyclo}
			\end{equation}
			where $m_e$ is the mass of electron. The observed polar diagram shown in \ref{fig:sci_cyclotron}a can be Fourier transformed into a sum of equivalent dipoles radiating at harmonics, $l$, of the relativistic gyrofrequency with intensities, $I_l$, 
			\begin{equation}
					I_l \propto \bigg(  \frac{v}{c} \bigg)^{2(l-1)}
			\end{equation}
			 These harmonics results in cyclotron emission to be of discrete energies. 
			 The spectra of the first 20 harmonics is shown in Figure \ref{fig:sci_cyclotron}b
			 The energy radiated for higher harmonics is very small and the maximum energy radiated in this case is from 1st harmonic.  			
			 For the cyclotron case, $\nu_c$ only depends on magnetic field B and not on the energy of the electron.
			 Equation \ref{eqn:gyro_cyclo}can be rearranged to express magnetic field strength required for the first harmonic to produce a photon of the energy E(in keV) as B(T) = 8.64 $\times$ 10$^6$ E (keV) \citep{Lei1997}.
			 The surface of a neutron star has a typical magnetic field of 10$^9$ T which means that cyclotron emission is likely to produce energies $<$ 100 keV \citep{Shrader1995}. 

			The polarization vector $\epsilon$ (electric field vector) of a cyclotron radiation lies in the plane described by acceleration vector \textbf{a} and the velocity vector of the photon. 
			For a distant observer if the line of sight is perpendicular to the magnetic field, the photons detected by the observer will be linearly polarized. 
			If the magnetic field is parallel to the line of sight, the photons detected will be circularly polarized. 
			If the magnetic field is at an angle $\theta$ between parallel and perpendicular to the line of sight, the photons detected will be elliptically polarized. 		
			 
		\subsubsection{Synchrotron Radiation}
		Synchrotron radiation is the relativistic case of the magneto-bremsstrahlung radiation. In the relativistic case $\gamma \to \infty$ , and the total radiation loss rate Equation \ref{eqn:general_radiation_rate} becomes 
			\begin{equation}
				-\frac{dE}{dt} = 2 \, \sigma_T\, c\, U_{mag} \gamma \sin^2 \alpha
			\label{eqn:radiation_rate_synchrotron}
			\end{equation}

\begin{figure}[hbtp]
 \centering
\begin{subfigure}[b]{0.85\textwidth}
 		 \includegraphics[width=1\linewidth]{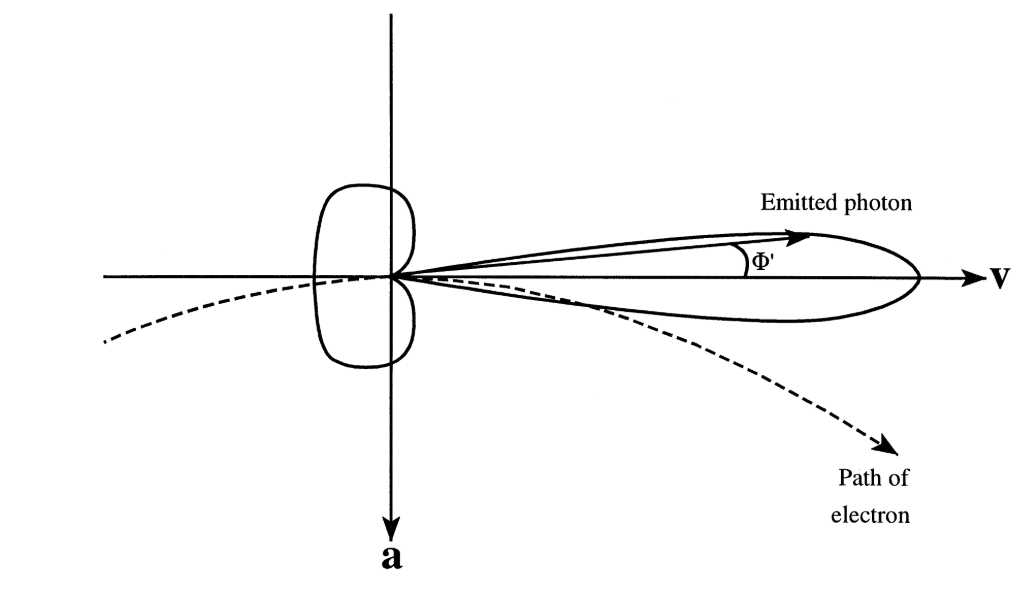}
		 \caption{}
\end{subfigure}    

 \begin{subfigure}[b]{0.60\textwidth}
 		 \includegraphics[width=1\linewidth]{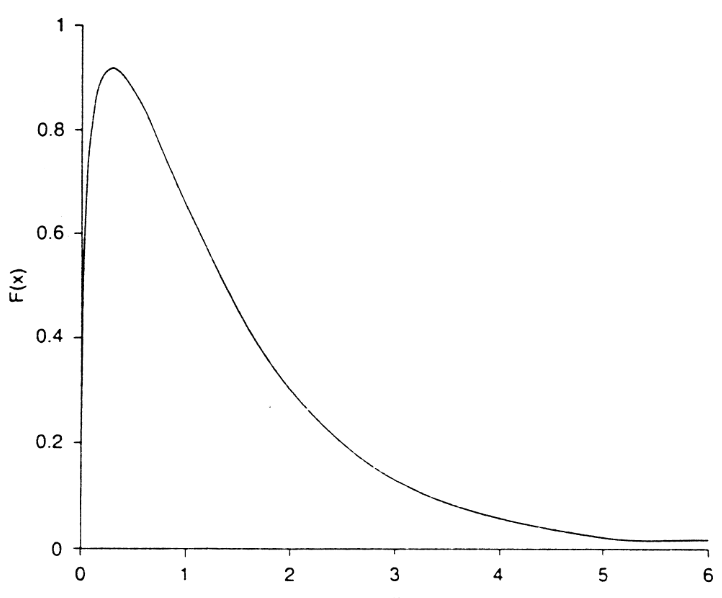}
		 \caption{}
\end{subfigure} 
  \caption{a) The dipole radiation emitted by a relativistic accelerated electron as transformed into the observer's frame of reference. b)The form of spectral energy distribution of a single electron by synchrotron radiation. \citep{Lei1997} }
\label{fig:sci_synchrotron}
\end{figure}	
			
	The radiation beamed is in the direction of the motion of the electron. In the rest frame of the electron, the radiation is similar to the dipole feature as in the cyclotron radiation but in observer's frame, the radiation is in the forward direction along the electron's velocity vector. The observer's frame of reference is shown in \ref{fig:sci_synchrotron}a. The synchrotron radiation spectrum emitted is a smooth continuum as compared to the discrete emission in cyclotron radiation.

		The radiation from a single electron is typically elliptically polarized but will tend towards linear polarization as the energy of the electron increases which is proportional to its velocity.  For the relativistic case of $ v \to c$, the pitch angle $\alpha$ is $90^\circ$ and the emission will be linearly polarized.
		For pitch angles between 90$^\circ$ and 0$^\circ$ the emissions will be elliptically polarized.
  	  	The net polarization is found by integrating over all of the electrons which contribute to the observed intensity. 
		The elliptical components are on either side of the line of sight so they cancel each other and the resultant polarization is linear \citep{Longair2011}. The perpendicular, parallel and total polarization are represented as
		\begin{equation}
				P_\perp(\omega) = \frac{\sqrt3 e^3 B \sin \alpha}{4 \pi m_e c^2} [ F(x) + G(x)]
		\label{eqn:sci_syn_perp}
		\end{equation}
		\begin{equation}
				P_\parallel(\omega) = \frac{\sqrt3 e^3 B \sin \alpha}{4 \pi m_e c^2} [ F(x) - G(x)]
		\label{eqn:sci_syn_para}
		\end{equation}
		\begin{equation}
				P(\omega) = P_\perp(\omega)+P_\parallel(\omega) =\frac{\sqrt3 e^3 B \sin \alpha}{2 \pi m_e c^2}F(x)
		\end{equation}		
	where  $F(x) = x \int_{x}^{\infty} K_{5/3}(z) dz$ and  $G(x) = x K_{2/3}(x)$, and $K_{5/3}$ and $K_{2/3}$ are modified Bessel functions, $x = 2\omega_r r / 3c\gamma^3$ , $\omega_r$ is the relativistic angular gyrofrequency ($\omega = 2 \pi \nu$) and r is the radius of curvature for the electron $r = v/\omega_r \, \sin \,\alpha$. 
	The energy radiated in the frequency range $\nu$ to $\nu + d\nu$ are from electrons with energies in the range $E$ and $E + dE$ so 
	\begin{equation}
		P(\nu)d\nu = \bigg(-\frac{dE}{dt}\bigg)N(E)dE		
		\label{eqn:sci_syn_pol}
  	\end{equation}
where,
		\begin{equation}
				N(E) dE= kE^{-p}dE
		\end{equation}
		\begin{equation}
				E = \gamma m_ec^2 =  \bigg( \frac{\nu}{\nu_g}\bigg)^{1/2} m_e \, c^2
		\end{equation}
		\begin{equation}
				dE =  \frac{m_e \, c^2}{2 \nu_g^{1/2}}\, \nu_g^{1/2} \,d\nu
		\end{equation}
		\begin{equation}
				-\frac{dE}{dt} = \frac{3}{4}\, \sigma_T\, c\, \bigg( \frac{E}{m_ec^2} \bigg)^2 \frac{B^2}{2\mu_0}
			\end{equation}
Substituting the above values in Equation \ref{eqn:sci_syn_pol} and expressed in terms of $\kappa, B, \nu$ and fundamental constants the emissivity can be expressed as
		\begin{equation}
				P(\nu)  = (constants) \kappa B^{(p+1)/2} \nu^{-(p-1)/2}
		\end{equation}
		 The emitted spectrum is represented by $J{\nu}\propto \nu^{-a}$ where a is the spectral index $a = (p-1)/2$. and p is the slope of electron energy spectrum. The spectral energy distribution for a single electron is shown in Figure \ref{fig:sci_synchrotron}b. For a beam of electrons, the fractional degree of polarization, $\Pi$, is defined as 
		\begin{equation}
						\Pi = \frac{P_\perp(\omega) - P_\parallel(\omega)}{P_\perp(\omega) + P_\parallel(\omega)}
						\label{eqn:sci_syn_lin_deg_a}
		\end{equation}
		Substituting the values of P form Equations \ref{eqn:sci_syn_perp}, \ref{eqn:sci_syn_para} we get
						\begin{equation}
						\Pi = \frac{G(x)}{F(x)} = \frac{p +1}{p + \frac{7}{3}}
						\end{equation}
						
			For the observed range of power-law indices, p,  from 1.5 to 5.0 for astrophysical sources that emit synchrotron radiations, the observed degree  of linear polarization is expected to range from approximately 65\% to 80\%. This is the maximum degree of linear polarization and any inhomogeneities in the structure of the magnetic field will result in the degree of linear polarization being reduced. 
		
		\subsubsection{Curvature radiation}
	Curvature radiation defines the radiation from an electron following a curved magnetic field that is accelerated by the curvature. If the radius of curvature is small, significant amount of magneto-Bremsstrahlung radiation will be emitted. In the relativistic limit for the curvature radiation we have, 
	\begin{equation}
	\nu_c = \frac{ \gamma^3 v}{2 \pi r_c}
	\end{equation}							
	where $r_c$ is the radius of curvature of electron's path. $v/r_c$ is the angular frequency associated with it. Magnetic poles of pulsars provide the necessary environment for emission via curvature radiation. The velocity $v$ of the electron is $~c$ in this case. The polarization feature is similar to that of synchrotron, except that the polarization vector of a curvature radiation is parallel to the magnetic field vector as compared to synchrotron where the local magnetic field is perpendicular to the polarization vector.
	
				\subsection{Inverse Compton Scattering}
\begin{figure}[!t]
\centering 
\includegraphics[width=0.8\textwidth]{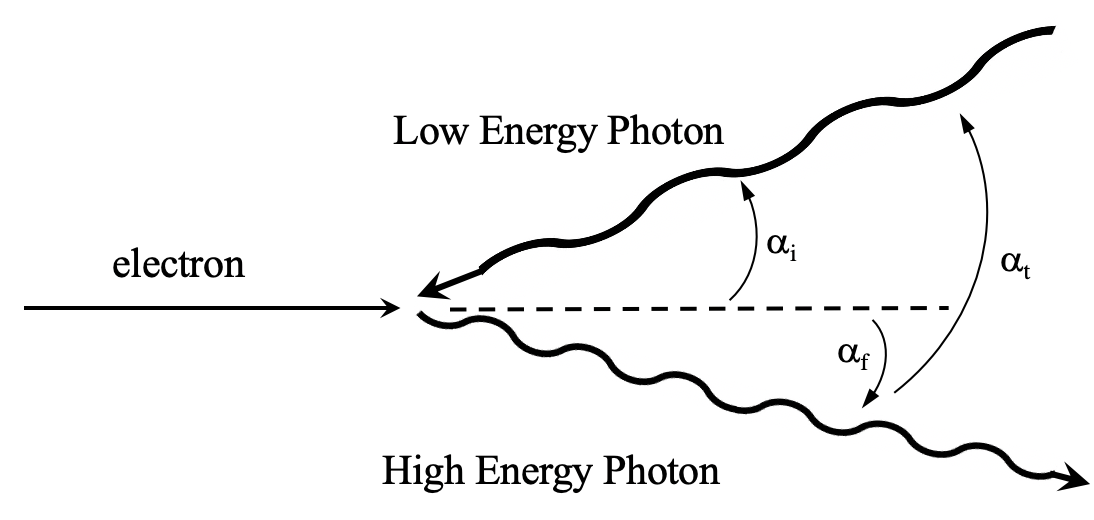}
\caption{Diagram showing the interaction of inverse Compton scattering. A high energy electron scatters a low energy photon. The electron transfers its energy to the scattered photon making it a high energy photon. }
\label{fig:sci_inverse_compton}       
\end{figure}

\begin{figure}[tbp]
\centering 
\includegraphics[width=0.8\textwidth]{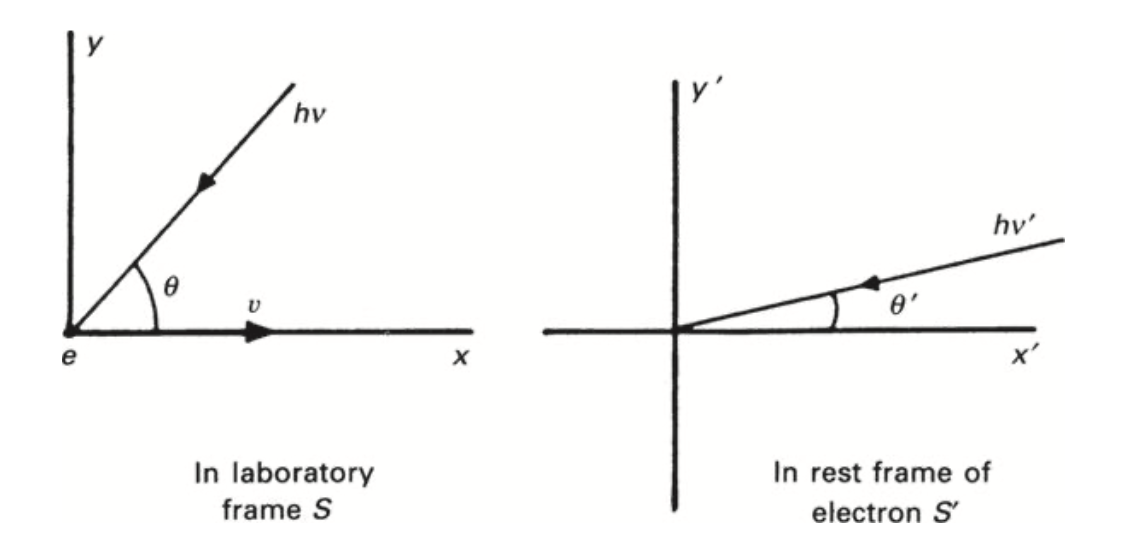}
\caption{Geometry of the inverse compton in the laboratory frame of reference S and in the electron's rest frame S' \citep{Longair2011}.}
\label{fig:sci_inverse_compton_a}       
\end{figure}

				Inverse Compton scattering occurs when a low energy photon scatters off a relativistic electron and gains kinetic energy from the electron. This interaction is shown in Figure \ref{fig:sci_inverse_compton}. The interaction in the rest frame of the relativistic electron and the laboratory frame is shown in Figure \ref{fig:sci_inverse_compton_a} \citep{Longair2011}.  In this case,  $\gamma \hbar \omega \ll m_e c^2$ and the energy of the photon is $\hbar \omega $ and the angle of incidence is $\theta$ in $S$, its energy in $S'$ using the relativistic doppler shift is 
			\begin{equation}
					\hbar \omega' = \gamma \hbar \omega [ 1+ (v/c) \cos \theta]
			\end{equation}

			The angle of incidence $\theta'$ in the frame $S'$ is related to $\theta$ in $S$ by the aberration formulae 
			\begin {equation}
				\sin \theta' = \frac{\sin \theta}{ \gamma [1+ (v/c) \cos \theta]}
			\end{equation}
			\begin {equation}
				\cos \theta' = \frac{\cos \theta + v/c }{[1 + (v/c) \cos \theta]}
			\end{equation}
for $\hbar \omega'  \ll m_ec^2$ the Compton interaction in the rest frame for electron electron , $S'$, is Thompson scattering so the energy loss rate for the electron in $S'$ is given by \citep{Longair2011}
\begin{equation}
		-  \bigg( \frac{dE}{dt} \bigg)' = \sigma_T cu_{rad}'
\end{equation}
where $u_{rad}'$ is the energy density of radiation in the rest frame of electron. The $u_{rad}'$ can be written as 
\begin{equation}
 u_{rad}' = [\gamma (1+ (v/c) \cos \theta) ]^2 u_{rad}
\end{equation}
and applying the Jacobian and the Lorentz transformation we can get $du_{rad}'$. Longair provides a detailed step by step algebra in the derivation of $du_{rad}'$ which ends up being \citep{Longair2011} 
\begin{equation}
	du_{rad}' = u_{rad} \gamma^2 [1+ (v/c) \cos \theta]^2 d\Omega = u_{rad}\gamma^2 [1+ (v/c) \cos \theta]^2 \frac{1}{2} \sin \theta d\theta
\end{equation}
where $d \Omega$ is the solid angle. So integrating over the solid angle, 
\begin{equation}
	u_{rad}' = u_{rad} \int_0^\pi{\gamma^2 [1+ (v/c) \cos \theta]^2 \frac{1}{2} \sin \theta d\theta} =  \frac{4}{3} u_{rad} \bigg( \gamma^2 - \frac{1}{4}\bigg)
\end{equation}
The relativistic invariant dictates that $(dE/dt) = (dE/dt)'$, so 
			\begin{equation}
					\bigg( \frac{dE}{dt} \bigg) = \bigg( \frac{dE}{dt} \bigg)' = \frac{4}{3} \sigma_T u_{rad} \bigg( \gamma^2 - \frac{1}{4}\bigg)
			\end{equation}			
			This is the energy gained by the photon field due to the low energy photon. For the total energy gained, the initial energy of the photon has to be subtracted. The rate at which the energy is removed from the field is $\sigma_T c u_{rad}$  so 
			\begin{equation}
			\bigg( \frac{dE}{dt} \bigg) =  \frac{4}{3} \sigma_T u_{rad} \bigg( \gamma^2 - \frac{1}{4}\bigg) -\sigma_T c u_{rad} = \frac{4}{3} \sigma_T c u_{rad} (\gamma^2 -1)
			\end{equation}
			Using the identity $(\gamma^2 -1 ) = (v^2/c^2)\gamma^2$, the loss rate becomes 
			\begin{equation}
				\bigg( \frac{dE}{dt} \bigg)_{IC} =\frac{4}{3} \sigma_T c u_{rad} \bigg( \frac{v}{c} \bigg)^2 \gamma^2
				\label{eqn:radiation_rate_inverse_compton}
			\end{equation}	
This energy loss rate from inverse Compton is similar to that of synchrotron radiation shown in Equation \ref{eqn:radiation_rate_synchrotron}. The spectral emissivity $I(\nu)$ is represented as
			\begin{equation}
			I(\nu) d\nu = \frac{3\sigma_T c}{16 \gamma^4} \frac{N(\nu_0)}{\nu_0^2} \nu \bigg[ 2\nu ln\bigg( \frac{\nu}{4 \gamma^2 \nu_0}\bigg) + \nu + 4\gamma^2\nu_0- \frac{\nu^2}{2 \gamma^2 \nu_0}\bigg] d\nu
			\end{equation}
			where the isotropic radiation field in the laboratory frame of reference S is assumed to be monochromatic with $\nu_0$ and $N(\nu_0)$ is the number density of photons. The maximum energy of the photon (for a head on head collision) is given by 
			\begin{equation}
					(\hbar \omega)_{max} = \hbar \omega \gamma^2 (1+v/c)^2 \approx 4 \gamma^2 \hbar \omega_0 
			\end{equation}
			The number of photons scattered per unit time is $\sigma_Tcu_{rad}/\hbar \omega$ so the average energy of the scattered photons can be written as 
			\begin{equation}
				\hbar \bar{\omega} = \frac{4}{3} \gamma^2 \bigg( \frac{v}{c} \bigg)^2 \hbar \bar{\omega}_0			
				\end{equation}
				This can be rewritten as 
				\begin{equation}
				E_{avg} = \frac{4}{3} \gamma^2 \bigg( \frac{v}{c} \bigg)^2  E_0
				\end{equation}
      			where $E_0$  is the incident photon energy. Using this equation an electron with $ \gamma = 1000$ can scatter optical photons ($\nu_0 = 4 \times 10^{14} $Hz) up to gamma rays ($ 4 \times 10^{20} Hz \sim 1.6 MeV$). 
			
			The energy loss rate Equation \ref{eqn:radiation_rate_inverse_compton} for the inverse compton is similar to that of synchrotron radiation Equation \ref{eqn:radiation_rate_synchrotron}. Hence the spectral form for inverse Compton will be similar to that of synchrotron radiation. The radiation spectral index of  $a = (p-1)/2$ where p is the spectral index of the electron energy spectrum. 
Inverse Compton scattering can produce polarized flux from an initially unpolarized flux. However, it can also be responsible for depolarizing a polarized flux. Compton scattering and inverse Compton scattering works on the principle that the photons prefer to scatter at an angle perpendicular to its polarization vector.  Compton polarimetry is extensively discussed in section  \ref{sec:compton_polarimetry}.
  The degree of linear polarization  depends on the vectors of the incident photons being aligned. For an unaligned source, the distribution of polarization angles of the photons are isotropic hence the inverse Compton scattering also ends up being isotropic and results in a non-polarized beam.

\section{Pulsars}

\begin{figure}[!t]
\centering 
\includegraphics[width=0.7\textwidth]{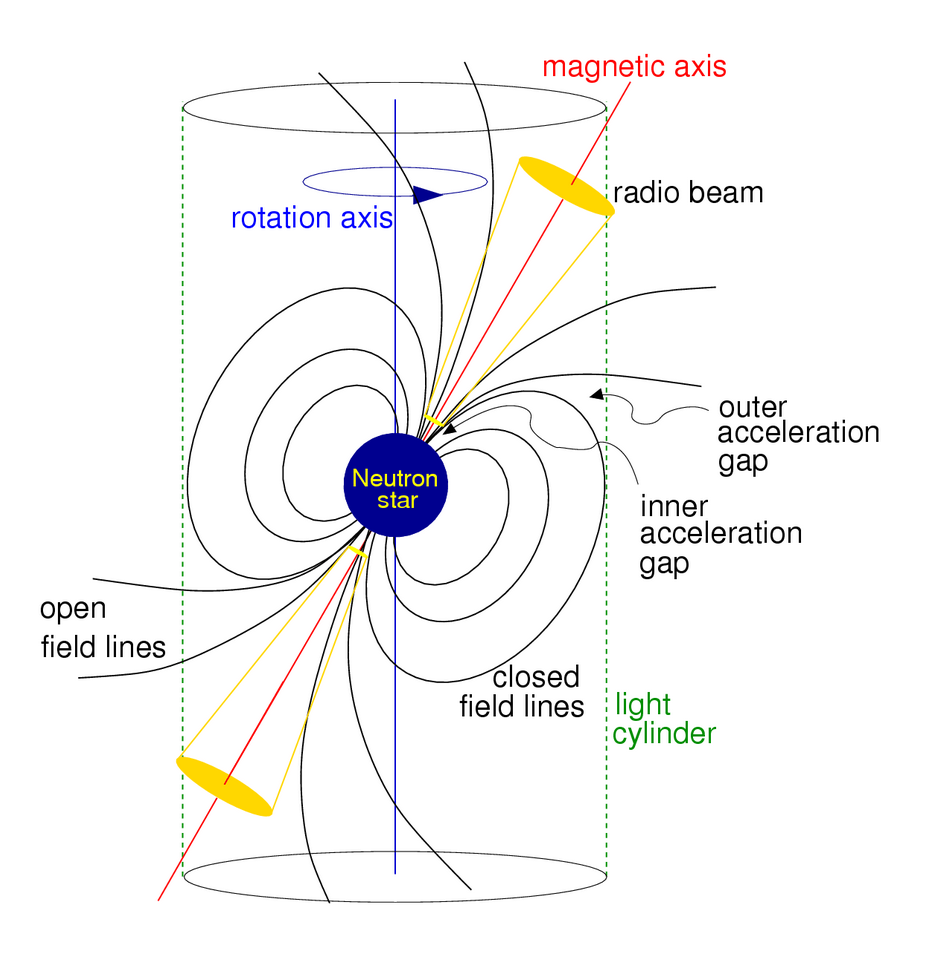}
\caption{Schematic drawing of a pulsar (\citep{Lorimer2012}). The magnetic axis is at some angle $\alpha$ with the rotational axis. The radiation beam is continuously emitted along the conical beam. It is detected when the beam is directed towards the observer and the pulses are noticed due to the rotation. The strong magnetic field accelerates the particle and light cylinder is defined at a radius r where the speed of the particles reaches speed of light. }
\label{fig:sci_pulsar_drawing}       
\end{figure}

One of the astrophysical sources that has the potential of emitting polarized gamma ray photons using the aforementioned emission mechanism are pulsars. Pulsars are rapidly rotating neutron stars formed in supernova explosions. They emit periodic pulses hence the name pulsar. A schematic drawing of a pulsar is shown in Figure \ref{fig:sci_pulsar_drawing}. The  magnetic field axis is at an angle to the rotational axis. The pulsar emits radiation continuously in the form of conical beams above the magnetic poles. Therefore the pulses are observed only when the beam points towards the observer. The angular momentum powers the pulsar so as the pulsar ages the rotation period slows down \citep{Lei1997}. 

\begin{figure}[tbp]
\centering 
\includegraphics[width=0.5\textwidth]{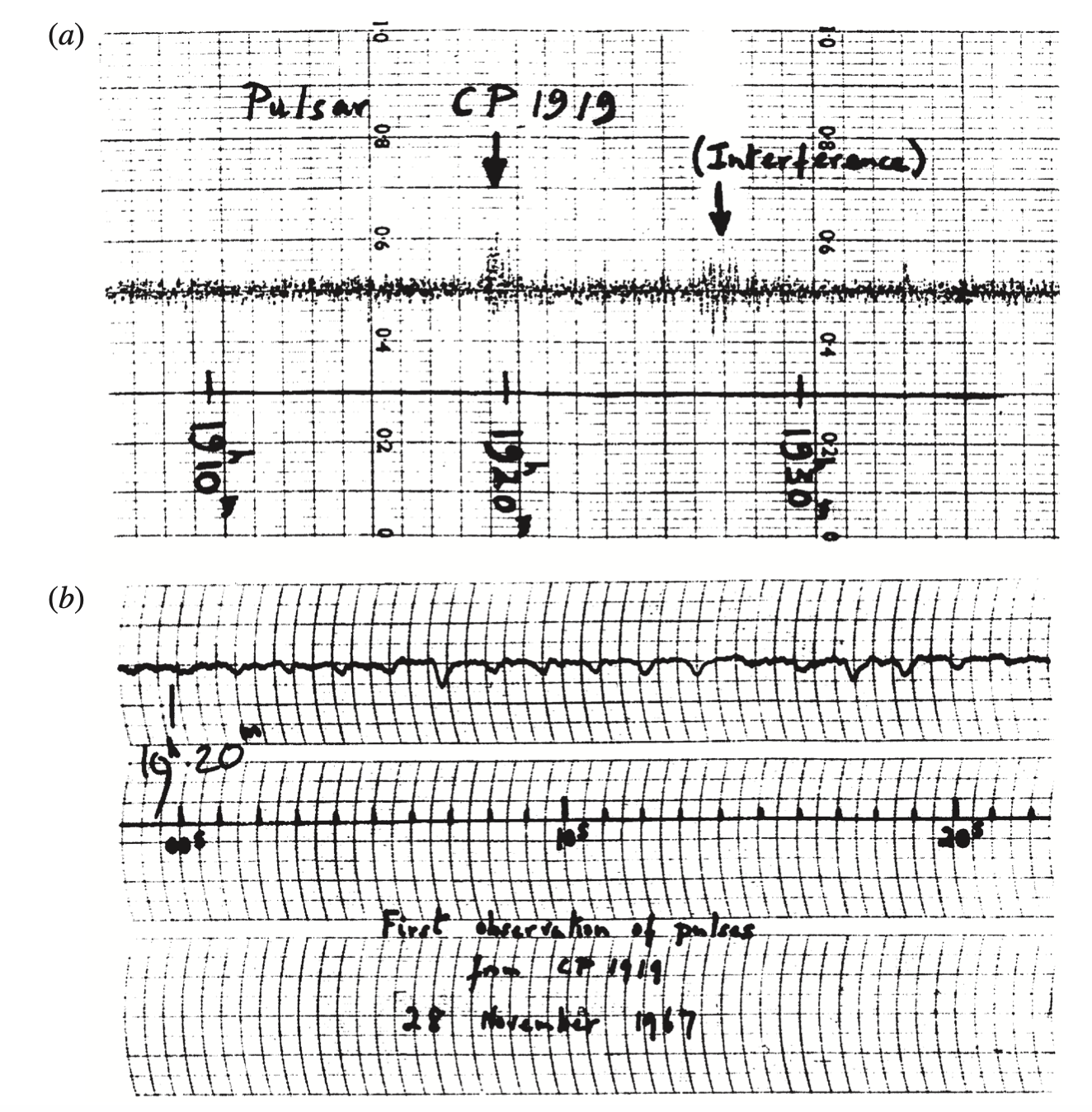}
\caption{The discovery records of the first pulsar discovered, PSR 1919+21. a)The scintillation source where the pulses were observed and labelled CP1919. b) The signal for the PSR 1919+21 observerved with a shorter time-constant showing the 1.33s period. \citep{Longair2011,Hewish1968}}
\label{fig:sci_pulsar_hewish}       
\end{figure}

Pulsars, at radio energies, were discovered by Hewish and Bell in 1967. 
During a sky survey, they discovered series of pulses with a pulse period of 1.33s. This is shown in Figure \ref{fig:sci_pulsar_hewish} \citep{Hewish1968}. These regular pulses appeared about 4 minutes ever solar day.  
It was determined that the signal came from outside our solar system and it was unofficially named "Little Green Men" (LGM-1) as the sources and emission mechanisms were unknown at the time \citep{Longair2011}. Since then hundreds of pulsars have been discovered at multiple wavelengths from radio to gamma rays. 

A pulsar is a rotating neutron star. We can model it as a magnetic dipole and retrieve the relevant equations for radiation and magnetic field. The radiated power from an inclined magnetic field can be represented as    
\begin{equation}
P_{rad} = \frac{2}{3} \frac{(\ddot{m_\perp})^2}{c^3}
\end{equation}
where m$_{\perp}$ is the component of the magnetic dipole moment perpendicular to the rotation axis \citep{Lorimer2012}. 

For this magnetic dipole, rotating with  angular velocity $\Omega$, the moment can be written as
	
\begin{equation}
m = m_0 \,exp(-i\Omega t)
\end{equation}
\begin{equation}
\dot{m} = -i\, \Omega\, m_0\, exp(-i\Omega t) 
\end{equation}
\begin{equation}
\ddot{m} = \Omega^2\, m_0\, exp(-i\Omega t) = \Omega^2\, m
\end{equation}
For a uniformly magnetized sphere with radius R and surface magnetic field strength B, the magnetic dipole moment is $m = BR^3$ \citep{Jackson1998}. A neutron star typically has a radius of 10 km and a mass of 1.4 M$_\odot$. Using the inclination angle of the B-field with respect to the rotation axis, $\alpha$, the magnetic dipole radiation becomes
\begin{equation}
P_{rad} = \frac{2}{3} \frac{(\ddot{m_\perp})^2}{c^3}=  \frac{2}{3} \frac{m_\perp^2 \, \Omega^4}{c^3} = \frac{2\Omega^4}{3c^3}(BR^3\, \sin\alpha)^2 = \frac{2}{3c^3} \bigg( \frac{2\pi}{P}\bigg)^4 (BR^3\, \sin\alpha)^2
\label{eqn:sci_pul_rad_rad}
\end{equation}
The Rotational Kinetic energy for a rotating sphere is given by 
	\begin{equation}
			E_{rot} = \frac{1}{2} I \Omega^2 = \frac{ 2\pi^2 I}{P^2}
	\end{equation}
	where P is the period of the pulsar, $\Omega = 2\pi/P$, moment of inertia for a sphere is $I= 2/5 MR^2$. The change in angular momentum can be written as
	\begin{equation}
	d\Omega/dt = \dot{\Omega} = -2\pi \dot{P}/P^2 
	\label{eqn:ang_mom_rate}
	\end{equation}
	The pulsar is slowing down and the decrease in rotational energy is used to accelerate the protons and electrons from the surface of the star.
	 These accelerated charged particles dissipate this energy loss by radiating photons as the beam along the pulsar axis. 
	 This decrease in rotational energy, also the spin-down luminosity, can be written as 
	\begin{equation}
			\frac{d E_{rot}}{dt} = L_{spin}= \frac{d}{dt} \bigg(  \frac{ 2\pi^2 I}{P^2} \bigg) = \frac{-4\pi^2 I \dot{P}}{P^3}
			\label{eqn:sci_pul_rad_rot}
	\end{equation}
	As this decrease in rotational kinetic energy powers the pulsar (which defines the energy radiated by the pulsar), we can equate the radiated power from rotational (Equation \ref{eqn:sci_pul_rad_rot}) with radiation from magnetic dipole moment (Equation \ref{eqn:sci_pul_rad_rad}) to get B.	\begin{equation}
			\frac{4\pi^2 I \dot{P}}{P^3} = \frac{2}{3c^3} \bigg( \frac{2\pi}{P}\bigg)^4 (BR^3\, \sin\alpha)^2
	\end{equation}
	\begin{equation}
			B^2 =    \frac{4\pi^2 I \dot{P}}{P^3} 
	\end{equation}
	The period of pulsars range from millisecond to seconds. Values of $\dot E$ for pulsar ranges from $\approx$ 5 $\times$ 10$^{38}$ ergs s$^{-1}$ (for millisecond pulsars) down to 3 $\times$ 10$^{28}$  ergs s$^{-1}$ (for slower pulsars) \citep{Gaensler2006}. This results in the pulsar's magnetic field to be between 10$^8$ G (for millisecond pulsars) to 10$^{15}$ G (for magnetars) \citep{Gaensler2006}. 
	
\subsection{Crab}
\begin{figure}[tbp]
 \centering
\begin{subfigure}[b]{0.40\textwidth}
 		 \includegraphics[width=1\linewidth]{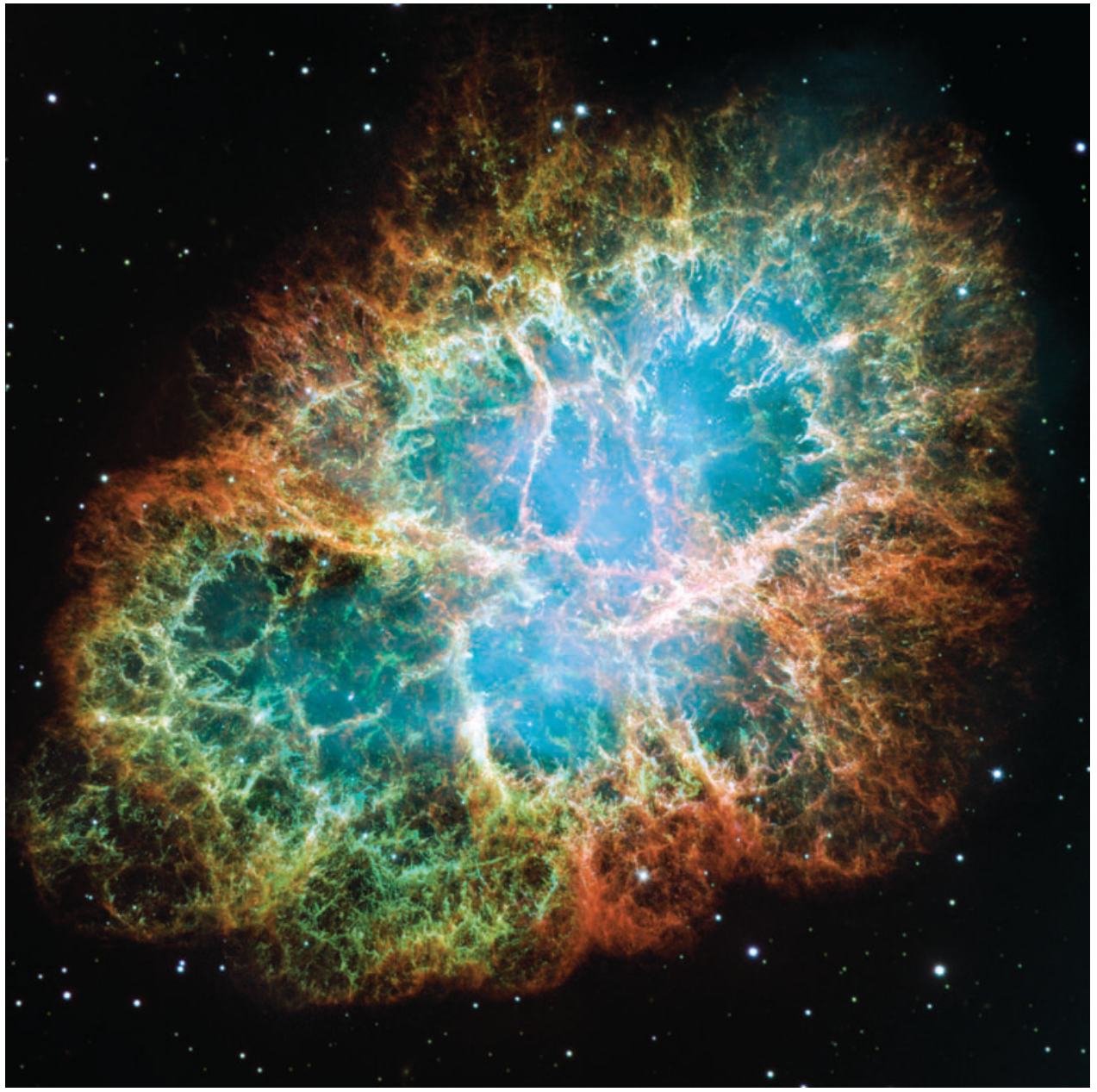}
		 \caption{}
\end{subfigure} \,
 \begin{subfigure}[b]{0.40\textwidth}
 		 \includegraphics[width=1\linewidth]{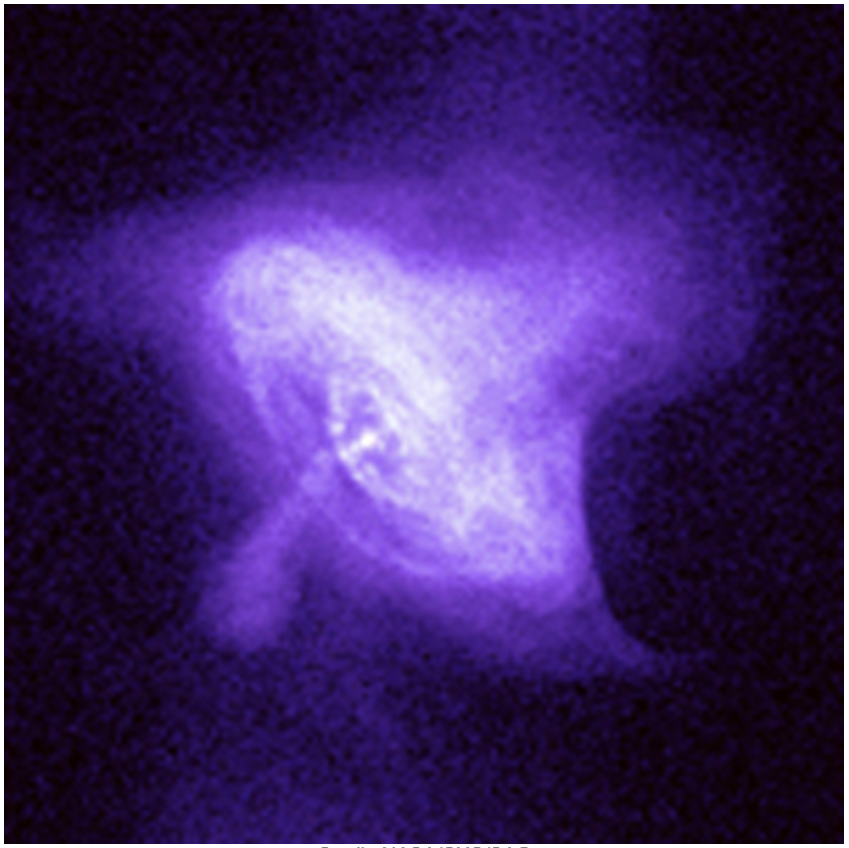}
		 \caption{}
\end{subfigure} 
  \caption{a) A composite Hubble Space Telescope (HST) image of the Crab Nebula. The synchrotron nebula is shown in blue. The thermal ejecta is shown in various shades of the red. The picture is of $6.8 \times 6.8$ arc min \citep{Hester2008}. b) X-Ray image taken by Chandra showing the center pulsar and nebula around it. This picture is for the central $2.5 \times 2.5$ arcmin (Picture Credit: NASA/CXC/SAO) }
\label{fig:sci_crab_a}
\end{figure}

The Crab Nebula is the remnant of a supernova explosion observed in 1054 AD by Chinese astronomers \citep{Clark1983}. It was rediscovered by John Bevis in 1731 \citep{Hester2008}. The name Crab was given to it by William Parsons, third Earl of Rosse, who observed and drew the Crab Nebula using his 72-inch reflecting telescope and claimed it looked like a crab \citep{Parsons1844}. A composite  Hubble Space Telescope image of the Crab Nebula, in the optical range, is shown in Figure \ref{fig:sci_crab_a} a \citep{Hester2008}. 

The supernova remnant consists of a pulsar engulfed by a nebula which can be seen in Figure \ref{fig:sci_crab_a}.  It is at a distance of $\sim$2 kpc \citep{Trimble1968}. 
The synchrotron nebula fills roughly on ellipsoidal volume ($\sim$30 pc$^3$) with a major axis of 4.4 pc and minor axis of 2.9 pc, tilted into the plane of the sky by 30$^\circ$ \citep{Hester2008,Lawrence1995}. 
The pulsar emits a pulse ~33 times in a second (rotation period P = 33 ms) and  $\dot{ \text{P}}$ = 4.21 $\times$ 10$^{-\text{13}}$ \citep{Hester2008}. 
Using these values, the moment of inertia for the Crab is given by $I = 2/5\, MR^2 = 1.12 \times 10^{45}$ gm cm$^2$. 
The spin-down luminosity (the rotational kinetic energy) can be calculated using the Equation \ref{eqn:sci_pul_rad_rot} as $ 5 \times 10^{38}$ erg s$^{-1}$ (130,000 $L_{\odot}$). 
The Crab can accelerate electrons and other particles to extremely high energies which can radiate X-rays and gamma rays. The radiative life time of these particles is very short so they have to radiate close to their acceleration zone. 


\begin{figure}[hbtp]
 \centering \,\,\,\,\,\,\,
\begin{subfigure}[b]{0.47\textwidth}
 		 \includegraphics[width=1\linewidth]{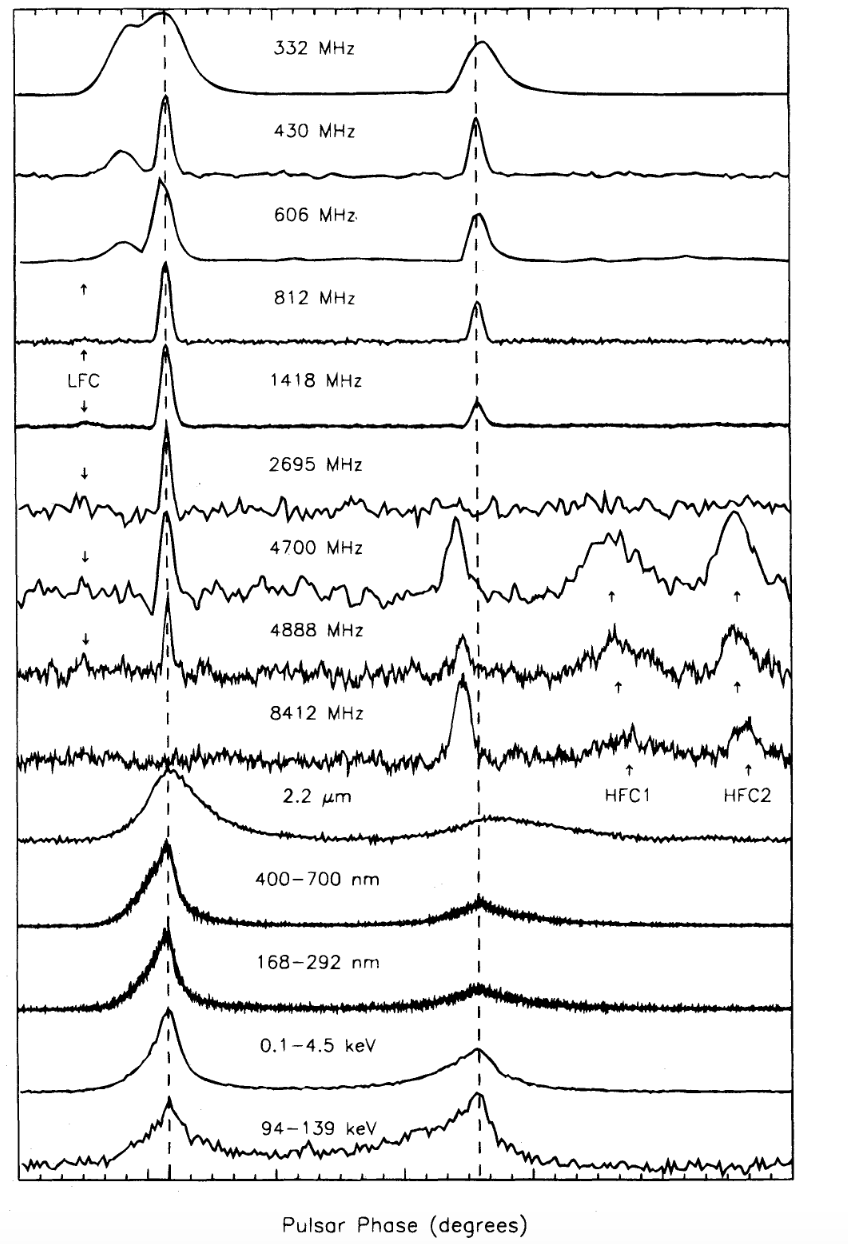}
		 \caption{}
\end{subfigure} 
 \begin{subfigure}[b]{0.52\textwidth}
 		 \includegraphics[width=1\linewidth]{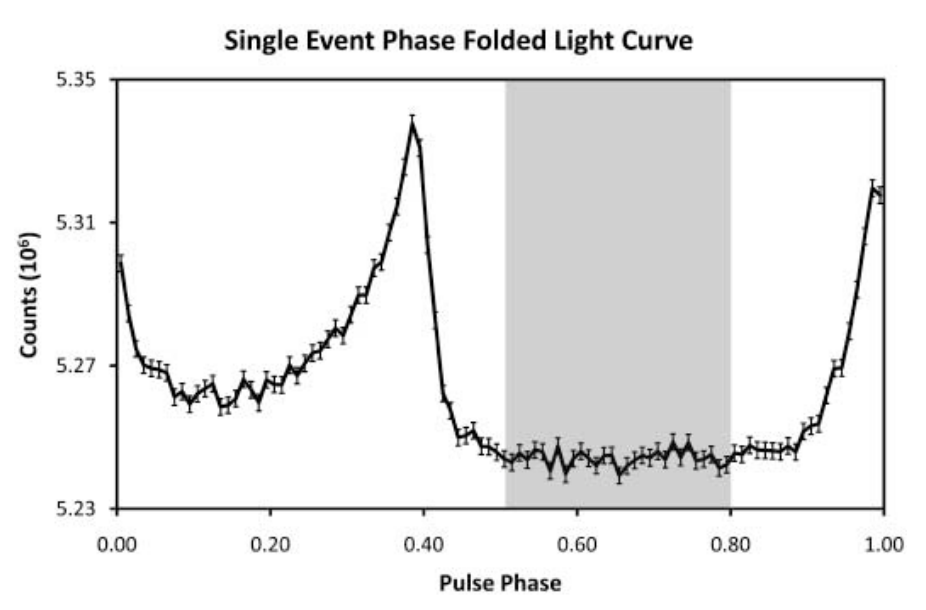}
		 \caption{}
\end{subfigure} 
  \caption{a) Pulse profile of Crab compiled for various energies upto 100 keV \citep{Moffett1996} b) Pulse profile of the Crab for energies 100 keV to 1 MeV \citep{Dean2008}. The peaks are the pulse phase which is attributed to the Crab pulsar and the emission beam. The shaded region is the off-pulse phase which corresponds to the Crab Nebula.}
\label{fig:sci_crab_pulse_profile}
\end{figure}

The Crab is observable from radio to high gamma energies. 
The pulse profile of Crab from radio to gamma energies is shown in Figure \ref{fig:sci_crab_pulse_profile}. 
The the peaks are referred to as as the pulse phase when the emission beam is pointing towards us.
The pulse data is used to extract information about the pulsar.
The off-pulse data (shaded portion on Figure \ref{fig:sci_crab_pulse_profile}b ) is presumed to be emissions from mainly the nebula.
The Crab Nebula and the pulsar follow a power law spectrum of  $\phi$= KE$^{-\Gamma}$ photons$/$cm$^2/$s$/$keV.
The off pulse spectra (representing the nebula) is well defined by $\Gamma$ = 2.23 $\pm$ 0.02 \citep{Massaro2006,Dean2008} and the pulsar component is defined by $\Gamma$ = 2.00 $\pm$ 0.08 \citep{Hameury1983}.
The polarization measurements of the Crab, done with various experiments in the gamma energy, is presented in section \ref{sec:compton_polarimeter_experiments}
There is still some debate about the emission region and the magnetic field structure responsible for the gamma ray emissions from the Crab Nebula and the pulsar.
The polarization measurement along with the spectral analysis of the radiation can aid in localizing the emission region of the photons, emission mechanisms and the magnetic field structure of these emission region \citep{Dean2008}. 
	
\clearpage

\chapter{Polarimetry}
\label{sec:polarimetry}

	Polarimetry is the measurement and analysis of the polarization of a transverse wave. 
	The polarization angle is defined as the orientation of the electric field vector \textbf{\textit{E}} of the source photon. 
	If all the photons observed have the same polarization angle then the radiation source is defined as a 100\% polarized source. 
	If only some photons have a preferred polarized angle then the radiation source is defined as partially polarized source.
	If the photons do not have any preferred polarization angle then the radiation source is defined as unpolarized source. 
	As discussed in previous chapter, the polarimetry, in addition to the spectral analysis, provides further knowledge of the emission mechanisms and the structural information of the observed source. 
	Our instrument does polarimetry based on Compton scattering. 
	This chapter introduces the Compton scattering process and Compton polarimetry.

\section{High Energy interaction with Matter}
\begin{figure}[tbp]
\centering 
\includegraphics[width=0.7\textwidth]{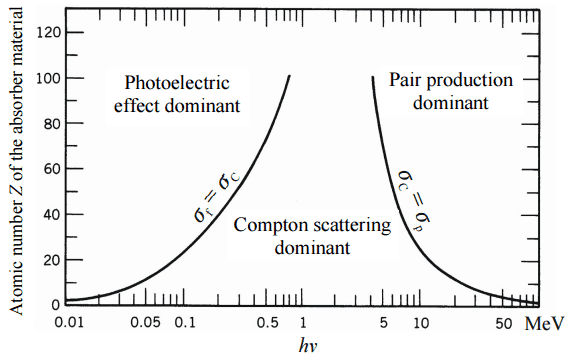}
\caption{Photon interactions at different energies and the atomic number of the material where the interaction occurs  \citep{Abdel-Rahman2010}.}
\label{fig:sci_interaction_attenuation}       
\end{figure}

   In the photon energy range of 1 keV to 100 MeV,  there are three main processes involved in the interaction of high energy photons with atoms, nuclei and electrons. 
   These are photoelectric absorption, Compton scattering and electron-positron pair production. 
   These three processes are used in the detection of high energy photons. 
   Relevant interaction processes for a detector depends on the energy of the photons to be detected and the material of the detectors.
   Figure \ref{fig:sci_interaction_attenuation} shows the dominant interaction process as a function of the atomic number and photon energy.
   Our instrument uses plastic and a CsI(Tl) detectors so we are mostly concerned with the Compton scattering and some with photoelectric absorption. 

		At low energy and high atomic number, the photons tend to interact with matter via photoelectric absorption.
		 If the energies of the incident photon $\varepsilon = h \nu$ are greater than the binding energy ($E_{bind}$)  of a bound electron, then that photon can eject electron with a kinetic energy of  
		\begin{equation}
				E_k = h \nu - E_{bind} 
		\end{equation}
		where $h$ is the Planck constant, $\nu$ is the photon frequency. 
		Detectors based on the photoelectric effect measure the energy of the emitted electron. 
		A mono-energetic beam of photons would result in the emission of photoelectrons having the same energy.
		The resulting spectrum would be a delta function.
		Photoelectrons tend to be ejected parallel to the electric field vector of the incident photon. 
		This fact can be exploited to measure the linear polarization of an incident photon beam by measuring the angular distribution of the photoelectrons \citep{Longair2011}. 

	At high energies, the photons interact with matter via pair production. 
	If a photon has energy greater than $2 m_e c^2$, pair production (electron/positron $e^+/e^-$ pair) can occur in the field of the nucleus. 
	Pair production cannot take place in free space as the momentum and energy cannot be conserved simultaneously.
	Therefore a third body such as nucleus  is required to absorb some of the momentum and energy \citep{Longair2011}. 
	The minimum energy for a photon to interact via pair production is $2 m_e c^2 = 1.022$ MeV. 
	The resulting $e^+/e^-$ pairs are produced in a plane that is parallel to the electric vector of the incident photon. 
	The distribution of these $e^+/e^-$ pair can be used to do polarimetry. 

\subsection{Compton Scattering}
	\label{sec:pol_comp_scattering}
\begin{figure}[tbp]
 \centering
    	\includegraphics[width=0.55\textwidth]{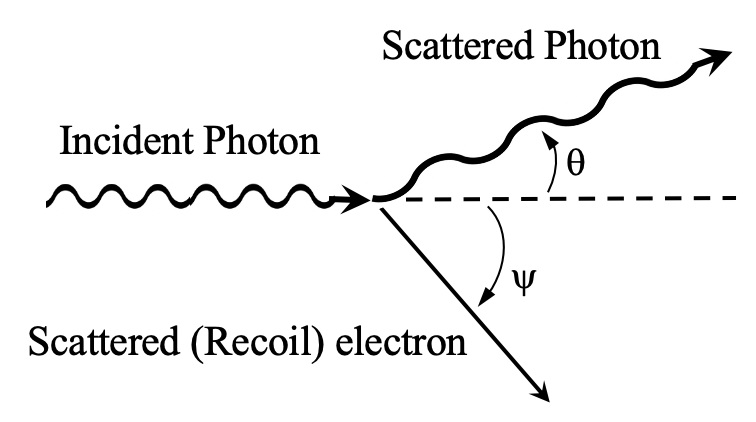}
    	\caption{Illustration of the Compton scattering process, showing the incident photon, the scattered photon and the scattered (recoil) electron. The angle $\theta$ and $\psi$ is the scatter angle for the photon and electron respectively. }
    	\label{fig:sci_comp_simple}
\end{figure}		
	
\begin{figure}[hbtp]
 \centering
    	\includegraphics[width=0.55\textwidth]{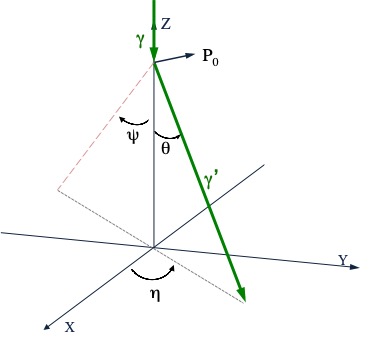}
    	\caption{Drawing of a Compton scattered vectors with various components. The green $\gamma$ is the incident photon of energy E  and $\textbf{P}_0$ is its polarization vector. Green $\gamma \,'$ is the scattered photon with energy E'. The $\theta$ is the Compton scatter angle for the photon, $\psi$ is the scatter angle for the electron and $\eta$ is the angle of the scattered photon in the azimuthal plane.}
    	\label{fig:sci_comp_pol_vector}
\end{figure}

		Compton scattering is a process in which a photon scatters off an electron. 
		A simple Compton scattering (for the case of free electron) is shown in figure \ref{fig:sci_comp_simple} where we can see the incident photon, the scattered photon and the scattered electron (recoil electron).
		For a Compton scatter event, the energy loss of the photon scatter by an electron is given by the Equation
		\begin{equation}
			\lambda'  - \lambda = \frac{h}{m_e c} (1-\cos \theta)
			\label{eqn:compton_scattering_a}
		\end{equation}
		$m_e$ is the mass of the electron, $\theta$ is the scattering angle of the photon. 
		$\lambda$ is the incident photon wavelength and $\lambda'$ is the scattered photon wavelength \citep{Compton1923}.
		Using, $ E = hc/\lambda$, we can  get the ratio of the scattered energy $E'$ to initial energy $E_o$ as 			
			\begin{equation}
					\epsilon = \frac{E'}{E_0}  = \frac{1}{1 + (E_0/m_e c^2) (1- \cos\theta)}
					\label{eqn:compton_scattering_b}
			\end{equation}
			The probability that a photon will scatter by an angle $\theta$ is given by the Klein-Nishina differential cross-section for a free electron at rest \citep{Lei1997,Longair2011}. 		
			For a polarized beam, the Klein-Nishina differential cross-section for a free electron at rest becomes
			\begin{equation}
					\frac{d \sigma_{KN,P}}{d \Omega} = \frac{1}{2} {r_0}^2 \epsilon^2 [ \epsilon + \epsilon^{-1} -2\, \sin^2 \theta\, \cos^2 \eta] 
					\label{eqn:compton_klein_nishina_pol}
			\end{equation}
			where $r_0$ is the classical electron radius, $\epsilon$ is the ratio $E'/E_0$,  $\theta$ is the compton scattering angle and $\eta$ is the azimuthal scattering angle \citep{Lei1997}. 
			This $\eta$ is the angle between the scattered photon and the polarization vector of the incident photon, \textbf{P}$_0$, as measured in a plane perpendicular to the incident photon direction.
			This can be seen clearly in the figure \ref{fig:sci_comp_pol_vector} which shows a typical scatter geometry. 
			The figure is comparable to figure \ref{fig:sci_comp_simple}, but now with a three-dimensional perspective. 
			
			Equation \ref{eqn:compton_klein_nishina_pol} can integrated over all possible values of $\eta$ to get the Klein-Nishina differential cross-section for unpolarized case as		
			\begin{equation}
			\frac{d \sigma_{KN,U}}{d \Omega} = \frac{1}{2} {r_0}^2 \epsilon^2 [ \epsilon + \epsilon^{-1} - \sin^2 \theta ] 
			\label{eqn:compton_klein_nishina_unpol}
			\end{equation}

\begin{figure}[hbtp]
 \centering
    	\includegraphics[width=0.70\textwidth]{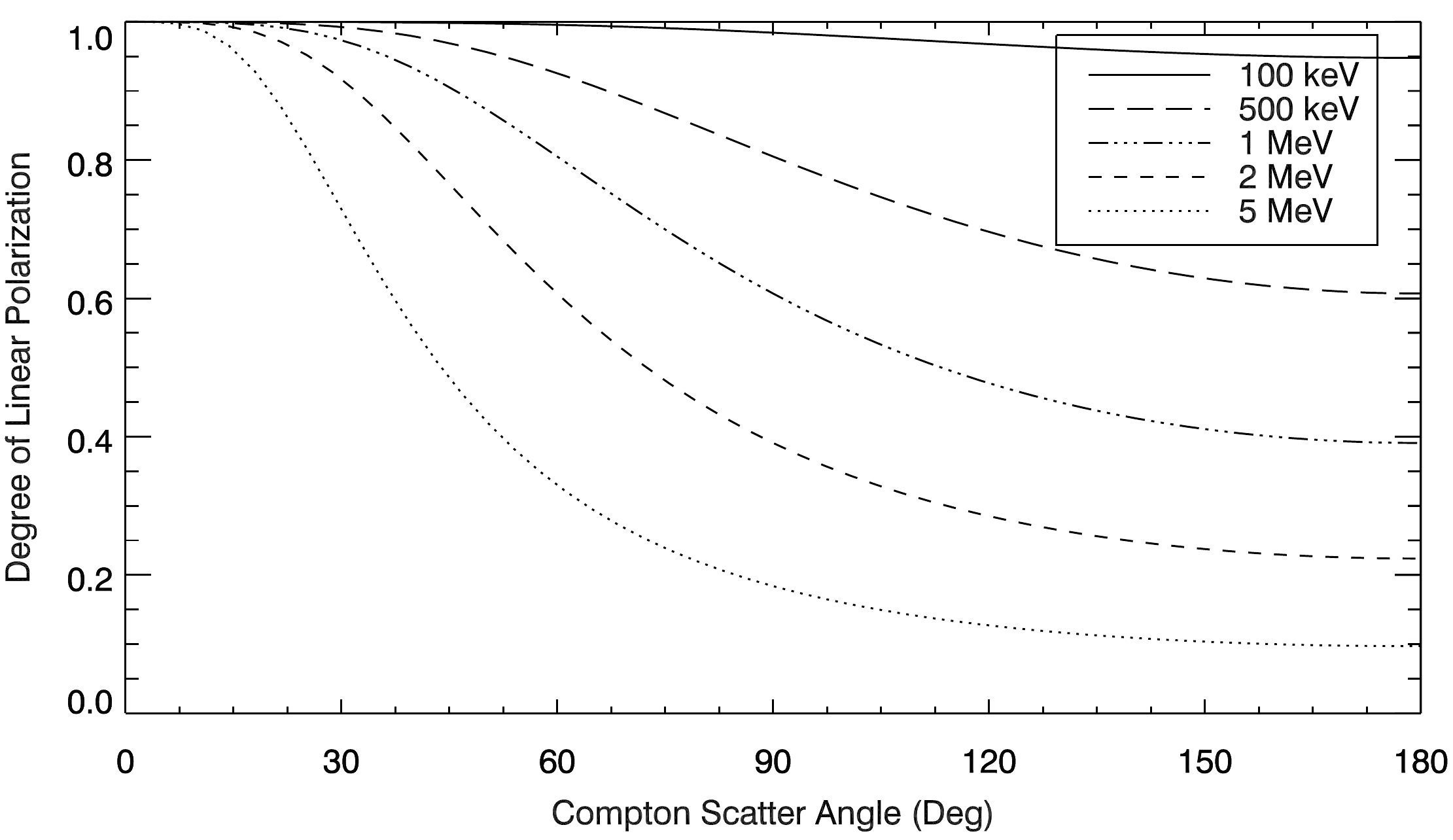}
    	\caption{Degree of linear polarization for a 100\% polarized photon beam versus Compton scattering angle $\theta$ for $\eta = 90^\circ$. The Compton scattering results in depolarization of a 100\% polarized beam. For example, a 100\% polarized photon beam with energy of 1 MeV would result in a fractional polarization of 60\% after being Compton scattered at an angle $\theta =$ 90$^\circ$. }
    	\label{fig:sci_comp_kn_lineardeg_pol}
\end{figure}

\begin{figure}[hbtp]
 \centering
    	\includegraphics[width=0.70\textwidth]{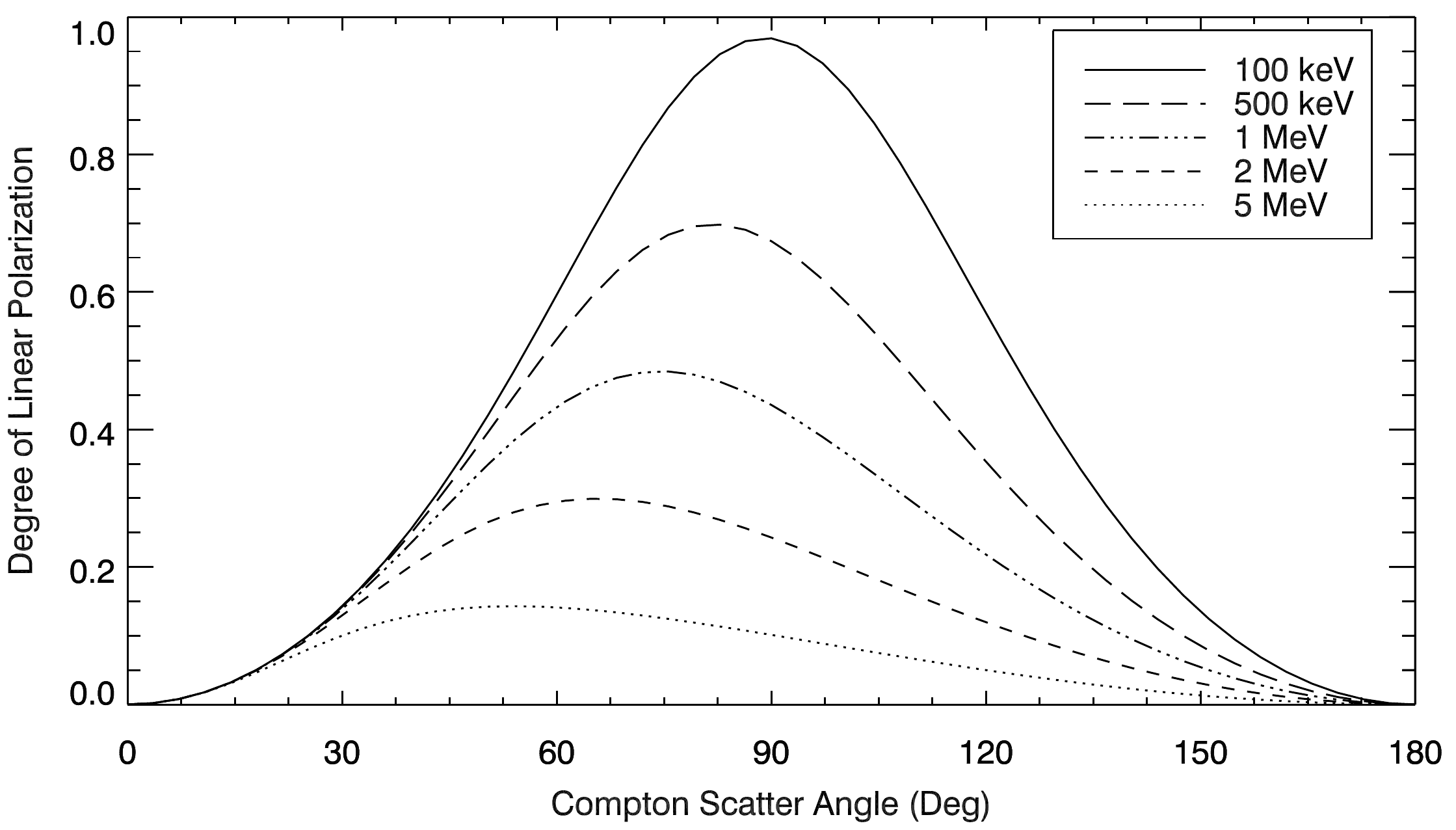}
    	\caption{Degree of linear polarization for various unpolarized incident energy beam($E_0$) versus Compton scattering angle. For example, a 100 keV incident photon beam Compton scattered at $90^\circ$ will have a fractional polarization of 98\%.}
    	\label{fig:sci_comp_kn_lineardeg_unpol}
\end{figure}

		Compton scattering can result in the polarization of an initially unpolarized beam, and it can result in the depolarization of an initially polarized beam. 
		For a fully polarized incident beam, the process of Compton scattering at an angle $\theta$ will produce a beam of reduced energy having a fractional polarization $\Pi_P$ (degree of linear polarization) given by
			\begin{equation}
				\Pi_P =2 \frac{1-\sin^2 \theta \, \cos^2\eta}{\epsilon + \epsilon^{-1} - 2\, \sin\theta \, \cos\eta}
				\label{eqn:gen_deg_lin_pol}
			\end{equation}
			The resulting polarization of a fully polarized incident beam at an azimuthal scattering angle of $\eta$ = 90$^\circ$ is shown in figure \ref{fig:sci_comp_kn_lineardeg_pol} for several different energies as a function of Compton scatter angle ($\theta$).
			For example the 5 MeV 100\% polarized beam gets depolarized to ~20\% at $\theta$ = 90$^\circ$.
 			The depolarization effect increases as the scattering angle $\theta$ increases from 0$^\circ$ to 180$^\circ$. 
			For the case of  an unpolarized beam, the process of Compton scattering at an angle $\theta$ will produce a beam with reduced energy fractional polarization $\Pi_U$ is given by
			\begin{equation}
				\Pi_U = \frac{\sin^2\theta}{\epsilon + \epsilon^{-1} - \sin^2\theta}
				\label{eqn:gen_deg_lin_unpol}
			\end{equation}
		This degree of linear polarization as a function of Compton scattered angle ($\theta$) for an unpolarized beam of photons is shown in figure \ref{fig:sci_comp_kn_lineardeg_unpol}. 
			The plot shows the degree of linear polarization expected for a photon with a specific initial energy (E$_0$) versus scattering angle ($\theta$). 
			A photon beam with 100 keV initial energy (E$_0$) Compton scatter at an angle of $\theta = 90^\circ$ will result in $\sim$ 98\% linearly polarized beam with energy $\sim$ 84 keV. 
			This method can be used to create a polarized source from an unpolarized source. 
			As will be discussed in section \ref{sec:ins_perf_polarization_measurements}, we have exploited this feature of Compton scattering to generate partially polarized photon sources in the lab. 
			For example, the 122 keV photons from Co-57, scattered at an angle of 90$^\circ$, will have an energy of 99 keV and a polarization level of ~95\%.
			
	\section{Compton Polarimetry}
	\label{sec:compton_polarimetry}
		
\begin{figure}[hbtp]
 \centering
    	\includegraphics[width=0.75\textwidth]{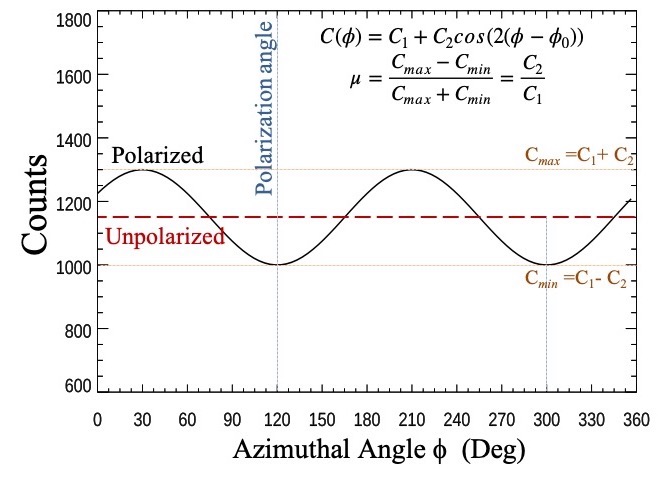}
    	\caption{Plot of counts vs azimuthal scatter angle for the ideal Compton polarimeter. The distribution is uniform for an unpolarized source (dashed red). For the polarized case, the distribution is modulated, as described by the functional form of equation \ref{eqn:pol_func_form} (black). The polarization angle is defined by the angles corresponding to minimum counts (C$_{min}$).}
    	\label{fig:sci_comp_ideal_compton}
\end{figure}

		The measurement of polarization based on Compton scattering exploits the fact that a photon prefers to Compton scatter at $90^\circ$ to their polarization vector (electric field vector  \textbf{\textit{E}}). 
		If these incident photons are polarized, the azimuthal distribution of the scattered photons is asymetric. 
		On the other hand, if these incident photons are not polarized, the azimuthal scatter distribution is uniform.
		An example of the distribution of counts versus the azimuthal angle $\eta$ is shown in figure \ref{fig:sci_comp_ideal_compton}. 
		The polarized plot (black line) displays the asymmetry observed due to polarization. 
		If the source was not polarized, then the asymmetry would be gone resulting in the unpolarized (red dashed-line). 
		\citet{Metzger1950} exploited this feature of Compton scattering photons to measure the degree of linear polarization for the first time. 
		Numerous Compton polarimeters have since been constructed based on same principle.
				
\begin{figure}[hbtp]
 \centering
    	\includegraphics[width=0.55\textwidth]{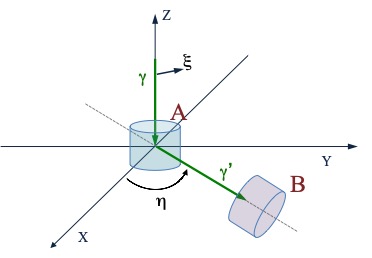}
    	\caption{Drawing of a setup for a simple Compton polarimeter. We have the incident photon $\gamma$ (with the polarization vector of $\xi$) and scattered photon $\gamma \, '$.  The incident photon is scattered from detector A to B. The Compton scatter angle $\theta$ is set to $90^\circ$ for this setup. The detector A is fixed and the detector B would be moved to different values of $\eta$ along the X-Y plane.}
    	\label{fig:sci_comp_pol_basic}
\end{figure}

		A polarization measurement requires a measure of the azimuthal scattering angle for a large number of photons. 
		This can be achieved by placing a number of detectors in the geometry shown in figure \ref{fig:sci_comp_pol_basic}. 
		The photon would be scattered at detector A to detector B where it will be absorbed. 
		The detectors are setup so that the Compton scatter angle is fixed to $90^\circ$.
		This corresponds to the Compton scatter angle where asymmetries in the azimuthal scatter angle are most pronounced.  		We can see the incoming photon (with Energy E), along Z axis, scattered off of detector A to detector B in figure \ref{fig:sci_comp_pol_basic}. 
		Detector A would be fixed.
		Detector B would be moved, within the X-Y plane, to different values of $\eta$ to observe the azimuthal distribution of counts.

		For a polarized source, the Klein-Nishina differential cross-section is defined by the Equation \ref{eqn:compton_klein_nishina_pol}.
			 	Using the trigonometric identity of $ \cos(2u) +1 = 2\cos^2u$, the  Equation becomes
		%
		\begin{equation}
			\frac{d \sigma_{KN,P}}{d \Omega} = \frac{1}{2}\ {r_0}^2 \epsilon^2 [ \epsilon + \epsilon^{-1} -\sin^2\theta \, ( \cos(2 \eta) +1)]
			\label{eqn:KN_comp_pol_1}
		\end{equation}
		%
			
		%
		We can introduce $\eta = \phi -\phi_0$ to include the phase-shift in azimuthal angle and rearrange the Equation to get the functional form
		\begin{equation}
			\frac{d \sigma}{d \Omega} = C(\phi) = C_1  + C_2\  \cos(2(\phi - \phi_0))
			\label{eqn:pol_func_form}
		\end{equation}
		where $C_1$ and $C_2$ are constants which are functions of $\epsilon$ (energy) and the angle $\theta$.  
		Equation \ref{eqn:KN_comp_pol_1} represents the azimuthal distribution of the scattered photons for a fixed energy ($\epsilon$) and Compton scatter angle ($\theta$). The azimuthal scatter angle distribution we observe from the setup is shown in Figure \ref{fig:sci_comp_ideal_compton}.
		In the polarized case, the distribution is nonuniform and for the unpolarized case, the distribution is uniform. 
				In the setup shown in Figure \ref{fig:sci_comp_pol_basic},  $\theta$ fixed to $90^\circ$ so the above Equation becomes
		%
		\begin{equation}
			\frac{d \sigma_{KN,P}}{d \Omega} = \frac{1}{2}\ {r_0}^2 \epsilon^2 [ \epsilon + \epsilon^{-1} - 2\cos^2 \eta]
			\label{eqn:KN_comp_pol}
		\end{equation}
			
		The effectiveness of the polarimeter response is measured by the polarization modulation factor $\mu$ and is given by 
		\begin{equation}
			\mu = \frac{N_\perp - N_\parallel}{N_\perp + N_\parallel}
			\label{eqn:modulation_1}
		\end{equation}
		where $N_\perp$ and $N_\parallel$ represents the counting rates in the azimuthal plane (XY-plane) \citep{Suffert1959}.
		Since the polarized photons prefer to scatter at $90^\circ$ to its polarization vector, the $N_\perp$ represents the maximum counts and $N_\parallel$ represents the minimum counts. 
			The modulation factor ($\mu$) represents the response of our instrument to a polarized source. $\mu_{100}$ represents the modulation response for a 100\% polarized source with our instrument. The $\mu_{100}$ represents the Figure Of Merit (FOM) of the polarimeter.
		The modulation factor can be represented as
		\begin{equation}
			\mu = \frac{C_{max}- C_{min}}{C_{max}+ C_{min}} 
			\label{eqn:modulation_2}
		\end{equation}
		
		\noindent where the $C_{max}$ and $C_{min}$ represents the maximum and minimum counts for a modulation profile represented by the Equation \ref{eqn:pol_func_form}. 
		Figure \ref{fig:sci_comp_ideal_compton} shows an ideal Compton polarimeter's modulation profile, the corresponding functional form, and the $C_{max}$ and $C_{min}$. 
		The values of $C_{max}$ and $C_{min}$  can also be written as $C_{max} = C_1+C_2$ and $C_{min} = C_1-C_2$. 
		The modulation factor can now be represented as 
		 \begin{equation}
			\mu = \frac{C_{max}- C_{min}}{C_{max}+ C_{min}} = \frac{(C_1+C_2)- (C_1-C_2)}{(C_1+C_2) +(C_1-C_2)} = \frac{2C_2}{2C_1}=\frac{C_2}{C_1}
			\label{eqn:modulation_3}
		\end{equation}
		The modulation factor $\mu$ can be used to get the degree of linear polarization.
		The degree of linear polarization $\Pi$ is the ratio of $\mu$ to $\mu_{100}$. 
		The $\mu_{100}$ is the modulation factor for a 100\% polarized source for the instrument \citep{Lei1997}.
		\begin{equation}
		\Pi = \frac{\mu}{\mu_{100}} =\frac{1}{\mu_{100}}\ \frac{C_2}{2C_1+C_2}
			\label{eqn:deg_lin_pol}
		\end{equation}
		In practice $\mu_{100}$ is generally derived from simulations. 

		%
		%
		We focus this section on emission mechanisms that are important for the 50-500 keV energy range of GRAPE. 
	This includes Bremsstrahlung, magneto-Bremsstrahlung, curvature radiation and inver Compton. 
	These are summarized here from \citet{Lei1997} and \citet{Longair2011}.
	
		\subsection{Minimum Detectable Polarization (MDP)}
		\label{sec:mdp}
		This section provides relevant equations in the derivation of the Minimum Detectable Polarization (MDP). 
		MDP defines the polarization sensitivity of the measurement. 
		These are summarized here from \citet{Weisskopf2006}.
		
		The observed signal through which the modulation profile is extracted is usually mixed with noise. 
		The mean of these signals can be defined by $\bar{S}$ with a variance of $\sigma^2$. 
		The functional form of the modulated signal due to polarization defined by Equation \ref{eqn:pol_func_form}) can be rewritten as
		\begin{equation}
				S = \bar{S}[ 1+ a_0\ \cos(2 (\phi-\phi_0))]
		\end{equation}
		where $\bar{S} = C_1$ and $ a_0 = C_2/C_1$ \citep{Weisskopf2006}. The $a_0$ and $\phi_0$ are the true modulation and the polarization angle. 
		The probability $p(a,\phi)$ of measuring a particular amplitude of modulation a and phase $\phi$ is given by \citet{Strohmayer2013,Weisskopf2006} as
		\begin{equation}
		P(a,\phi) = \frac{N\bar{S}^2a}{4\pi \sigma^2} \text{exp} \bigg[- \frac{N\bar{S}^2}{4 \sigma^2}(a^2 + a_0^2 - 2aa_0\cos(\phi-\phi_0)\bigg]
		\label{eqn:prob_a_phi}
		\end{equation}
		N is the total number of data points and therefore $N\bar{S}$ gives the total counts $C_T$. 		
		The probability of measuring a particular amplitude $a$ independent of $\phi$ is given by integrating Equation \ref{eqn:prob_a_phi} over the possible range of $\phi$ that is $0$ to $2\pi$ gives us
		\begin{equation}
		P(a) = \int_{-\pi/2}^{\pi/2}P(a,\phi) \, d\phi = \frac{N\bar{S}^2a}{2 \sigma^2} exp \bigg[- \frac{N\bar{S}^2}{4 \sigma^2}(a^2 + a_0^2) \bigg] I_0 \bigg[ \frac{N\bar{S}^2 a a_0}{2\sigma^2}\bigg] 
		\label{eqn:prob_a}
		\end{equation}
		This distribution is also called the Rice distribution where $I_0$ is the zeroth order modified Bessel function \citep{Rice1944}. 
		In case of no polarization, the modulation $a_0 = 0$. The probability function represented by Equation \ref{eqn:prob_a_phi} simplifies to 
		\begin{equation}
		P(a,\phi)_{UnPol} =  P(a=0) =\frac{N\bar{S}^2a}{4\pi \sigma^2} \text{exp} \bigg[- \frac{N\bar{S}^2}{4 \sigma^2}a^2 \bigg]
		\label{eqn:prob_a_phi_Unpol}
		\end{equation}
		The data from various polarimeters can be assumed as Poisson distributed data. For Poisson statistics $\sigma^2 = \bar{S}$ \citep{Knoll2010}. The total counts represented by $N \bar{S}$  = $C_T$. 
		The probability of having no modulation ($a_0 =0 $) becomes
		\begin{equation}
				P(a=0) =\frac{N\bar{S}^2a}{4\pi \sigma^2} \text{exp} \bigg[- \frac{N\bar{S}^2}{4 \sigma^2}a^2 \bigg] = \frac{C_T\,a}{4\pi } \text{exp} \bigg[- \frac{C_T \,a^2}{4 } \bigg]
		\label{eqn:prob_no_amp}
		\end{equation}

		Any modulation present would be defined by $a>0$. We would like to know the amplitude which has the probability of being more than 0 (chance for unpolarized source) \citep{Weisskopf2006}. In general, the probability that $x_1 \leq x \leq x_2$ is defined by
		$P(x_1 \leq x \leq x_2) = \int_{x1}^{x2} p(x)dx$. Similarly for our case, the probability to have a modulation value (a') greater than a given amplitude a is given by. 
		\begin{equation}
		P(a' \geq a) = \int_{a}^{\infty}P(a') da' = \frac{C_T\,a}{4\pi } \text{exp} \bigg[- \frac{C_T \,a^2}{4 } \bigg]
= \text{exp}  \bigg[- \frac{C_T}{4}a^2\bigg]
		\label{eqn:prob_a_gt_a'}
		\end{equation}		
	    With this Equation we can retrieve the amplitude that has only 1\% probability of occurring as
	    	\begin{equation}
		a_{1\%} = \frac{4.29}{\sqrt{C_T}}
		\label{eqn:prob_a1}
		\end{equation}
		This probability is in terms of mean total counts where $C_T = C_S + C_B$ where $C_B$ is background counts and $C_S$ is source counts. The modulation described in Equation \ref{eqn:prob_a1} can be expressed in terms of fraction of source counts ( $C_S/C_T$ ) as 
		\begin{equation}
		a_S = \frac{a_{1\%}}{C_S/C_T} = \frac{4.29}{\sqrt{C_S+C_B}} \frac{{C_S+C_B}}{C_S}= \frac{4.29}{C_S} {\sqrt{C_S+C_B}}
				\label{eqn:prob_a_s}
		\end{equation}
		Each of the polarimeter instruments might not respond fully to a 100\% polarized source. 
		The modulation factor for a 100\% polarized source would be the ideal response of the instrument.
		As discussed previously, $\mu_{100}$ is calculated using simulation and the minimum detectable polarization at the 99\% confidence level, $MDP_{99}$ is 
		 \begin{equation}
		MDP_{99} = \frac{a_{S}}{\mu_{100}} = \frac{4.29}{\mu_{100}\ C_S}{\sqrt{C_S+C_B}}
				\label{eqn:mdp}
		\end{equation}
		
	%
	%
	\section{Compton Polarimeter Experiments}
	\label{sec:compton_polarimeter_experiments}
		
		Multiple Compton polarimeters have been constructed using the aforementioned principles to do gamma ray polarimetry.
		\citet{Mcconnell2017} and \citet{Tatischeff2019} provides a good list and a detailed discussion on several of  these experiments. 
		I have listed some of the selected ones and provided a quick overview of how they achieved Compton polarimetry.
		

\begin{figure}[!ht]
 \centering
\begin{subfigure}[b]{0.55\textwidth}
 		 \includegraphics[width=1\linewidth]{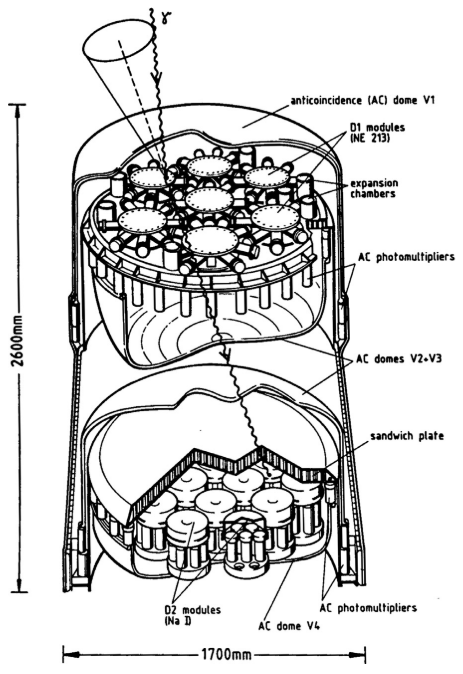}
		 \caption{}
\end{subfigure}
\begin{subfigure}[b]{0.4\textwidth}
 		 \includegraphics[width=1\linewidth]{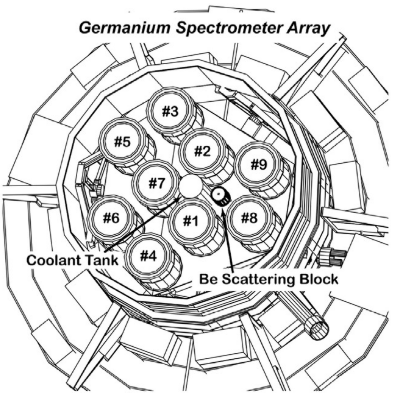}
		 \caption{}
\end{subfigure}
  
  \caption{(a) COMPTEL onboard the CGRO. The imaging telescope was exploited to do compton polarimetry using the  position of D1 and D2 detectors and the azimuthal scattering angle  \citep{Schoenfelder1993, Lei1996}.(b)RHESSI instrument with its9 germanium detectors and the scattering block that is used to get an azimuthal scattering histogram to do polarimetry  \citep{Lin2002,Smith2002,McConnell2002}. }
\label{fig:sci_comp_exp_2}

\end{figure}    
		The Compton Imaging Telescope (COMPTEL) was one of the experiments onboard the Compton Gamma Ray Observatory (CGRO). 
		COMPTEL was primarily, as the name suggests, an imaging telescope capable of imaging in the 0.75 - 30 MeV range. 
		Figure \ref{fig:sci_comp_exp_2}a shows the COMPTEL experiment \citep{Schoenfelder1993}. 
		COMPTEL consisted of two detector arrays. The top array of detectors, labeled D1, consists of 7 cylindrical modules of liquid scintillators NE 213A. Each of these module is 27.6 cm in diameter and 8.5 cm thick. The total area of the D1 detector array is $\sim$4188 cm$^2$. 
		The bottom array of detectors, labeled D2, consists of 14 cylindrical NaI(Tl) blocks of 7.5 cm thickness and 28 cm in diameter, covering a total area of 8620 cm$^2$. D1 and D2 are separated by a distance of 1.5m. An incident photon from an astrophysical source would Compton scatter from D1 to D2. In the default double scattering mode (DSM), photons triggering both D1 and D2 are recorded. The energy and direction from the interaction sites within each of the triggered detectors are used to create an event circle to do imaging. The locations of the triggered D1 and D2 detectors can also be used to create an azimuthal scatter angle histogram. \citet{Lei1996} attempted to measure polarization of a GRB using COMPTEL from 750 keV to 1.5 MeV but the results were inconclusive. COMPTEL was not optimized to operate as a polarimeter as it was primarily an imaging experiment. Hence, its polarimetry capabilities were limited.

		The Ramaty High-Energy Solar Spectroscopic Imager (RHESSI) is shown in Figure \ref{fig:sci_comp_exp_2}b. 
		RHESSI was built to image solar photons in the range of 3 keV to 17 MeV \citep{Lin2002}. 
		The spectrometer consisted of 9 germanium detectors, one behind each  Rotating Modulation Collimators (RMC) \citep{Smith2002}. 
		A small Be scattering block was used to scatter the incoming solar photons to the rear segments of nearby Ge detectors.
	     The position angles are exploited to generate an azimuthal scattering histogram to do polarimetry. 
	     \citet{McConnell2002} attempted to measure the polarization of GRB using this approach but was met with only limited success.

			\begin{figure}[!t]
 \centering
\begin{subfigure}[b]{0.4\textwidth}
 		 \includegraphics[width=1\linewidth]{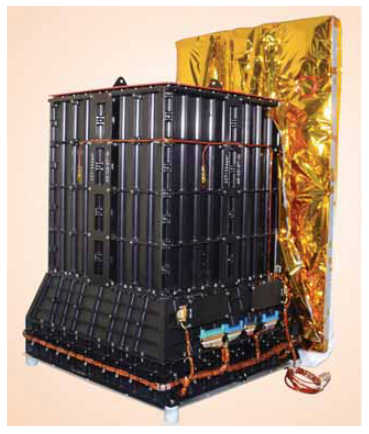}
		 \caption{}
\end{subfigure} 
 \begin{subfigure}[b]{0.4\textwidth}
 		 \includegraphics[width=1\linewidth]{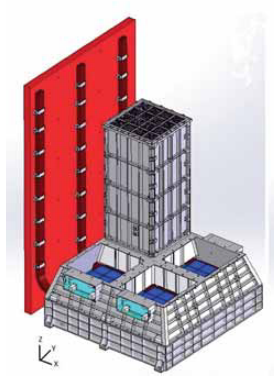}
		 \caption{}
\end{subfigure} 
  \caption{a) Figure of fully assembled Cadmium Zinc Telluride Imager (CZTI) before integrating in the spacecraft \citep{Bhalerao2017}. b) Layout of the CZTI. The four identical quadrants of 16 CZT detector modules make up the CZTI. Only three of the four quadrants are hidden for clarity \citep{Bhalerao2017}.}
\label{fig:sci_comp_exp_astrosat}
\end{figure}
		Cadmium Zinc Telluride Imager (CZTI) is one of the four instruments onboard AstroSat (multi wavelength space observatory) launched by Indian Space Research Organization (ISRO). The CZTI is shown in Figure \ref{fig:sci_comp_exp_astrosat}. The CZTI instrument is divided into four identical quadrants. Each quadrant has a Coded Aperture Mask at top and 16 CZT detector modules. These modules are sensitive to the 20-200 keV X-rays. The overall dimensions of the payload are 482 mm $\times$ 458 mm $\times$ 603.5 mm. The X-ray photon in the energy range 10-100 keV deposits the full energy in the CZT. Above $\sim$ 100 keV, the pixellated detectors have the ability to measure polarization via Compton polarimetry \citep{Bhalerao2017}.  \citet{Vadawale2018} used this feature to measure polarization of the Crab in the 100-380 keV band. They report a Polarization Fraction (PF) of 32.7$\pm$5.8\% and a Polarization Angle (PA) of 143.5$\pm$2.8$^\circ$ for the total Crab. For the off-pulse (representing the Crab Nebula), they report a PF of 39.0$\pm$10\% and a PA of  140.9$\pm$3.7$^\circ$ \citep{Vadawale2018,Bhalerao2017}. 

		POLAR is a dedicated for Gamma Ray Bursts (GRBs) polarimeter that took data on the second Chinese space laboratory Tiangong-2 (TG-2). It is a Compton polarimeter and is optimized for 15-500 keV energy range. The Compton polarimeter is made up of 64-scintillator in a 8x8 array read by the Multi-Anode Photo-Multiplier Tube (MAPMT) (similar to that used in GRAPE). Polarimetry is achieved by analyzing the Compton scatter events between these detectors. The design provided a wide field of view and the Figure of Merit ($\mu_{100}$) $\approx$ 35\%. The telescope was switched off in 2017 due to failure of its power distribution unit. POLAR measured the light curves of 55 GRBs and 5 of these were bright enough in the 10-1000 keV to do polarization measurements. POLAR-2, a successor of POLAR has been selected to be installed on board the China Space Station (CSS) officially with a launch schedule in 2024 \citep{Li2019,Mcconnell2017,Tatischeff2019}. 
		
\begin{figure}[!t]
 \centering
\begin{subfigure}[b]{0.4\textwidth}
 		 \includegraphics[width=1\linewidth]{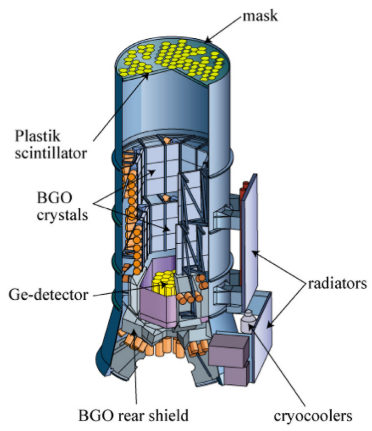}
		 \caption{}
\end{subfigure}
\begin{subfigure}[b]{0.4\textwidth}
 		 \includegraphics[width=1\linewidth]{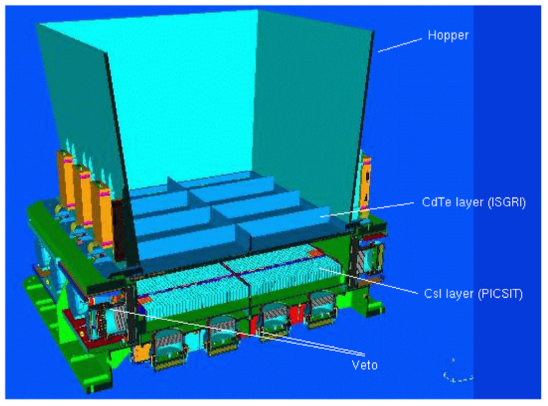}
		 \caption{}
\end{subfigure}
  \caption{Figure of the two instruments on board INTErnational Gamma-Ray Astrophysics Laboratory (INTEGRAL). a) Spectral Image on board Integral (SPI) and b) Imager on Board the INTEGRAL Satellite (IBIS). The figures are taken from the European Space Agency website: \href{https://www.cosmos.esa.int/web/integral/instruments-spi}{\textit{\underline{ https://www.cosmos.esa.int/web/integral/instruments-spi}}} and \href{https://www.cosmos.esa.int/web/integral/instruments-ibis}{\textit{\underline{ https://www.cosmos.esa.int/web/integral/instruments-ibis}}}.  }
\label{fig:sci_comp_exp_3}

\end{figure}

		The INTErnational Gamma-Ray Astrophysics Laboratory (INTEGRAL) is an European Space Agency (ESA) space telescope to observe gamma rays.
		 It was launched in 2002 and is still operational. 
		Two instruments on this telescope provided imaging, spectroscopy, and polarimetry.
		 The Spectrometer on INTEGRAL (SPI) and the Imager on Board the INTEGRAL Satellite (IBIS). 
		SPI was designed to operate from 20 keV to 8 MeV \citep{Vedrenne2003}. 
		This instrument is shown in Figure \ref{fig:sci_comp_exp_3}a. 
		SPI consists of an array of 19 hexagonal germanium detectors at the bottom of this instrument. 
		A hexagonal coded aperture mask is located 1.7m above the detection plane in order to image large regions of the sky. 
		Polarimetry can be achieved by selecting the Compton scattering event between the Ge-detectors. 
		\citet{Dean2008} measured the polarization of the Crab Nebula (only the non-pulse region) using SPI in 0.1 MeV to 1 MeV range. 
		They reported a polarization of 46 $\pm$ 10\% and the polarization angle was estimated to be 124.0$^\circ$ $\pm$ 0.1$^\circ$.


		IBIS is a coded-aperture instrument that provides imaging capabilities between 15 keV and 10 MeV. 
		IBIS is composed of two detection planes, each of which can operate independently or in coincidence with each other. Both use the same coded mask for imaging. The top layer is the INTEGRAL Soft Gamma Ray Imager (ISGRI) and the bottom one is the Pixellated Imaging CsI Telescope (PiCsIT). They are separated by 90mm.
	ISGRI is sensitive between 15 keV and 1 MeV. It consists of 8 modules of 64 $\times$ 32 CdTe pixels for a total of 16384 pixels, and each pixels measures 4$\times$4 mm$^2$ in area. \citep{Lebrun2003,Labanti2003,Ubertini2003}. 
 		The PiCsIT also has 8 modules each with 32 $\times$ 16 pixels for a total of 4096 CsI pixels. Each of these measuring $8.4 \times 8.4 mm$ in area and $3 cm $deep. 
		Operated in Compton mode, photons can be detected that Compton scatter from ISGRI and subsequently absorbed in PiCsIT.  
		The azimuthal scatter angle histogram can be created using the position of the hit pixels in ISGRI and PiCsIT. 
		\citet{Forot2008} used this approach to measure the polarization of the Crab Nebula from 200 to 800 keV. 
		They measured the polarization angle of 120.6$^\circ$ $\pm$ 7.7$^\circ$ and a polarization of $>$72\% at 95\% confidence for the Crab Nebula (off pulse region). 
		For the whole Crab observation (all phases), the polarization angle measured was 100$^\circ$ $\pm$ 11$^\circ$ and a polarization of 47$_{-0.13}^{+0.19}$\%.

\begin{figure}[!t]
 \centering
\begin{subfigure}[b]{0.4\textwidth}
 		 \includegraphics[width=1\linewidth]{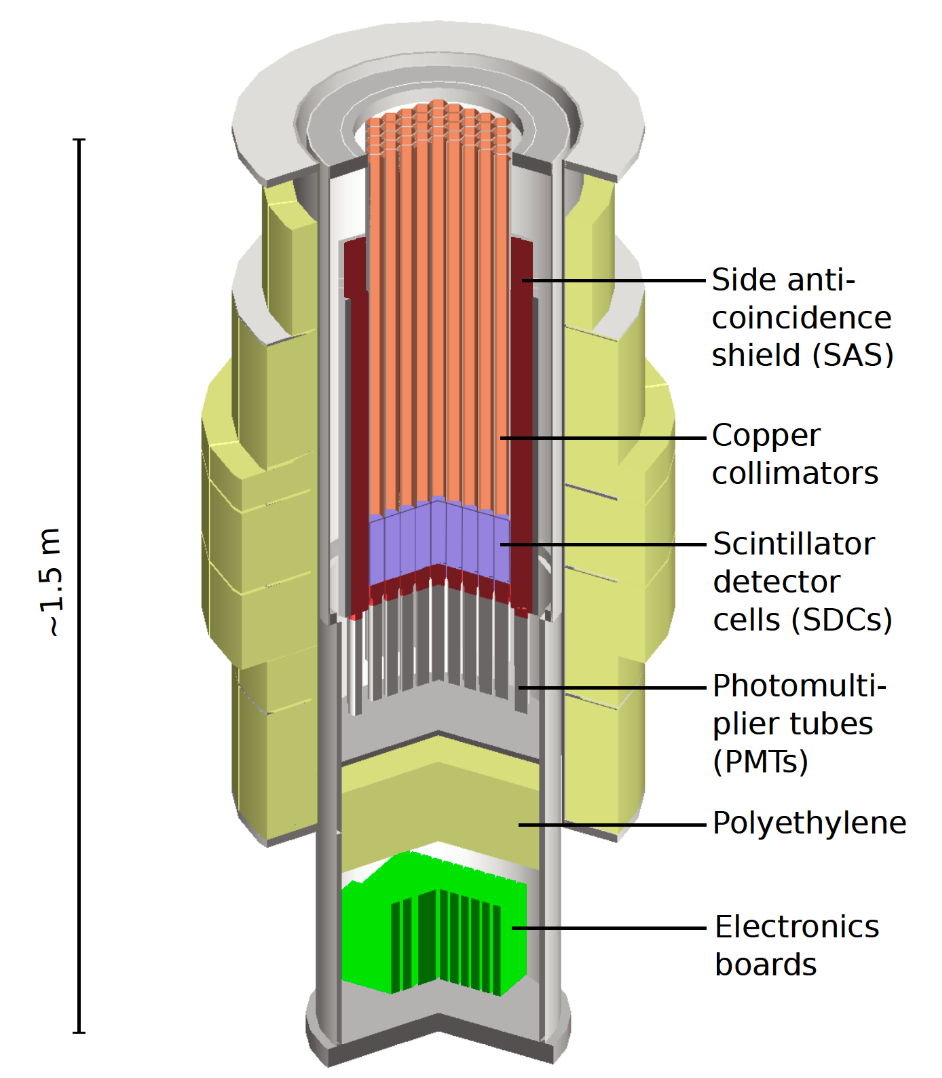}
		 \caption{}
\end{subfigure}   
 \begin{subfigure}[b]{0.5\textwidth}
 		 \includegraphics[width=1\linewidth]{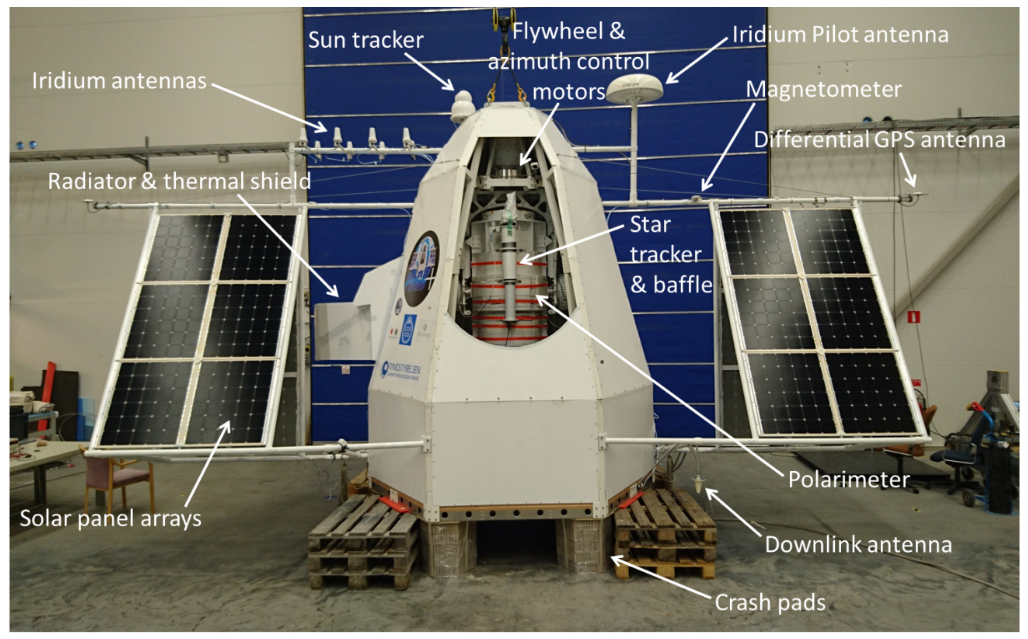}
		 \caption{}
\end{subfigure} 
  \caption{The PoGO+ instrument and the payload. a) The PoGO+ instrument which shows the closely packed hexagonal grid of of the scintillator detectors cells, collimators and the PMTs. b) The payload containing the instrument.  \citep{Chauvin2016,Chauvin2017}}
\label{fig:sci_comp_exp_4}
\end{figure}
		
		PoGO+ was a balloon borne Compton polarimeter that flew in 2016. 
		It performed polarimetry using a detector array consisting of closely packed 61 hexagonal plastic scintillators. 
		Each of these plastic scintillator rods are 3cm wide and 20 cm long.
		The instrument is shown in Figure \ref{fig:sci_comp_exp_4}. 
		The incoming photon would travel through the copper collimators and hit the scintillator array. 
		Events which involved the two of the scintillator elements (two hit events) were used to get an azimuthal scatter angle histogram and determine the polarization fraction. 
		The instrument rotated about the zenith direction to remove the systematic effects in the measurements.
		\citet{Chauvin2017} measured the polarization of Crab Nebula (off pulse) to be 20.9 $\pm$ 5.0\% using the data from PoGO+ in the 20-160 keV energy range.

\begin{figure}[hbtp]
 \centering
\begin{subfigure}[b]{0.75\textwidth}
 		 \includegraphics[width=1\linewidth]{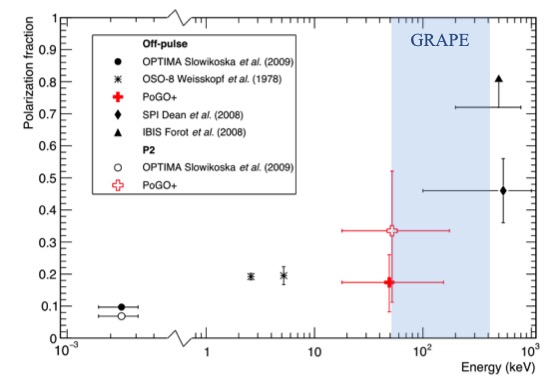}
		 \caption{}
\end{subfigure}
 \begin{subfigure}[b]{0.75\textwidth}
 		 \includegraphics[width=1\linewidth]{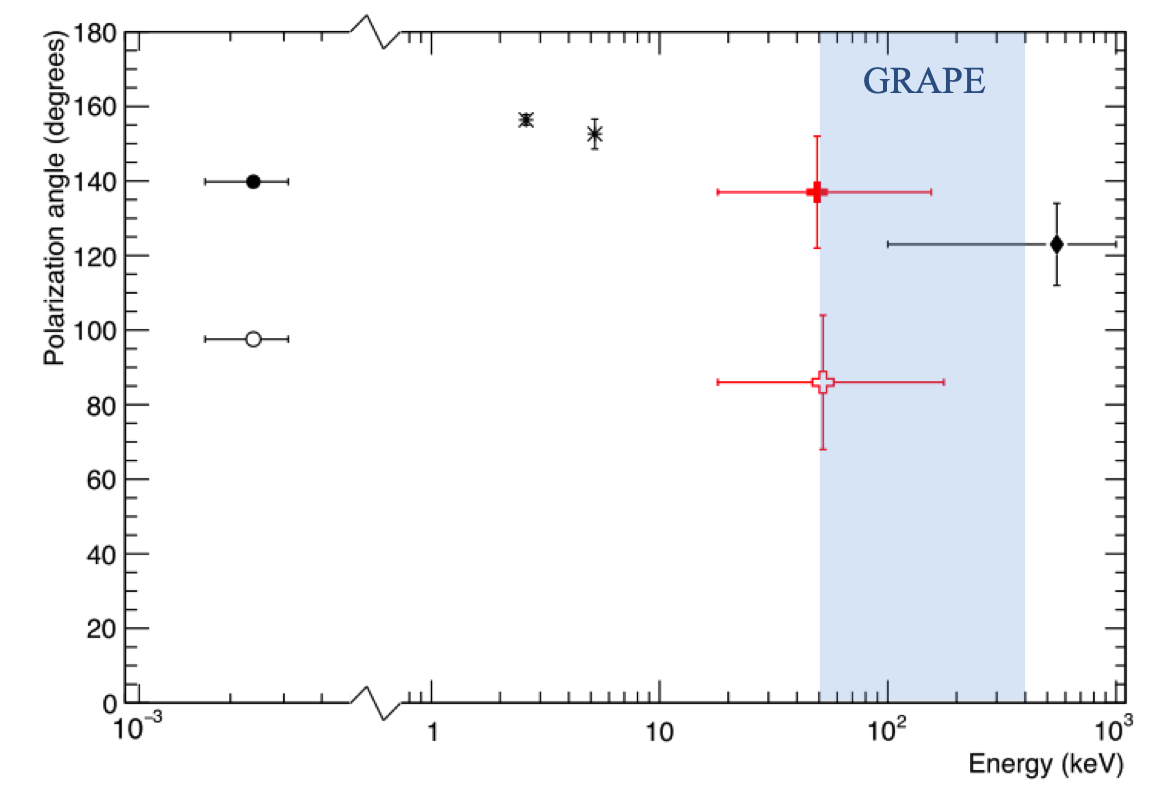}
		 \caption{}
\end{subfigure} 
  \caption{Comparison of  a) Polarization Fraction (PF) and b) Polarization Angle of the Crab from different polarimetric studies \citep{Chauvin2017}. The shaded region labelled GRAPE is added to this plot which represents the operational energy range of GRAPE and the measurement done by GRAPE would contribute a data point in those regions.}
\label{fig:sci_compiled_result}
\end{figure}

			\citet{Chauvin2017} also compiled the results from other experiments that did a polarization measurement of the Crab. These are shown in Figure \ref{fig:sci_compiled_result}. The shaded region (labeled GRAPE) represents the energy range of GRAPE (50-300 keV). The polarization measurement made by GRAPE would provide additional data points in those regions.

\chapter{GRAPE Instrumentation}
\label{sec:grape_instrumentation}
The Gamma RAy Polarimeter Experiment (GRAPE) is a balloon borne polarimeter optimized for the 50-300 keV energy range.
It was initially flown in the fall of 2011 from Fort Sumner, New Mexico and again in the fall of 2014 from the same location.
The second balloon flight incorporated several improvements to the payload, as outlined below. 
This thesis focuses on data from the 2014 ballon flight.
	
	\section{Polarimeter Module}
	\label{sec:polarimeter_module}
\begin{figure}[tbp]
 \centering
    	\includegraphics[width=0.85\textwidth]{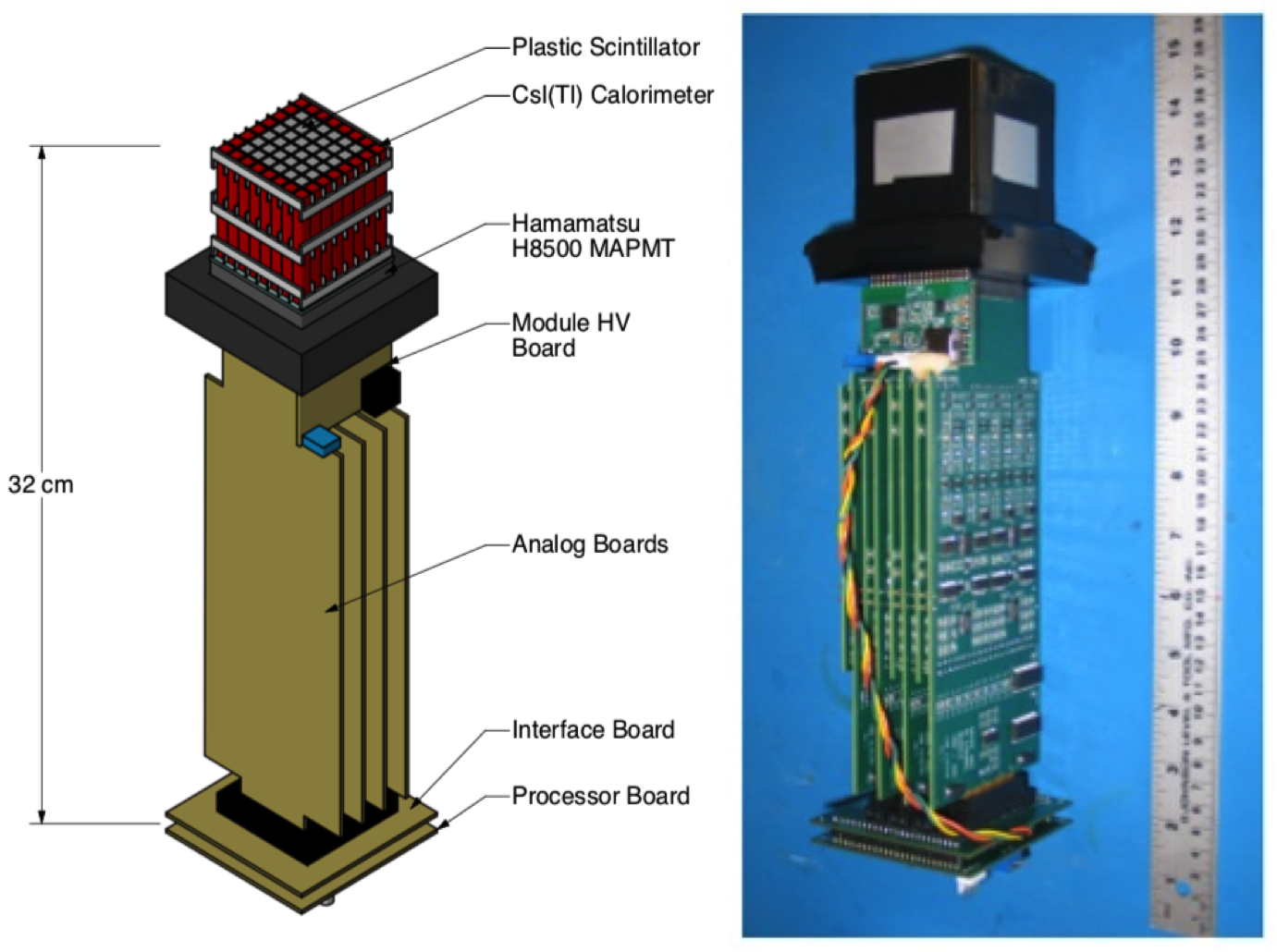}
    	\caption{A diagram of a module (left) alongside a fabricated module (right). We see the scintillator array, MAPMT, HV board, analog boards, the processor board and the interface board that makes up a module.}
    	\label{fig:ins_module_block}
\end{figure}

\begin{figure}[!ht]
 \centering
\begin{subfigure}[b]{0.45\textwidth}
 		 \includegraphics[width=1\linewidth]{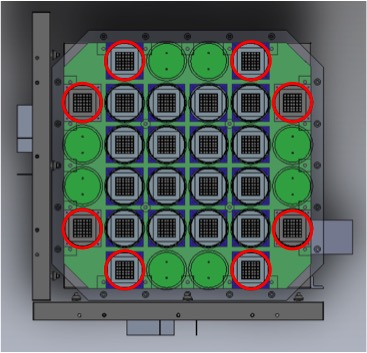}
		 \caption{}
\end{subfigure} \,\,\,
 \begin{subfigure}[b]{0.45\textwidth}
 		 \includegraphics[width=1\linewidth]{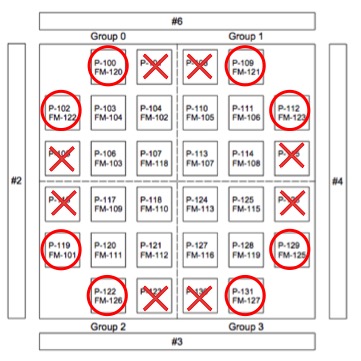}
		 \caption{}
\end{subfigure} 
  \caption{Module array for the 2014 GRAPE payload. The left is top view from simulation and right is a drawing with position and module numbers for each of the modules. The center 4 $\times$ 4 array as the 2011 configuration. Eight more modules (2 on each corner represented by red circles) were added for the 2014 configuration. The X's denote the unoccupied module locations.}
\label{fig:ins_module_array}
\end{figure}

GRAPE is comprised of multiple independent polarimeter modules. 
A polarimeter module is the smallest detector block that is capable of doing polarization measurements.
It consists of an array of scintillator elements (scintillator array) which are placed on top of a single 2-inch multi-anode photomultiplier tube (MAPMT).  
Beneath each MAPMT, and contained within the MAPMT footprint, are the high voltage board, four analog boards (16 channels each), a processing board, and a module interface board.
Each module connects to the Module Interface Board (MIB; see section \ref{sec: mib_electronics})
An example of a polarimeter module is shown in Figure \ref{fig:ins_module_block}. 
There were 16 polarimeter modules (in a  4$\times$4 grid) on the 2011 GRAPE payload. 
As an improvement, 8 more modules were added to the 2014 GRAPE payload, which resulted in the total number of polarimetry modules to be 24, as shown in Figure \ref{fig:ins_module_array}. 

			\subsection{Scintillator Array}
			\label{sec:scintillator_array}
The scintillator array consists of 64 scintillator elements in an 8 $\times$ 8 grid. 
This array consists of two different kinds of scintillators, plastic and calorimeters.
Plastic scintillators are organic, low Z (atomic number) scintillators.
These were selected to maximize the probability of Compton scattering the incident photons.
The calorimeters are inorganic, high Z scintillators. 
These were selected to maximize the absorption of the scattered photons.


The plastic elements are Eljen Technology's EJ-204 plastic scintillators.
These elements are characterized by fast timing (0.7 ns rise time and 1.8 ns decay time) and an emission spectrum that peaks at 408 nm (providing a good spectral match to traditional bi-alkali photocathodes). 
The calorimeter elements are CsI(Tl) scintillator crystals from Saint Gobain. 
The CsI(Tl) was chosen for its good light output and its relatively low cost.  
The primary decay time is $\sim$1000 ns. 
The scintillator arrays were assembled in-house and the low hygroscopic tendencies of CsI(Tl) made it easier to handle than CsI(Na) which was used in previous iteration of GRAPE \citep{Connor2012}. 

\begin{figure}[!ht]
 \centering
    	\includegraphics[width=0.40\textwidth]{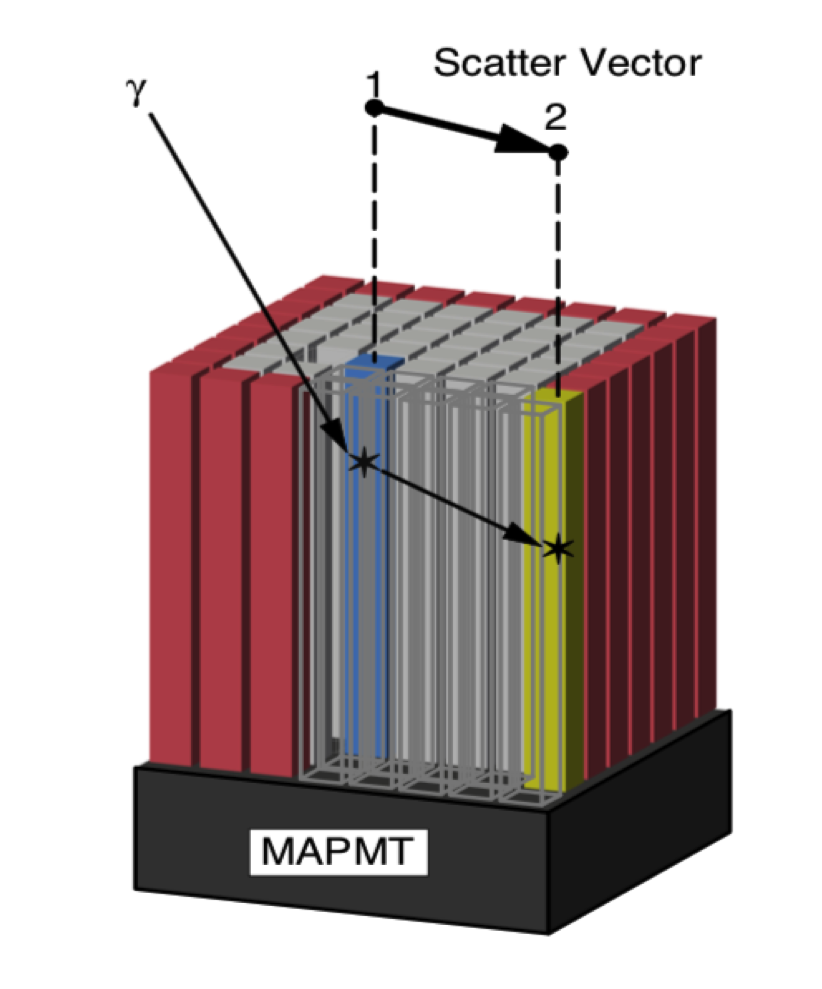}
    	\caption{A diagram of the scintillator array. There are 64 elements. There are 36 plastic elements (grey) surrounded by 24 calorimeter elements (red). It is also shows a $\gamma $ ray scattering from a plastic (blue) to a calorimeter element (yellow). The scatter vector defines the azimuthal scatter angle.}
    	\label{fig:ins_scint_array_block}
\end{figure}

Each scintillator element (both plastic and calorimeter) are rectangular element in shape, with dimensions of 5 mm $\times$ 5 mm $\times$ 50 mm. 
These element are arranged in a grid of 64 (8 $\times$ 8) where 36 plastics (6 $\times$ 6) are surrounded by 28 calorimeters which forms the scintillator array. 
The scintillator arrays were assembled using two alignment grids (made of Delrin\textsuperscript{\textregistered}) to hold the elements in proper geometric relationship with the Multi-Anode Photo Multiplier Tube (MAPMT) so that each element was aligned with a single anode of the MAPMT. 
The plastic scintillator was diamond-milled and the CsI(Tl) surfaces were polished.
Each scintillator element was individually wrapped (on all four sides and the top) in a layer of VM2000 reflective material (now marketed as Vikuiti\textsuperscript{\texttrademark} Enhanced Specular Reflector or ESR) to maximize the light collection.
A high temperature molding process was employed to provide a good fit of the VM2000 to the shape of each scintillator element.  
In addition to maximizing the light collection, the wrapping also served as a means to provide optical isolation for each element \citep{Ertley2014,Connor2012}.

			\subsection{Multi-Anode Photo Multiplier Tube (MAPMT)}
			\label{sec:ins_MAPMT}

\begin{figure}[tbp]
     			\centering
    				\includegraphics[width=0.6\textwidth]{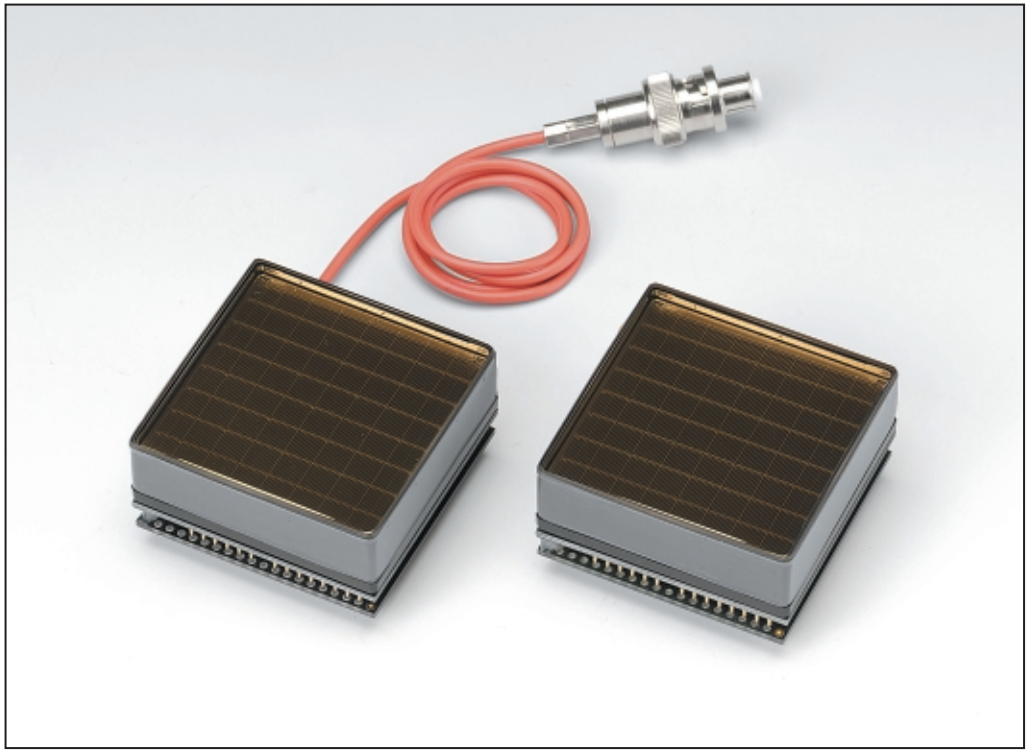}
    				\caption{Hamamatsu 12-stage H8500 series. The H8500C is the left one with cable input which we use. It is 52  mm$^2$, and has 8 $\times$ 8 multianodes.}
    				\label{fig:ins_mapmt_hamamatsu}
				\end{figure}

Each GRAPE module employs a Hamamatsu H8500C MAPMT, which is shown in Figure \ref{fig:ins_mapmt_hamamatsu} (left). 
This MAPMT is compact with minimal dead space.
It is 52$\times$52 mm$^2$ in area and 28 mm in depth.
It has 64 (8 $\times$ 8) independent anodes with 5mm anodes  arranged on a pitch of 6 mm.
The MAPMT (Hamamatsu H8500) reads out both plastic and calorimeter elements simultaneously. 
A high voltage of $~$1100 V is supplied to the MAPMT. 
The scintillator array is attached to the front end of this MAPMT using an optical cement (St. Gobain Crystals BC-600).
Each scintillator element (either plastic or CsI(Tl)) is co-aligned for readout by a single anode, so that each of the 64 anodes is designed to read out one of the 64 scintillator elements in the scintillator array. 
There is a $~$1.5 mm thick glass window at the front end of the MAPMT.
The light from the scintillators has to pass through this glass window to reach its respective anode in the MAPMT. 
This setup allows for some level of optical crosstalk that is not avoidable.
This optical crosstalk is in addition to the 3\% electronic crosstalk that is already present in the MAPMT.      
Recognizing the potential for crosstalk issues, a Pulse-Shaped Discriminator (PSD) circuit was devised and implemented to partially deal with this issue.

			\subsection{Pulse-Shaped Discrimination}
			\label{sec:ins_psd}
			
		The Pulse-Shaped Discrimination (PSD) seeks to identify signals in an anode whose timing characteristics do not correspond to those associated (or expected) for that anode. PSD takes advantage of the large difference in pulse decay times between plastic and CsI(Tl) to reject crosstalk between elements of different scintillator types. 
(It should be noted that crosstalk between materials of the same scintillator can not be rejected this way).     
The plastic has a short decay time of $~$1.8 ns and the CsI(Tl) has an average decay time of 1000 ns. 
Scintillator logic signals (triggers) are generated at the end of the signal decay. 
The large difference in decay time means that, for a coincidence event involving two different scintillator types, the calorimeter trigger is generated well after the plastic trigger (about 600 ns later). 
The plastic triggers were delayed by 600 ns so that the signals from a true coincidence between plastic and CsI(Tl) would generate simultaneous signals. 
A coincidence window of 180 ns was defined for detecting a true coincidence between the CsI(Tl) trigger and the delayed plastic trigger. 
			
\begin{figure}[hbtp]
\centering
\includegraphics[scale=0.5]{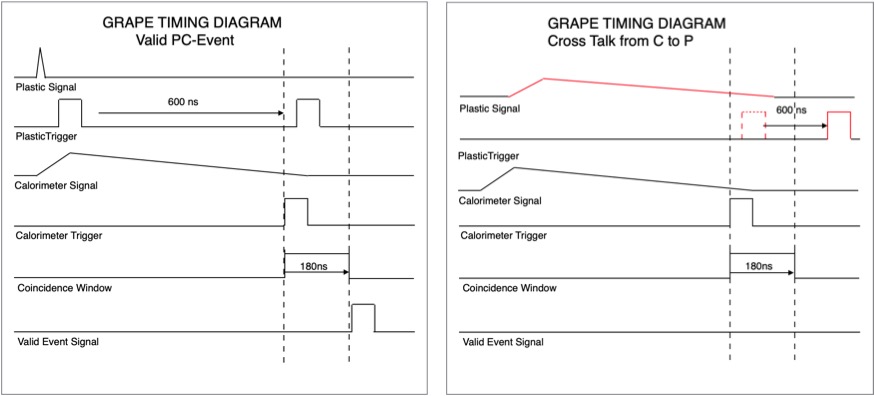}
\caption{PSD timing plots showing plastic and calorimeter triggers and signals along with coincidence window. Scintillator logic signals (triggers) are generated at the end of the signal decay. 
(Left) a: Shows a valid PC event, where the delayed plastic and calorimeter trigger is within coincidence window.  
(Right) b: Shows the plastic trigger due to crosstalk being rejected as the delayed trigger does not lie within the coincidence window }
\label{fig:ins_timing_plots}       
\end{figure}

To demonstrate how the PSD circuitry operates, signal timing plots are shown in Figure \ref{fig:ins_timing_plots}. 
Two cases are shown, one for a valid plastic-calorimeter (PC) coincidence (Figure \ref{fig:ins_timing_plots}a) and one for coincidence generated by crosstalk from calorimeter to plasic(Figure \ref{fig:ins_timing_plots}b). 
The plastic trigger is delayed by 600 ns to match the calorimeter trigger; generation time of the CsI(Tl) trigger. 
The calorimeter trigger starts the coincidence window. 
For a valid coincidence events the CsI(Tl) trigger is generated just as the decayed plastic trigger arrives. 
Right after the coincidence window closes, a VALID signal is generated. 
The VALID signal is generated for any event with at least one plastic and one calorimeter elements involved. 
Events are recorded only when associated with a VALID signal. 

To see how events generated by a crosstalk are rejected, consider Figure \ref{fig:ins_timing_plots}b. 
Here crosstalk has occurred from a calorimeter event to an adjacent plastic element. 
Since it is a crosstalk event, the signal from the plastic anode mimics the signal from the calorimeter anode. 
Both logic signals (triggers) are generated simultaneously. 
Since the plastic trigger is delayed by 600 ns, the plastic trigger does not fall in the coincidence window. 
VALID signal is not generated in this case.  

This PSD technique can only identify crosstalk between different types of scintillator elements.
It cannot identify crosstalk between elements of same type. 
Additionally, due to the difference in light yield and threshold values, a crosstalk from plastic to calorimeter is very unlikely. Therefore, the PSD circuit specializes in detecting crosstalk from calorimeter elements to plastic elements.

			\subsection{Module Electronics }		
			\label{sec:Module_Electronics}	
		
\begin{figure}[hbtp]
\centering
\begin{subfigure}[b]{0.95\textwidth}
 		 \includegraphics[width=1\linewidth]{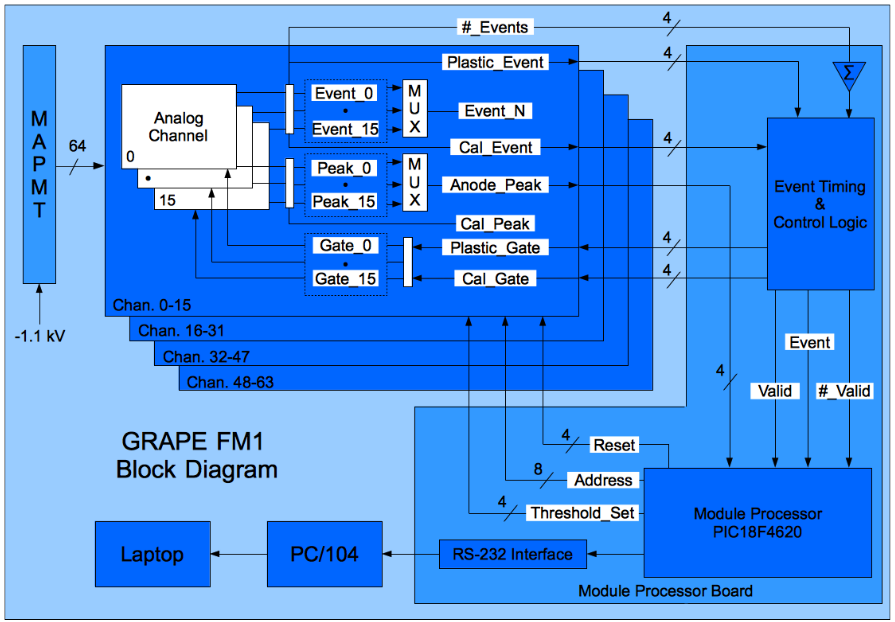}
		 \caption{}
\end{subfigure}
\\
\begin{subfigure}[b]{0.95\textwidth}
 		 \includegraphics[width=1\linewidth]{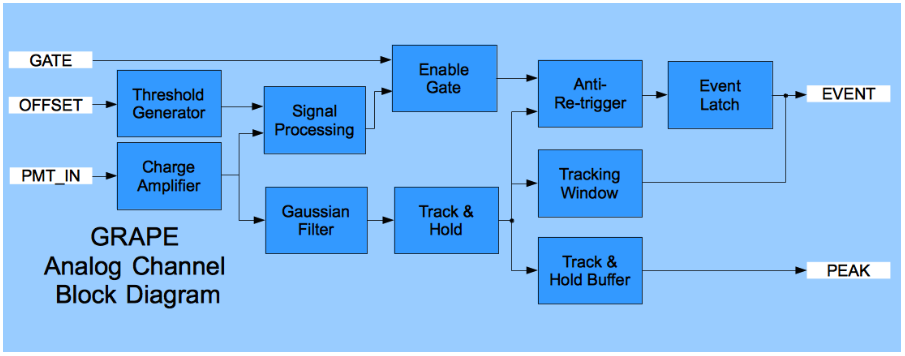}
		 \caption{}
\end{subfigure}
  
  \caption{A block diagram of module level electronics.(a) shows the full electronics setup. (b) focuses on the the analog block.}
\label{fig:ins_elec_block}
\end{figure}

Each of the anodes from the MAPMT goes to one of the four 16-channel readout boards. 
Each analog channel includes a charge pre-amplifier, a constant-fraction discriminator (CFD), a pulse height discriminator (PSD) and Gaussian shaping filter.
A block diagram of the analog board is shown in Figure \ref{fig:ins_elec_block} (b). 
Total power was 1.8 W for each 64-channel module.  
The design was intentionally developed without the use of custom ASICs to assure flexibility with evolving balloon instrument configurations while minimizing the cost of future modifications.
The analog boards fits into an interface board that connects to the processor board at the base of each module assembly.
The processor board includes a PIC18F4620 microprocessor that controls the readout of the data. 
The processor includes an RS-232 interface through which the data can be sent to the instrument computer.
Commands from the instrument computer are acted upon by the microprocessor to control the operational state of the module. 
The microprocessor was programmed to provide various modes that serve different functionalities for calibration, flight and housekeeping. 
The modes are discussed in section \ref{sec:ins_perf_module_calibration}.
The microprocessor also stores the threshold data (which controls the flow of signals in the analog board) and processes each event.

The output data type and format is depended on the programmed mode which were selected from the connected PC terminal. 
For calibration of individual modules, an ASCII format and an E mode where every data was printed out and recorded. 
For calibration of the whole instrument, flight mode was used which gave us data in binary form and had a limit of 8 anodes per event. 
This mode was ran during the flight. 
The important modes are further explained in chapter \ref{sec:insrument_performance}.
Along with these recording modes, housekeeping modes were ran periodically to get the rates to verify the workings of the modules. 
The rates were primarily used to identify the lower threshold value to avoid electronic noises for each of the scintillators.
Additionally the processor also identified and defined the events into different event classification at hardware level.

			\subsection{Event Classification}
			\label{sec:event_classification}
Each event is classified by the type of anode element(s) involved in that event.
C is used to denote a calorimeter element and P is used to denote a plastic element. 
The events are classified at two separate levels, the hardware and the software. 
The hardware level event classification happens in the onboard hardware when anode elements are triggered.
During the analysis of the data, the events are reclassified once energy calibrations and energy thresholds are applied.

The hardware event classification is defined by the anodes triggered during the event.
The hardware threshold was set to $\sim$6 keV for plastics and $\sim$20 keV for calorimeters.
The hardware threshold is set for the pulse-height channel (which is converted to energy using the energy calibration) so the nearest channel corresponding the aforementioned values.
The events are classified into three groups at hardware level, which are C, CC and PC.  
C defines an event where only one calorimeter element is triggered. 
CC defines an element where 2 or more calorimeter elements are triggered.
PC defines an event where there are at least 1 plastic and 1 calorimeter elements triggered. 
The hardware did not recorded events with only plastic elements triggered. 

The software event classification happens at the analysis of the post-flight analysis of the data after threshold is applied. 
A software threshold, higher than the hardware threshold, is defined so that the elements have a common threshold. 
The software threshold was set to 10 keV for plastic elements and 40 keV for CsI(Tl) calorimeter elements. 
The software event classification uses the exact number of anode elements to categorize on event. 
A C event represents a single calorimeter element above the software threshold.  
A CC event represents 2 calorimeter elements and a CCC event represents 3 calorimeter elements above the software threshold. 
A PC event involves exactly 1 plastic and 1 calorimeter element and a PPC event is defined as an event with 2 plastic elements and 1 calorimeter element above the software threshold. 
Because of the software threshold, the software event classification can result in PP or PPP events. For example, a PPC event where the calorimeter's energy is less than the software threshold of 40 keV can result in the event being classified as PP, PPP, etc at software level.

			\begin{figure}[tbp]
     			\centering
    				\includegraphics[width=0.35\textwidth]{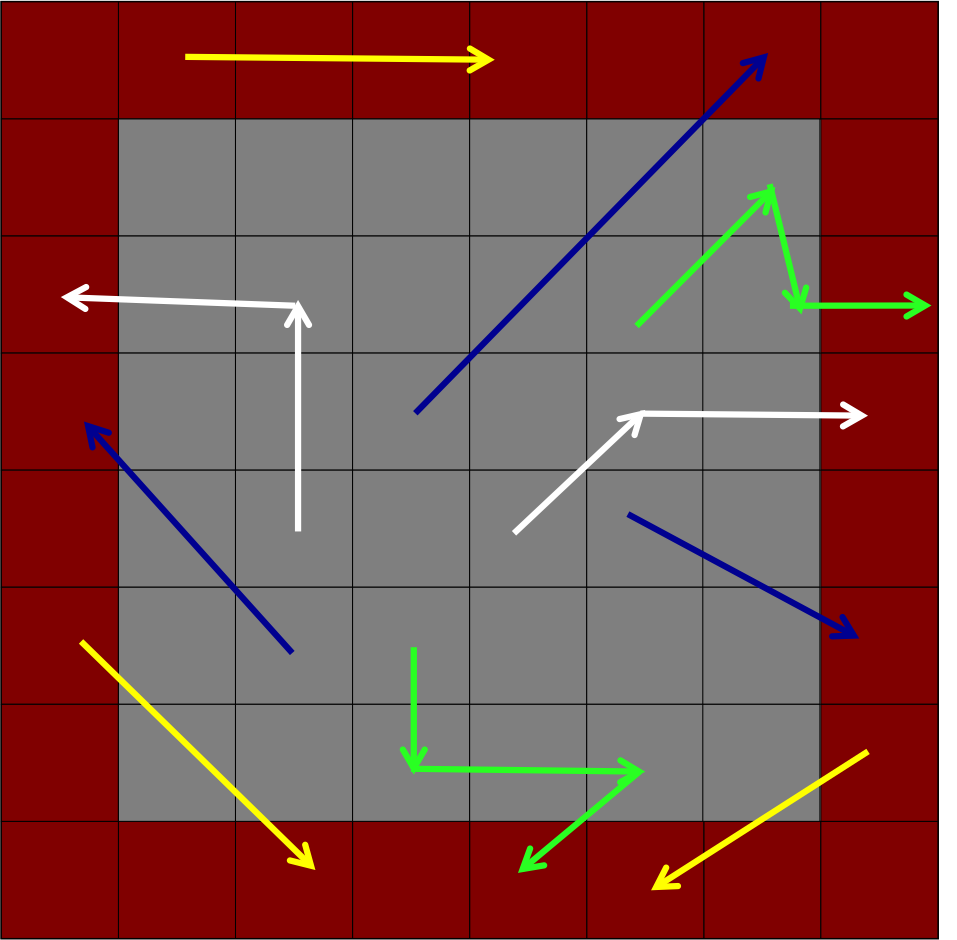}
    				\caption{Top view of the scintillator array grid (grey blocks are plastic and red blocks are calorimeters.  Various example of scattering event types are shown here: CC event (Yellow), PC(Blue), PPC(White) , PPPC(Green). Single C events are not scattered.}
    				\label{fig:ins_evt_classes}
				\end{figure}

An ideal PC event would be an incident photon scattering off a plastic element and completely being absorbed by a calorimeter. 
The high probability of scattering from the low-Z plastic and high probability of being fully absorbed in the high-Z calorimeters makes these PC events most ideal for polarimetry.
Therefore, for the remainder of the document, the analysis is mainly focused on PC events. 
					\subsubsection{Event Type}
					\label{sec:event_types}
					 					
\begin{figure}[tbp]
     			\centering
    				\includegraphics[width=0.35\textwidth]{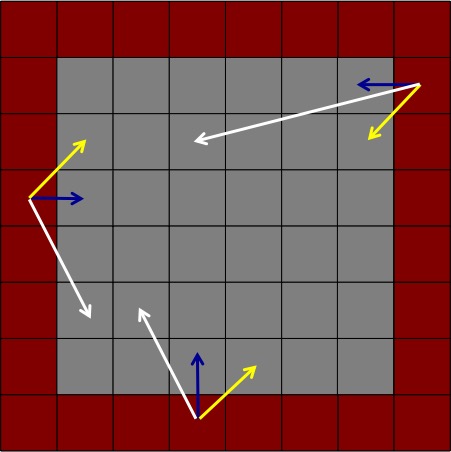}
    				\caption{Top view of the scintillator array grid displaying various event types. The examples  shown here are of PC events. Type 1 (white) are the non adjacent, type 2 (blue)  are the adjacent anodes sharing a side and type 3 (yellow) are adjacent anodes sharing a corner. }
    				\label{fig:ins_evt_type}
				\end{figure}

An event type is defined for events with two triggered anodes. 
The event type provides additional information on the spatial relationship between the anodes involved and the susceptibility of the event to the optical crosstalk. 
Type 1 events are those in which the two anodes are clearly separate from each other. 
Type 2 events are those in which the two anodes are adjacent to one another, sharing a side between them. 
Type 3 events are those in which the two adjacent anodes share a corner between them. 
These are shown in Figure \ref{fig:ins_evt_type}

			\subsection{Crosstalk}
			\label{sec:ins_crosstalk}

There are 2 types of crosstalk that we must consider. 
As mentioned in section \ref{sec:ins_MAPMT}, the MAPMT comes with an intrinsic electronic crosstalk of $~$3\%
The electronic crosstalk is electronic noise or interference that makes its way from the electronics of one anode to the electronics of an adjacent anodes.

\begin{figure}[tbp]
     			\centering
    				\includegraphics[width=0.75\textwidth]{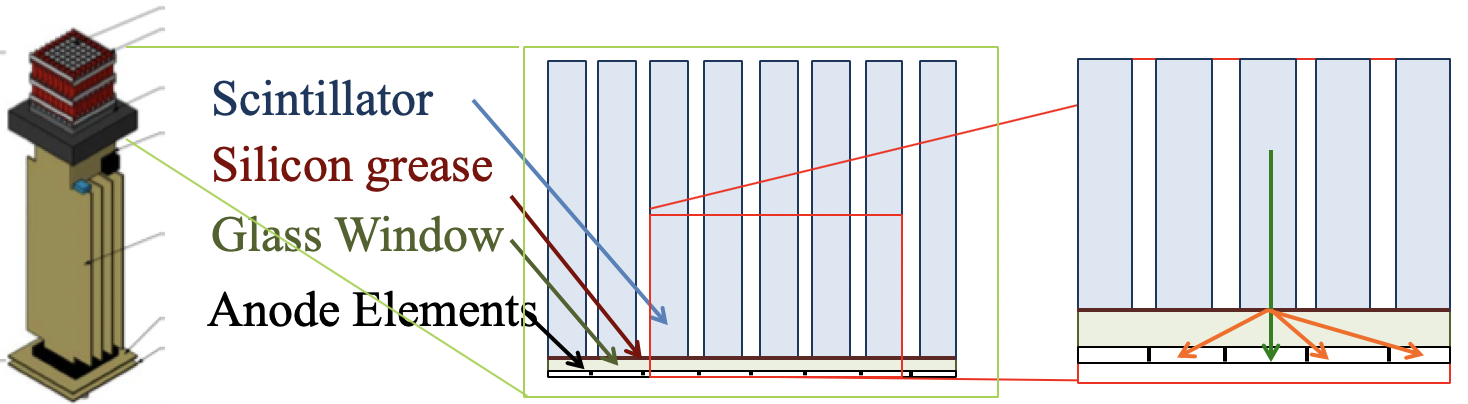}
    				\caption{An example of crosstalk. An incoming photon and the corresponding anode in MAPMT shown in green. The triggering of other anodes in MAPMT due to crosstalk in orange.}
    				\label{fig:ins_ct_example}
				\end{figure}

The second type of crosstalk is known as optical crosstalk. Ideally, in the absence of crosstalk, the scintillation light from each of the scintillator elements corresponds to a single anode in the MAPMT.
In practice the light spreads out laterally as it leaves the base of the scintillator.
Some of that light makes it to adjacent anodes.
This phenomenon is known as optical crosstalk and is shown in Figure \ref{fig:ins_ct_example}.
The enlarged view on the right shows the potential paths of the scintillation light (green), some of which can make its way into adjacent anodes (red). 
This light received in the adjacent anodes is treated as light from its respective element by the MAMPT anode.
The adjacent anodes will trigger if the deposited signal is higher than the threshold energy.
The triggering of these anodes, due to crosstalk, leads to misclassification of events.        

Misclassification of events simply means that a particular event can be classified incorrectly if additional anodes are triggered by crosstalk.
For example, a C event can be misclassified as PC event if it triggers a plastic due to crosstalk. 
Another example would be a PC event misclassified as PPC event if an additional plastic is triggered due to crosstalk. 
Misclassification of events is a huge problem as it provides false statistics to our data.            
There is no direct way of dealing with crosstalk between adjacent scintillators of the same material (plastic to plastic or calorimeter to calorimeter).
However, the PSD technique has been designed to deal with the crosstalk from different scintillator types.         
	
	\section{Instrument Assembly}
	\label{sec:instrument_assembly}
	
\begin{figure}[tbp]
 \centering
\begin{subfigure}[b]{0.67\textwidth}
 		\includegraphics[width=1\textwidth]{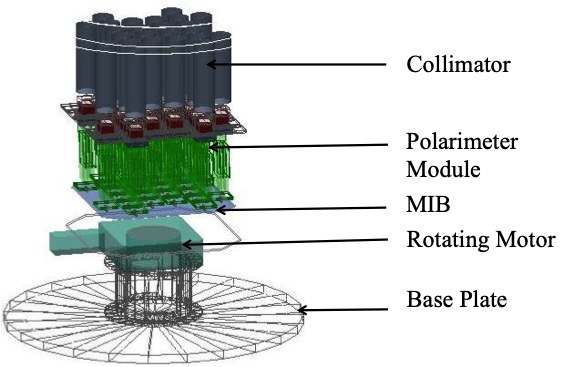}
		 \caption{}
\end{subfigure}    
\begin{subfigure}[b]{0.27\textwidth}
 		 \includegraphics[width=1\linewidth]{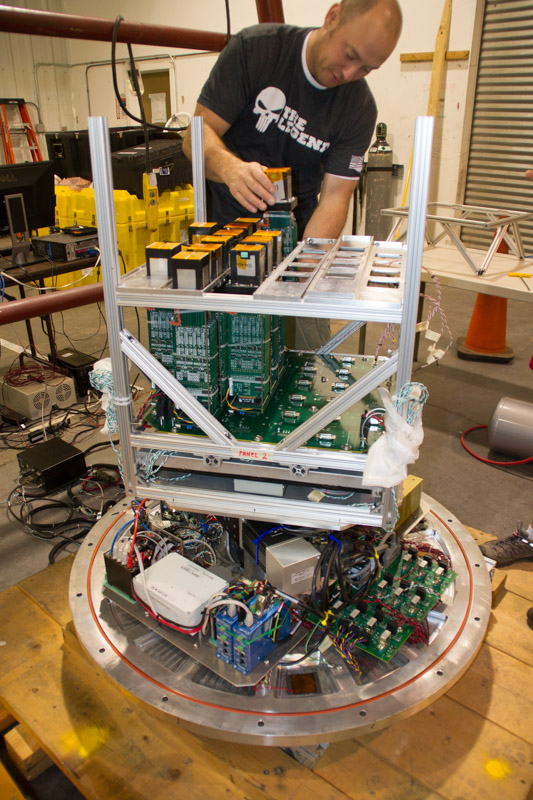}
		 \caption{}
\end{subfigure} 
  \caption{a) A skeleton model of GRAPE instrument assembly generated through GEANT4 toolkit. We can see the collimator array, polarimeter module array, MIB and rotating motor. The polarimeter module array is placed on the MIB. The 6 shields encloses the setup and sits on top of the rotating motor. For visibility of the collimators and modules, the shields are not shown here. This instrument assembly sits on rotating motor. b) GRAPE being assembled. We can see the MIB, module array, the electronics on top of the pressure vessel base plate.}
\label{fig:ins_assembly}
\end{figure}

A total of 24 independent polarimeter modules form the polarimeter module array (Figure \ref{fig:ins_module_array}).	
In front of each module is a cylindrical Pb-Al collimator that is designed to allow flux from a limited Field of View (FoV) to reach the module. 
Each polarimeter module plugs into the Module Interface Board (MIB).
The assembly of MIB, polarimeter array and collimator is enclosed by a six-sided shield assembly consisting of both active and passive shielding materials. 
Collectively, this assembly is referred to as the Instrument Assembly, as  shown in Figure \ref{fig:ins_assembly}.
		
			\subsection{Collimator Array}
			\label{sec:collimator}

Each of the 24 modules has its own collimator which is cylindrical in shape. 
Each collimator consists of an Al tube (7.6 cm OD with 1.2 mm thick walls) lined on the outside with 0.4 mm of Pb and extending 25 cm in front of the top surface of each polarimeter module.
This provides a $\sim$10$^\circ$ FoV for each modules.
There are two aluminum plates on top of the module array with circular holes for the collimators to be supported.
The collimators fit into these circular hole.
This arrangement of 24 collimators, one for each module, is referred to as the Collimator Array.

			\subsection{Shielding}
 			\label{sec:instrument_shielding}
		
The polarimeter array, the collimator array, the MIB and its electronics are completely enclosed on all sides by 6 anti-coincidence (AC) panels.
This arrangement can be seen in Figure \ref{fig:ins_assembly}. 
Each of the panels is composed of both passive and active shielding.
The active shielding consists of a 6-mm thick sheet of plastic scintillator (Eljen EJ-204 ), contained in a rigid, light aluminum housing. 
This plastic sheet is split into two equal pieces and a wavelength shifting (WLS) bar is placed between them. 
The WLS bar redirects the light produced in the plastic shields to two PMTs to read out the sum signal from each scintillator panel.
This active shield design with WLS bar was a technique developed for the FiberGlAST project.
This design is 99$\%$ efficient at detecting minimum ionizing particles \citep{Pendleton1999}. 
Charged particles detected by the active shields would generate a coincidence signal which is used to reject (veto out) the background events. 

The passive shield consists of a layer of 4.24 mm of Pb and a layer of 0.8 mm of Sn.  
The shield lies is on the inner side of each AC panel, facing the detectors.    
The passive shielding layers are attached to the inner walls of the AC panels, with the Sn facing the detector array. 
The 4.24mm layer of Pb was was an upgrade from 0.8mm from GRAPE 2011 flight.
Passive shields serve to absorb high-energy atmospheric photons from the sides and from below the payload.
The top panel has a cut out of 338.5 $\times$ 338.5 mm$^2$ for the module array to allow photons to enter from the forward direction. 
The 6 AC panels forms an enclosed box (cuboid shape) and is defined as the instrument assembly. 
			
			\subsection{Module Interface Board (MIB) and Instrument Electronics}
			\label{sec: mib_electronics}

The Module Interface Board (MIB) is a rigid printed a circuit board with a field programmable gate array (FPGA) and 32 module interface connectors.
We only utilized 24 of these connectors for the 2014 GRAPE balloon flight. 
The MIB placement is seen in Figure \ref{fig:grp_ins_e}.
The MIB is used to supply the modules with power, provide mechanical support for the modules, and serve as an interface between the module detectors and the Science Data Computer (SDC).          
                                                                                                                                                                                                                                                                                                                                                                                                                                                                                                                                                                                                                                                                                                                                                                                                                                                                                                                                                                                                                                                                                                                                                                                                                                                                                                                                                                                                                                                                                                                                                                                                                                                                                                                                                                                                                                                                                                                                                                                                                                                                                                                                                                                                                                                                                                                                                                                                                                                                                                                                                                                                                                                                                                                                                                                                                                                                                                                                                                                                                                                                                                                                                                                                                                                                                                                                                                                                                                                                                                                                                                                                                                                                                                                                                                                                                                                                                                                                                                                                                                                                                                                                                                                                                                                                                                                                                                                                                                                                                                                                                                                                                                                                                                                                                                                                                                                                                                                                                                                                                                                                                                                                                                                                                                                                                                                                                                                                                                                                                                                                                                                                                                                                                                                                                                                                                                                                                                                                                                                                                                                                                                                                                                                                                                                                                                                                                                                                                                                                                                                                                                                                                                                                                                                                                                                                                                                                                                                                                                                                                                                                                                                                                                                                                                                                                                                                                                                                                                                                                                                                                                                                                                                                                                                                                                                                                                                                                                                                                                                                                                                                                                                                                                                                                                                                                                                                                                                                                                                                                                                                                                                                                                                                                                                                                                                                                                                                                                                                                                                                                                                                                                                                                                                                                                                                                                                                                                                                                                                                                                                                                                                                                                                                                                                                                                                                                                                                                                                                                                                                                                                                                                                                                                                                                                                                                                                                                                                                                                                                                                                                                                                                                                                                                                                                                                                                                                                                                                                                                                                                                                                                                                                                                                                                                                                                                                                                                                                                                                                                                                                                                                                                                                                                                                                                                                                                                                                                                                                                                                                                                                                                                                                                                                                                                                                                                                                                                                                                                                                                                                                                                                                                                                                                                                                                                                                                                                                                                                                                                                                                                                                                                                                                                                                                                                                                                                                                                                                                                                                                                                                                                                                                                                                                                                                                                                                                                                                                                                                                                                                                                                                                                                                                                                                                                                                                                                                                                                                                                                                                                                                                                                                                                                                                                                                                                                                                                                                                                                                                                                                                                                                                                                                                                                                                                                                                                                                                                                                                                                                                                                                                                                                                                                                                                                                                                                                                                                                                                                                                                                                                                                                                                                                                                                                                                                                                                                                                                                                                                                                                                                                                                                                                                                                                                                                                                                                                                                                                                                                                                                                                                                                                                                                                                                                                                                                                                                                                                                                                                                                                                                                                                                                                                                                                                                                                                                                                                                                                                                                                                                                                                                                                                                                                                                                                                                                                                                                                                                                                                                                                                                                                                                                                                                                                                                                                                                                                                                                                                                                                                                                                                                                                                                                                                                                                                                                                                                                                                                                                                                                                                                                                                                                                                                                                                                                                                                                                                                                                                                                                                                                                                                                                                                                                                                                                                                                                                                                                                                                                                                                                                                                                                                                                                                                                                                                                                                                                                                                                                                                                                                                                                                                                                                                                                                                                                                                                                                                                                                                                                                                                                                                                                                                                                                                                                                                                                                                                                                                                                                                                                                                                                                                                                                                                                                                                                                                                                                                                                                                                                                                                                                                                                                                                                                                                                                                                                                                                                                                                                                                                                                                                                                                                                                                                                                                                                                                                                                                                                                                                                                                                                                                                                                                                                                                                                                                                                                                                                                                                                                                                                                                                                                                                                                                                                                                                                                                                                                                                                                                                                                                                                                                                                                                                                                                                                                                                                                                                                                                                                                                                                                                                                                                                                                                                                                                                                                                                                                                                                                                                                                                                                                                                                                                                                                                                                                                                                                                                                                                                                                                                                                                                                                                                                                                                                                                                                                                                                                                                                                                                                                                                                                                                                                                                                                                                                                                                                                                                                                                                                                                                                                                                                                                                                                                                                                                                                                                                                                                                                                                                                                                                                                                                                                                                                                                                                                                                                                                                                                                                                                                                                                                                                                                                                                                                                                                                                                                                                                                                                                                                                                                                                                                                                                                                                                                                                                                                                                                                                                                                                                                                                                                                                                                                                                                                                                                                                                                                                                                                                                                                                                                                                                                                                                                                                                                                                                                                                                                                                                                                                                                                                                                                                                                                                                                                                                                                                                                                                                                                                                                                                                                                                                                                                                                                                                                                                                                                                                                                                                                                                                                                                                                                                                                                                                                                                                                                                                                                                                                                                                                                                                                                                                                                                                                                                                                                                                                                                                                                                                                                                                                                                                                                                                                                                                                                                                                                                                                                                                                                                                                                                                                                                                                                                                                                                                                                                                                                                                                                                                                                                                                                                                                                                                                                                                                                                                                                                                                                                                                                                                                                                                                                                                                                                                                                                                                                                                                                                                                                                                                                                                                                                                                                                                                                                                                                                                                                                                                                                                                                                                                                                                                                                                                                                                                                                                                                                                                                                                                                                                                                                                                                                                                                                                                                                                                                                                                                                                                                                                                                                                                                                                                                                                                                                                                                                                          
The MIB continuously cycles through each of the 32 module ports (even though we only use 24 of them) checking to see if a valid event is ready to be to be read out. 
Valid events are read out with a time resolution of 20 $\mu$s. 
The MIB cycle time (102.4 $\mu$s) plus the additional processing of the data in the MIB results an absolute time accuracy of 125 $\mu$s. 

\begin{sidewaysfigure}[hbtp]
 \centering
    	\includegraphics[width=\columnwidth]{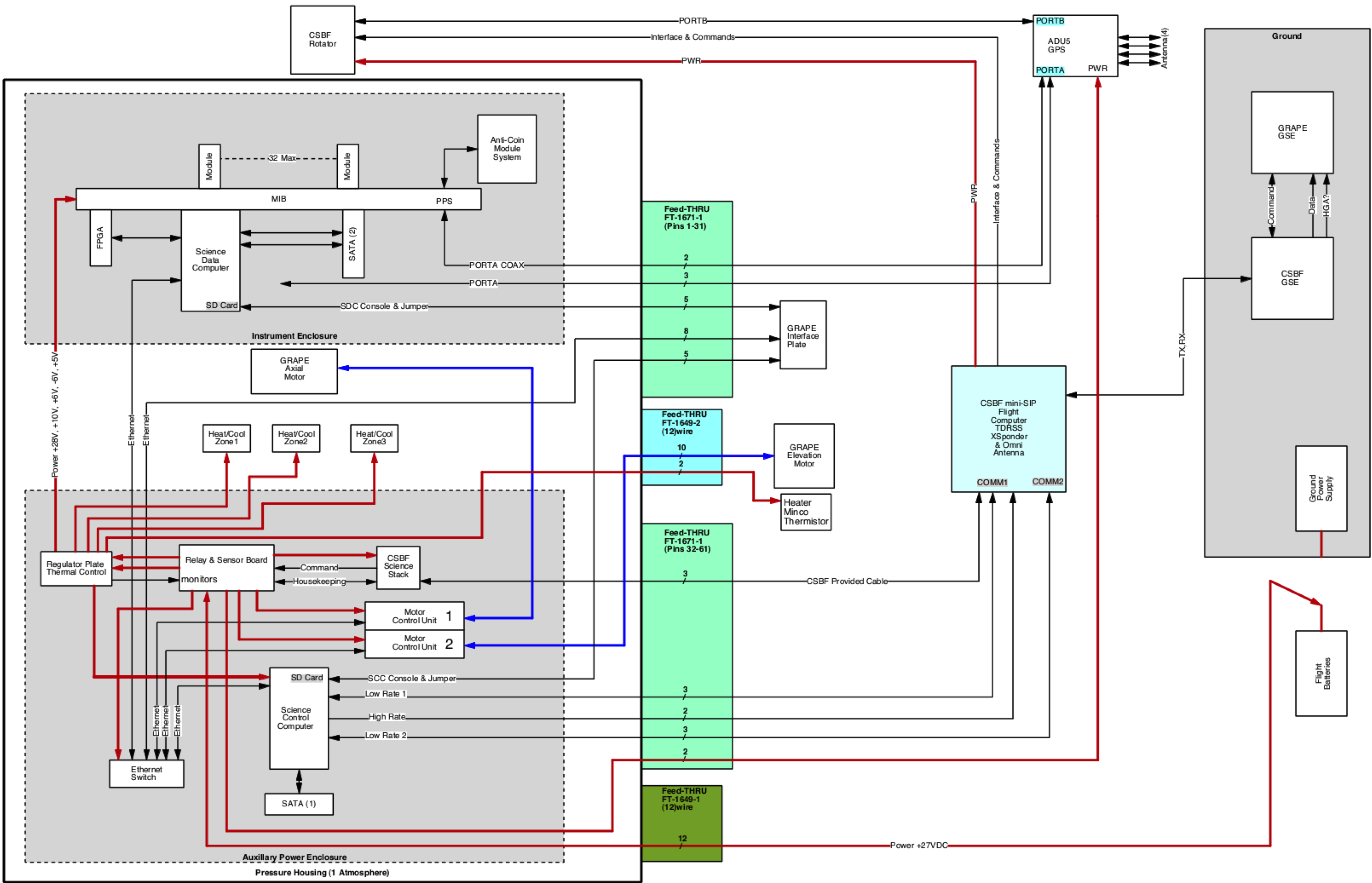}
    	\caption{The combined electronic block diagram showing SCC, SDC, GSE terminals, etc and the connections and the data flow. }
    	\label{fig:grp_flt_elec_sys_block}
\end{sidewaysfigure}

The electronics enclosed within the instrument assembly is labeled as the Instrument Electronics. 
A block diagram of the Instrument Electronics can be seen in the top left of Figure \ref{fig:grp_flt_elec_sys_block} labeled as instrument enclosure. 
The MIB with the plugged modules, FPGA, SDC computer and a SATA hard drive associated with the SDC are within the instrument enclosure.
The Anti-Coincidence (AC) Module System also communicates with the MIB to generate an AC flag for the events that coincide with the detection of charged particles from the AC panels. 
The AC are described briefly in section \ref{sec:instrument_shielding}.
The SDC communicates with the ADU5 GPS antenna that receives the GPS time.
The events are time tagged using this GPS time. 
We use the ADU5 GPS receiver (and the antenna), model \#: 800952, produced by Thales Navigation.
The data acquisition of science and housekeeping data from each module was routed through the MIB to the SDC. 
The control of modules (thresholds, power, etc.) was also routed from the SDC through the MIB to the respective modules. 
The event data and module housekeeping data is stored in the SATA drive associated with the SDC.

	\section{Rotation table}
	\label{sec:rotationtable}
\begin{figure}[tbp]
 \centering
    	\includegraphics[width=0.8\textwidth]{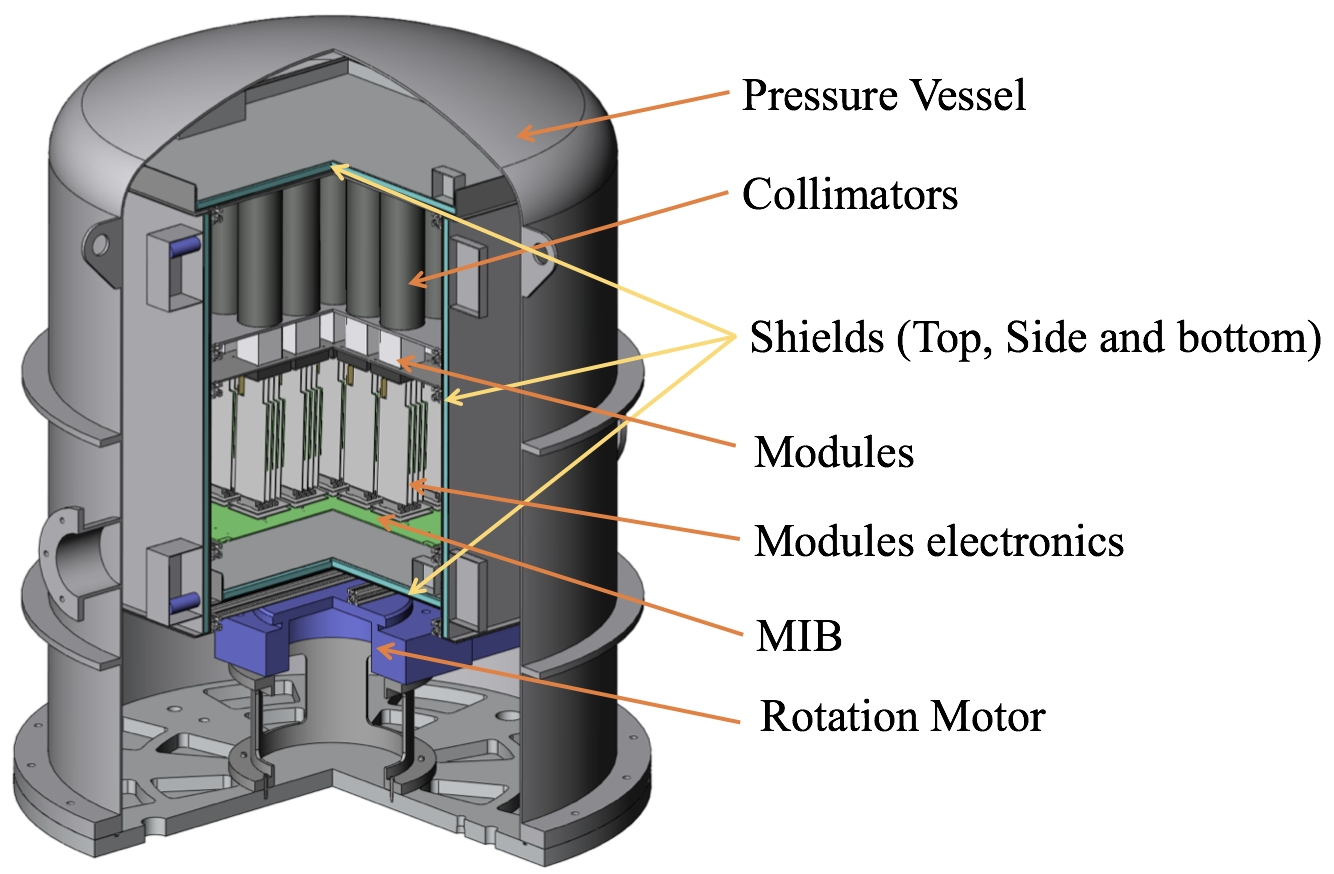}
    	\caption{A model picture showing pressure vessel, collimators, shields, rotation motor and various other components of GRAPE assembly.}
    	\label{fig:grp_ins_e}
\end{figure}

The instrument assembly sits on top of a rotation motor.
This motor enables the rotation of the instrument about the pointing axis. 
This rotation of the instrument helps to remove effects of anisotropies in the polarization response.
The rotation motor can be seen in Figure \ref{fig:grp_ins_e} and the pointing axis goes through center of the motor and the center of the instrument assembly. The rotation happens at 4$^\circ$ increments (steps) and a full rotation of  360$^\circ$ is called a sweep. 
Once a sweep is completed in a clockwise direction, the next sweep occurs in an anti-clockwise direction.
This prevents the cables from tangling if rotation continued in one direction. 
The rotation is halted at each step for a set time which we refer to as the dwell time. 
The dwell time (observation time at a fixed angle) was set at 5s.
It took 3s to rotate from one step to next, so each sweep took 720s (12 mins).       
The dwell time is one of the table parameters that can be set by user command. 
During the 2011 GRAPE flight, sweeps continued even during a change in pointing direction.
As an improvement for the 2014 flight, an option to pause at the end of the sweep was added to the set of user commands. 
Once the pause command was sent, the rotation paused at the end of the sweep so that the instrument could be pointed to a new target before resuming a new sweep.
This allowed us to observe a new source from the beginning of a sweep. 
The recording of the data did not occur during a pause of the rotation motor.
An encoder in the axial drive system recorded the time-tagged orientation of the rotation parameters as part of the housekeeping data. 

\subsection{Baseplate Electronics}
	
Additional electronics are placed outside the instrument assembly on top of the pressure vessel base plate. 
This can be seen in Figure \ref{fig:ins_assembly} b. 
These electronics include the telemetry interface, rotation motor, SCC, CSBF Science Stack and power relays. 
The Science Control Computer (SCC)  handled the housekeeping data which included thermistors, module rate data, shield rate data, rotation table data. 
It also processed the ground commands received from the Colombia Scientific Balloon Facility (CSBF) mini-SIP (Support Instrument Package). 
The SCC also prepared data for the telemetry stream provided by the mini-SIP. 
Any module commands received by the mini-SIP from the ground station were routed via the SCC to the SDC to the MIB and then to the respective module.
The mini-SIP computer had an Omni antenna that used to send telemetry data to ground station.
Data from the SDC (located in the MIB) was routed to the SCC and stored on the SCC-SATA drive.
The housekeeping data was also stored on the SCC-SATA drive.
For calibration and pre-flight operations, a direct ethernet connection was established to transfer data.
A block diagram of the SCC can be seen in Figure \ref{fig:grp_flt_elec_sys_block} on the bottom left shaded region labelled auxiliary power enclosure. 
During flight, the telemetered data was used to monitor the status of the experiment using Ground Station Equipment (GSE) computers.

%

					\subsubsection{GSE Computers}
					\label{sec:gse_computers}
					
\begin{figure}[tbp]
 \centering
    	\includegraphics[width=0.95\textwidth]{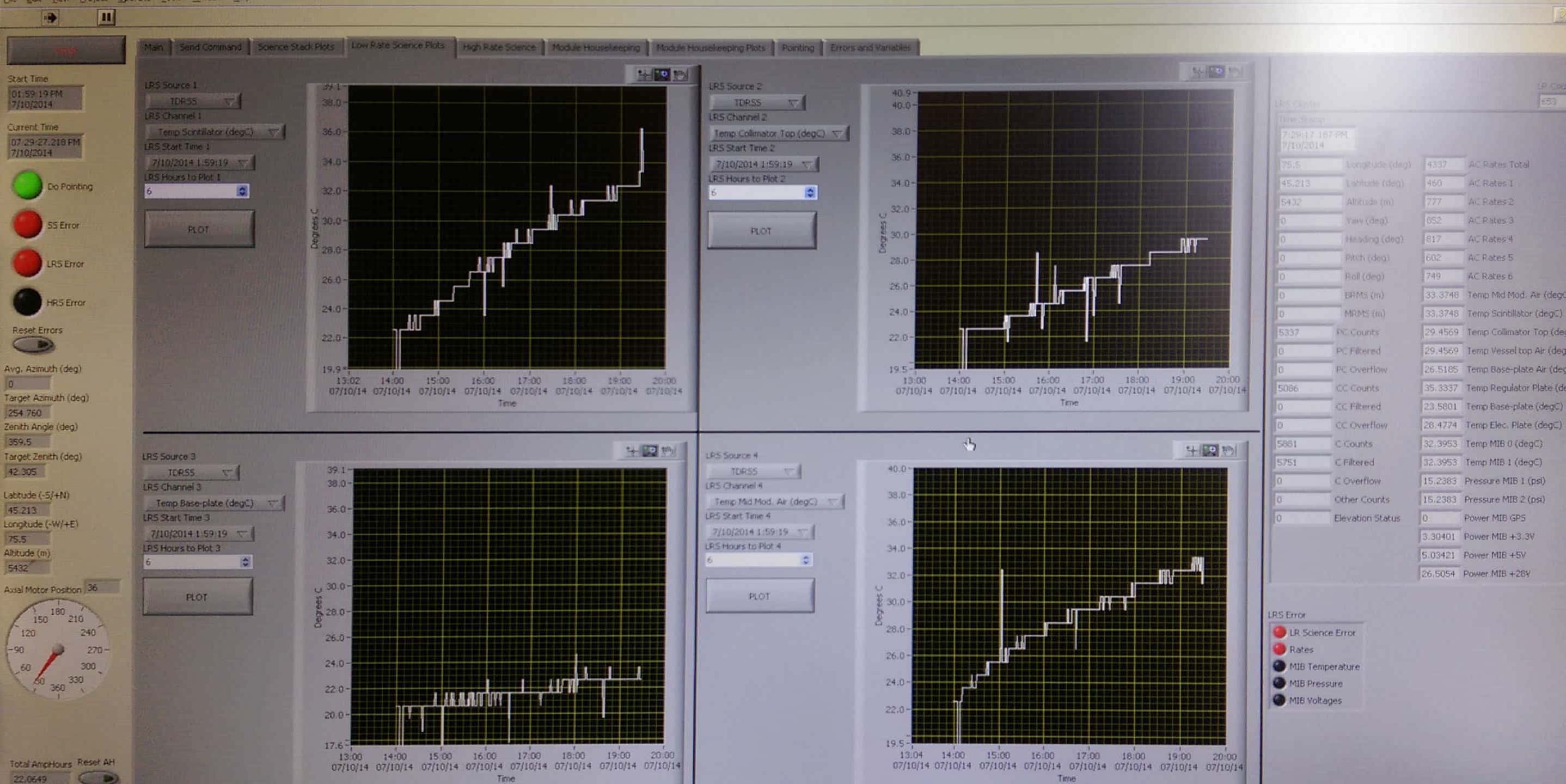}
    	\caption{The GSE LabVIEW low rate science (LRS) tab. We can see the LRS data on the right and some graphs. We can also see the various other tabs present in the GUI. There were various options in the drop down menu which allowed us to focus various units of the experiment to monitor.}
    	\label{fig:grp_gse_tab_lrs}
\end{figure}

GSE computers were set up with a LabVIEW Graphical User Interface (GUI). 
The GUI had 9 tabs to display various data and diagnostic information.  
They were labelled Main, Send Command, Science Stack Plots, Low Rate Science Plots, High Rate Science, Module Housekeeping, Module Housekeeping Plots, Pointing, errors and variables. 
One of the tabs, Low Rate Science Plots, is shown in Figure \ref{fig:grp_gse_tab_lrs}.
The Main tab updated every 30s and represented approximately the last hour of data. 
The Send Command tab was used to record and send commands to GRAPE. 
This tab was also used to  monitor rates and adjust the detector thresholds as needed.
The command for the rotation parameters was also sent via this tab. 
The Science Stack, Module Housekeeping, Low Rate Science, High Rate Science and other tabs were mostly diagnostic data that were monitored for abnormalities in the instrument. 
The pointing tab was used to monitor our pointing parameters for a target source and a command could be sent if it needed to be adjusted or to point to a new target.
We could adjust the zenith from this tab and the azimuthal direction was done via CSBF GSE. 
All these different tabs helped us to monitor the instrument during the flight .

	\section{Pressure Vessel}
	\label{sec:ins_pressure_vessel}	
	
	The instrument assembly, the rotation motor and the supporting electronics are all sencased in the pressure vessel (PV). 
	This can be seen in Figure \ref{fig:grp_ins_e}. 
	The PV is a two-piece aluminum assembly consisting of a cylindrical sidewall with an upper dome and a separate flat flange base.
	 It was maintained at 1 atm of pressure throughout the flight. 
	 Instrument integration, test and debug activities were conducted with the PV's sidewall and top dome structure removed to facilitate access. 
	 All instrument hardware is supported from the base plate, which also has ports for electrical wires and gas feed-throughs. 
	 The PV's top dome element is formed with 2-3 mm thick aluminum. 
	 Mounting brackets for the pressure vessel are located on the sidewalls. 
	 The two parts of the PV are connected with a series of screws around the perimeter of the flange. 
	 An O-ring was used to insure a tight seal so that the PC maintained the 1 atm pressure. 
	 The PV resides in a sturdy aluminum frame that forms the gondola.
			
	\section{Gondola}
	\label{sec:gondola}
\begin{figure}[tbp]
 \centering
    	\includegraphics[width=0.9\textwidth]{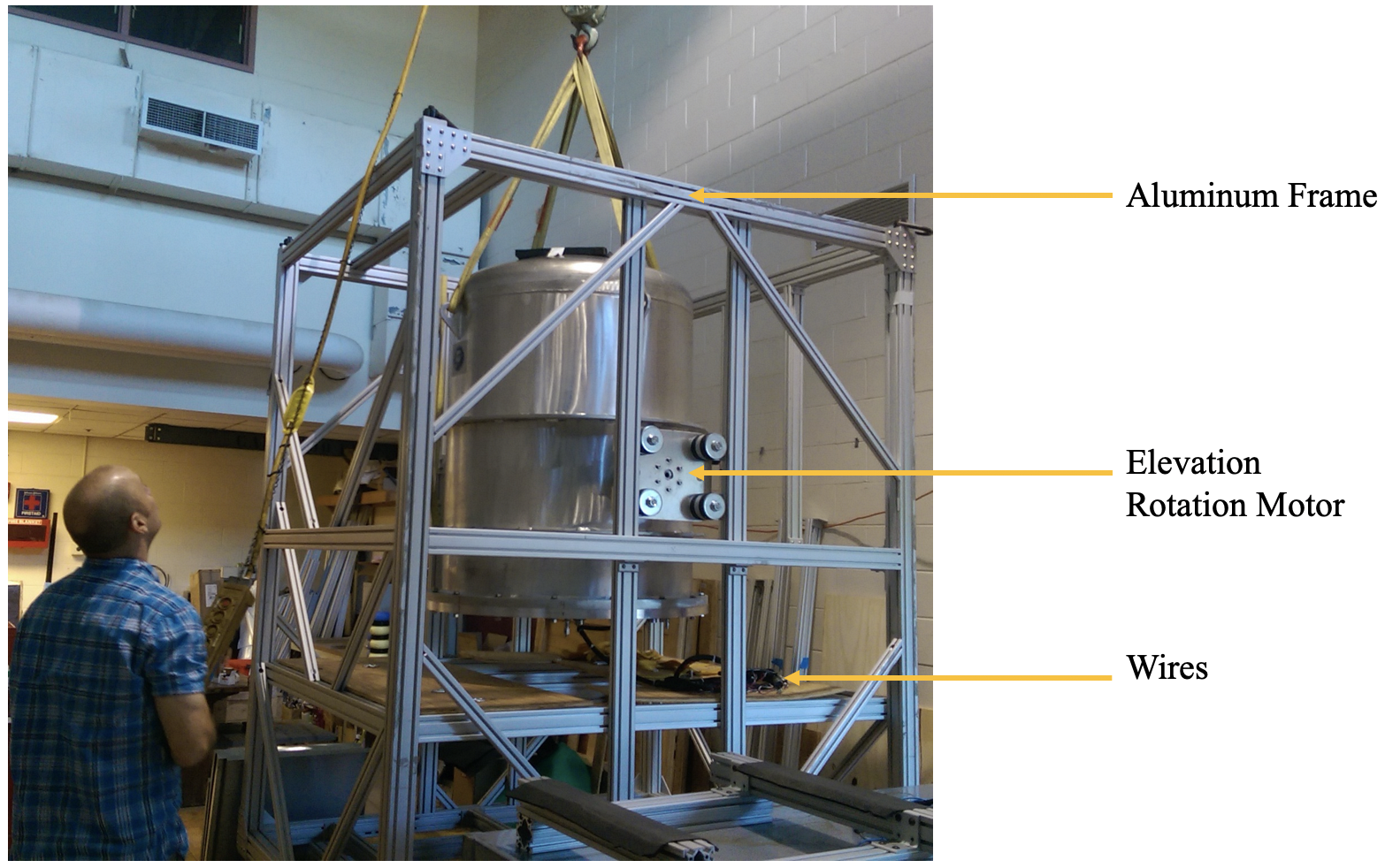}
    	\caption{The gondola with GRAPE instrument (inside the pressure vessel). We can see the wires that supply the power. We also notice the elevation rotation motor. The yellow ropes are of the mechanical crane as this picture was taken as the assembly was done. Instrument pointing is discussed in section \ref{sec:flt_pointing}.}
    	\label{fig:grp_ins_frame}
\end{figure}

	The PV was attached to an extended Aluminum frame (80/20 modular Al framing).
	The aluminum frame is sturdy and light which suits our needs for the balloon flight. 
	The PV is attached to the frame on two points on each side of the pressure vessel (180$^\circ$) apart. 
	One of the attachment is an elevation motor and another one is a bearing that allows the rotation. 
	This can be seen in Figure \ref{fig:grp_ins_frame} where we can see the pressure vessel, aluminum frame and the rotation motor.
	This motor provides pointing of the pressure vessel (and the instrument) in zenith angle. 
	The instrument pointing is further discussed in section \ref{sec:flt_pointing}.
	The Al frame is referred to as the gondola. 	
	For the calibration and ground observations, the powers were directly supplied and the data was directly received to the GSE computers.
	The power supplied to the instrument was through +28V batteries and distributed to the instrument through the relay board. 
For flight, the batteries were present in the gondola and for pre-flight, a regular battery unit was used to provide the +28V supply.
	The additional CSBF electronics, batteries and the rotator that enables the azimuthal pointing associated with the balloon flight are discussed in chapter \ref{sec:balloon_campaign}.
	
\chapter{Instrument Performance}
\label{sec:insrument_performance}

		The instrument performance defines the ability of our instrument to do various measurements.
		The expected performance of the instrument is determined via simulation of the instrument.
		Flight background is simulated using the parameters determined by the flight plan and eventually compared to the measured flight background. 
		Simulation is also used to define the instrument response and determine the polarimetry sensitivity.
		This section gives a detailed description of the simulations performed and also presents the calibration data.

	\section{Simulation}
	\label{sec:ins_perf_simulation}
	The instrument simulation for GRAPE is performed using the GEANT4 toolkit.
	The GEANT4 toolkit is used for simulating the passage of particles through matter and is applicable to a large number of projects and experiments in high energy physics, astrophysics, medical physics and radiation protection \citep{Allison2016, Agostinelli2003, Allison2006}. 
	GRAPE 2014 is simulated in GEANT4.10 (current version of simulations are in GEANT4.10.5). 
		
			\subsection{GEANT4}
			\label{sec:ins_perf_geant4}
			GEANT4 requires three main components to perform a simulation.
			The user must specify these three components of the simulation: 1) an instrument description; 2) a list of particle interaction; and 3) a description of the incident beam. 
			These 3 files are mandatory user classes and are required (at minimum) by GEANT4 to perform a simulation. 
			Additional user classes can be called depending on the specific needs of the experiment.
			\subsubsection*{Mass Model}
			\label{sec:ins_pref_mass_model}
		

\begin{figure}[hbtp]
 \centering
\begin{subfigure}[b]{0.4\textwidth}
 		 \includegraphics[width=1\linewidth]{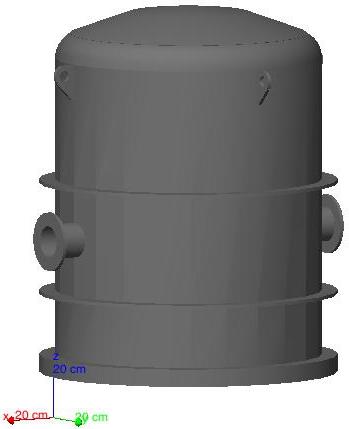}
		 \caption{}
\end{subfigure}
\begin{subfigure}[b]{0.4\textwidth}
 		 \includegraphics[width=1\linewidth]{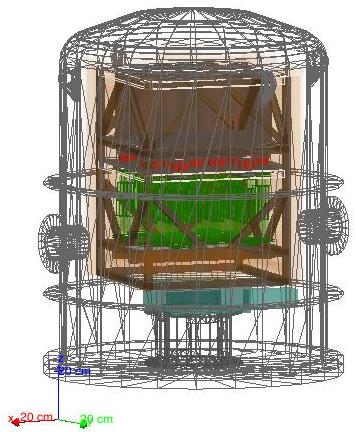}
		 \caption{}
\end{subfigure}

\begin{subfigure}[b]{0.40\textwidth}
 		 \includegraphics[width=1\linewidth]{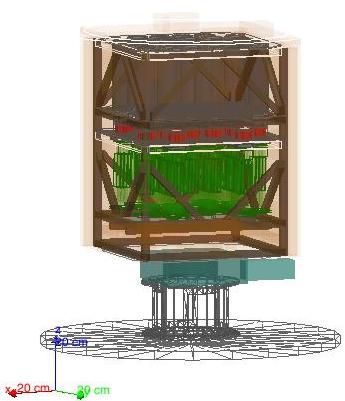}
		 \caption{}
\end{subfigure}
\begin{subfigure}[b]{0.40\textwidth}
 		 \includegraphics[width=1\linewidth]{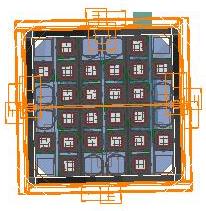}
		 \caption{}
\end{subfigure}

  \caption{Instrument Description (Mass Model) for GRAPE. (a) Picture outlining the pressure vessel we saw in section \ref{sec:ins_pressure_vessel}. (b) Outlines the instrument assembly with skeleton of pressure vessel. (c) Outline showing the instrument assembly and the rotation motor. (d) This shows the instrument from the top view where we can see the modules and shields. }
\label{fig:sim_geo_model_1}

\end{figure}

\begin{figure}[!ht]
 \centering
\begin{subfigure}[b]{0.48\textwidth}
 		 \includegraphics[width=1\linewidth]{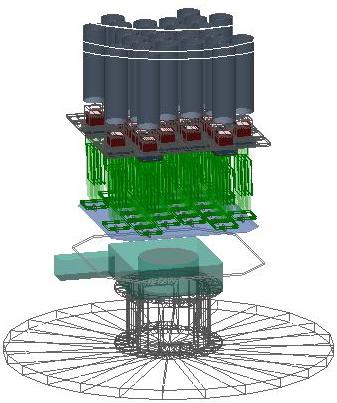}
		 \caption{}
\end{subfigure}
\begin{subfigure}[b]{0.48\textwidth}
 		 \includegraphics[width=1\linewidth]{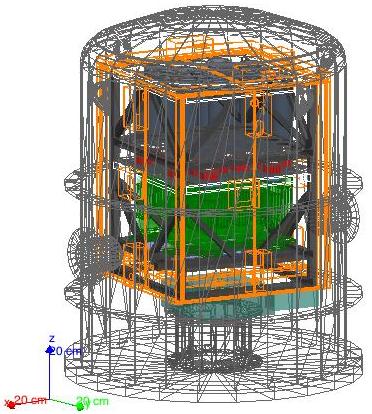}
		 \caption{}
\end{subfigure}

\caption{Further instrument description (mass model) for GRAPE. (a) The shields and pressure vessel are removed to see the collimator and module array (b) The overall skeleton figure of the payload.}
\label{fig:sim_geo_model_2}

\end{figure}

			The instrument geometry, the first mandatory component, is defined by what is often referred to as the mass model.
			The mass model includes the descriptions (measurements, shape and material) of the various components that makes up the instrument. 
			These components include all the detectors, shields, pressure vessel, beams, collimators, screws, bolts, etc. 
			Anything in the instrument that can interact with the particle needs to be described here. 
			We also have to include the air which fills up the pressure vessel since the particles can interact with the air molecules. 
			The mass model for GRAPE is shown in Figure \ref{fig:sim_geo_model_1} and \ref{fig:sim_geo_model_2}. 
			These Figures show various components of the instrument to give a better understanding of the mass model.
			GEANT4 requires that these modeled components do not overlap with each other. 
			Hence these components have to be modeled and arranged carefully. 
			GEANT4 has inbuilt checks for overlaps. 
			After every modification, an overlap check was conducted. 
			We also need to define the components whose interaction with the particles (hits) are to be recorded. 
			These components are defined as sensitive detectors. 
			In our simulation, the scintillators (plastic and CsI) and the active shields are the sensitive detectors. 
			
			\subsubsection*{Physics List}
			\label{sec:ins_pref_physics_list}
			
			The second mandatory component is the description of the interaction physics for the particles being simulated.
			These definitions are in physics list modules that are called within the simulations. 
			We can import the pre-defined models, combine multiple models or create a new ones. 
			The decay physics list (G4DecayPhysics) and the standard electro-magnetic list (G4EmStandardPhysics) were always invoked for the GRAPE simulations.
			The G4DecayPhysics defines all the particles and their decay physics. 
			The G4EmStandardPhysics is the default list which covers most physics from 0-100 TeV for photons, up to 1 PeV for electrons and positrons. 
			Electro-Magnetic (EM) interactions of charged hadrons and ions up to 100 TeV are also covered by this list.
			Additionally, it also defines the electro-magnetic physics for muons, pions, kaons, protons, anti-protons etc. 
			It covers physics of multiple Coulombs scattering, pair production, Compton scattering, Bremsstrahlung, etc. 
			A complete list of these processes can be found in the GEANT documentation. 
				
			Additional lists were used depending on special cases of the simulations. 
			For polarized runs, G4EmLivermorePolarizedPhysics list was invoked.
			This physics list specializes in low energy polarized electromagnetic interactions for electrons and photons. 
			This includes all low energy processes like photo-electric effect, Compton scattering, Rayleigh Scattering, gamma conversion, bremsstrahlung and ionization and more. 
			This model also includes the regular Livermore physics list by default.
			This model was used for the simulation of gammas, electrons and positrons. 
			For neutrons and protons (the hadronic particles) the QGSP\_BERT\_HP model was additionally invoked. 
			This physics description contains a list of basic physics that applies to the quark gluon string model for high energy interaction of protons, neutrons, pions, Kaons and nuclei. 
			It uses the BERTini cascade model for primary protons, neutron, pions and Kaons below $\sim$10GeV. 
			This is further improved by including the data driven High Precision (HP) neutron package (NeutronHP) to transport neutrons below 20 MeV down to thermal energies. 


\begin{figure}[hbtp]
 \centering
\begin{subfigure}[b]{0.24\textwidth}
 		 \includegraphics[width=1\linewidth]{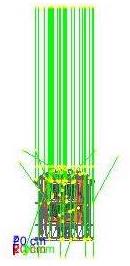}
		 \caption{}
\end{subfigure}
\begin{subfigure}[b]{0.59\textwidth}
 		 \includegraphics[width=1\linewidth]{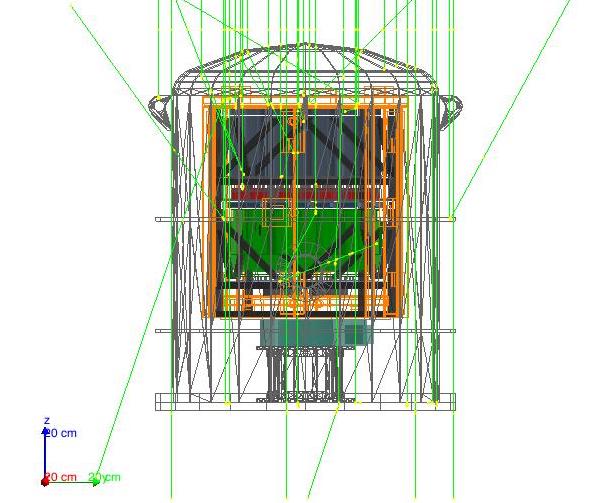}
		 \caption{}
\end{subfigure}

\begin{subfigure}[b]{0.26\textwidth}
 		 \includegraphics[width=1\linewidth]{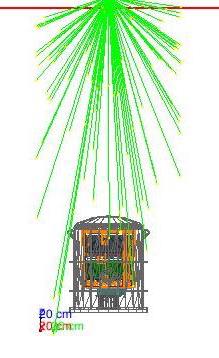}
		 \caption{}
\end{subfigure}
\begin{subfigure}[b]{0.46\textwidth}
 		 \includegraphics[width=1\linewidth]{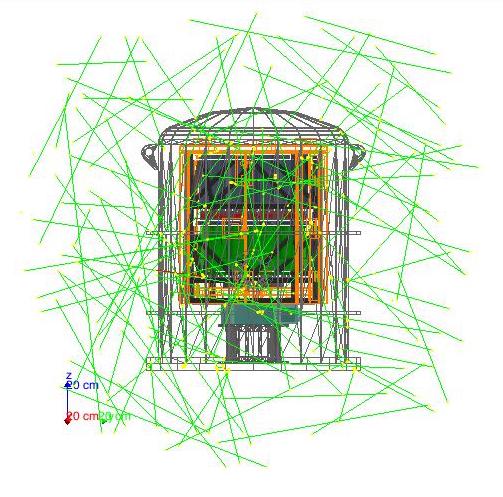}
		 \caption{}
\end{subfigure}
  \caption{Examples of various source descriptions. a) Plane parallel beam: where the particles are incident downward to the instrument from a plane (for distant sources) b) This is a closer view of the plane parallel run where we can see some of the particles interacting with the instrument. c) An example of a point source run (flood run). d) An example of a source description for a background simulation run. The source particles are generated in the surface of an imaginary sphere enclosing the instrument. This setup replicates the instrument in a radiation environment (at flight altitude) from various input fluxes form literature.  }
\label{fig:sim_src_desc_1}

\end{figure}
			
			\subsubsection*{Source Description}
			\label{sec:ins_perf_sim_src_desc}
			The third mandatory component is the description of the incident particle beam. 
			The particle type, energy and geometry of the particle beam must be provided.
		    The type of particle is generally a photon (gamma ray) unless we are doing a background run. 
		    For the background, the particle simulated were gammas, electrons, positrons, neutrons and protons.
		    The energy of the particle can be either mono-energetic or some type of spectral distribution. 
		    A spectral distribution can be provided as some sort of functional form or through a text file. 
		    The text file description is used mainly for the background simulation. 

		   The geometry of the incident particle depends on the simulation. 
		   There are mainly three forms of source geometry used in our simulations.
		   They are plane-parallel beams (for distant sources), point source runs (for calibration sources) and isotropic distributions (for  the background runs).
		   Examples of these source geometries are shown in Figure \ref{fig:sim_src_desc_1}.
		   
		    A plane parallel run is shown in Figure \ref{fig:sim_src_desc_1} (a). 
		    The particles are incident downward (in z direction)  as a circular beam covers the instrument cross-section.
		    This description is used to generate the instrument response in terms of  the response matrix and the effective area. 
		    Figure \ref{fig:sim_src_desc_1}(b) is a zoomed in picture which shows a few interactions (scattering and hits) in the detector. 
		    
		    A point source run also called a "flood run" is shown in Figure \ref{fig:sim_src_desc_1} (c).
		    Here a particle is isotropically radiating at a given distance. 
		    In our simulation, the distance is along z-direction. 
		    This description is used for simulation of the calibration runs.
		    For point source runs, we can make the simulation efficient by fixing the angle of incident to cover the instrument rather than isotropically radiating where a lot of generated photons are wasted.
		    
			Figure \ref{fig:sim_src_desc_1} (d) shows a typical distribution of particles during background simulation.
			This description defines the instrument in a radiation environment which is analogous to GRAPE during the flight. 
			The source particles are generated in the surface of an imaginary sphere enclosing the instrument. 
			The example shown in Figure \ref{fig:sim_src_desc_1} (d) shows particles bombarding at the instrument from all angles. 
			The description of the background particles  as a function of energy and angle is taken from the literature.
			Some particles (primary electron, positron or proton) the distribution is defined such that the particles are only bombarding from the upper hemisphere. 
			This is further explained in section \ref{sec:ins_perf_bgd_sim}.
			
\begin{figure}[hbtp]
 \centering
 	\includegraphics[width=0.95\textwidth]{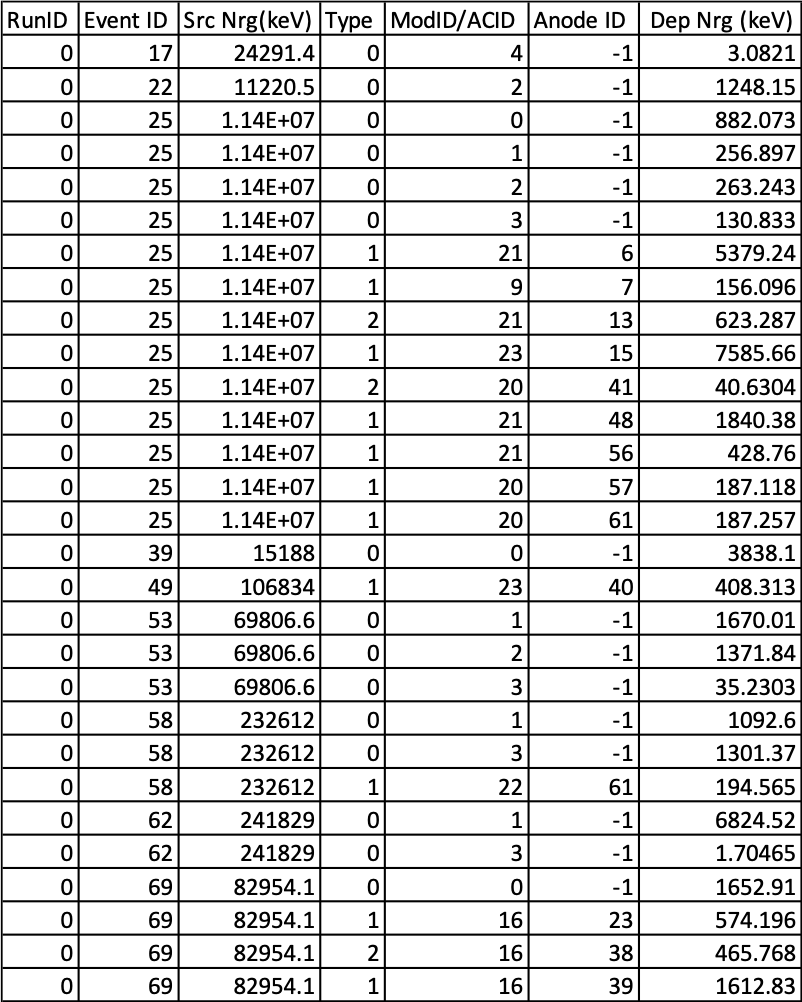}
    	\caption{Example of the output from GEANT4 simulation. This is an example of version 8, which is used for simulation of non-hadronic particles.}
    	\label{fig:sim_csv_file_8}
\end{figure}		

\begin{figure}[hbtp]
 \centering
 	\includegraphics[width=0.95\textwidth]{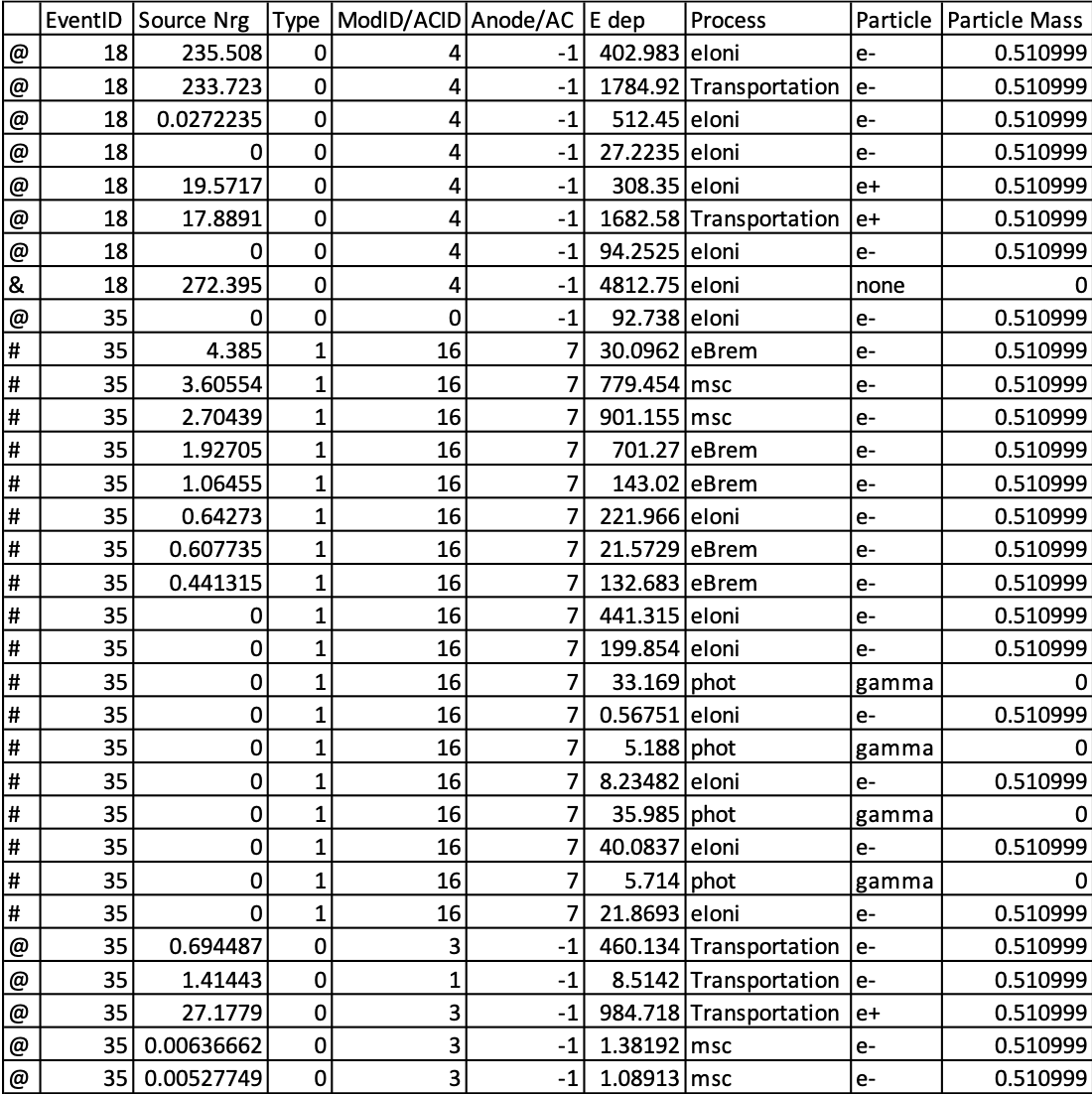}
    	\caption{Example of the output from GEANT4 simulation. This is an example of version 8.1, which is used for simulation of hadronic particles. This has more information than the version 8 which are needed for processing the hadronic processes. }
    	\label{fig:sim_csv_file_8_1}
\end{figure}		


			\subsubsection*{Simulated Raw Output}
			
			Each of the GEANT4 simulations generate a comma separated values (which has a file extension of .csv) output file.  
			An example  of the data in one such file is shown in Figure \ref{fig:sim_csv_file_8}. 
			This is a column file with 7 columns.  
			The columns, from left to right, are run id (run number), event id (Event no.), Src Energy (Energy of the incoming particle), Detector type (0 is for Anti-Coincidence (AC) shields, 1 is for Calorimeter and 2 is for Plastics), the Module/AC ID (the identification number for the module or the AC ), Anode ID (Anode identification number, -1 if this trigger was AC), Dep Nrg (Deposited energy).
			 Each row represents one interaction. 
			 In an event, there could be multiple interactions so for each event processing, all the interactions that belong to the event are processed accordingly.
			 This output was for non-hadronic simulations. 
			 A sample output for the hadronic simulations is shown in Figure \ref{fig:sim_csv_file_8_1}.
			 This file has additional information that is needed to process these hadronic interactions. 
			 The additional information includes the interaction process, particle type and  particle mass. 
			 This additional information is used for converting the electron equivalent energy from heavier ions and protons when necessary. 
			 Section \ref{sec:ins_perf_elec_equi_energy} discusses the electron equivalent process. 
			 The initial run number is not that important, as we have one file per run,  so this column is replaced with various markers that aid in processing of the events.
			 Ideally, this version could have been used for non-hadronic simulations too.
			 The non-hadronic simulations were performed first, and the output was modified for the hadronic simulations so we ended up having two different versions. 
			
			\subsection{Simulated Output Processing}
			\label{sec:ins_perf_sim_out}
			
\begin{figure}[hbtp]
 \centering
 	\includegraphics[width=0.75\textwidth]{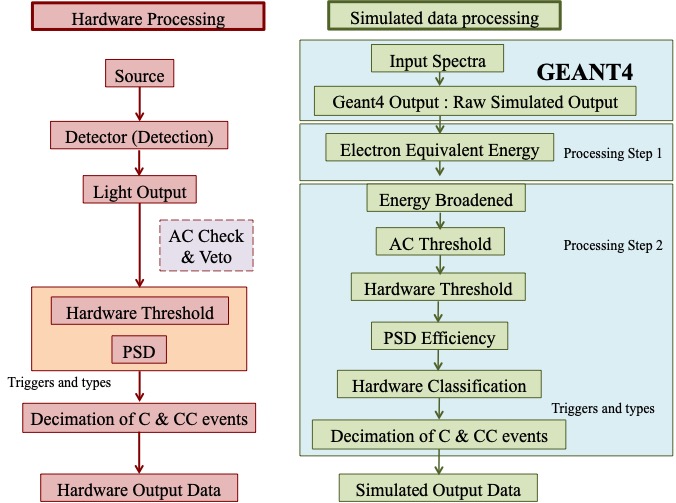}
    	\caption{A flow diagram comparing the key steps in the hardware processing (left) and the simulated data processing (right) designed to replicate the hardware processing. }
    	\label{fig:sim_flow_process}
\end{figure}

			The raw GEANT4 output represents an ideal detector that measures precise energy deposits.
			However, in reality our instrument measures the energy loss with some level of uncertainty (as defined by the instrument response). 
			The raw GEANT4 output must be processed to replicate these uncertainties following the hardware processing of the data.
		 The key processes are shown in Figure \ref{fig:sim_flow_process}. 
			The left side shows the key steps (processes) in the hardware processing and the right side shows the key steps for the simulated data processing.
			
			The hardware processing is shown in the left hardware side of Figure \ref{fig:sim_flow_process}.
			The incoming photon is detected via scintillation in our instrument detector.
			The photon loses its energy in the detector and generates scintillation photons.
			The light output is directly proportional to the energy and the proportionality is measured during calibration.
			Hardware threshold values are set for each of the scintillator elements. 
			These hardware threshold values are determined during calibration and stored in the Programmable Integrated Chip (PIC) processor of each module.
			An anode only creates a trigger signal if the deposited energy is higher than the hardware threshold value set for that anode.
			The triggered anodes also go through the Pulse-Shaped Discrimination (PSD) circuit when applicable. 
			The event is classified into various event classes and event types after these checks.
			For event classes of C and CC, a decimation value of 5 was set which meant that only 1 out of 5 of these events were recorded.
			The decimation value of 5 was only set after the whole instrument was assembled and was not set for single module measurements done in the lab.
			This value of 5 was used in 2011 flight.
			Therefore it was set when the instrument was assembled and the decimation value was 5 during the 2014 flight as well. 
			The Anti-Coincidence (AC) shield data is processed in the Science Data Computer (SDC) and creates a veto flag when there is an AC signal associated with the event.			
			The output after these processing is defined as the hardware output 
			
			The simulated data processing is designed to replicate the key steps of the hardware processing described above.
			These steps are shown in the right side in Figure \ref{fig:sim_flow_process}. 
			For simulated data processing, the AC threshold and rejection of vetoed events is done earlier as it makes the code more efficient. 
			The decimation, event classification and PSD are modeled according to the description in the hardware processing steps.
			Energy broadening and the electron equivalent energy are the two processes that are added to the simulated data processing that are not listed in the hardware processing.
			
			\subsubsection*{Energy Broadening}
			\label{sec:ins_perf_energy_broadening}
			
			The simulation reports the precise absorption of energy. 
			For a mono-energetic photon beam, it would report a delta function.   
			In reality, the detectors generate a Gaussian distribution with some width.
			This phenomena is integrated in our simulation via Gaussian broadening. 
			A Gaussian function is defined by the peak energy E$_\text{0}$ and width $\sigma$. 
			Gaussian broadening determines a $\sigma_{\text{E0}}$ for a given energy E$_\text{0}$ and randomly shifts the energy based on a Gaussian distribution of random numbers. 
			The resolution for the calorimeter is defined for 10\% at 662 keV as 
			\begin{equation}
					\sigma_{Cal} = \frac{0.1}{2.35}\sqrt{662. \cdot E_0}
					\label{eqn:gauss_broad_cal_sig}
			\end{equation}
			where, FWHM = $2 \sqrt{2 \ln 2} \sigma \approx$ 2.35 $\sigma$. 
			For plastic, which has a broader energy peak, we use a wider broadening given by
			\begin{equation}
					\sigma_{Pla} = \sqrt{0.5692 + (2.837 \times E_0)}
					\label{eqn:gauss_broad_pla_sig}
			\end{equation}
			These conversion values were used from \citet{Ertley2014} which used the 2011 calibration data for the calorimeter elements and plastics.
			The type of scintillator elements are still the same  and the 16 polarimeter modules are reused for GRAPE 2014.
			Even the new modules are made up of same scintillator elements therefore this calculation of $\sigma$ is still applicable. 
			The energy broadening $E_{broad}$ is defined as a function $\sigma$ by
			\begin{equation}
					E_{broad} = \text{Random}_G(\sigma,E_0)
					\label{eqn:gauss_broad}
			\end{equation}			
			Here $\text{Random}_G$ is a gaussian distributed random number. 
			\begin{table}[tbp]
\begin{center}
\caption{ $\sigma$ for the plastic and calorimeter element using the Gaussian broadening. E$_0$ is the input energy and the respective grid shows the calorimeter and plastic $\sigma$ as well as the corresponding FWHM calculation.}
\label{table:sim_gauss_broad}
 \begin{tabular}{||c | c| c | c |c||}
  \hline
  &\multicolumn{4}{c||}{ Energies in keV}\\
  \hline
E$_0$  & 50 &100 &200&500 \\ [0.5ex] 
 \hline
 \hline
$\sigma_{Cal}$  &7.7& 10.9 & 15.5& 24.5\\ 
 \hline
 FWHM$_{Cal}$  &18.1& 25.6 & 36.4& 57.5\\ 
 \hline
 \hline
$\sigma_{Pla}$& 11.9 & 16.9& 23.8 & 37.7\\ 
 \hline
  FWHM$_{Pla}$  &28.0& 39.7 & 55.9& 88.5\\ 
 \hline
  \hline
 \end{tabular}
 \end{center}
\end{table}

			Typical values of $\sigma$ from Equation \ref{eqn:gauss_broad_cal_sig} and \ref{eqn:gauss_broad_pla_sig} are shown in table  \ref{table:sim_gauss_broad}. 
						

\begin{figure}[tp]
 \centering
\begin{subfigure}[b]{0.49\textwidth}
 		 \includegraphics[width=1\linewidth]{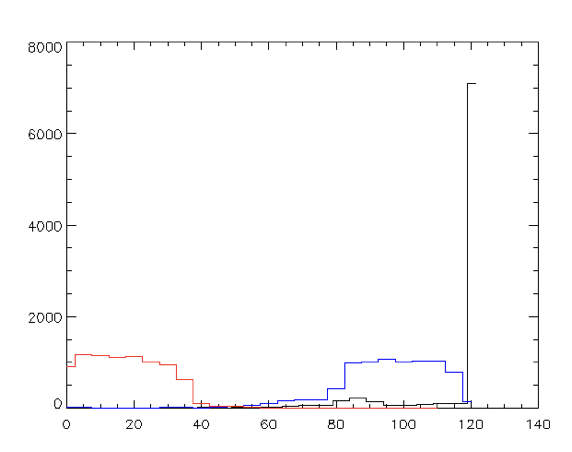}
		 \caption{}
\end{subfigure}
\begin{subfigure}[b]{0.49\textwidth}
 		 \includegraphics[width=1\linewidth]{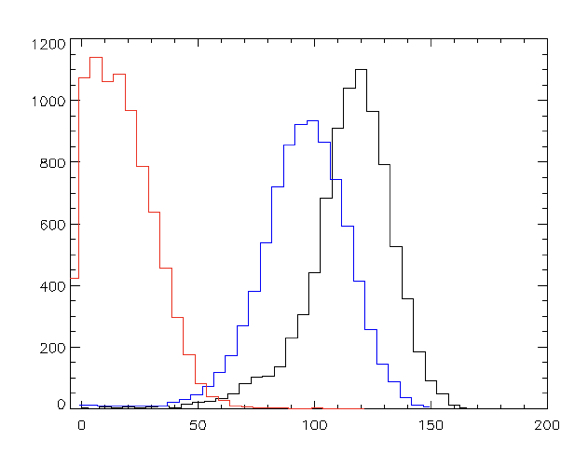}
		 \caption{}
\end{subfigure}

  \caption{An example of energy broadening for a PC event of 120 keV energy. The black is the PC event and the P and C are the plastic and calorimeter that makes the PC event. a) Un-broadened energy loss spectra. We see a clear deposit of, almost all, 120keV. b) Gaussian broadened energy loss spectra replicating our instrument.}
\label{fig:sim_ener_broad}

\end{figure}
			An example of this energy broadening is shown in Figure \ref{fig:sim_ener_broad}. 
			The left Figure \ref{fig:sim_ener_broad}(a) is an un-broadened simulated raw output for a simulation of a 122 keV source.
			This Figure represents an energy loss spectra for a PC data (black).
			The red and blue are the Plastic(P) and the Calorimeter(C) data, respectively.
			The P and C spectra are not delta functions.
			The initial energy of the photon beam is 122 keV and during the PC event, the photon splits the energy between the two elements during the scattering. 
			This split in energy is not predetermined hence the energy is not a delta function. 
			However, the total energy deposited has to equal the initial energy so the total energy in the un-broadened case is almost a delta function centered at 122 keV.

			Figure \ref{fig:sim_ener_broad}(b) represents the corresponding energy broadened spectra. 
			The energy broadening is done to each plastic and calorimeter energy and the broadened energies are added to determine the total energy for that PC event.
			The total number of counts, the area under the graph of each of these plots, are same as the same number of P and C data are needed to form PC data. 
		
			\subsubsection{Electron Equivalent Energy}
			\label{sec:ins_perf_elec_equi_energy}
			
\begin{figure}[tbp]
 \centering
    	\includegraphics[width=.90\textwidth]{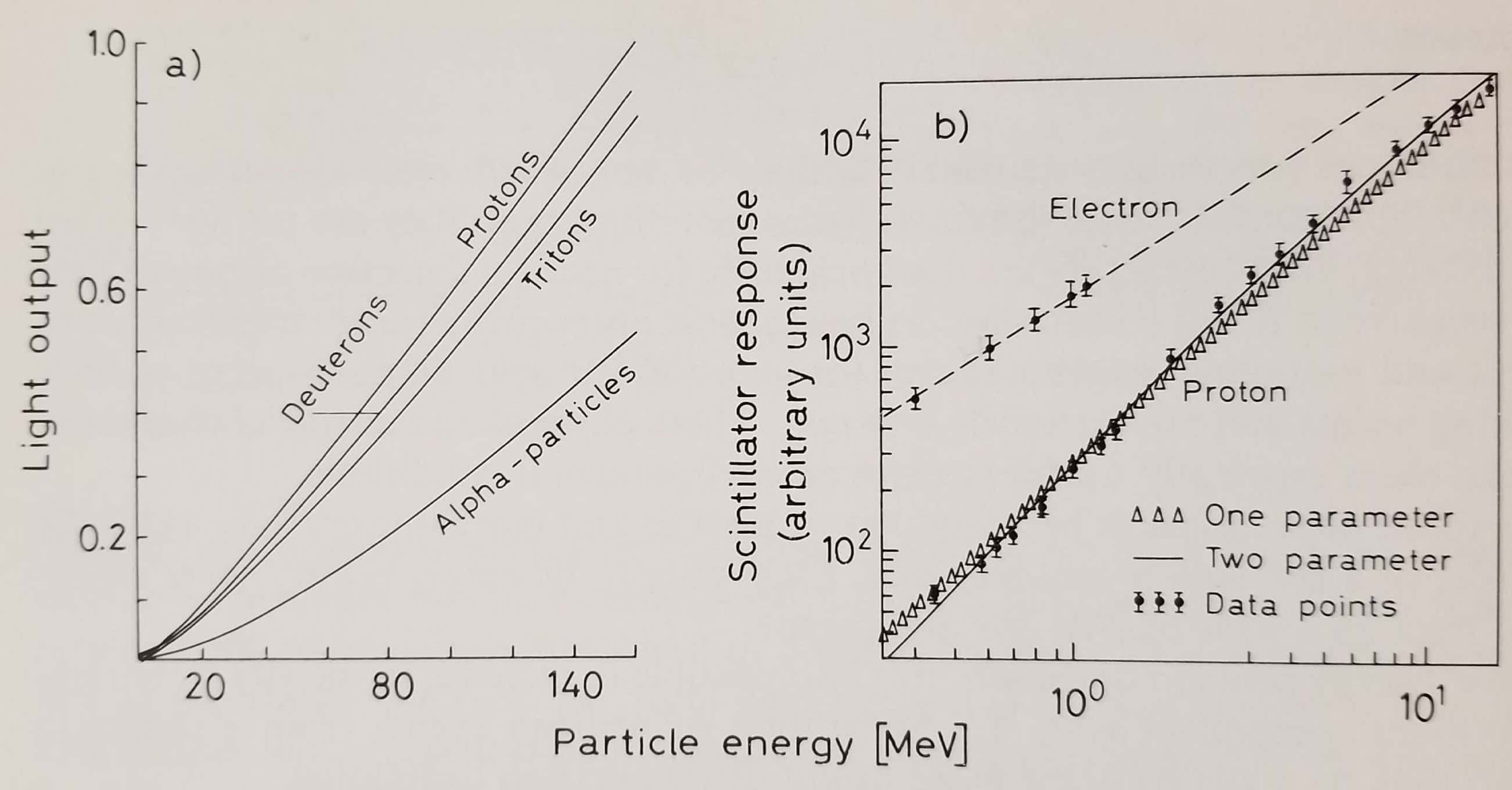}
    	\caption{Response of a plastic scintillator (NE102 similar to our plastic scintillator) to different particles. (a) from \citet{Gooding1960} (b) from \citet{Craun1970} \citep{Leo1994}. Our scintillator elements are calibrated using the electron's light output. The light yield from the heavier particles is read as it is from an electron. This process is replicated in simulations using the response of our plastic scintillators for different particles (taken from spec-sheet \href{https://eljentechnology.com/products/plastic-scintillators}{\textit{\underline{https://eljentechnology.com/products/plastic-scintillators}}}). }
    	\label{fig:sim_ele_eqiv_ener}
\end{figure}

			The incident photon scatters the electrons in the detector and the energy loss of these electrons results in the generation of the scintillation light.
			Only a fraction of the Kinetic Energy (KE) lost by a charged particle is converted to the florescent energy in the organic scintillators. 
			The number of photons (light yield) produced depends on the material of the scintillator and the incident energy. 
			The light yields don't always have a linear dependence, but this is calibrated. 
			The initial energy and this dependence is determined in calibration.
			Our instrument is calibrated for the KE loss of an electron. 
			For a given KE loss, heavier ions produce lower light yields (LY) as compared to the electrons.
			The response of a plastic scintillator to different particles are shown in Figure \ref{fig:sim_ele_eqiv_ener}.
			The instrument is calibrated using photons, so the light output (pulse height) to energy calibration is based on electrons. 
			When the heavier ions generate this lower light yield, $\text{LY}_{\text{ion}}$ for an energy E$_{\text{ion}}$, the calibrated instrument assumes it to be electron and uses this light yield as if it was of the electron (LY$_{\text{ion}}$ = LY$_{\text{ele}})$ and reports the energy that was calibrated for the electron E$_{\text{ele}}$. 
			For heavier particles, the energy must be converted to electron equivalent energy deposit to recreate the response of the instrument.
			The scintillator's response (light output vs energy) to the different particles are retrieved from spec-sheet provided by the manufacturer.

			Similar process is used for the electron equivalent energy process in CsI scintillator (using the relative light output for different particles in the CsI scintillator).
			For version 8 of our GEANT simulation that deals with electrons, positrons and photons, we do not have to worry about the electron equivalent energy. 
			For hadronic simulations in version 8.1, the physics of each of the interactions are recorded and electron equivalent energy conversion is done when necessary.  
		 	Both of these outputs are processed similarly for analysis and comparisons apart from the electron equivalent energy conversion. 
			With inclusion of these processes (energy broadening and electron equivalent energy), our simulated data processing is comparable to the hardware processing.

	\section{Instrument Response}
	\label{sec:ins_perf_instrument_response}		
	
	Our instrument does not directly measure the photon flux vs. energy. 
	Our instrument measures counts as a function of energy loss which is the result of the the source flux interacting with our instrument.
	This interaction of the flux with our instrument can be understood as the instrument response. 
	The instrument response can be viewed as a function that takes in the input flux and outputs what our instrument measures. 
	The instrument's measured counts (C) vs. instrument channel (I) (which is related to energy loss) can be represented as 
	\begin{equation}
			C(I) =  \int f(E) R(I,E) dE	
			\label{eqn:instrument_response_detailed}		
	\end{equation}
 where R(I,E) is the response of the instrument, f(E) is the input flux and E is the energy.
 	Our instrument's measurement include background measurements along with source. 
	The instrument data (D) with background (B), the C(I) can be defined by 
 	\begin{equation}
 				C(I)  = D(I) - a_{factor} B(I)
				\label{eqn:bgd_sub}
	\end{equation}
	here the a$_{\, \text{factor}}$ is the scale factor to normalize the background data to the instrument data. 
	The response R(I,E) is a continuous function of  E and defines the instrument's response to an incoming photon of energy E. 
	Therefore, R(I,E) is proportional to the probability that an incoming photon with E will be detected in channel I.
	This response R(I,E) is retrieved from simulation where the instrument is simulated for incoming photons of energy E.
	To simulate this, the continuous function of R(I,E) is changed to discrete form.
	The discretization of the R(I,E) is represented by 
	\begin{equation}
		R_D (I,J) = \frac{\int^{E_J}_{E_{J-1}} R(I,E) dE}{E_J - E_{J-1}}
		\label{eqn:discrete_response_function}
	\end{equation}
	Here J is the bin number of the discrete response. E$_\text{J}$ is the energy for the J$^{\text{th}}$ bin. 
	The R$_{\text{D}}$(I,J) represents the instrument's response (what energy loss bins in the instrument (I) gets triggered) by the flux of energy E$_{\text{J-1}}$ to E$_{\text{J}}$. 
	The response R$_{\text{D}}$(I,J) has two components, the response matrix R$_{\text{P}}$(I,J) and the effective area A$_{\text{D}}$(J) and they can be represented as 
	\begin{equation}
	R_D(I,J) = R_P(I,J) \times A_D(J)
	\label{eqn:resp_func_resp_matrix_eff_area}
	\end{equation}
	
	\subsection{Response Matrix}
	\label{sec:ins_perf_response_matrix}		
\begin{figure}[!ht]
\centering
\includegraphics[width=0.85\linewidth]{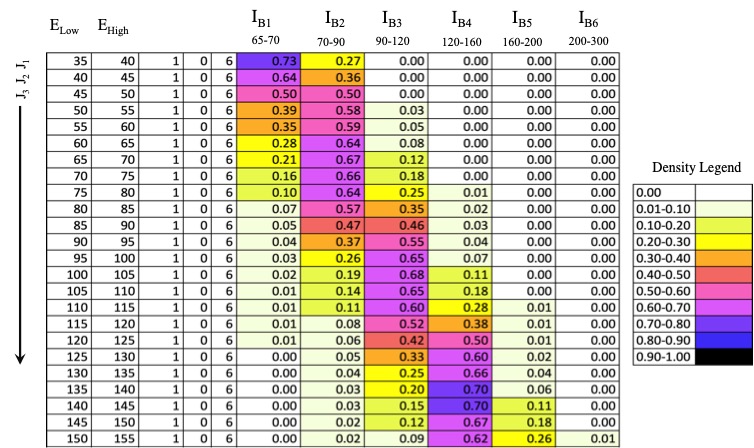}
  \caption{An example of the response matrix file. We see the rows represent the bins of the incident energy J. The first row is the first incident energy bin (J$_1$) which represents 35 keV to 40 keV. The columns labelled I$_{\text{BN}}$ represents our instrument's response bins. The energy range of those instrument bins are labelled right below them are in units of keV. The density values are color coded and the legend for these values are shown on right. The response for an ideal case would be only diagonal, in reality it would be some spread about the diagonal as seen here.}
\label{fig:sim_resp_matrix}
\end{figure}

	The response matrix R$_{\text{P}}$(I,J) is a probability density function that defines the probability for an incoming photon of energy E$_{\text{J}}$ to be detected in the instrument channel I. 
	It is a matrix of  dimension N $\times$ M where N is the number of energy bins (I) in our instrument or in our measured pulse height spectrum and M is the number of bins for the incident energy (J). 
	The R$_{\text{P}}$(I,J) is a probability density function with a normalization of 1 for the J$^{th}$ row .  
	For GRAPE, we run a simulation  of mono-energetic runs from 1 keV to 1000 keV at 1 keV intervals. 
	Each run is processed as discussed in section \ref{sec:ins_perf_sim_out} and a response matrix is generated after collecting the data into various bins.
	The matrix dimensions depend on the analysis.
	Figure \ref{fig:sim_resp_matrix} shows an example of response matrix file used for the deconvolution of energy spectra. 
	We can see the indices J$_{\text{M}}$ (each row), that defines the incident energy bins. 
	Here we have 5 keV bins for the incident photon energy (J). 
	The two column labeled, E$_{\text{Low}}$ and E$_{\text{High}}$ represents the lower and the higher values of the energy bin for the J.
	Each columns labelled I$_{\text{BN}}$ is the N$^{\text{th}}$ bin in our energy loss spectra (generated by our instrument). 
	The energy ranges associated with our instrument binning is labelled right below them. 

	\subsection{Effective Area}
	\label{sec:ins_perf_effective_area}		

\begin{figure}[hbtp]
 \centering
\begin{subfigure}[b]{0.8\textwidth}
 		 \includegraphics[width=1\linewidth]{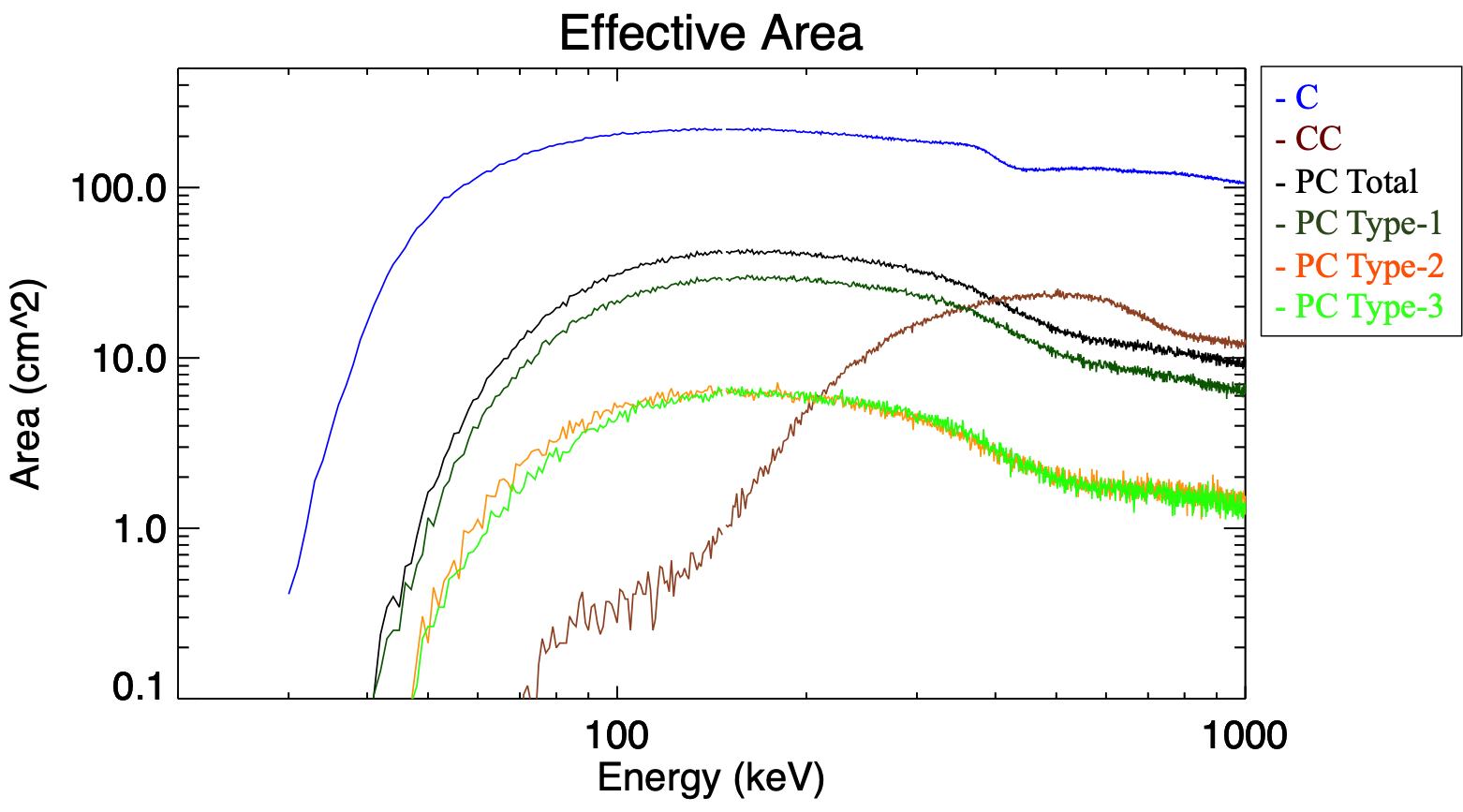}
		 \caption{}
\end{subfigure}   

 \begin{subfigure}[b]{0.8\textwidth}
 		 \includegraphics[width=1\linewidth]{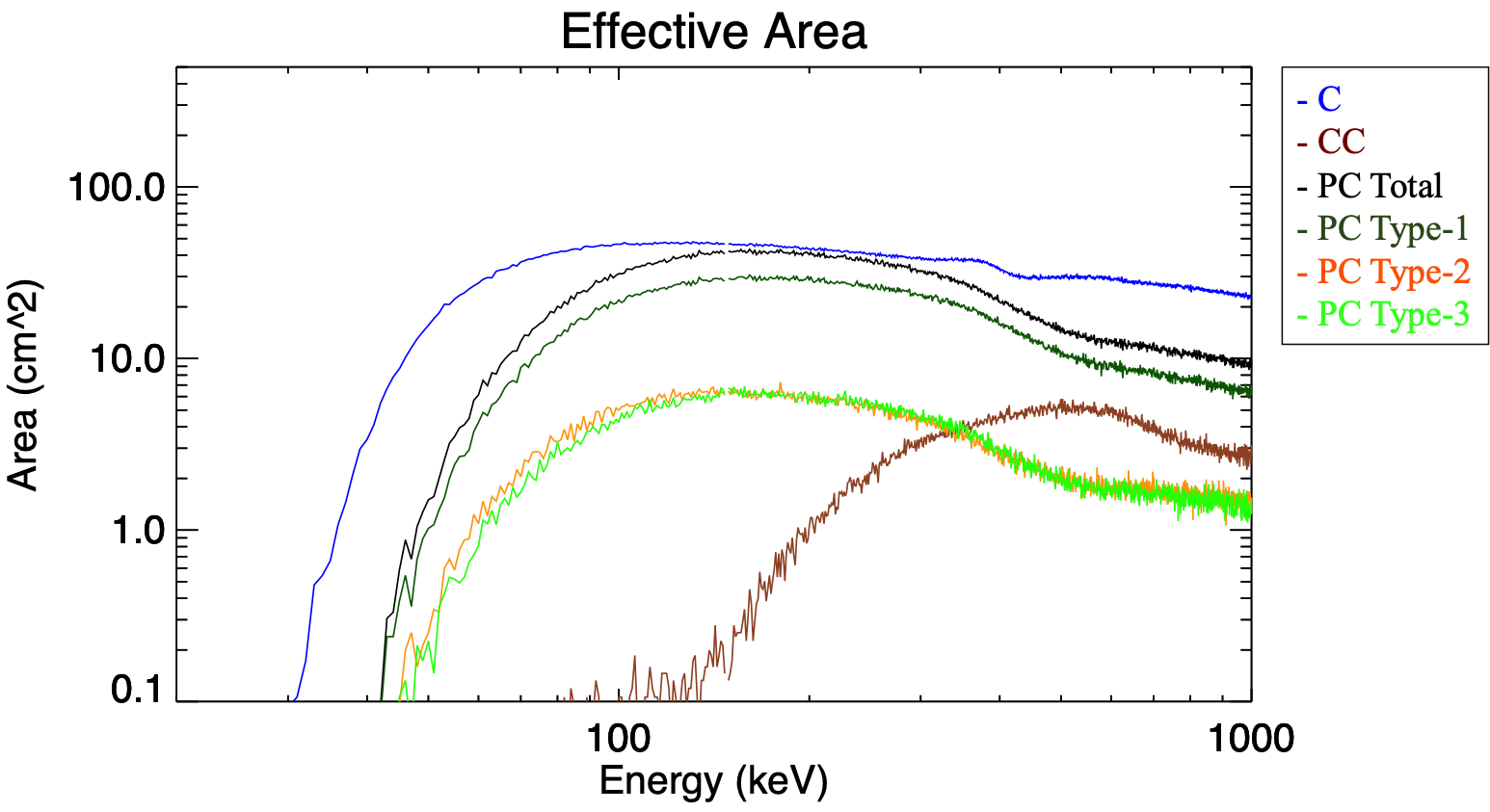}
		 \caption{}
\end{subfigure} 
  \caption{The effective area for different event types. a) This represents effective area without decimation. b) This represents effective area with a decimation value of 5. The C and CC that are affected by decimation are reduced by a factor of 5 but the PC remains intact as it is not affected by the decimation.}
\label{fig:sim_resp_effective_area}
\end{figure}

	The second part of the instrument response is the effective area A$_\text{D}$(J). 
	This can be understood as the sensitive area of our instrument.
	It is defined by the ratio of number of incident versus the detected  counts multiplied by the area of the incidence.
	The mono-energetic runs from 1 keV to 1000 keV was also used to determine the effective area of GRAPE. 
	It is a function of energy, has units of area and has the dimension of $N$ (number of incident energy bins) as in the response matrix.  	
	We decimate C and CC events at our hardware level and is set at a value of 5. 
	Therefore the effective area for C and CC are affected by the decimation but not the PC events. 
	This is shown in Figure \ref{fig:sim_resp_effective_area} a and \ref{fig:sim_resp_effective_area} b where we can see the effective area of the various event types with and without decimation. 
	For our analysis, we only look at the effective area with decimation. 
	

	\section{Background Simulation}
	\label{sec:ins_perf_bgd_sim}
	
	Background simulation refers to simulation of the instrument background at flight altitudes. 
	The instrument background comes from the radiation environment to which the payload is exposed to at these altitudes. 
	The radiation environment consists of several components.
	These include the relevant radiation components consist of photons (gamma rays), and various sub atomic particles (protons, electron, positrons and neutrons). 
	These components can come directly from deep space and are defined as cosmic rays (primary component).
	Some of these components are generated by this cosmic ray interactions in the atmosphere which are defined as atmospheric components (secondary components). 
	The parameterized spectra of each components is retrieved from the literature and used to create the particle description file for our simulations.  
	The relevant components and their parameterized definitions retrieved from the literature are outlined here. 

			\subsection{Background components}

			\subsubsection{Gammas}
			\label{sec:ins_perf_inputspec_gamma}
			
			The parameterized definition of gamma component of the background was retrieved from \citet{Gehrels1985}.
			This definition includes both the cosmic diffuse (primary) and the atmospheric gammas (secondary). 
			Gehrels used the data from balloons flown from Palestine, Texas at an atmospheric depth of 3.5 g/cm$^2$ ($\sim$ 39 km in altitude)  to define the gamma ray flux.
			The combined total gamma ray flux (cosmic and the atmospheric component) and corrected it  to the depth of 3.5 g/cm$^2$.
			The power-law forms were applied to 0.02 - 10 MeV energy range and the total gamma ray flux were defined for the four zenith angle ranges as shown in table \ref{table:sim_desc_gamma_flux}.
			 \begin{table}[h]
 \begin{center}
 \caption{ The gamma ray flux definitions used for simulations retrieved from \citet{Gehrels1985}. The total flux refers to the total gamma ray flux incident on the instrument.}
\label{table:sim_desc_gamma_flux}
\begin{tabular}{c c c}
\textbf{Zenith Angle} & \textbf{Form} & \textbf{Total flux}\\
0$^\circ$  - 65$^\circ$  & 0.052  E$^{-1.81}$  ph/s/cm$^2$/sr/MeV & 15200   ph/s/m$^2$/sr \\
65$^\circ$ - 95$^\circ$  & 0.085  E$^{-1.66}$  ph/s/cm$^2$/sr/MeV & 16800   ph/s/m$^2$/sr \\
95$^\circ$  - 130$^\circ$  & 0.14  E$^{-1.50}$  ph/s/cm$^2$/sr/MeV & 18900   ph/s/m$^2$/sr \\
130$^\circ$ - 180$^\circ$  & 0.047  E$^{-1.45}$  ph/s/cm$^2$/sr/MeV & 5710   ph/s/m$^2$/sr \\
\end{tabular}\\

\end{center}
\end{table}
			
			These input spectra from the are shown in Figure \ref{fig:sim_bgd_input_flux_desc} in red.
			A separate simulation is conducted for each of these angular intervals. 
			These definitions are used to generate four separate input flux file (text file).
			The simulated particles radiate from the surface of a sphere that encloses the instrument as shown in Figure \ref{fig:sim_src_desc_1}(d).
			Each of the four angular intervals for gammas would be used to generate photons from the surface of the sphere over the corresponding angle range. 
			The output from these four simulations are normalized accordingly and summed to get a total gamma ray energy loss spectra (section \ref{sec:ins_perf_sim_normalization}).
			The normalized total gamma ray energy loss spectra is shown in Figure \ref{fig:sim_bgd_output_noct}.

			\subsubsection{Primary Charged Particles}
			\label{sec:ins_perf_inputspec_primaries}
			
\begin{figure}[hbtp]
\centering
\includegraphics[width=0.70\linewidth]{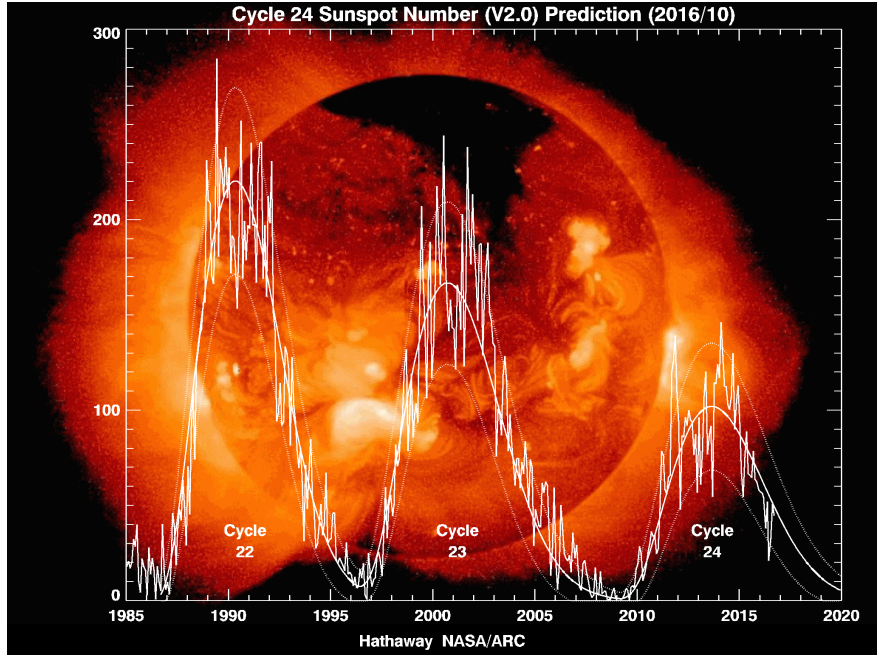}
  \caption{Figure showing number of sunspots corresponding to solar cycle. 2014 was solar maximum so the corresponding value  of 1200 MV was used. \href{http://solarscience.msfc.nasa.gov/predict.shtml}{\textit{\underline{http://solarscience.msfc.nasa.gov/predict.shtml}}} }
\label{fig:sim_bgd_input_flux_sol_mod}

\end{figure}

			The primary particles refer to galactic cosmic rays that originate outside our solar system. 
			Their spectra are known to be affected by solar activity and Earth's geomagnetic field.
			The spectra of these primary particles are typically modeled by a power law in rigidity defined in interstellar space, unaffected by solar modulation by
			\begin{equation}
					\text{Unmod}(E_k) = A \bigg[ \frac{R(E_k)}{ GV }\bigg]^{-\alpha}
					\label{eqn:ins_perf_pri_un_mod}
			\end{equation}
			where $E_k$ and $R$ are the kinetic energy and rigidity of the particles, respectively \citep{Mizuno2004}.
			The $\alpha$ and $A$ are the spectral index and normalization values respectively.  
			The definitions from  \citet{Gleeson1968} are used for the solar modulation which modifies the  Equation \ref{eqn:ins_perf_pri_un_mod} as			
 			\begin{equation}
					Mod(E_k) =  Unmod(E_k+Ze\phi) \times \frac{(E_k + Mc^2)^2 - (Mc^2)^2}{(E_k + Mc^2 + Ze\phi)^2 - (Mc^2)^2}
					\label{eq:ins_perf_pri_mod}
			\end{equation}
 			where $e$ is the magnitude of the charge, $Z$ is the atomic number, $M$ is the particle mass and c is the speed of light. 
			Since we are only interested in positrons, electrons and protons for our experiment, $Z$ and $e$ have the value 1. 
			The parameter $\phi$ is the solar modulation that depends on the solar activity and varies from $\sim$400 MV (solar  minimum) to $\sim$1200 MV (solar maximum). 
			The solar activity is shown in Figure \ref{fig:sim_bgd_input_flux_sol_mod} which plots the number of sun spots per year. 
			The number of sunspots can be used to determine the solar modulation during the time of the flight.
			

			The influence of the Earth's geomagnetic field was modeled using the Alpha Magnetic Spectrometer (AMS) measurements done by \citet{Alcaraz2000a}. Using this data, \citet{Mizuno2004} found the reduction factor to be
		
		 	\begin{equation}
					\frac {1}{ 1+ (R / R_{cut})^{-r} }
					\label{eq:pri_rfac}
		 	\end{equation}	
			where R$_{cut}$ is the cut off rigidity and calculated via the following Equation assuming the Earth's magnetic field to be dipole and is defined by \citep{Alcaraz2000a} 
		 	\begin{equation}
					R_{cut} = 14.9 \times \bigg(1 + \frac{h}{R_\text{Earth}}\bigg)^{-2.0} ( \cos \theta_M)^4 \, GV
					\label{eq:rcut}
			\end{equation}	
			where  $\theta_M $ is the geomagnetic latitude and h is the altitude. 
			Combining the above definitions, the general form of primary cosmic ray spectrum can be defined as		
			\begin{equation}
			 \text{Pri}(E_{k}) = A \times \bigg[ \frac{E_k + Ze\phi}{ GV }\bigg]^{-a} \times \frac{(E_k + Mc^2)^2 - (Mc^2)^2}{(E_k + Mc^2 + Ze\phi)^2 - (Mc^2)^2} \times \frac {1}{ 1+ (R / R_{cut})^{-r} }
			\label{eq:ins_perf_pri_main} 
			\end{equation}
		     One thing to note is that this function will create a kink around the geomagnetic cutoff.         
		     For 2014 GRAPE, we use h $\approx$ 39  km from our flight plan (with knowledge of GRAPE 2011 flight) and $\theta_M$ = 43.53$^\circ$N for Fort Sumner, NM. 
			This results in $R_{cut} =$ 4211 MV. 
			The solar activity was at the highest cycle, from Figure \ref{fig:sim_bgd_input_flux_sol_mod}, so $\phi \approx $ 1200 MV and the value of $ze\phi \approx $1200 MeV. 
			There are three primary particles to simulate and they are protons, electrons and positrons. 
			Primary protons, electrons and positrons and we need to determine the normalization value of A and the power law index $\alpha$ from literature to define these components.

			\subsubsection*{Cosmic Protons (Primary)}
			\label{sec:ins_perf_inputspec_pri_prot}
			\addcontentsline{toc}{subsubsection}{\hspace{2.5cm} Cosmic Proton}
			
			The Alpha Magnetic Spectrometer (AMS) and Balloon-borne Experiment with Superconducting Spectrometer (BESS) measurements were used to get an accurate spectrum of primary protons \citep{Alcaraz2000c,Sanuki2000}. 
			BESS measured the fluxes at $\approx $ 37 km. 
			AMS measured flux at energies below and above the geomagnetic cutoff. 
			The experimental data from these experiments are fitted for proton to get a normalization of A= 23.9 counts s$^{-1}$  m$^{-2}$ sr$^{-1}$ MeV$^{-1}$, and $\alpha$ = 2.83. 
			The flux of the primary protons at balloon altitude (far above the Pfotzer maximum) suffers some attenuation due to their interaction with the air it passes through. 
			The nuclear interaction length in air is 90.0 g /cm$^2$. 
			For GRAPE 2014, an altitude of 39 km which corresponds to 3.5 g cm$^2$, results in 96.1\%  probability that the flux will reach our instrument. 
			This method was used to scale the normalization constant of the cosmic ray proton spectrum to our altitude. 
			Additionally, the particles at oblique angle would have to travel through more of the atmosphere. 
			Therefore, a factor of $1/\cos\, \theta_Z$ (where $\theta_Z$ is the zenith angle) was used to scale the depth to get the effective atmospheric depth. 
			This scaling continues until $ \cos \theta_Z=$ 0.2, which is at $\theta_Z \approx$ 78$^\circ$.
			The scaling of $ \cos \theta_Z=$ 0.2 is used from 78$^\circ$ till 90$^\circ$.
			Below the horizon, it is considered to be $0$. 
			The resulting proton input flux is shown in Figure \ref{fig:sim_bgd_input_flux_desc} in solid blue. 
			The output from the simulation is normalized and shown in \ref{fig:sim_bgd_output_noct}.            
			
			\subsubsection*{Cosmic Electron and Positron (Primaries)}
			\label{sec:ins_perf_inputspec_pri_elec}
			\addcontentsline{toc}{subsubsection}{\hspace{2.5cm} Cosmic Electron and Positron}
			
			A combined spectrum of primary electron and protons spectra was modeled by \citet{Mizuno2004} using a compilation of data by \citet{Webber1983}.  This model has normalization A= 0.7 c s$^{-1}$ m$^{-2}$ sr$^{-1}$ MeV$^{-1}$ and a power law index $\alpha$ of 3.3 which can be represented as
			\begin{equation}
				\text{Unmod} (E_k) = 0.7 \bigg[  \frac{R(E_k)}{GV}\bigg] ^{-3.3}  \text{c s}^{-1} \text{m}^{-2} \text{sr}^{-1} \text{MeV}^{-1}
			\label{eq:ins_perf_pri_elec_posi} 
			\end{equation}

		    This data referred to a combination of positrons and electrons \citep{Longair2011,Webber1983}. 
			The primary electrons and positrons have the same spectral index $\alpha$ but have a different normalization constant A. 
		    There has been several experiments that measured the positively charged fraction, $e^+ / (e^+ + e^-)$. 
		    \citet{Golden1994} measured this to be 0.078 $\pm$ 0.016 between 5 and 50 GeV which is used to get the constants for electron as A$_{e^-}$ = 0.65 and positron as A$_{e^+}$ = 0.055. 
		    
		    The attenuation for the electron and positron particles has to take into account the energy loss due to ionization and bremsstrahlung in the atmosphere. 
		    The radiation length for these particles is 36.6 g cm$^{-2}$ so the transmission at 3.5  g cm$^{-2}$ is $e^{-\frac{3.5\  g\ cm^{-2}}{3.6.6\  g\ cm{^{-2}}} }= 0.91$ \citep{Mizuno2004}. 
		    This value was used for the attenuation of  these particles at the balloon altitude. 
		    
		    The oblique angles have the same effects and forms as that of proton flux and it scales with $1/\cos \theta_Z$.
		    We can see this input flux on Figure \ref{fig:sim_bgd_input_flux_desc} and is presented in solid green and magenta.
		    The output of this simulation is normalized and presented in Figure \ref{fig:sim_bgd_output_noct} with rest of the components.            
		    
			\subsubsection{Secondary Charged Particles}
			\label{sec:ins_perf_inputspec_secondaries}
		The high energy primary particles (cosmic particles) interact within the atmosphere to produce many secondary particles (atmospheric particles).
		The primary electron and positron can undergo an electromagnetic shower to create atmospheric secondaries. 
		The electrons and positrons can radiate photons via bremsstrahlung radiation.
		The resulting bremsstrahlung photons can undergo the pair-production again if the photons have high enough energy.  
		This avalanche of gamma rays and electron-positron pairs is referred to as an electromagnetic shower. 
		The energy is halved after every iteration so after two iteration of this process the photons will have energy of $E_0 /4$ and $E_0/8$ after third iteration.
		The cascade effect is higher as the particles move deeper into the atmosphere.
		At an altitude of $\sim$ 70 kft ($\sim$20 km), the flux of atmospheric secondaries reaches a maximum (the Pfotzer maximum).
		
\begin{figure}[!t]
 \centering
    	\includegraphics[width=0.7\textwidth]{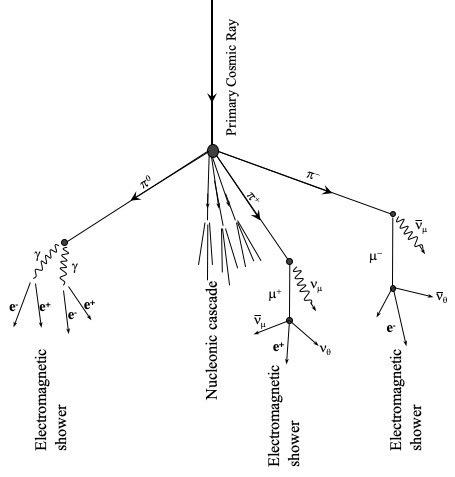}
    	\caption{Schematic diagram showing various process for creation of secondaries in the atmosphere due to primary components \citep{Longair2011}}
    	\label{fig:sim_bgd_input_flux_cascade}
\end{figure}

		A high energy cosmic-ray proton can interact strongly with the nucleus of an atmospheric particle to primarily generate $\pi^+, \pi^-, \pi^0$.
		This shown in Figure (\ref{fig:sim_bgd_input_flux_cascade}).
		In the aftermath of this interaction, the nucleus is left in an highly excited state. 
		It can lose energy through nuclear evaporation which produces neutrons. 
		The secondary nucleons and the charged pions which have sufficient energy will continue to multiply through this fashion until the energy per nucleon drops below that is required of pion production, that is about 1GeV. 
		This is called a nucleonic cascade.
		The excited nucleus can further evaporate with more spallation fragments release $\gamma$ rays in the subsequent de-excitation of the nuclei to its ground state. 
		Each of these secondary particles are capable of creating another collision inside the same nucleus which can create a mini-nucleon cascade.
		The pions generated can also further lead to electromagnetic shower. 
							
			\subsubsection*{Atmospheric Proton (Secondary)}
			\label{sec:ins_perf_inputspec_atm_prot}
			\addcontentsline{toc}{subsubsection}{\hspace{2.5cm} Atmospheric Proton}
			
			\citet{Mizuno2004} modeled the secondary proton spectra at balloon altitude using the Gamma-Ray Large Area Space Telescope (GLAST) Balloon Flight Engineer Model (BFEM) data at R$_{cut}\sim$ 4.5 GV. 
			The atmospheric protons are separated into the downward and upward components. 
			At energies below our R$_{cut}$ the downward atmospheric proton flux is defined by 
			\begin{equation}
			0.17 \bigg(\frac{E_k}{100\ MeV}\bigg)^{-1.0} \text{counts s}^{-1} \text{m}^{-2} \text{MeV}^{-1}
			\label{eq:ins_perf_sec_proton_down_1} 
			\end{equation}
			and above the geomagnetic cutoff $R_{cut}$ this flux becomes a power-law function
			\begin{equation}
			0.236 \bigg(\frac{E_k}{GeV}\bigg)^{-2.83} \text{counts s}^{-1} \text{m}^{-2} \text{MeV}^{-1}
			\label{eq:ins_perf_sec_proton_down_2} 
			\end{equation}
			
			The atmospheric depth of GLAST BFEM was 3.8 g cm$^{-2}$. The downward flux is known to be proportional to the atmospheric depth \citep{Verma1967,Abe2003}.  Therefore, the downward flux is scaled appropriately to 3.5 g cm$^{-2}$ for GRAPE using the simple ratio $\frac{3.5}{3.8}$. 
			The upward proton flux does not change very much with depth \citep{Verma1967} so it is used as it is.
			The upward proton flux is modeled as 
			\begin{equation}
			0.17 \bigg(\frac{E_k}{100\ MeV}\bigg)^{-1.0} \text{counts s}^{-1}\text{m}^{-2} \text{MeV}^{-1}
			\label{eq:ins_perf_sec_proton_up_1} 
			\end{equation}
			below 100 MeV and above the 100 MeV it is modeled as
			\begin{equation}
			0.17 \bigg(\frac{E_k}{100\ MeV}\bigg)^{-1.6} \text{counts s}^{-1} \text{m}^{-2} \text{MeV}^{-1}
			\label{eq:ins_perf_sec_proton_up_2} 
			\end{equation}
			
			For the downward fluxes, to include the angular dependence, the flux is multiplied by $1/\cos\theta$ ($\theta =$ zenith angle) till $\theta=60^\circ$. 
			And  a factor of 2 from $ 60^\circ \leq \theta \leq 90^\circ$.
			For the upward fluxes, the zenith angle is substituted to nadir angle. 
			The $1/\cos\theta$ factor is continuous so for our simulation, we chose 6 angular bins  of $10^\circ$ bins to $60^\circ$ and the seventh bin to be from $60^\circ - 90^\circ$.
			We simulate each of these separately so there are 7 separate simulations for upward secondary protons and 7 more for downward secondary protons. 
			For each of the 7 simulations for upward and downward components, the input spectra is the same but they differ in the angular description.
			The input spectra of the secondary upward proton is shown in Figure  \ref{fig:sim_bgd_input_flux_desc}  in dashed blue and the secondary downward proton is shown in Figure \ref{fig:sim_bgd_input_flux_desc} in dashed cyan. 
			The output from the simulations are collected, normalized and added to get the atmospheric proton spectra and is shown in Figure  \ref{fig:sim_bgd_output_noct}.
			
			\subsubsection*{Atmospheric Electrons and Positrons (Secondaries)}
			\label{sec:ins_perf_inputspec_atm_elec_posi}
			\addcontentsline{toc}{subsubsection}{\hspace{2.5cm} Atmospheric Electron and Positron}		
		 
\begin{figure}[!t]
 \centering
    	\includegraphics[width=0.9\textwidth]{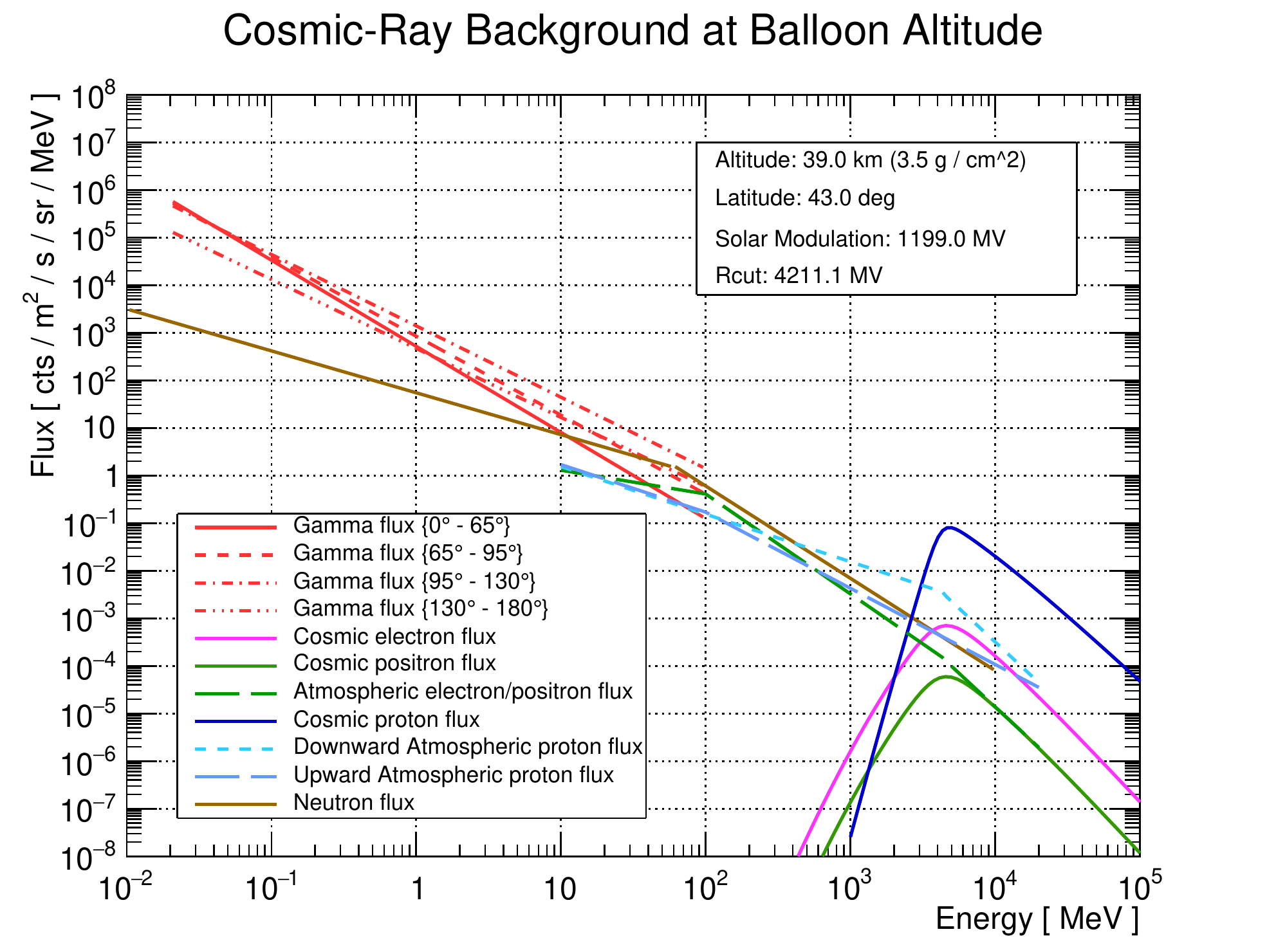}
    	\caption{Input flux description of the components that contribute to the background. The definitions are derived from literature and adjusted to our flight parameters. These input fluxes are used in the GEANT4 simulations of these components.}
    	\label{fig:sim_bgd_input_flux_desc}
\end{figure}

			 \citet{Barwick1998} used a balloon borne experiment, the High-Energy Antimatter Telescope (HEAT) also flown from Ft. Sumner, to measure atmospheric positrons and electrons.
			 He showed that the positron fraction is nearly $50\%$ at R$_{cut}$ = 4.5 GV. 
			 Data from \citet{Alcaraz2000c} showed that above this geomagnetic latitude, the spectral shapes are almost identical.        
			 Therefore for simplicity we used the same spectral definition for both atmospheric electrons and positrons.
			 For these atmospheric electrons and positrons, we have upward and downward components. 
			 Measurements from \citet{Duvernois2001} found that the upward fluxes of these components are similar to that of the downward fluxes. 
			 Hence the same spectral shape is used for upward and downward fluxes for both the atmospheric electrons and positrons.
			 Mizuno et. al. compiled these results to model these fluxes as 
			 \begin{equation}
			0.41\bigg(\frac{E_k}{100\ MeV}\bigg)^{-0.5} \text{counts s}^{-1} \text{m}^{-2} \text{MeV}^{-1}
			\label{eq:ins_perf_atm_elec_1} 
			\end{equation}
			for 10 MeV- 100 MeV and
			 \begin{equation} 
				0.41\bigg(\frac{E_k}{100\ MeV}\bigg)^{-0.21} \text{counts s}^{-1} \text{m}^{-2} \text{MeV}^{-1}
			\label{eq:ins_perf_atm_elec_2} 
			\end{equation}
			for 100 MeV to our R$_{cut}$ = 4.2GV and from R$_{cut}$ = 4.2GV to 10 GeV as 
			 \begin{equation} 
				6.289\bigg(\frac{E_k}{100\ MeV}\bigg)^{-2.83} \text{counts s}^{-1} \text{m}^{-2} \text{MeV}^{-1}
			\label{eq:ins_perf_atm_elec_3} 
			\end{equation}
			These fluxes are shown in Figure \ref{fig:sim_bgd_input_flux_desc} in dashed green. 
			The same angular distribution as that of the atmospheric protons is assumed due to lack of experimental data at large zenith angle.  		
			The output from the simulation of these components are normalized and added to generate the atmospheric positron flux and electron flux and shown in Figure \ref{fig:sim_bgd_output_noct}.
			
			\subsubsection{Neutrons}
			\label{sec:ins_perf_inputspec_neutrons}
			
			The secondary neutrons are created by the interaction of charged primary particles in the atmosphere  as described in the section \ref{sec:ins_perf_inputspec_secondaries} \citep{Simpson1951}. \citet{Armstrong1973} describes the neutron spectra at different atmospheric depths at geomagnetic latitude of 42$^\circ$ N. The spectra is a power-law described as        
			\begin{equation}
				f_i(E) =A_iE^{\alpha_i}
			\end{equation}
		where A$_i$ is in counts s$^{-1}$ cm$^{-2}$ keV$^{-1}$. This description is inferred from \citet{Armstrong1973}and \citet{Simpson1951} and divided into three energy range as shown in table \ref{table:sim_input_flux_neutron} 
			\begin{table}[hbtp]
\begin{center}
\caption{Neutron description in the three energy range as inferred from \citet{Armstrong1973}.}
\label{table:sim_input_flux_neutron}
 \begin{tabular}{||c | c |c||}
  \hline
Energy Range   & $A_i$ &$\alpha_i $\\ [0.5ex] 
 \hline
 0.01 $eV$ $-$ 0.1 $eV$ & 7.96E+04  & 1.0\\ 
 \hline
 0.1 $eV$ $-$ 60  $MeV$ & 2.4E-03  & -0.88\\
  \hline
  60 $MeV$ $-$ 10  $GeV$ & 3.023E+02  & -1.94\\
  \hline
 \end{tabular}
\end{center}
\end{table}

			For GRAPE, the first energy bin is not relevant as the events generated are lower than the GRAPE threshold. 
			The other two neutron flux is shown in Figure \ref{fig:sim_bgd_input_flux_desc} as a solid brown line. 
			The simulated neutron background is shown in Figure \ref{fig:sim_bgd_output_noct}. 
			
			\subsubsection{Normalization}
			\label{sec:ins_perf_sim_normalization}
			The absolute normalization relates the simulations to the real world.
			This relation can be shown by the equation
			\begin{equation}
					\text{X}_{r} = \text{X}_{s} \cdot \frac{N_r}{N_s}
					\label{eqn:ins_perf_sim_norm}
			 \end{equation}
			 where X$_{r}$ and X$_{s}$ represents the real and simulated results, respectively. N$_r$ is equivalent to the expected number of real events and N$_s$ represents the number of simulated events. 
			The N$_r$ is in units of events s$^{-1}$ and N$_s$ is in units of events. 
			The X$_{s}$ is in units of counts so the real events are in units of counts s$^{-1}$.
			The  X$_{s}$ and N$_s$ are directly retrieved from simulation. 
			The N$_r$ is calculated from the input flux descriptions. 
			
\begin{figure}[!t]
\centering
\includegraphics[width=1\linewidth]{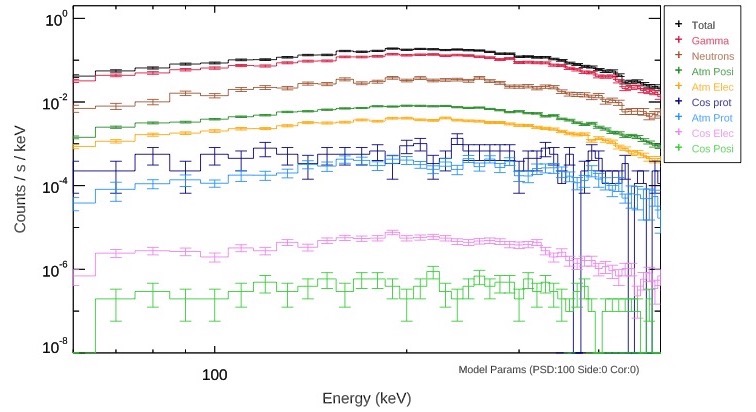}
  \caption{Plot showing the output of the simulation for various components. Each components have been normalized and summed when necessary and presented here.  The summation of all the components are shown in black. The most important components, that contribute the most are the gammas (red), neutrons (brown) and then the atmospheric positron (dark green).}
\label{fig:sim_bgd_output_noct}
\end{figure}

			The normalized simulated components are shown in Figure \ref{fig:sim_bgd_output_noct}.
			Each simulation is normalized separately. 
			The normalized components are summed together and represented in the plot. 
			For example, the red gamma represents the four gamma simulations normalized and summed together. 
			The errors from the simulated output treated as Poisson statistics. 
			The error is propagated accordingly when the components are normalized and added together. 
			The plot also shows the summation of all the normalized simulated components in black and is labelled as 'Total'. 

		\section{Polarization Measurements}
		\label{sec:ins_perf_polarization_measurements}

		A polarization measurement for a partially polarized beam is used to show our instrument's ability do polarimetry.	 
		As discussed in chapter \ref{sec:polarimetry}, a polarized source was created by Compton scattering photons from an unpolarized source. 
		A plastic block is used to Compton scatter photons from a sealed radioactive source. 
		As an example, $^{57}$Co emits unpolarized photons with an energy of 122 keV.
		Scattering those photons at 90$^\circ$ results in a beam of partially polarized photons at $\sim$98 keV.
		The $\sim$ 98 keV beam is expected to be $\sim$95\% linearly polarized (Figure \ref{fig:sci_comp_kn_lineardeg_unpol}). 
		
\begin{figure}[hbtp]
\centering 
\includegraphics[width=0.6\textwidth]{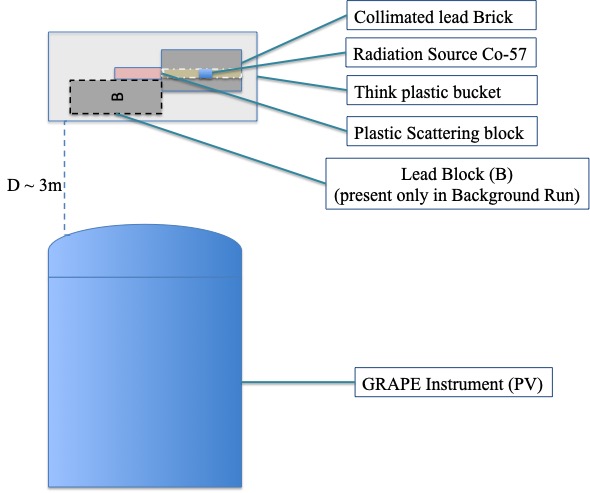}
\caption{Drawing of the setup to create a linearly polarized gamma source. a) This figure displays the generalized setup. The lead block labelled 'B' is removed for the source run and only included in the background run. The source is placed in the collimated lead block and scattered off of the plastic. The setup is $\sim$ 3m away from the GRAPE PV. }
\label{fig:sim_pol_setup}       
\end{figure}

\begin{figure}[hbtp]
 \centering
 	\includegraphics[width=0.6\textwidth]{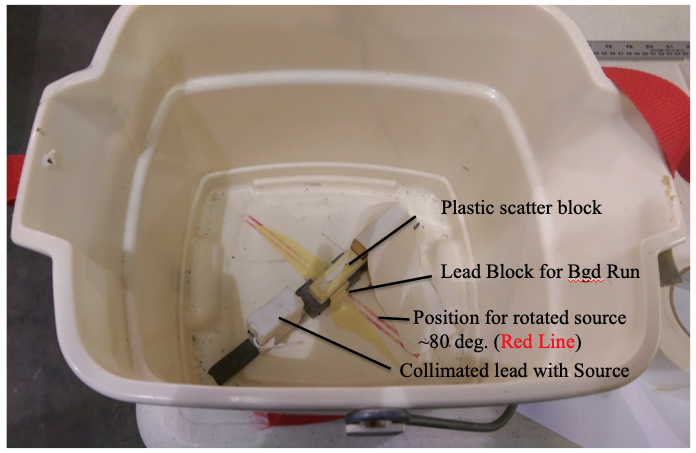}
    	\caption{Picture of the setup with the bucket that was used to create the polarized source for our UNH and FTS runs. This bucket was then raised using a crane pulley and suspended during various runs. This shows the setup for the polarized background run. For the polarized source run, we remove the lead block at the bottom and do the run. There are few wooden blocks and lead blocks used to level the collimated source and scattering plastic block. Any feature arising from these are also present in the background run which gets eliminated when we subtract the background. }
    	\label{fig:sim_pol_fts_setup}
\end{figure}

\begin{figure}[htbp]
 \centering
\begin{subfigure}[b]{0.40\textwidth}
 		 \includegraphics[width=1\linewidth]{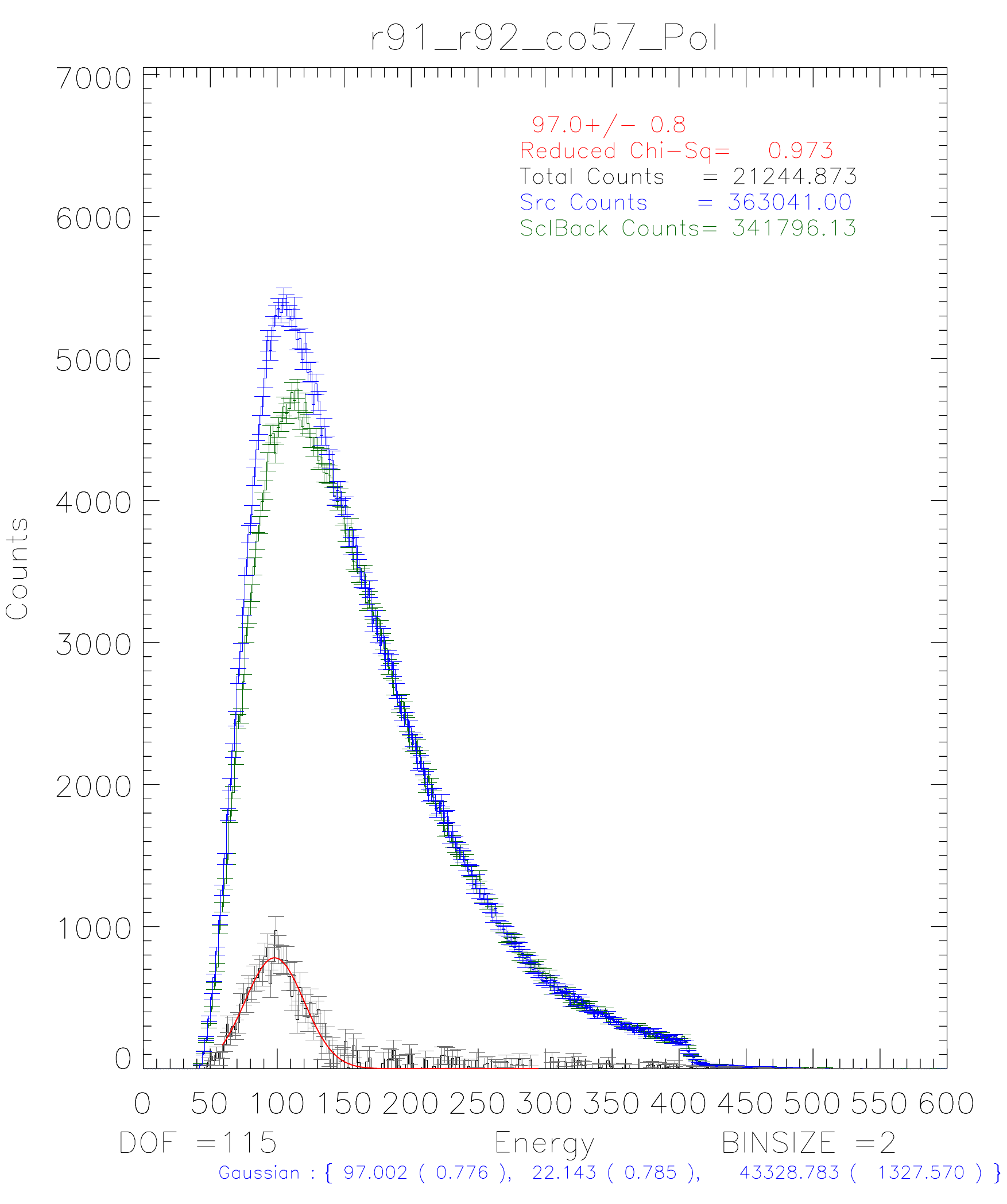}
		 \caption{}
\end{subfigure} \; \; \; 
\begin{subfigure}[b]{0.41\textwidth}
 		 \includegraphics[width=1\linewidth]{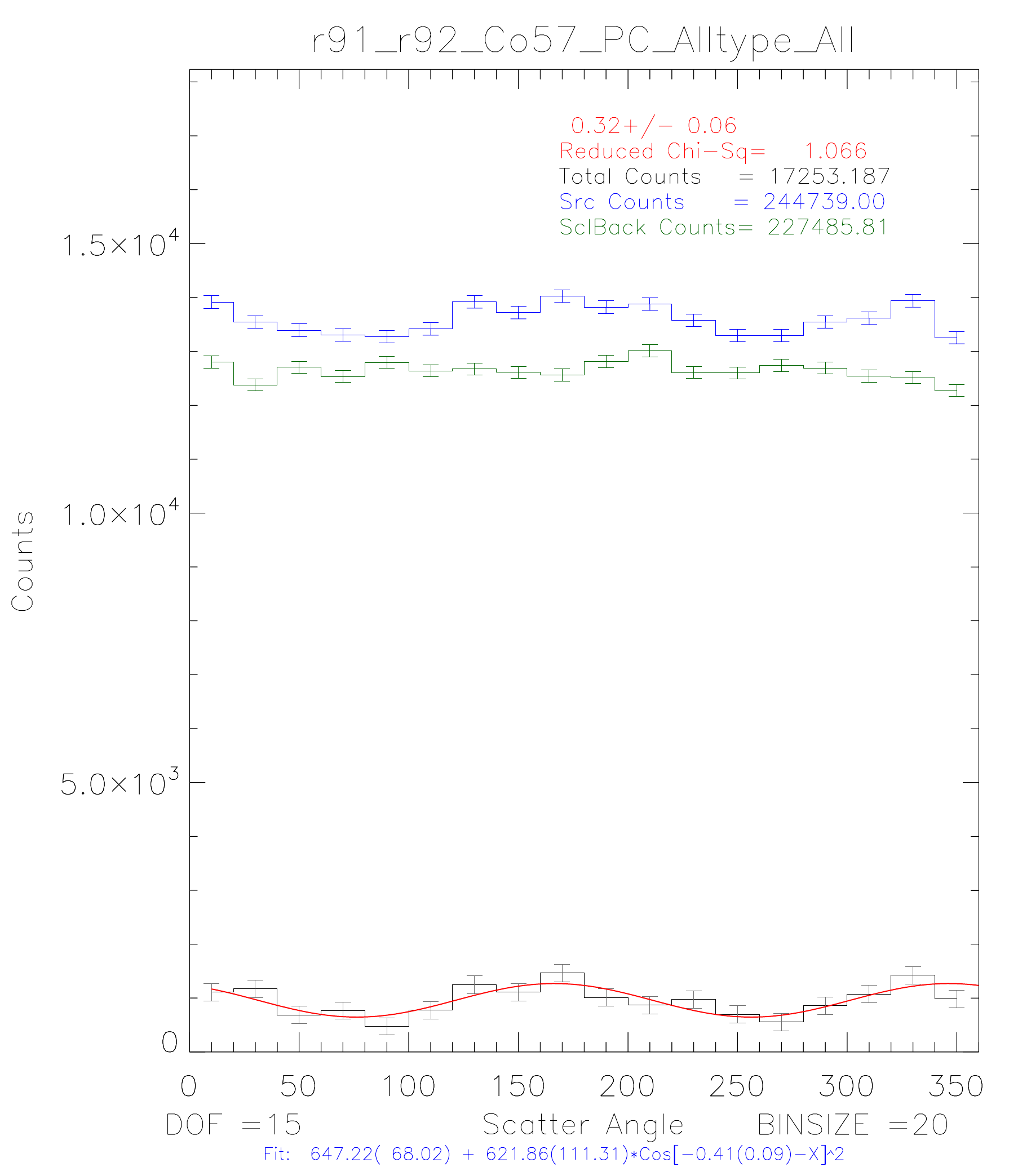}
		 \caption{}
\end{subfigure} 

\begin{subfigure}[b]{0.41\textwidth}
 		 \includegraphics[width=1\linewidth]{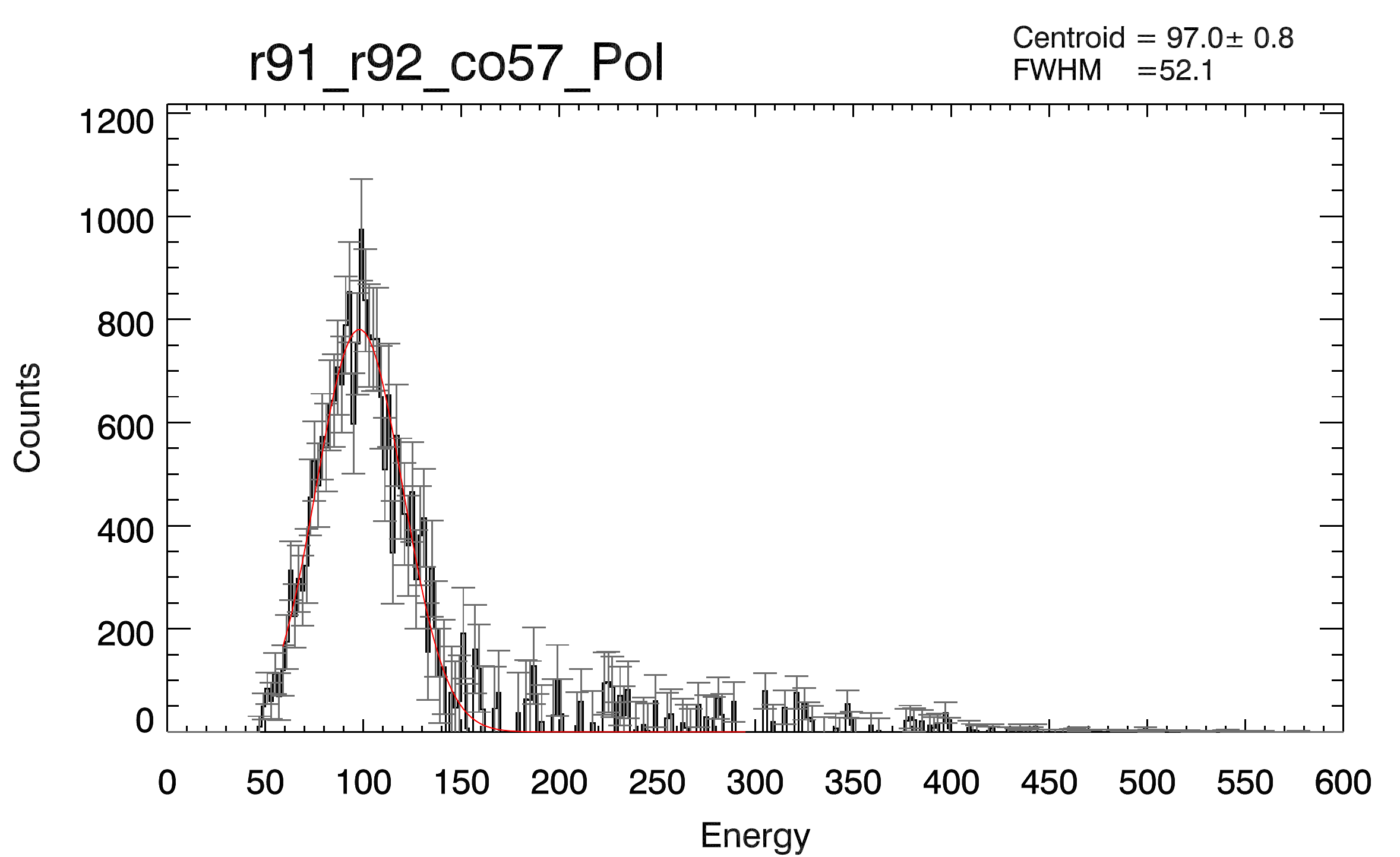}
		 \caption{}
\end{subfigure} \; \; \;
 \begin{subfigure}[b]{0.4\textwidth}
 		 \includegraphics[width=1\linewidth]{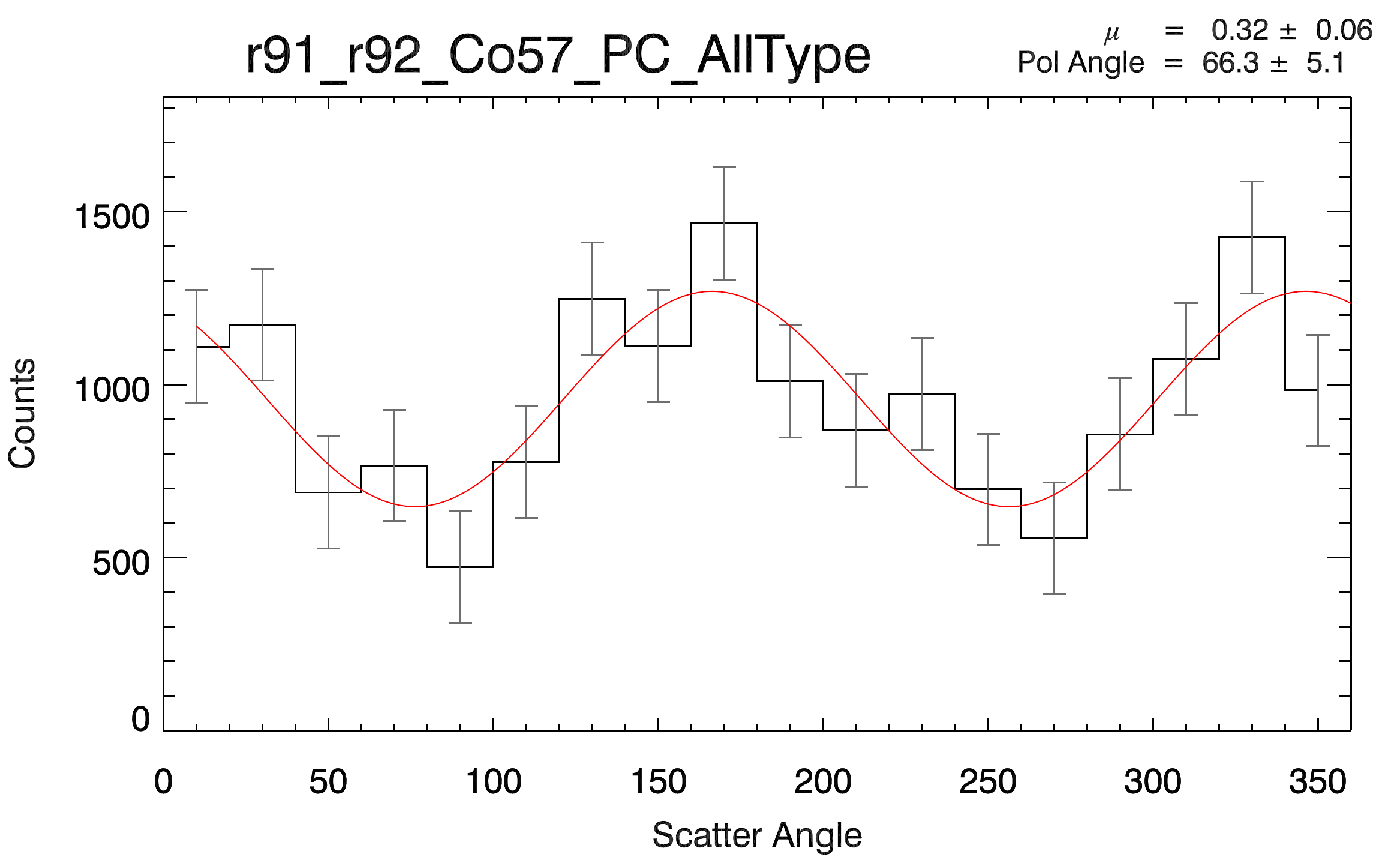}
		 \caption{}
\end{subfigure} 

  \caption{Plot of the polarized Co-57 via compton scattered at $\sim$90$^\circ$. This is summed from all the modules. Initial energy of the source is 122 keV and the 90$^\circ$ compton scattered is about 98 keV. The energy spectra for the scattered source is shown in figure (a) and (c). The scattering angle histogram is shown in (b) and (d) for PC events of all types.}
\label{fig:sim_pol_unh_co57_a}

\end{figure}

\begin{figure}[htbp]
 \centering
\begin{subfigure}[b]{0.60\textwidth}
 		 \includegraphics[width=1\linewidth]{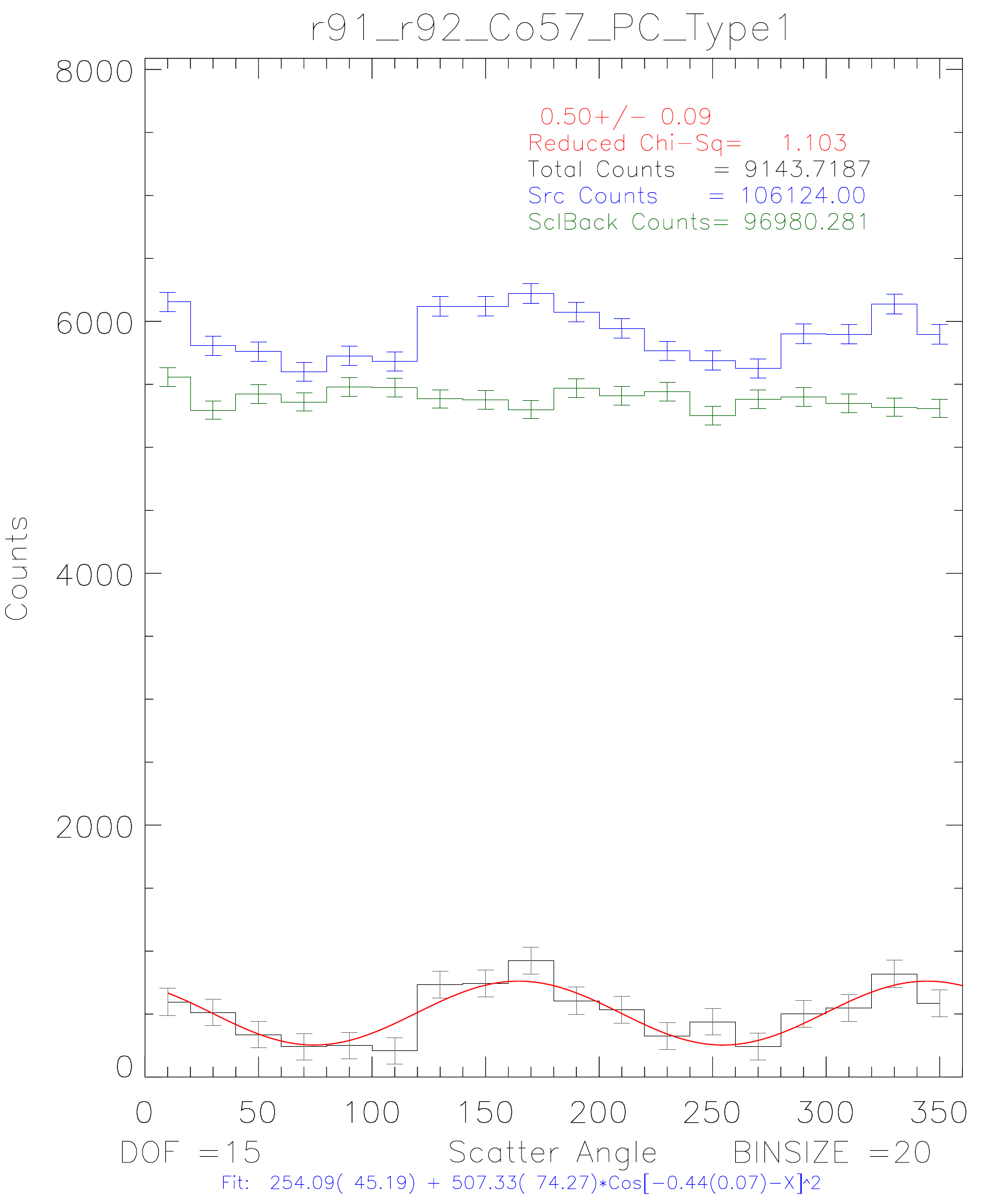}
		 \caption{}
\end{subfigure} \; \; \;     

 \begin{subfigure}[b]{0.60\textwidth}
 		 \includegraphics[width=1\linewidth]{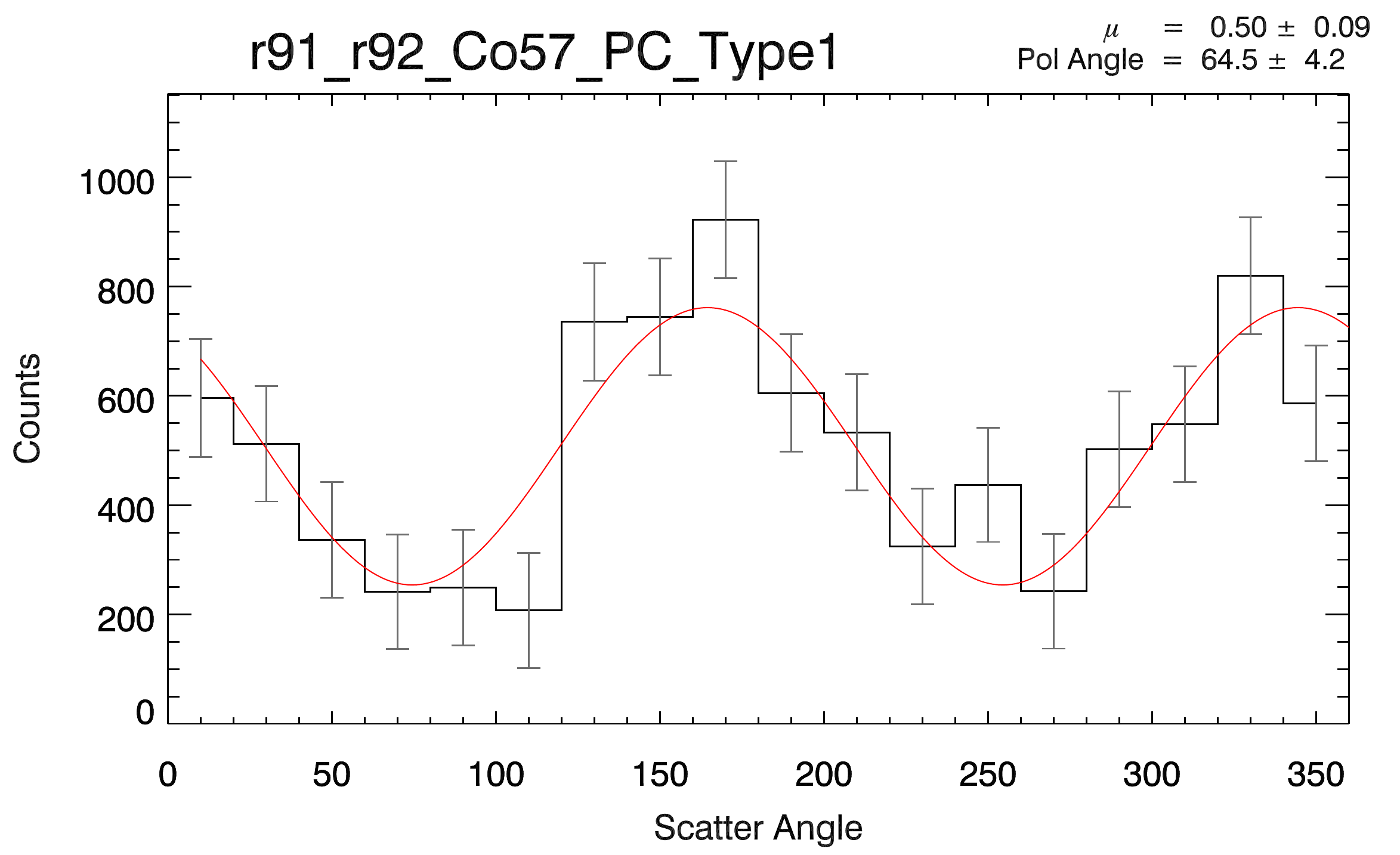}
		 \caption{}
\end{subfigure} 
  \caption{Scatter angle histogram plot of polarized Co-57 PC Type-1 events. This data is from all the modules combined. }
\label{fig:sim_pol_unh_co57_t1}
\end{figure}

		The lab setup for this scattering is shown in Figure \ref{fig:sim_pol_setup} and in \ref{fig:sim_pol_fts_setup}. 
		The source is placed in the collimated lead brick and scattered through the plastic at 90$^\circ$. 
		The lead block 'B' (in Figure \ref{fig:sim_pol_setup} is only present at the background run and removed for the source run. 
		This setup is used on the fully assembled GRAPE at University of New Hampshire (UNH) and at Fort Sumner (FTS) as this setup could not be ran individual modules due to time constraints. 
		The fitted energy peak, at UNH, is shown in Figure \ref{fig:sim_pol_unh_co57_a} a and \ref{fig:sim_pol_unh_co57_a}c and the respective scattered angle histogram is shown in \ref{fig:sim_pol_unh_co57_a} b and \ref{fig:sim_pol_unh_co57_a} d. 
		The data in blue is from the source run and the green is from the background run.
		The livetime of each module is measured separately and used to maximize the data from that module.
		Each of the collected data is poisson statistics so the error is the square root of the counts. 
		The data presented is  the sum of normalized data from all the modules with the error propagated accordingly.
		The black is the background subtracted  spectrum from the source run and the red is the fit to this data. 
		The error bars for this background subtraction is also propagated accordingly. 
		The scatter angle histogram for the PC Type-1 (non-adjacent) events is shown in Figure \ref{fig:sim_pol_unh_co57_t1}.
		The modulation for all PC event types was measured to be $\mu = 0.32 \pm 0.06$ and for PC Type-1 was  $\mu = 0.50 \pm 0.09$.
		The PC Type-1 has a higher modulation factor for two reasons.
		First, being further away, there is smaller uncertainty in the scatter angle (larger lever arm) that would tend to generate a more well defined modulation. 
		Larger uncertainty (smaller lever arm, as in case of the adjacent anodes) would tend to flatten the modulation.
		Second, the effect of crosstalk is significantly lower for the Type-1 events as they are further away. 
		The crosstalk results in misclassified event (for example a C event can be misclassified as PC event). 
		As there is no preferred direction for the crosstalk, the misclassified event would also flatten the modulation.
		 

\begin{figure}[hbtp]
 \centering
\begin{subfigure}[b]{0.42\textwidth}
 		 \includegraphics[width=1\linewidth]{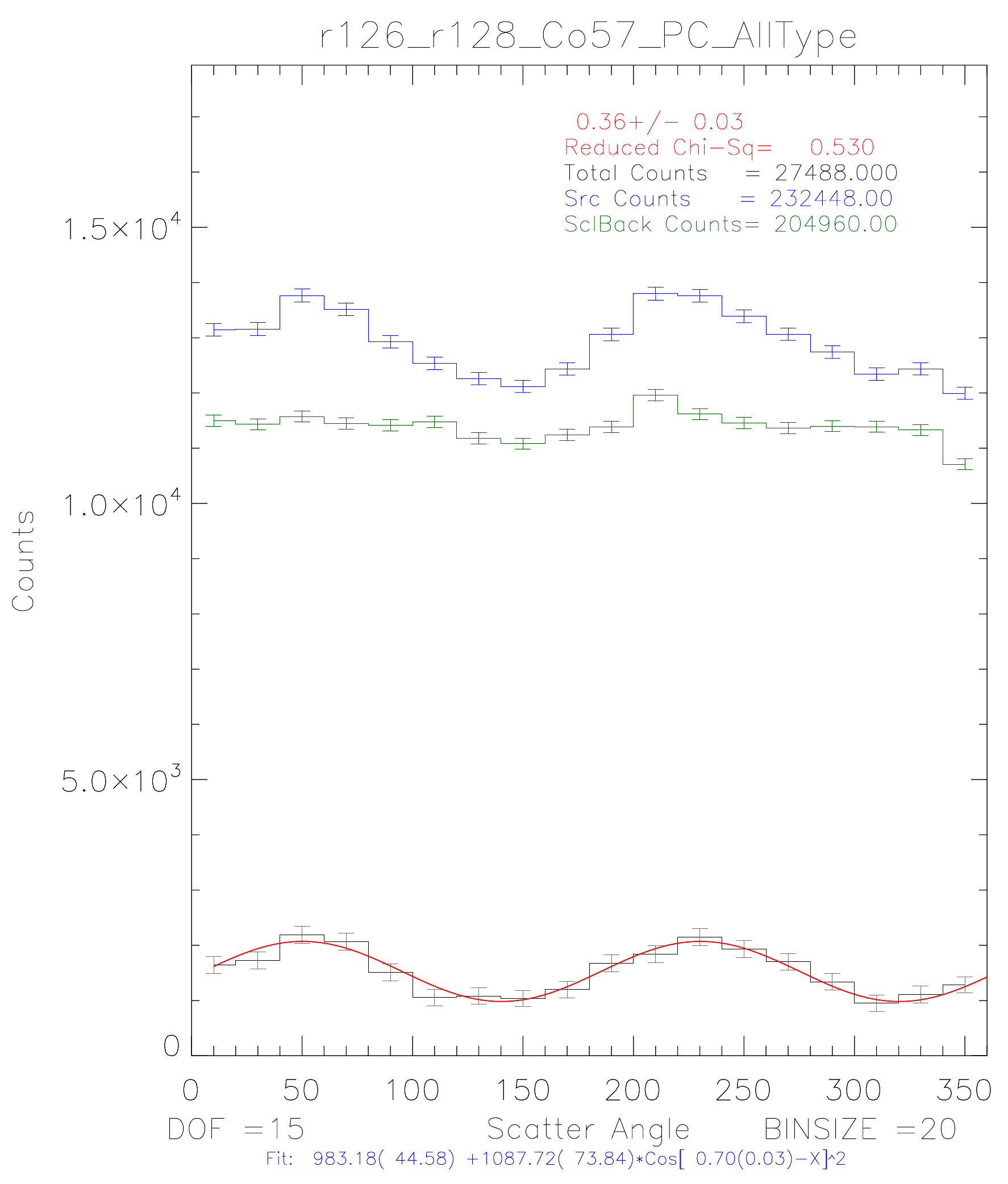}
		 \caption{}
\end{subfigure} \; \; \; 
\begin{subfigure}[b]{0.41\textwidth}
 		 \includegraphics[width=1\linewidth]{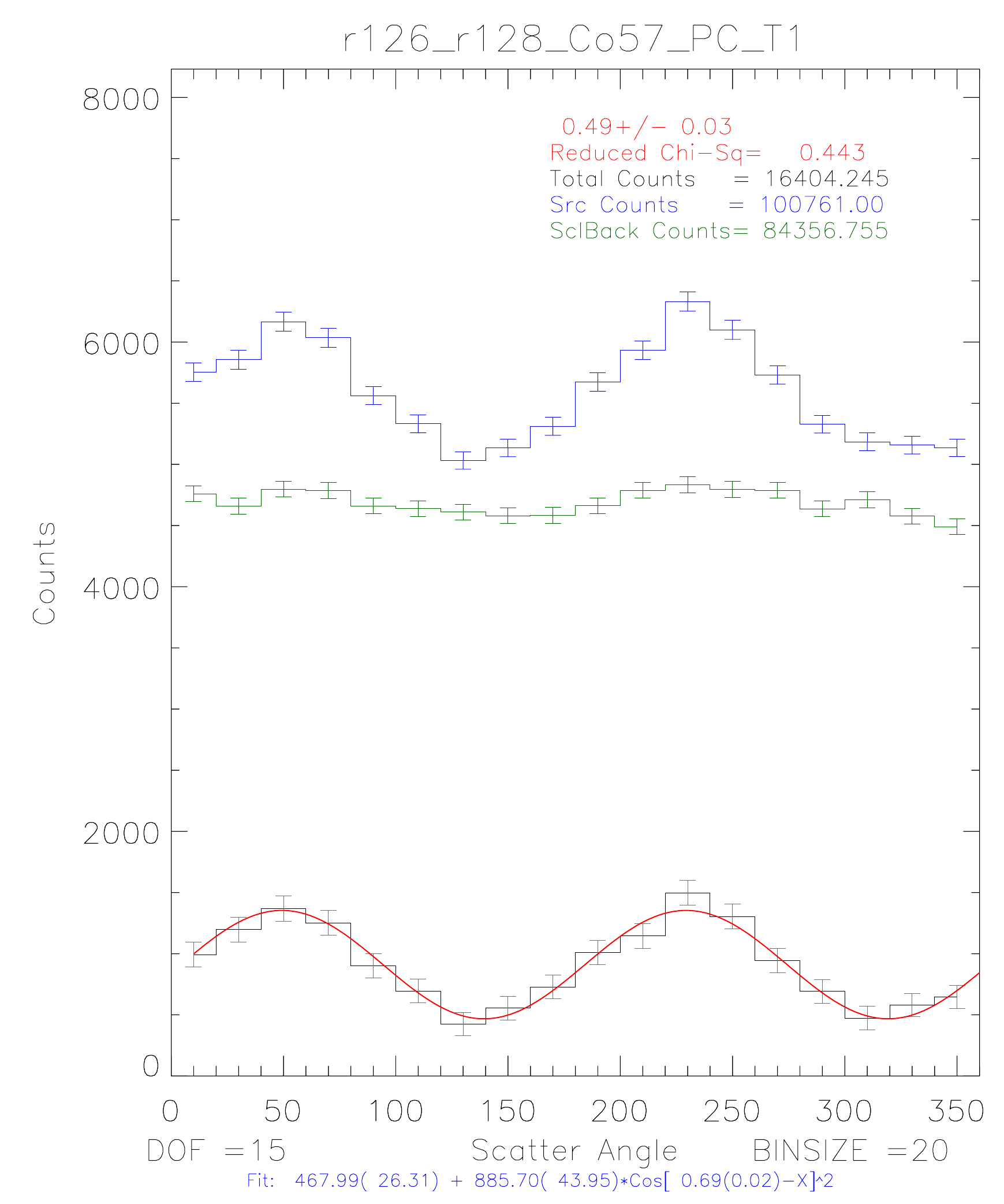}
		 \caption{}
\end{subfigure} 

\begin{subfigure}[b]{0.41\textwidth}
 		 \includegraphics[width=1\linewidth]{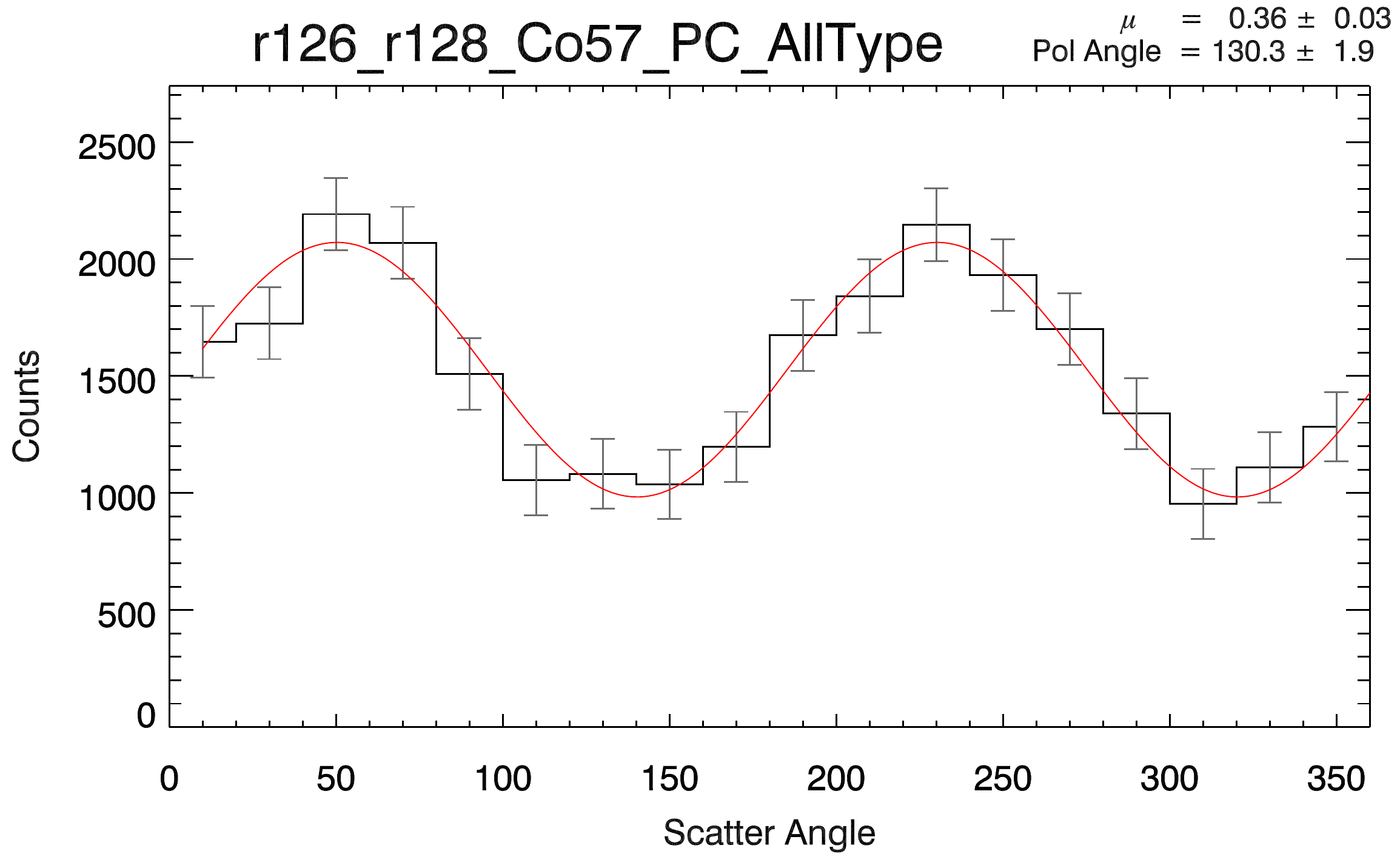}
		 \caption{}
\end{subfigure} \; \; \;
 \begin{subfigure}[b]{0.41\textwidth}
 		 \includegraphics[width=1\linewidth]{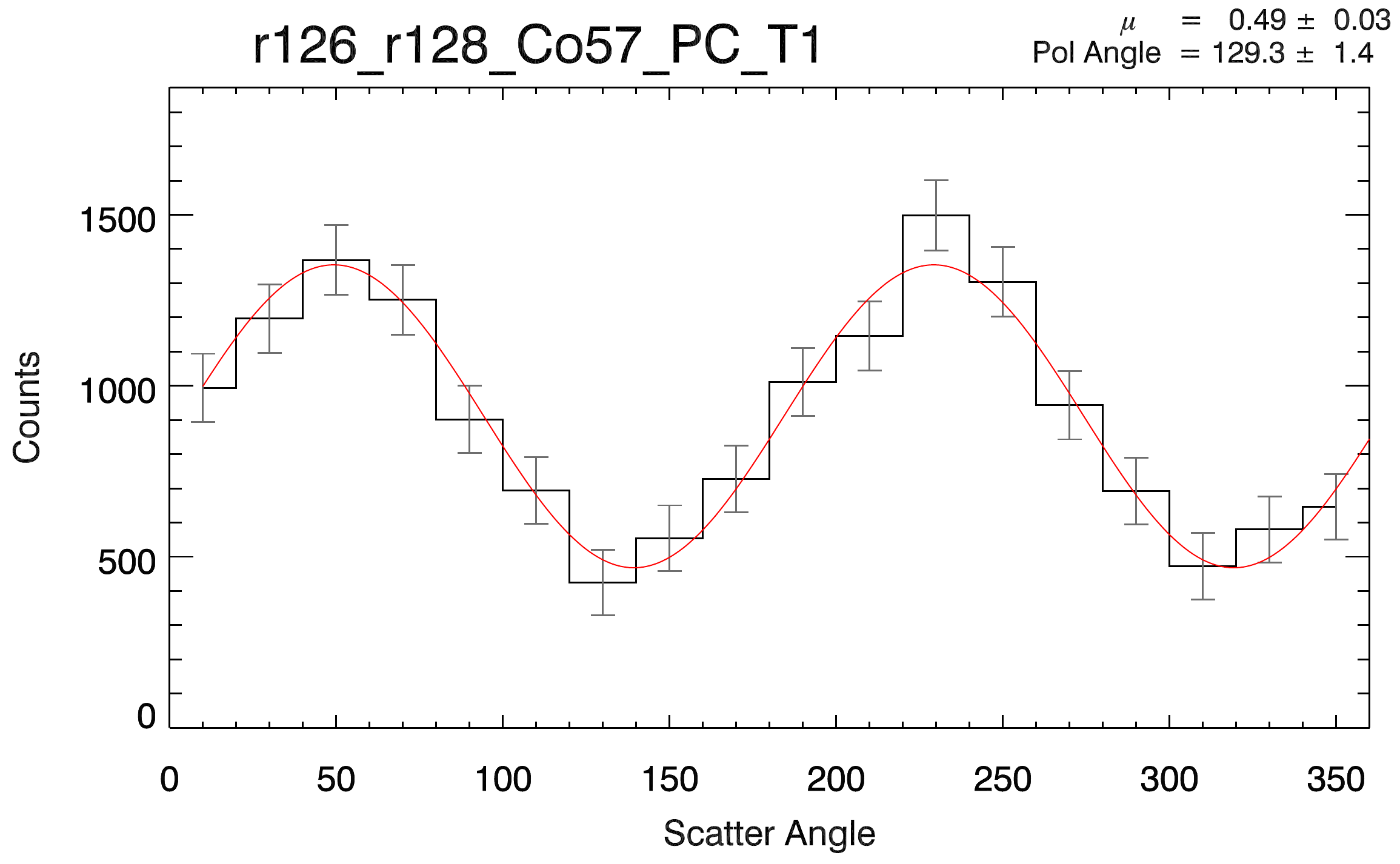}
		 \caption{}
\end{subfigure} 

  \caption{Plot of the polarized Co-57 via compton scattered at $\sim$90$^\circ$ at Fort Sumner similar to UNH.  Initial energy of the source is 122 keV and the 90$^\circ$ compton scattered is about 98 keV energy. The scatter angle histogram is plotted here. Plot (a) and (c) represents PC events of all types . Plots (b) and (d) represents PC events Type 1 (Non-Adjacent) Events. The polarization angle here is $\sim$130$^\circ$.}
\label{fig:sim_pol_fts_co57_a}

\end{figure}

		Similar runs were performed at Fort Sumner during pre-flight calibrations. 
		Those data are shown in Figure \ref{fig:sim_pol_fts_co57_a}. 
		The modulation for this run for all PC event types was $\mu = 0.35 \pm 0.03$ and for PC type-1 $\mu = 0.49 \pm 0.09$.
		The modulation factor is slightly higher for Type-1 PC as compared to all types PC events as seen previously. 

\begin{figure}[hbtp]
 \centering
\begin{subfigure}[b]{0.415\textwidth}
 		 \includegraphics[width=1\linewidth]{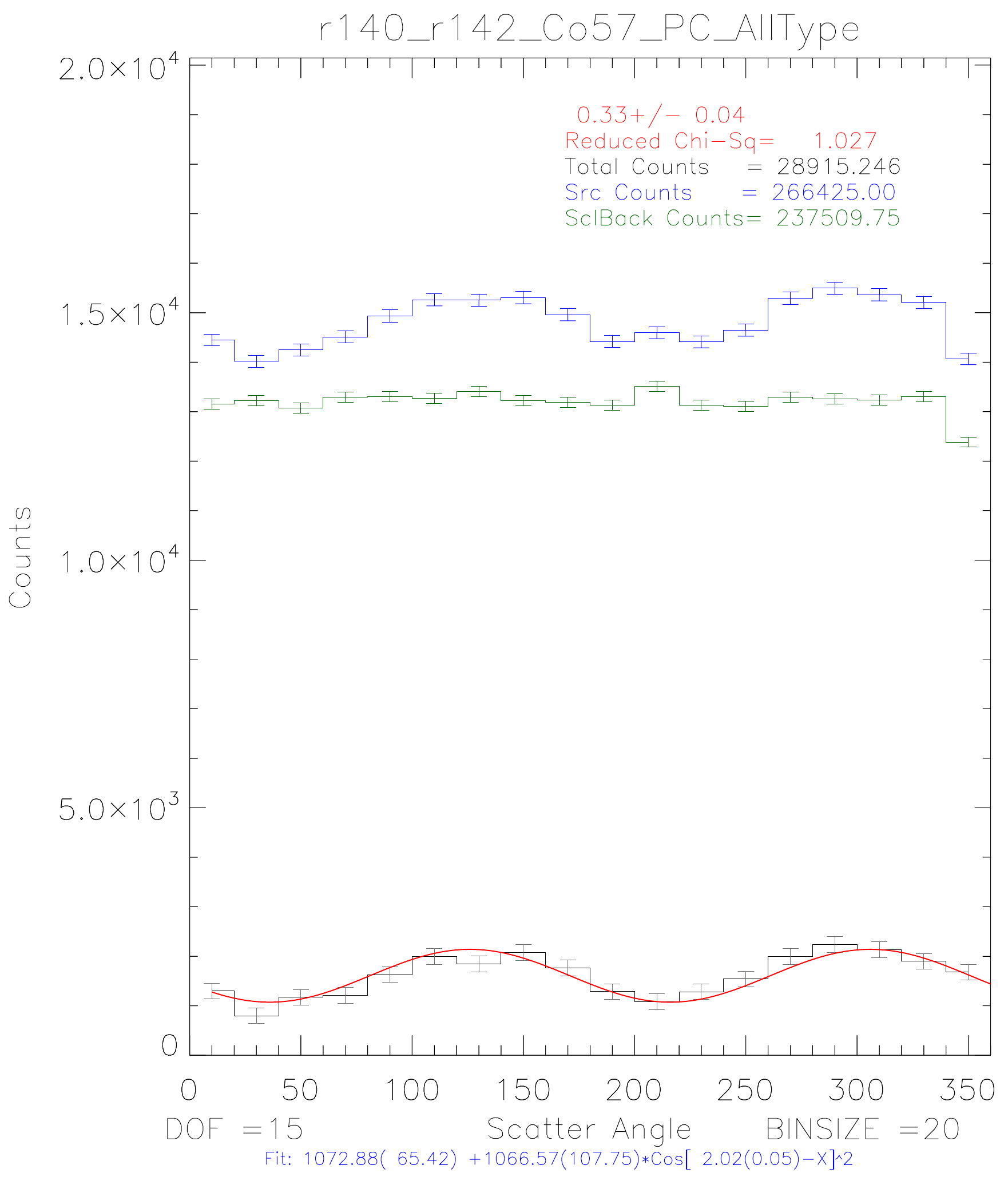}
		 \caption{}
\end{subfigure} \; \; \; 
\begin{subfigure}[b]{0.41\textwidth}
 		 \includegraphics[width=1\linewidth]{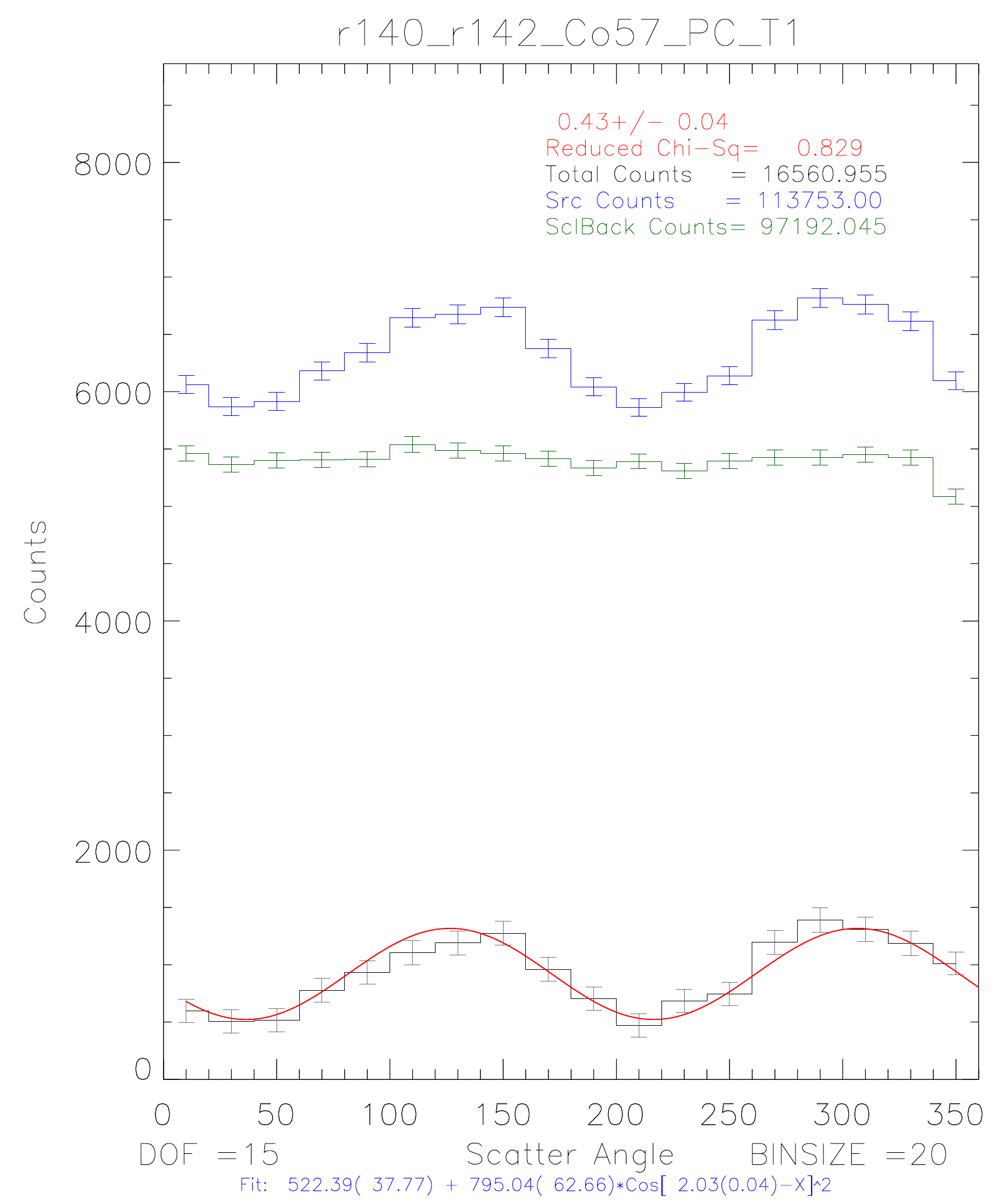}
		 \caption{}
\end{subfigure} 
 \; \; \;
\begin{subfigure}[b]{0.41\textwidth}
 		 \includegraphics[width=1\linewidth]{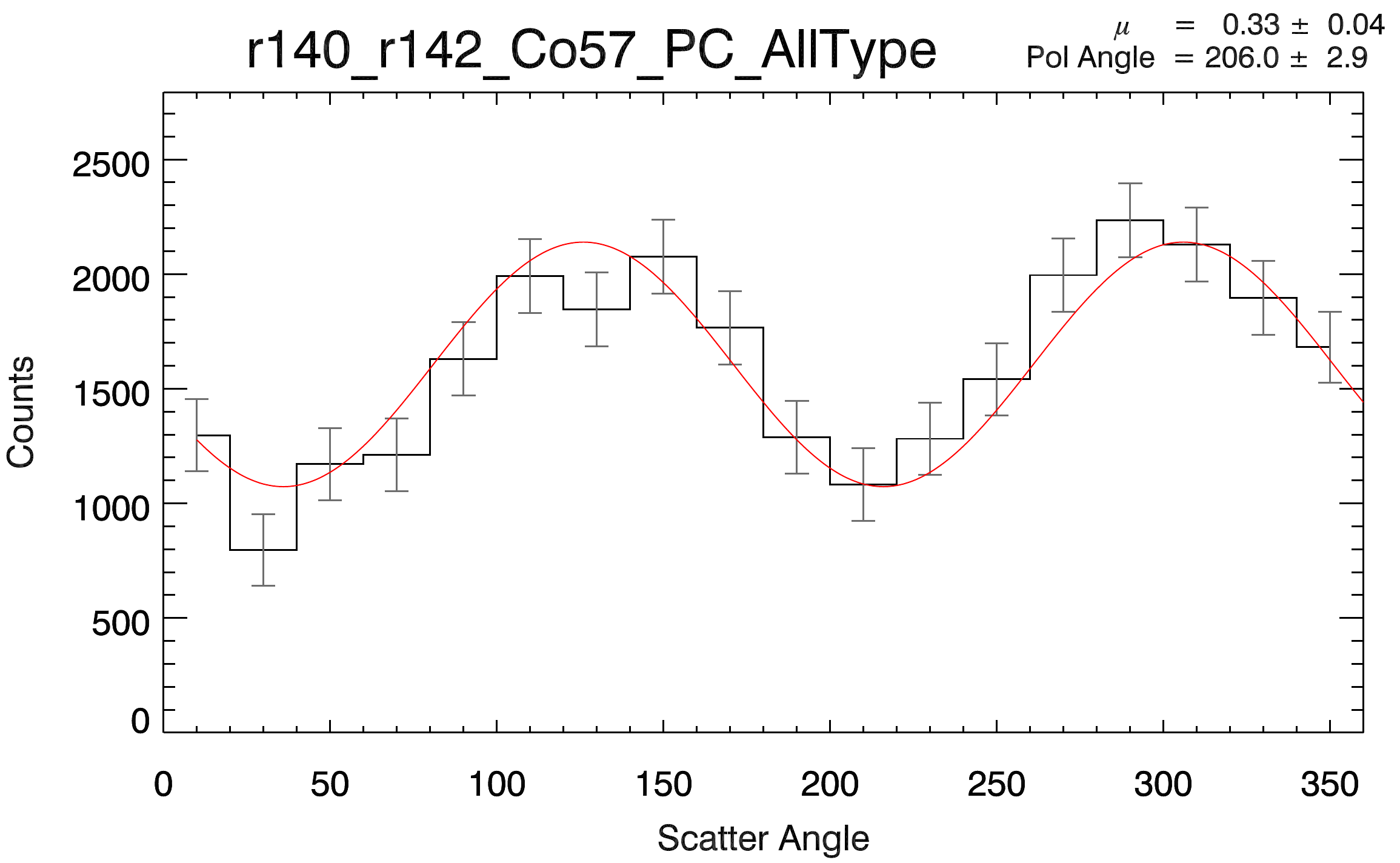}
		 \caption{}
\end{subfigure} \; \; \; 
 \begin{subfigure}[b]{0.41\textwidth}
 		 \includegraphics[width=1\linewidth]{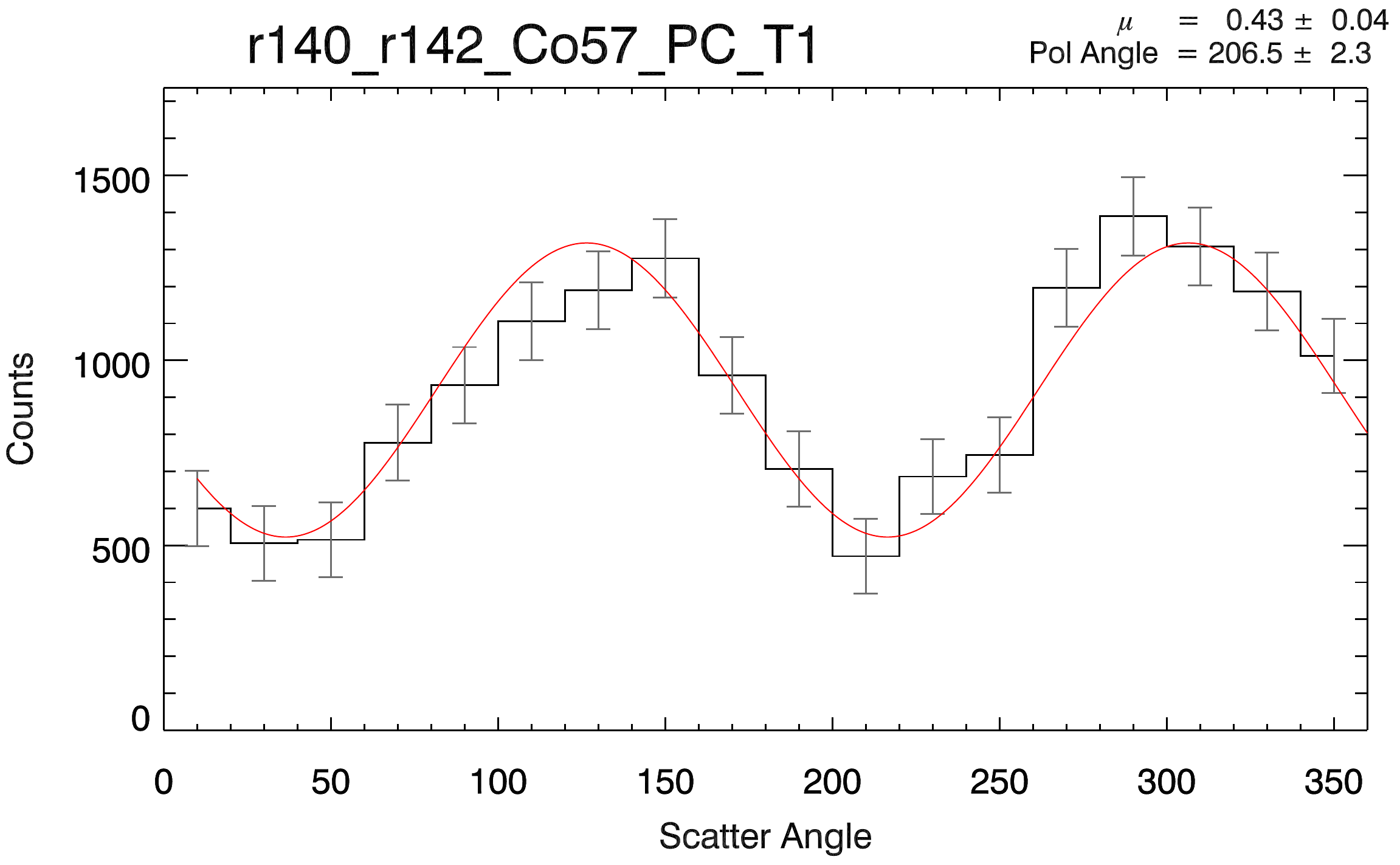}
		 \caption{}
\end{subfigure} 

  \caption{Plot of a different polarized Co-57 run at Fort Sumner. The source setup is rotated to verify whether we get a similar polarization angle change in our measurements. The scatter angle histogram is plotted here. Plot (a) and (c) represents PC events of all types . Plots (b) and (d) represents PC events Type 1 (Non-Adjacent) Events. The polarization angle here is $\sim$207$^\circ$. }
\label{fig:sim_pol_fts_co57_b}

\end{figure}

		We further experimented with rotating the source and observing whether the new polarization angle reflected this change. 
		The setup is shown in Figure \ref{fig:sim_pol_fts_setup} which replicates the drawing in Figure \ref{fig:sim_pol_setup}. 
		In this Figure, the marked red line is the rotated position for the rotated polarized source. 
		Since the bottom of the bucket was a rectangle, the rotated angle was measured to be $\sim$80$^\circ$. 
		The data for this rotated polarized run is shown in Figure \ref{fig:sim_pol_fts_co57_b}. 
		The polarization angle was measured to be $\sim$207$^\circ$. 
		This change is about 80$^\circ$ which validates our instrument's capability to measure the change in polarization angle. 
		
		The degree of linear polarization for a measurement is determined by the equation \ref{eqn:deg_lin_pol} which is the ratio of the modulation factor of the instrument and the modulation for the 100\% polarized source.
	\[ \Pi = \frac{\mu}{\mu_{100}} \]
	where the ${\mu}$ is measured and $\mu_{100}$ is determined from simulations. 
		For the measurements mentioned above, we need to determine the 100\% polarized from simulation. 
		
\begin{figure}[tp]
\centering 
\includegraphics[width=0.7\textwidth]{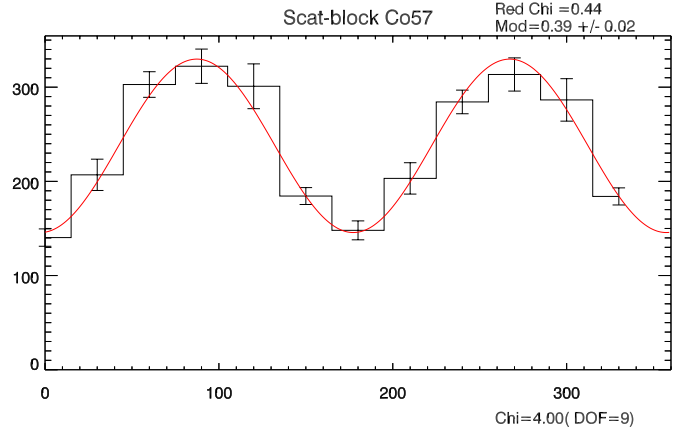}
\caption{A scatter angle histogram of a 100\% polarized 99 keV photons. The polarized run was corrected with the unpolarized run for the simulation to correct the anisotropy of the instrument. The $\mu_100 = 0.49 \pm 0.01$.  }
\label{fig:ins_perf_co57_100_pol}       
\end{figure}

		100\% polarized photons of 99 keV was simulated to determine the $\mu_{100}$ for the above measurements. 
		This is shown in Figure \ref{fig:ins_perf_co57_100_pol}. The modulation factor for this 100\% polarized source for all PC event types is measured to be $\mu_{100} = 0.49 \pm 0.01 $. 
		Therefore the measured linear degree of polarization (where $\mu \approx 0.33$) would be $\sim 0.33/0.40 \approx 0.83$ ( or 83\%). 
		Although this is not the result we initially expected, the setup used to generate the 99 keV polarized in the lab is not ideal. 
	The geometry results in a broader distribution of scatter angles. Unfortunately, we never simulated this. 
		It can be seen from Figure \ref{fig:sci_comp_kn_lineardeg_unpol}, the maximum fractional polarization is achieved at $\sim$90$^\circ$ Compton scatter angle for the unpolarized beam of 122 keV.
		The polarization fraction lowers as the Compton scatter angle diverges from $\sim$90$^\circ$.
		The measured 83\% suggests that the scatter angles are between $\sim$80$^\circ$ and $\sim$100$^\circ$.
	
		\subsubsection{Polarization Sensitivity}	
		\label{sec:ins_perf_pol_sensitivity}	
		Polarization sensitivity of an instrument is defined by the Minimum Detectable Polarization (MDP) as given by equation \ref{eqn:mdp} as
	\[MDP_{99} = \frac{a_{S}}{\mu_{100}} = \frac{4.29}{\mu_{100}\ C_S}{\sqrt{C_S+C_B}}\]
	where C$_S$ is the total number of source counts, C$_B$ is the total number of background counts and $\mu_{100}$ is the modulation factor for a 100\% polarized source (determined from simulations). 
	The MDP is a characteristic of the measurement.
	For the run shown in Figure \ref{fig:sim_pol_fts_co57_a}, C$_S$ = 232448, C$_B$ = 204960 and with the  $\mu_{100} = 0.49$, the MDP = 0.025. 
	This measurement was characterized as being able to measure a minimum polarization of  0.025 (or 2.5\%). 
	The MDP for a measurement decreases (the sensitivity increases) for longer observation times.
	 The lab measurements were done over longer periods of time and the radiation sources are strong as compared to the astrophysical sources.
	 Hence, the MDP for the lab measurements ends up being pretty low. 
	 This is not the case for astrophysical measurements done from balloon experiments as the counts are pretty low as compared to radiation sources in lab. 
	 The MDP and  $\mu_{100}$ of the Crab observations for flight data are presented in chapter \ref{sec:polarization_analysis}.
	

	\section{GRAPE Energy Calibration}
	\label{sec:ins_perf_calibration}
	GRAPE is made up of 24 modules, each with 64 scintillator elements.
	The incoming photons produce scintillations in the detectors and the amount of scintillation is proportional to the incident energy. 
	The collected light is used to generate a pulse-height spectrum (counts vs channels). 
	We use known radiation sources to determine the energy calibration which translates pulse height channel to energy loss in the scintillator.
	GRAPE was calibrated using various radiation sources at University of New Hampshire (UNH) and also before the flight at Fort Sumner (FTS). 
 	We do calibration of these detectors at module level first and then verify these at instrument level (after the modules are integrated in the pressure vessel).

			\subsection{Module level calibration at UNH}
			\label{sec:ins_perf_module_calibration}
	
			Module level calibration provided calibration information for each scintillator element (both plastic and calorimeters). 
			Each module is made up of 64 scintillator elements of which 36 are plastic and 28 are calorimeter. 
			We uses sources of known energy to get a series of pulse-height histogram for each anode element.
			The pulse-height histograms provide a peak channel number which corresponds to the energy of the radiation. 
			This constitutes of a calibration point (the energy and the corresponding peak channel number). 
		 	This process is repeated for several of known radiation sources to get multiple calibration points. 
			These calibration points are fitted to get the necessary parameters to convert the pulse-height channels to energy.

\begin{figure}[hbtp]
 \centering
\begin{subfigure}[b]{0.85\textwidth}
 		 \includegraphics[width=1\linewidth]{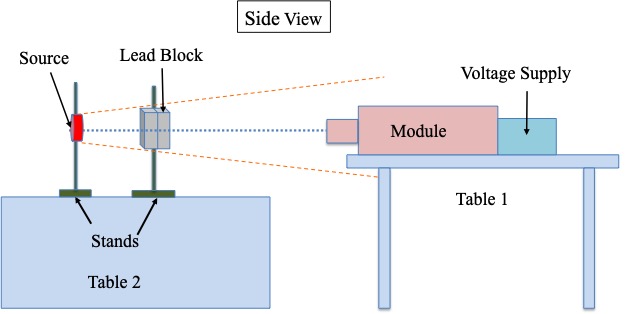}
		 \caption{}
\end{subfigure}
\begin{subfigure}[b]{0.85\textwidth}
 		 \includegraphics[width=1\linewidth]{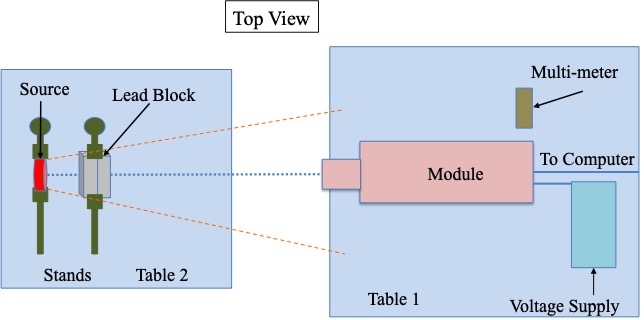}
		 \caption{}
\end{subfigure} 
  \caption{Drawing of a setup for a module level energy calibration. Module is placed on table1 with respective electronics and we see the radiation source and the lead block for the background run. For the source run, just the lead block is removed while leaving everything else the same. The sources were about half a meter to a meter away from the modules.}
\label{fig:sim_cal_mod_setup}

\end{figure}
			
			The typical setup of a module calibration run is shown in Figure \ref{fig:sim_cal_mod_setup} a. 
			The module and supporting electronics are placed on table.
			The source is placed some distance away on a separate table.
			A background run was conducted with a lead block between the source and the detector module. 
			The lead block obstructs the direct line of sight from the radiation source to the detector. 
			This prevents direct radiations to hit the detector but allows any other scattered radiation that contributes to the background.
			The setup in \ref{fig:sim_cal_mod_setup} b is for the background run.

\begin{figure}[hbtp]
 \centering
\begin{subfigure}[b]{0.45\textwidth}
 		 \includegraphics[width=1\linewidth]{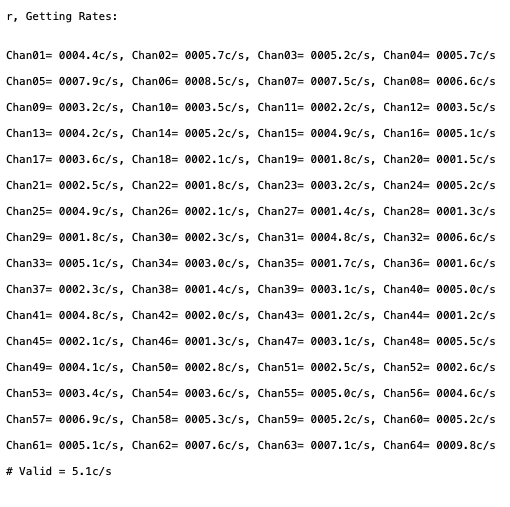}
		 \caption{}
\end{subfigure} \; \;     
 \begin{subfigure}[b]{0.45\textwidth}
 		 \includegraphics[width=1\linewidth]{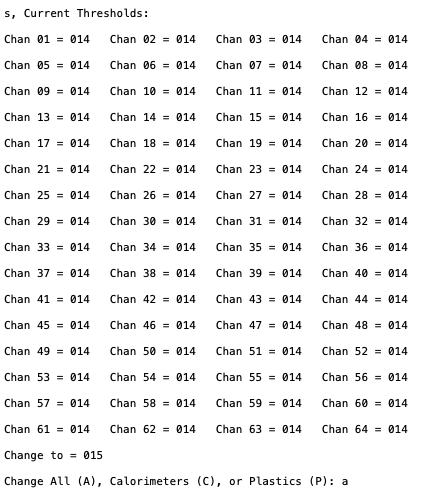}
		 \caption{}
\end{subfigure}       

\begin{subfigure}[b]{0.45\textwidth}
 		 \includegraphics[width=1\linewidth]{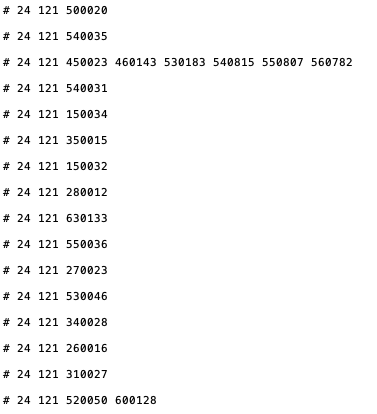}
		 \caption{}
\end{subfigure} 
  \caption{Examples of the various modes for each modules. (a) This is an example of rates R mode and (b) This is an example of set threshold S mode. (c) This is an example of calibration C mode.}
\label{fig:ins_pref_modes_a}
\end{figure}
			Module control and communication (data collection) is achieved via the Programmable Integrated Chip (PIC) present in each module electronics.
			Various operational modes are pre-programmed in the PIC processor.
			There are 4 modes that are important for data acquisition: the Flight mode, Calibration (C) mode, set threshold (S) mode and Rates (R) mode. 
			An example of these modes is shown in Figure \ref{fig:ins_pref_modes_a}. 
			These output of Flight and C mode and can set the output to be binary or ASCII file.
			The flight mode is used to generate binary data files and is used predominantly  during operation of the fully assembled GRAPE instrument. 
			For flight mode, the number of triggered anodes was limited to 8 anodes.
			The R mode is used to output the counting rates for each of the scintillator elements of the module.
			The S mode was used to set the threshold values for the scintillator elements. 
			The C (calibration) mode was created to output all the triggered anodes during an event so this mode does not have the limitation of only 8 anodes triggered. 
			The flight and C mode also provides the calculated fractional live time for each event. 
			The output from C mode is an ASCII file, that is used to generate pulse height spectra.			
					\subsubsection{Calibration sources}
					\label{sec:ins_perf_calibration_sources}

					Various radiation sources with known peak energy were used to calibrate detectors. 
					The energy ranges of plastics were from $\sim$6 to $\sim$200 keV and the calorimeter were from  $\sim$20  to   $\sim$400 keV. 
					The radiation sources used to calibrate our detectors at the module levels were $^{109}$Cd, $^{241}$Am and $^{133}$Ba. 
					$^{109}$Cd produces peaks at 22 keV and 88 keV. 
					For plastics, the 88 keV peak is not visible but we do see 22 keV (because of their relative intensities, 22 keV overshadows the 88 keV peak).
					For calorimeters, we observe either one or both peaks. 
					Therefore we get 1 calibration point for plastic and at least 1 for calorimeter from this Cd-109 run. 
					The $^{241}$Am produces a 60 keV peak which is seen by both plastics and calorimeters. 
					Poor energy resolution in the plastics makes them more difficult to calibrate . 
					Therefore get one calibration point for each of the scintillators from the $^{241}$Am run.
					

\begin{figure}[hbtp]
\centering
\begin{subfigure}[b]{0.8\textwidth}
 		 \includegraphics[width=1\linewidth]{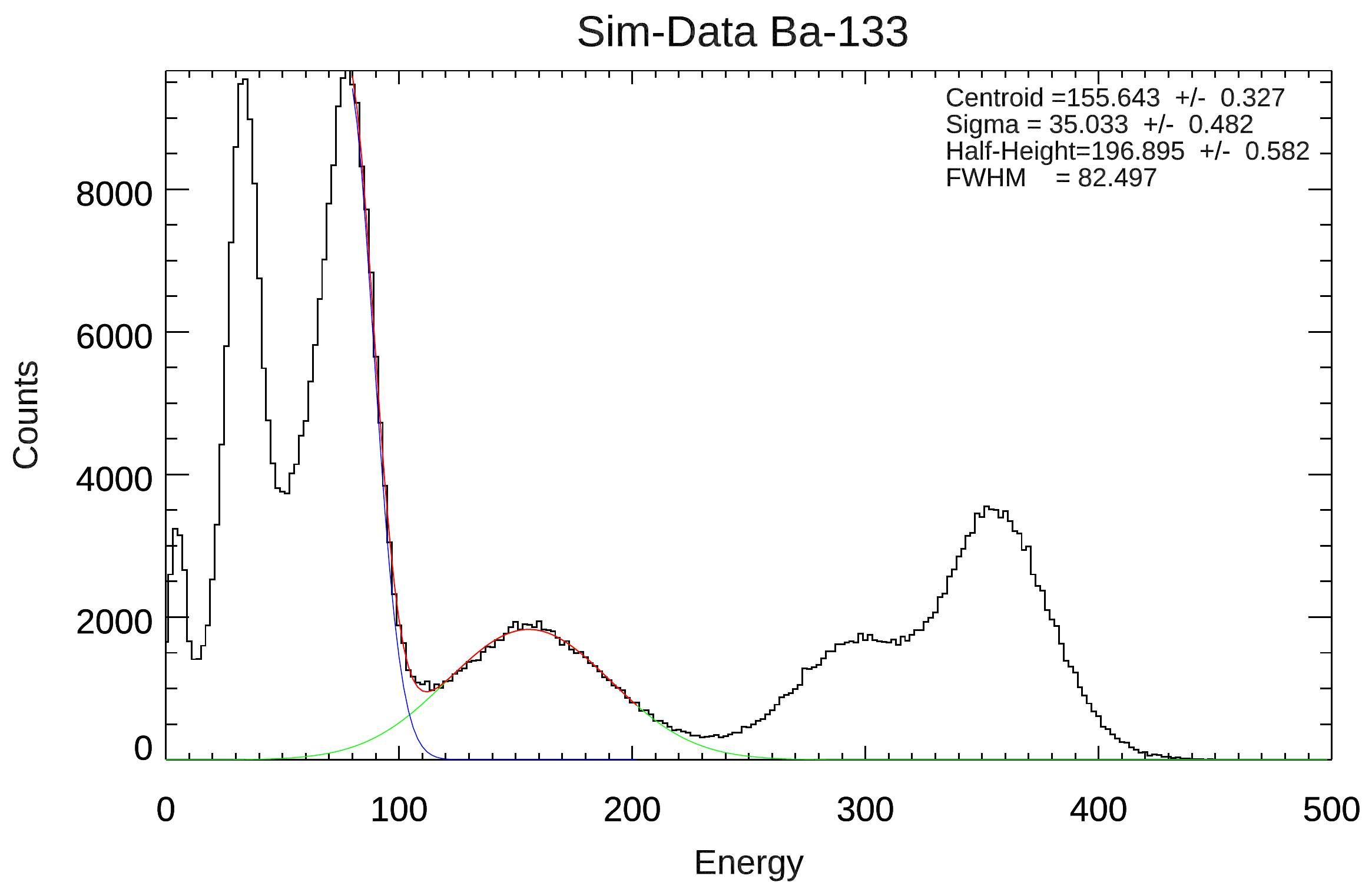}
		 \caption{}
\end{subfigure}
\begin{subfigure}[b]{0.8\textwidth}
 		\includegraphics[width=1\linewidth]{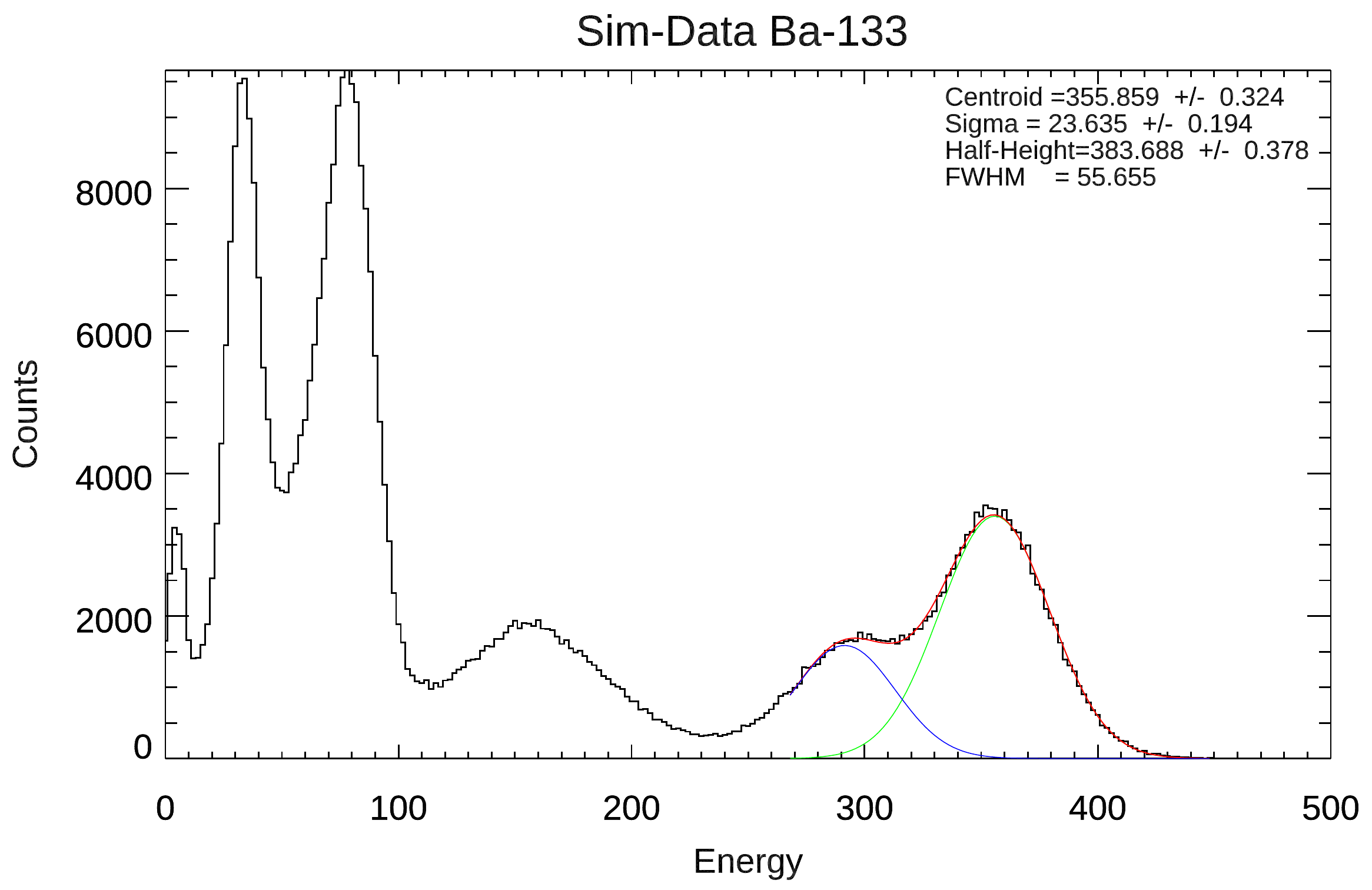}
		 \caption{}
\end{subfigure}

  \caption{Simulated response of a single calorimeter with Ba-133. We expect to observe these peaks in our calorimeters for the Ba-133 runs. We use the 155 keV and 356 keV peaks as calibration points for calorimeters. The green is the fitted gaussian in these plots.}
\label{fig:sim_cal_fit_ba133_cal_sim}

\end{figure}

\begin{figure}[!ht]
 \centering
    	\includegraphics[width=0.8\textwidth]{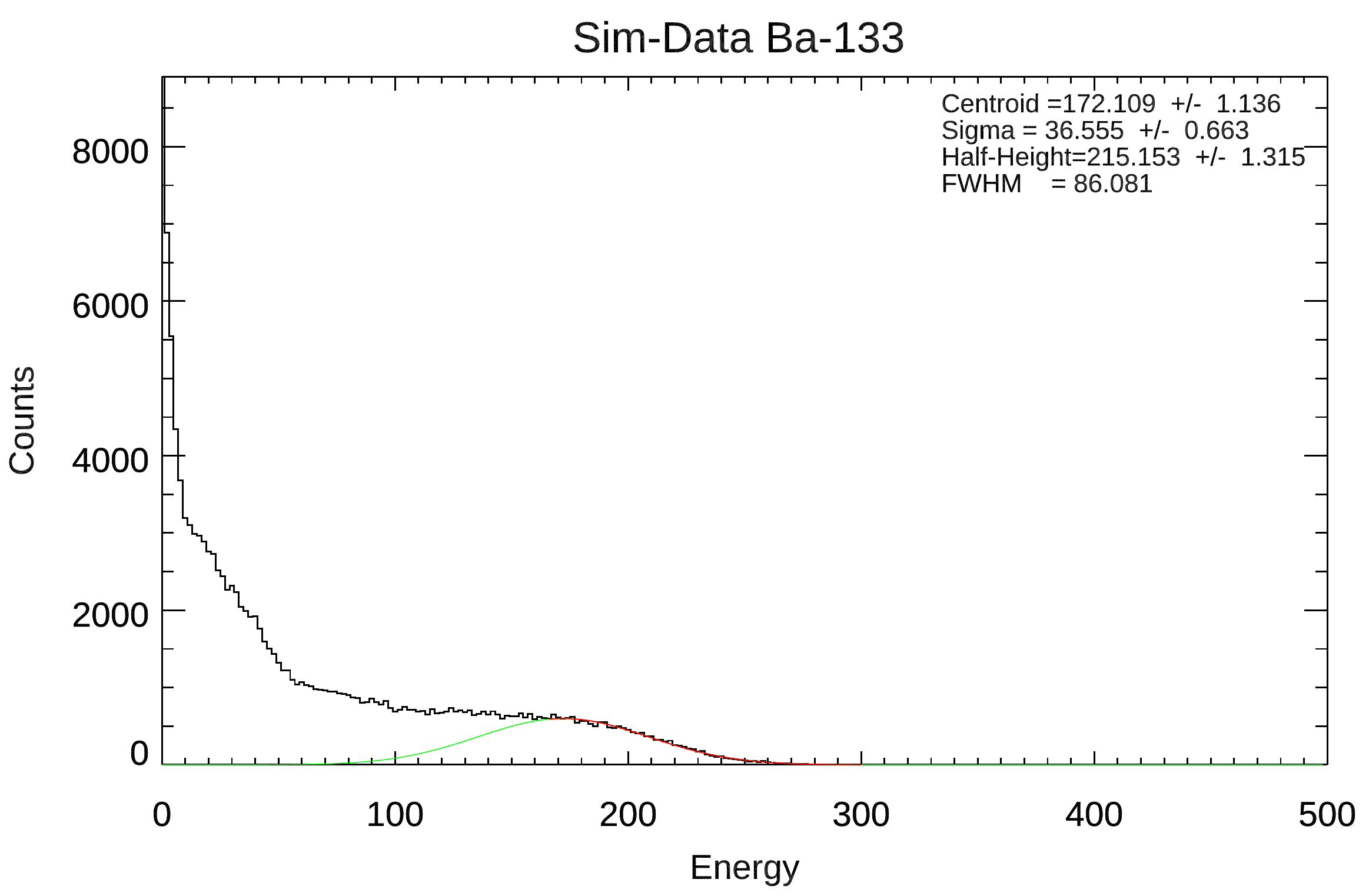}
    	\caption{Simulated response of a single plastic with Ba-133. We attempt to fit the compton edge with a gaussian function. Green is the full gaussian function and the selected data and fit is shown in red. We use the FWHM to get the energy for half-height which is a better calibration point. This half-height energy value is approximately 215 keV and is represented as Half-Height in the plot.}
    	\label{fig:sim_cal_fit_ba133_pla_sim}
\end{figure}		


					$^{133}$Ba is our third source used for calibration.
					$^{133}$Ba radiates gamma rays at  30 keV, 80 keV, 276 keV, 302 keV, 356 keV and 383 keV with relative intensities of 95\%, 34\%, 7\%, 18\%, 62\% and 9\%.
					A simulation of a single module yields calorimeter spectrum shown in Figure \ref{fig:sim_cal_fit_ba133_cal_sim}. 
					As we have calibration points for less than 100 keV from $^{241}$Am and $^{109}$Cd runs, we use $^{133}$Ba to get calibration points for energies higher than 100 keV.
					The $155\ keV$ and  $356\ keV$ calibration points, as seen in Figure \ref{fig:sim_cal_fit_ba133_cal_sim}, were selected from the $^{133}$Ba run for the calorimeters. 
					A plastic spectrum from a simulation of a single module for the $^{133}$Ba radiation source is shown in Figure \ref{fig:sim_cal_fit_ba133_pla_sim}. 
					A compton edge fit was used for plastics as shown in Figure \ref{fig:sim_cal_fit_ba133_pla_sim} to fit the plastic.
					For some plastic elements this point is skipped as this was beyond its range.
					We do 1 source run (with 1 background run) for $^{133}$Ba which provides us with 2 calibration points for calorimeter and 1 plastic point (in fortunate circumstances).

					The list of these runs and the run time is shown in Table \ref{table:sim_cal_mod_run}. 
					Each unique run, for each module, is colored separately and each of the runs consists of a source run followed by the background run. 
					Each $^{109}$Cd run needed 8 hrs.
					The $^{241}$Am ran for 8 hrs focusing the calorimeters and 12 hrs focusing the plastic. 
					The plastic run was scheduled for the night so it had a longer time block. 
					The $^{133}$Ba run took 12 hrs so each module took roughly 40 hrs to complete all the runs.
					We have 24 modules so a total time of 960 hrs is needed for the whole calibration runs. 
					Which ends up to be roughly more than 40 days of calibration runs for the 24 modules. 
					As these setups had to be changed manually by being present in the lab, some of these runs ran longer than shown in the table as they ran overnight. 
					Some runs ran over the weekend if arrangements could not be made to update the setup.
					\begin{table}[tbp]
\begin{center}
\caption{ Table showing various radiation sources and the total time for each run. Separate colors represent separate runs. Am241 is ran twice. Once for calorimeters by increasing the plastic threshold to max and then for plastics by raising calorimeter threshold to max.}
\label{table:sim_cal_mod_run}
   	\includegraphics[width=0.8\textwidth]{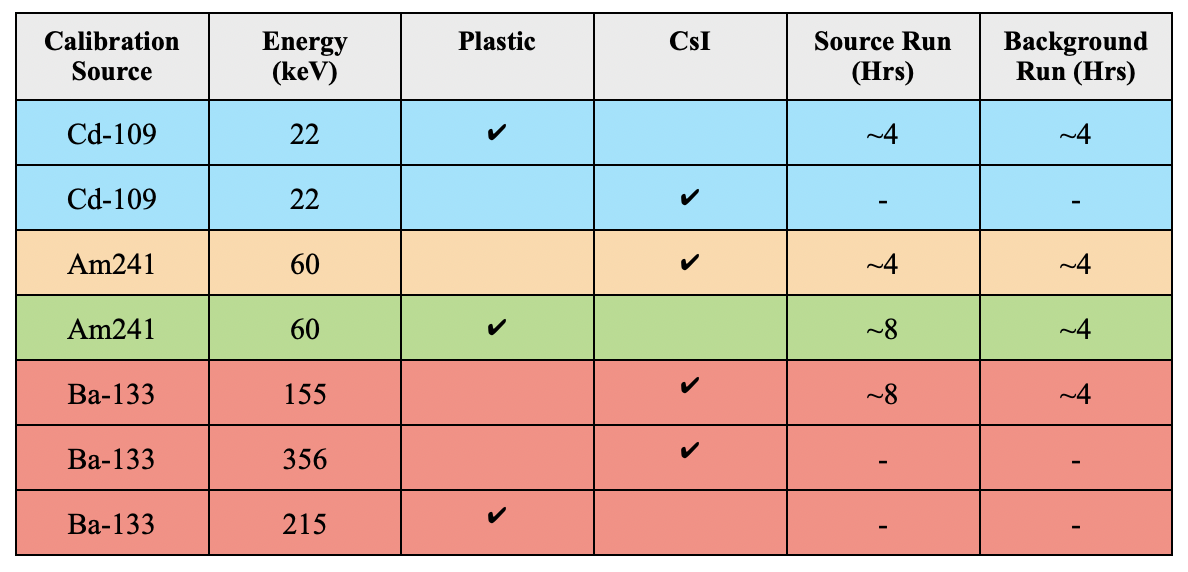}
  \end{center}
\end{table}


\begin{figure}[hbtp]
 \centering
\begin{subfigure}[b]{0.35\textwidth}
 		 \includegraphics[width=1\linewidth]{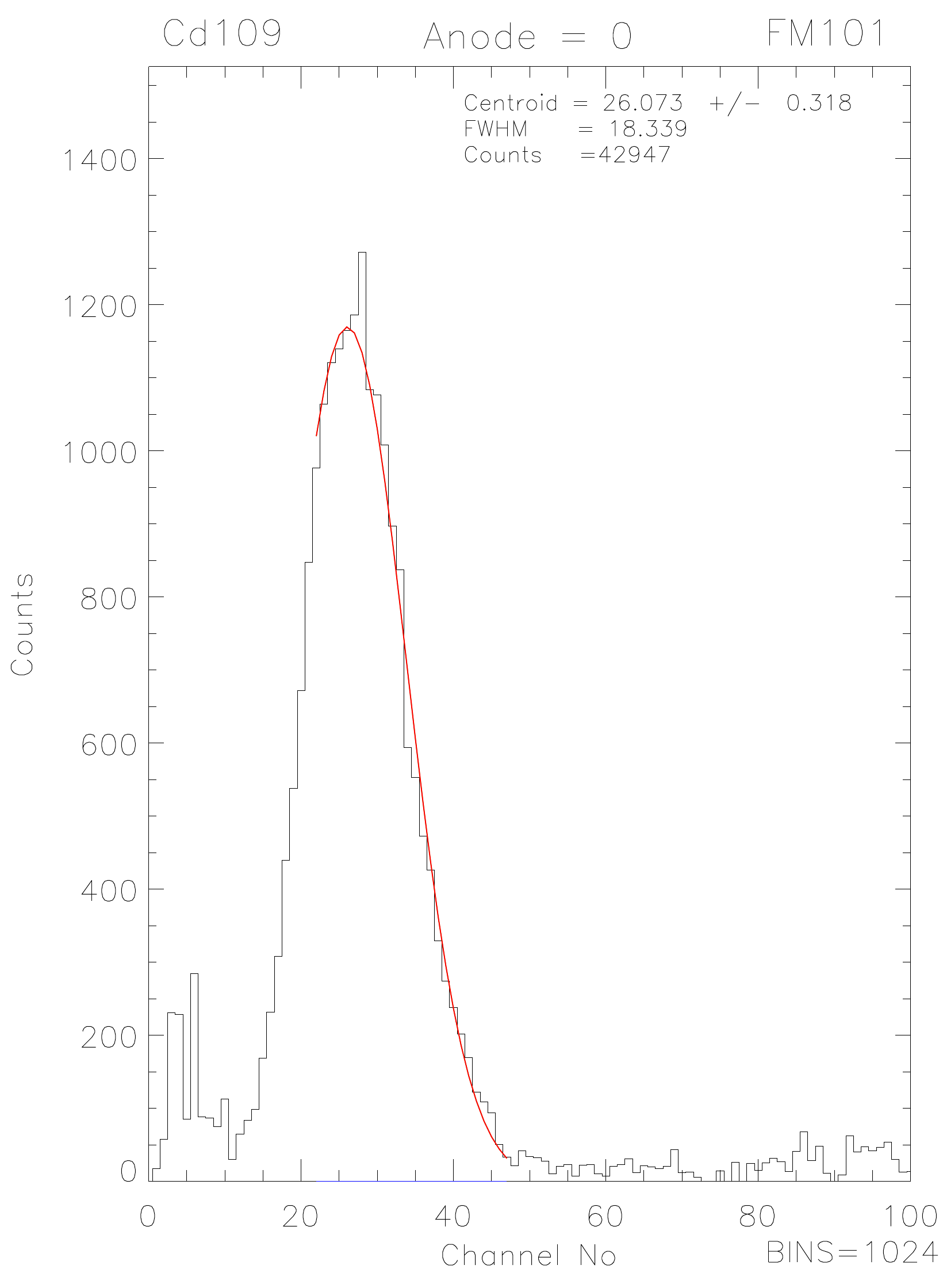}
		 \caption{}
\end{subfigure} \; \; \; 
\begin{subfigure}[b]{0.35\textwidth}
 		 \includegraphics[width=1\linewidth]{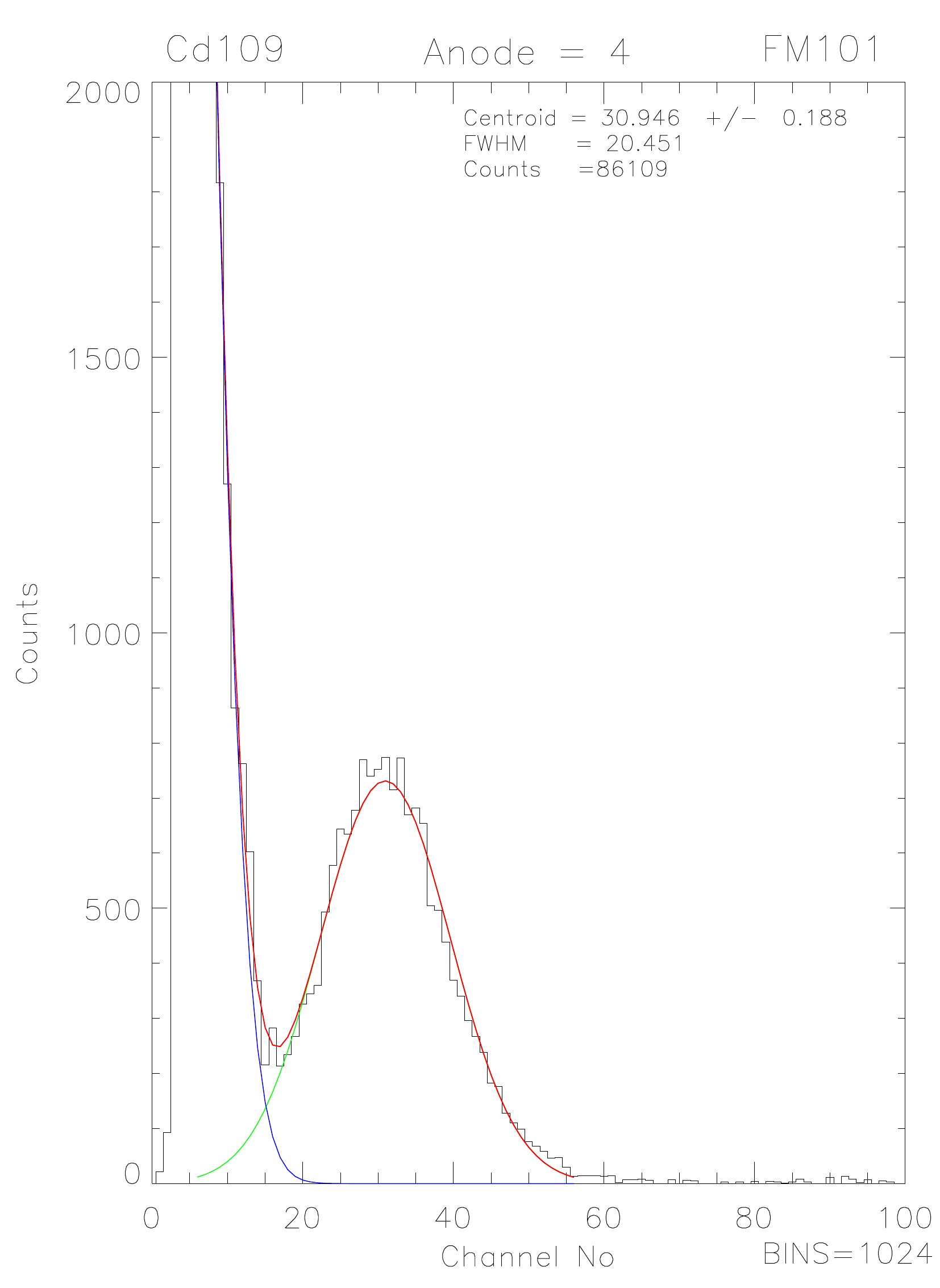}
		 \caption{}
\end{subfigure} 

\begin{subfigure}[b]{0.35\textwidth}
 		 \includegraphics[width=1\linewidth]{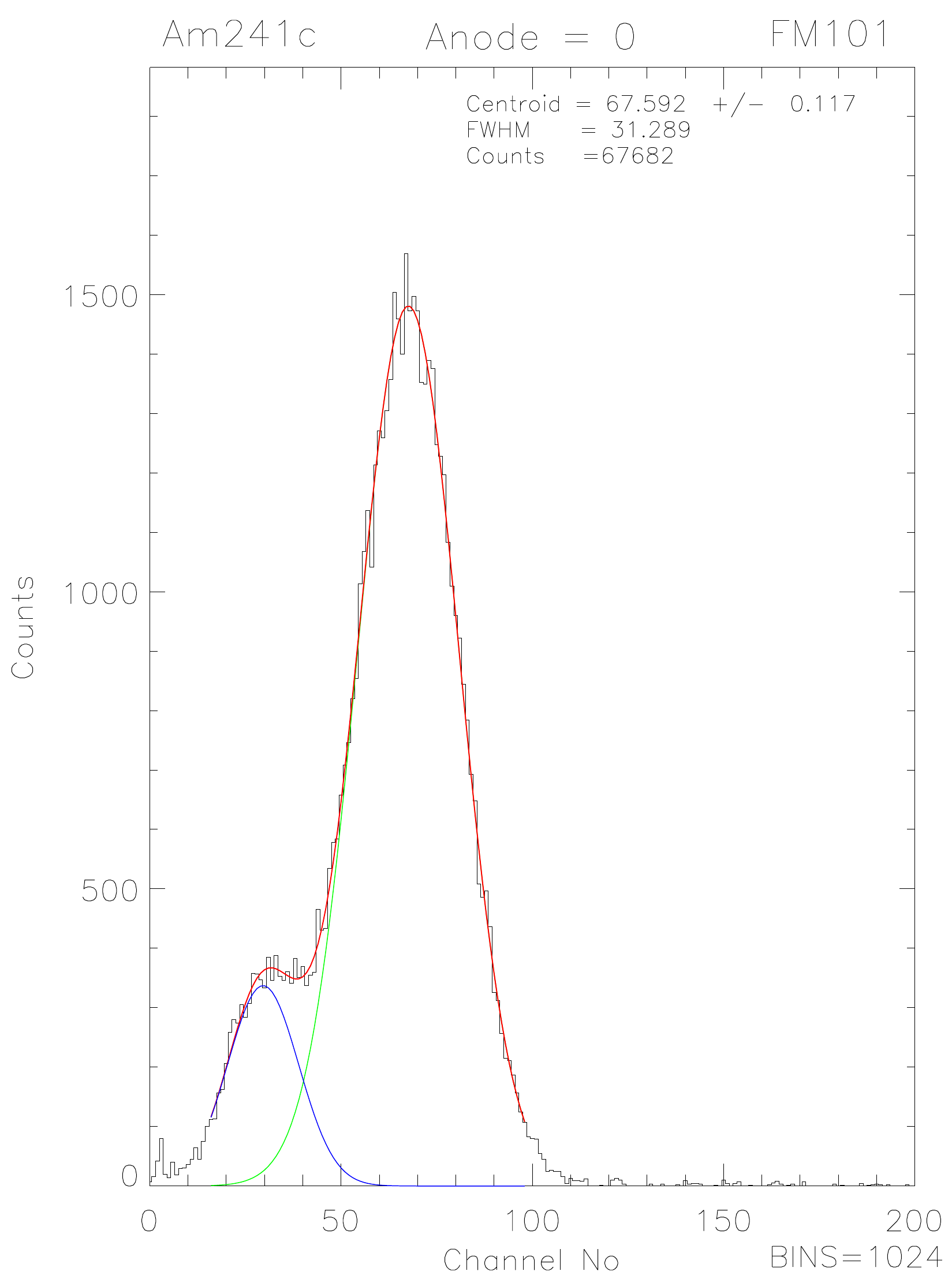}
		 \caption{}
\end{subfigure} \; \; \;
 \begin{subfigure}[b]{0.35\textwidth}
 		 \includegraphics[width=1\linewidth]{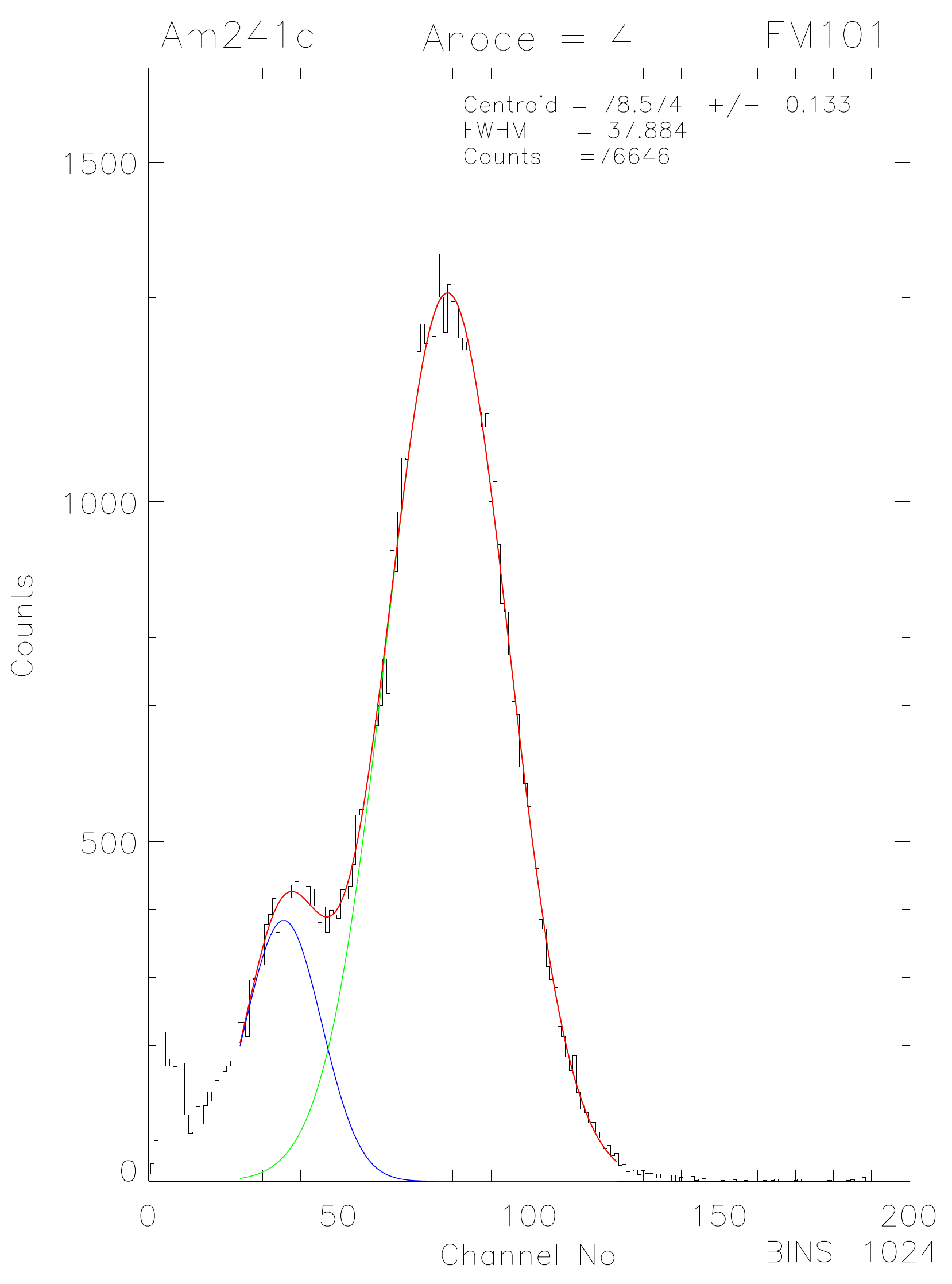}
		 \caption{}
\end{subfigure} 

  \caption{Calibration spectra for two different calorimeters. (a) and (b) are fits for $^{109}$Cd 22 keV peak for these calorimeters. (c) and (d) are fits for the $^{241}$Am 60 keV peaks.}
\label{fig:sim_cal_fit_cal_1}

\end{figure}

\begin{figure}[hbtp]
 \centering

\begin{subfigure}[b]{0.36\textwidth}
 		 \includegraphics[width=1\linewidth]{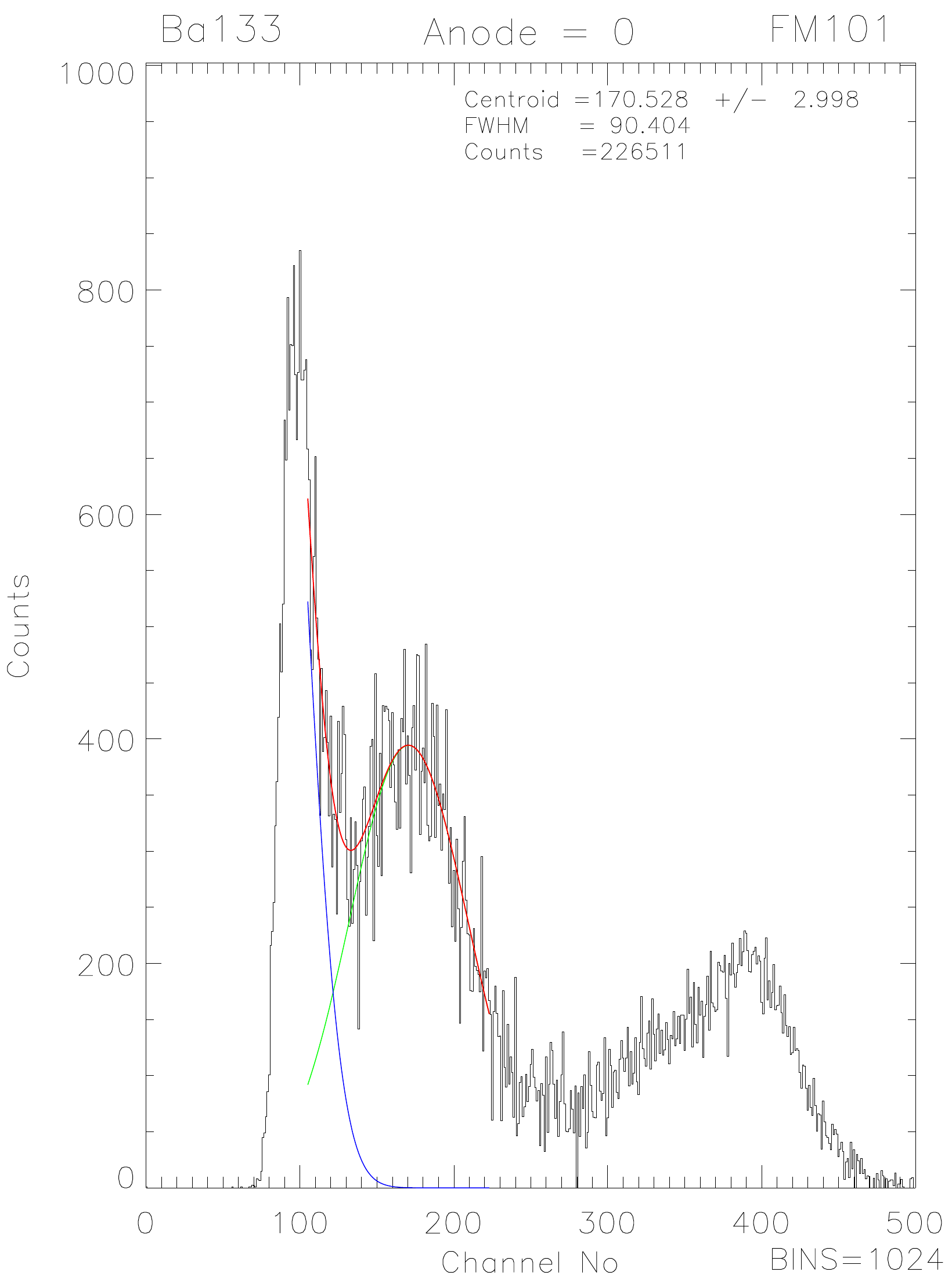}
		 \caption{}
\end{subfigure} \; \; \; 
\begin{subfigure}[b]{0.35\textwidth}
 		 \includegraphics[width=1\linewidth]{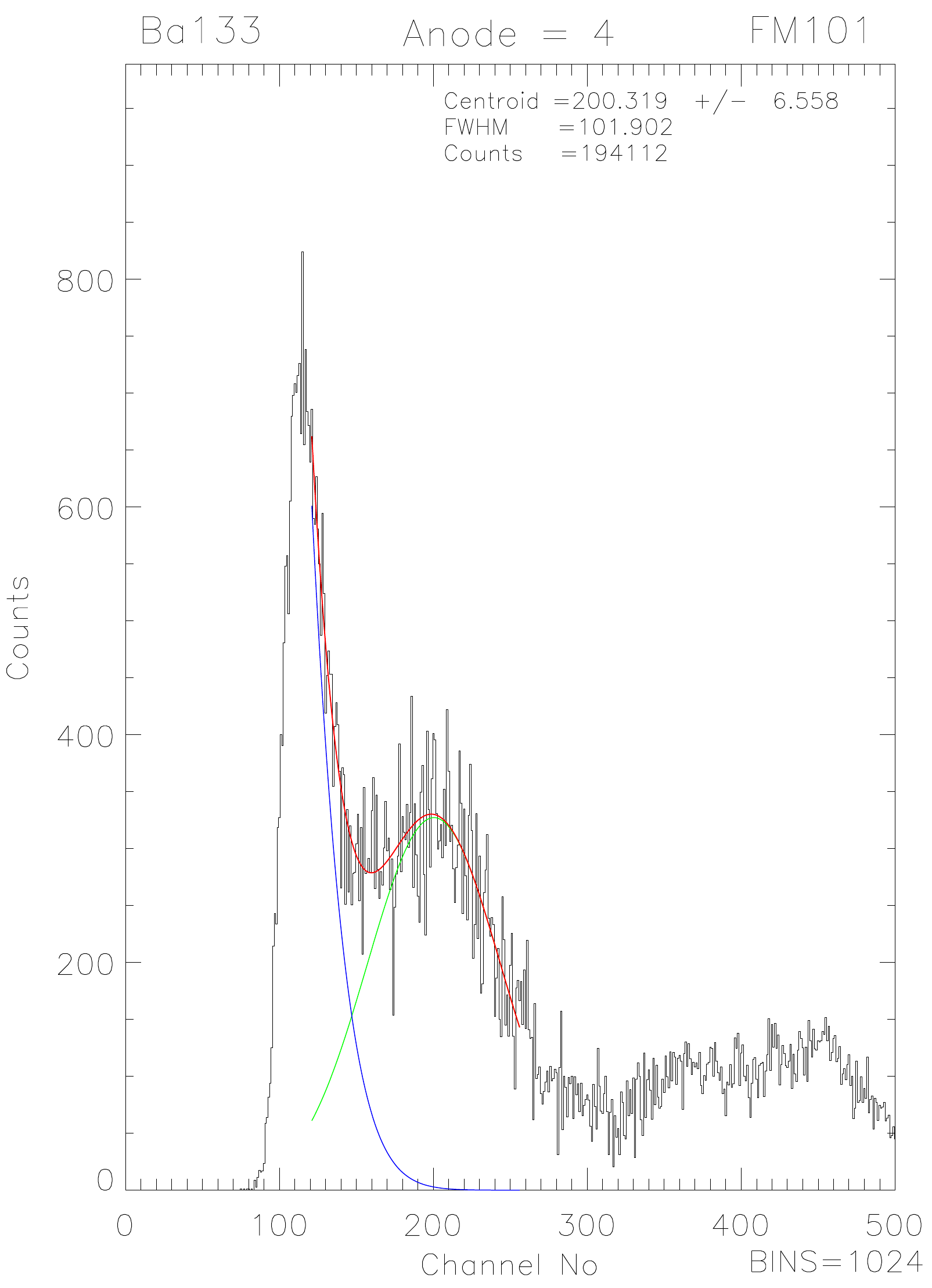}
		 \caption{}
\end{subfigure} 
 
\begin{subfigure}[b]{0.36\textwidth}
 		 \includegraphics[width=1\linewidth]{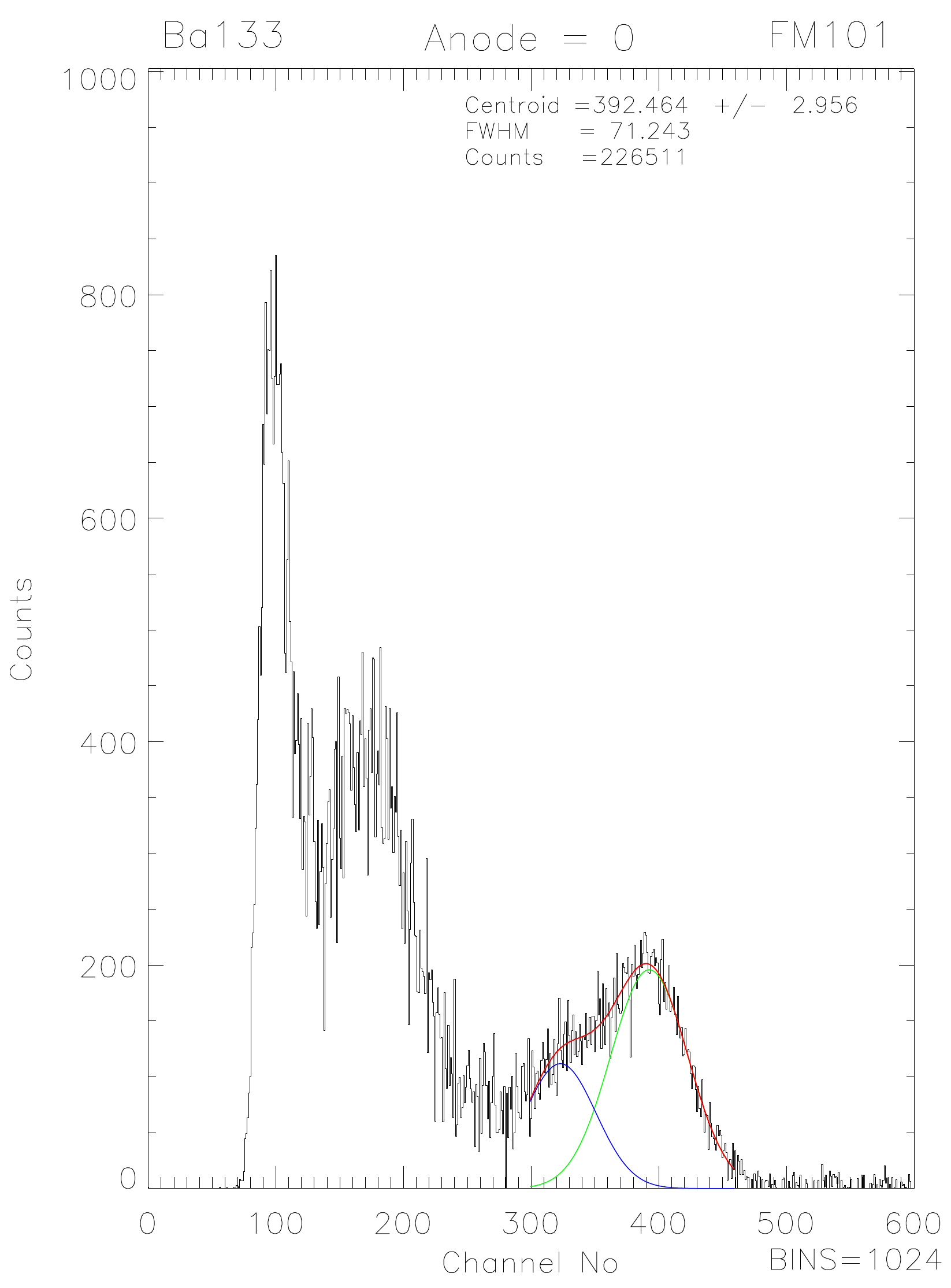}
		 \caption{}
\end{subfigure} \; \; \;  
\begin{subfigure}[b]{0.35\textwidth}
 		 \includegraphics[width=1\linewidth]{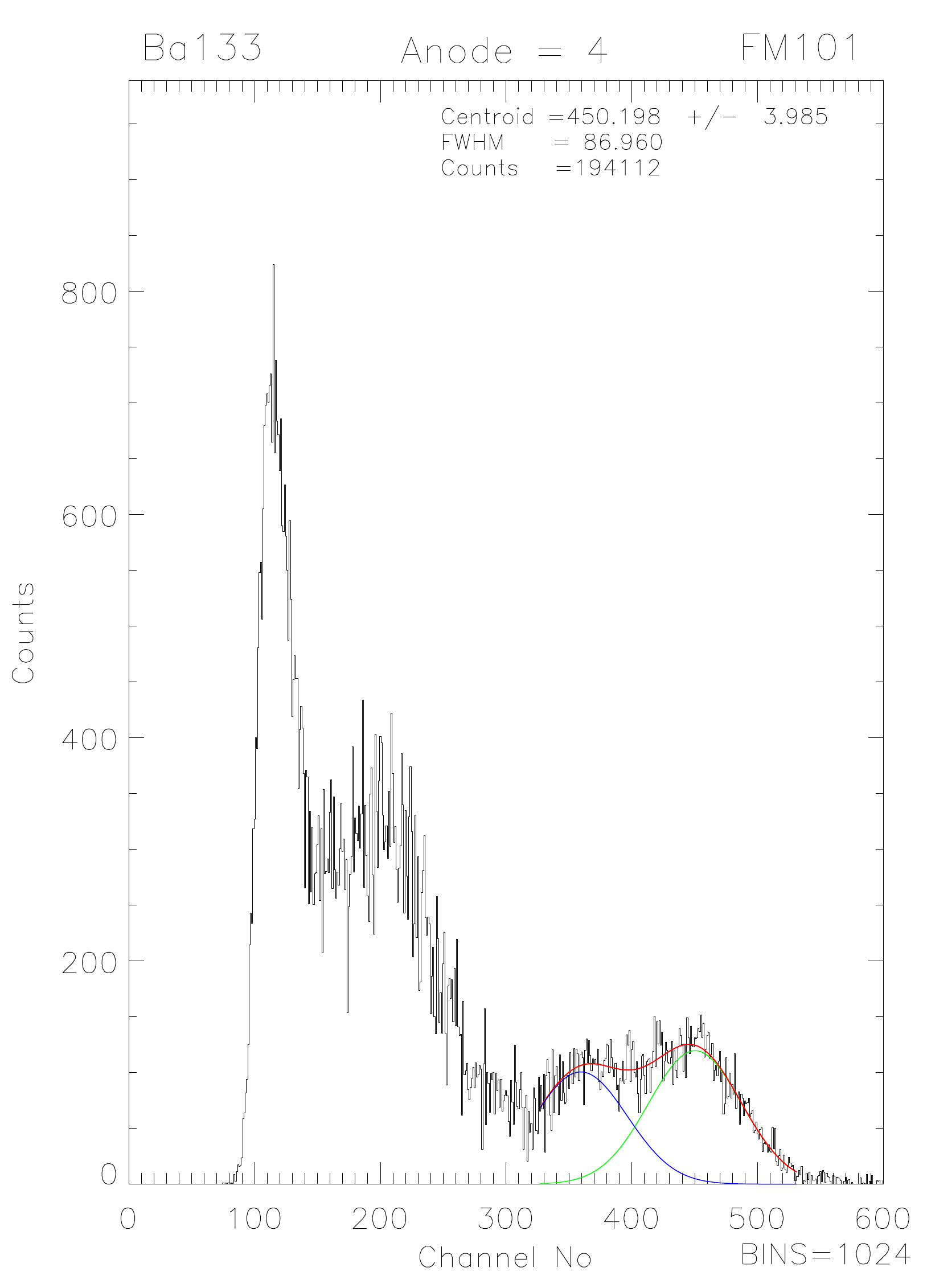}
		 \caption{}
\end{subfigure} 

  \caption{Additional calibration spectra for the two calorimeters. (a) and (b) are fits for$^{133}$Ba 155 keV peak for these calorimeters. (c) and (d) are fits for the $^{133}$Ba 356 keV peaks.}
\label{fig:sim_cal_fit_cal_2}

\end{figure}	
					
					\subsubsection*{\underline{Calorimeters}}
					\label{sec:ins_perf_calibration_calorimeters}
					
\begin{figure}[tbp]
 \centering
\begin{subfigure}[b]{0.35\textwidth}
 		 \includegraphics[width=1\linewidth]{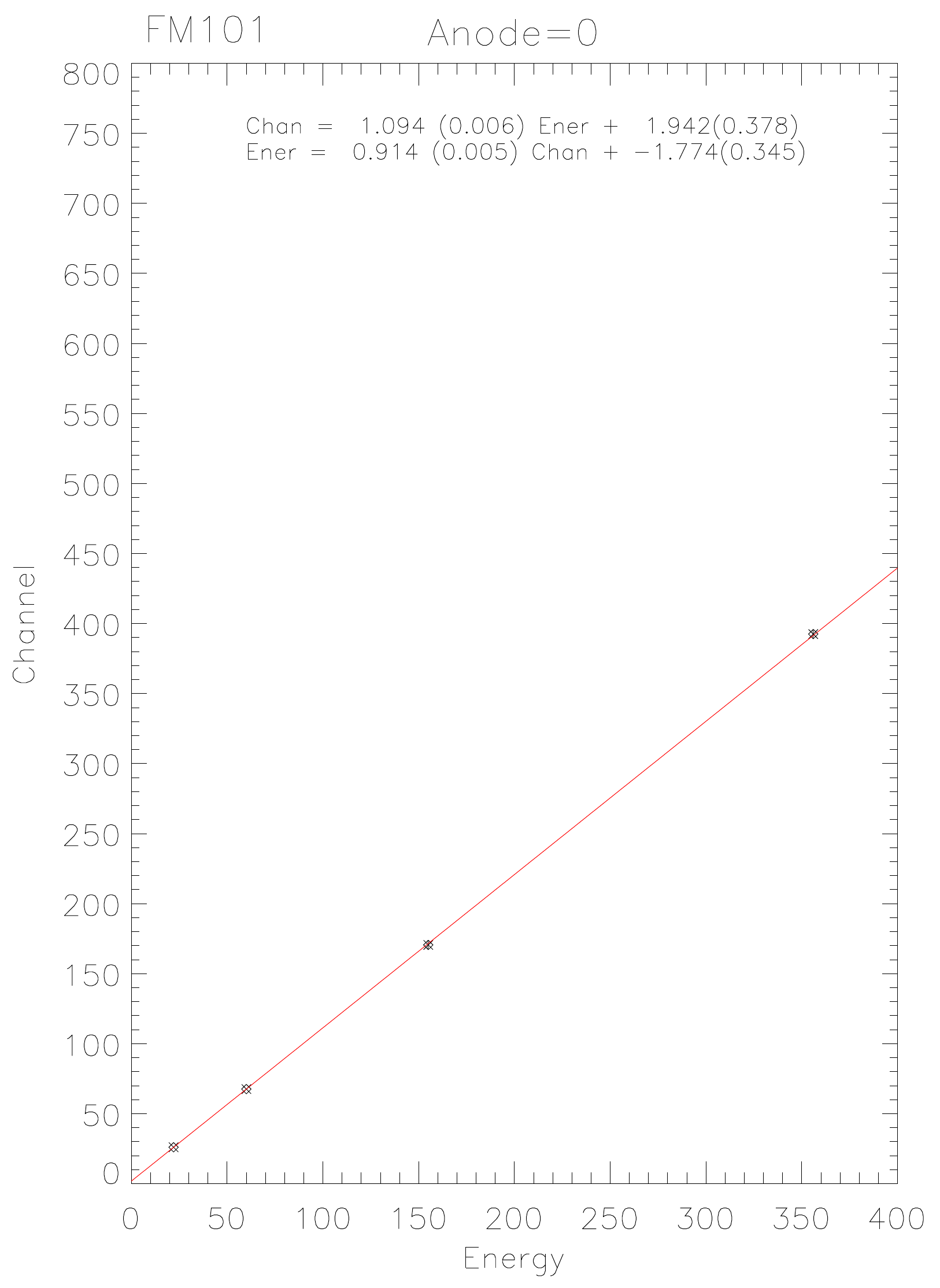}
		 \caption{}
\end{subfigure} 
\begin{subfigure}[b]{0.35\textwidth}
 		 \includegraphics[width=1\linewidth]{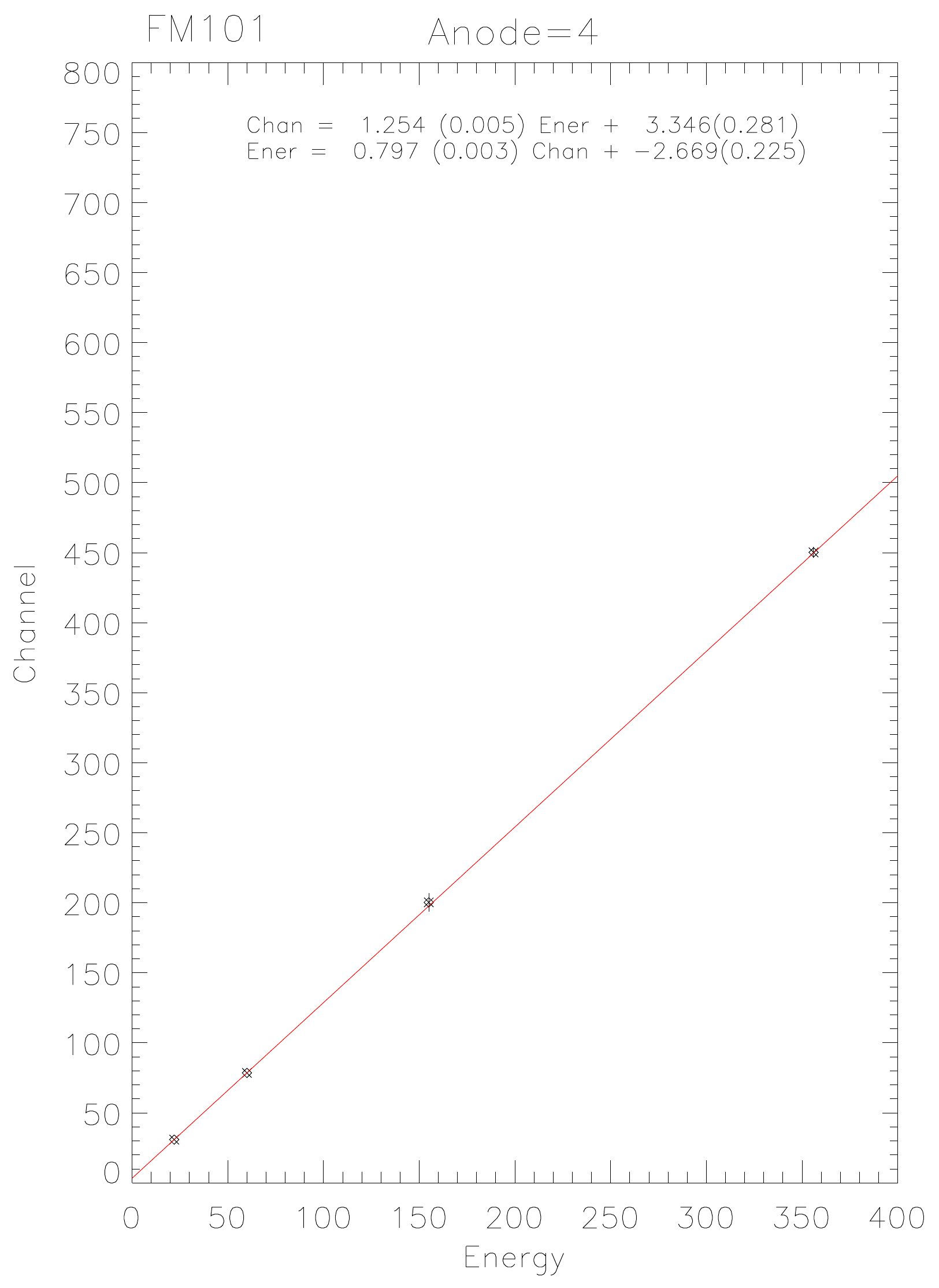}
		 \caption{}
\end{subfigure}

  \caption{Calibration curves for two different calorimeter elements.}
\label{fig:sim_cal_fit_cal_nrg}

\end{figure}

					Examples of background subtracted calibration spectra are shown in Figure \ref{fig:sim_cal_fit_cal_1} and Figure \ref{fig:sim_cal_fit_cal_2}.
					We see the $^{109}$Cd 22 keV,  $^{241}$Am 60 keV and $^{133}$Ba 155 keV and 356 keV fits for these anodes. 
					The resulting calibration curves for the two anodes are shown in Figure \ref{fig:sim_cal_fit_cal_nrg}. 
					These conversion parameters for each individual anodes are stored in a calibration file. 
					We generate one file for each individual module which has the calibration information for all 64 anodes (calorimeters and plastics) of that module.
					Any calorimeters that had less than two calibration point was flagged and the events associated with those anodes were rejected during processing. 
					There were only two calorimeter anodes (out of all the calorimeters in GRAPE) that were rejected.

					\subsubsection*{\underline{Plastics}}
					\label{sec:ins_perf_calibration_plastics}
						
\begin{figure}[hbtp]
 \centering
\begin{subfigure}[b]{0.35\textwidth}
 		 \includegraphics[width=1\linewidth]{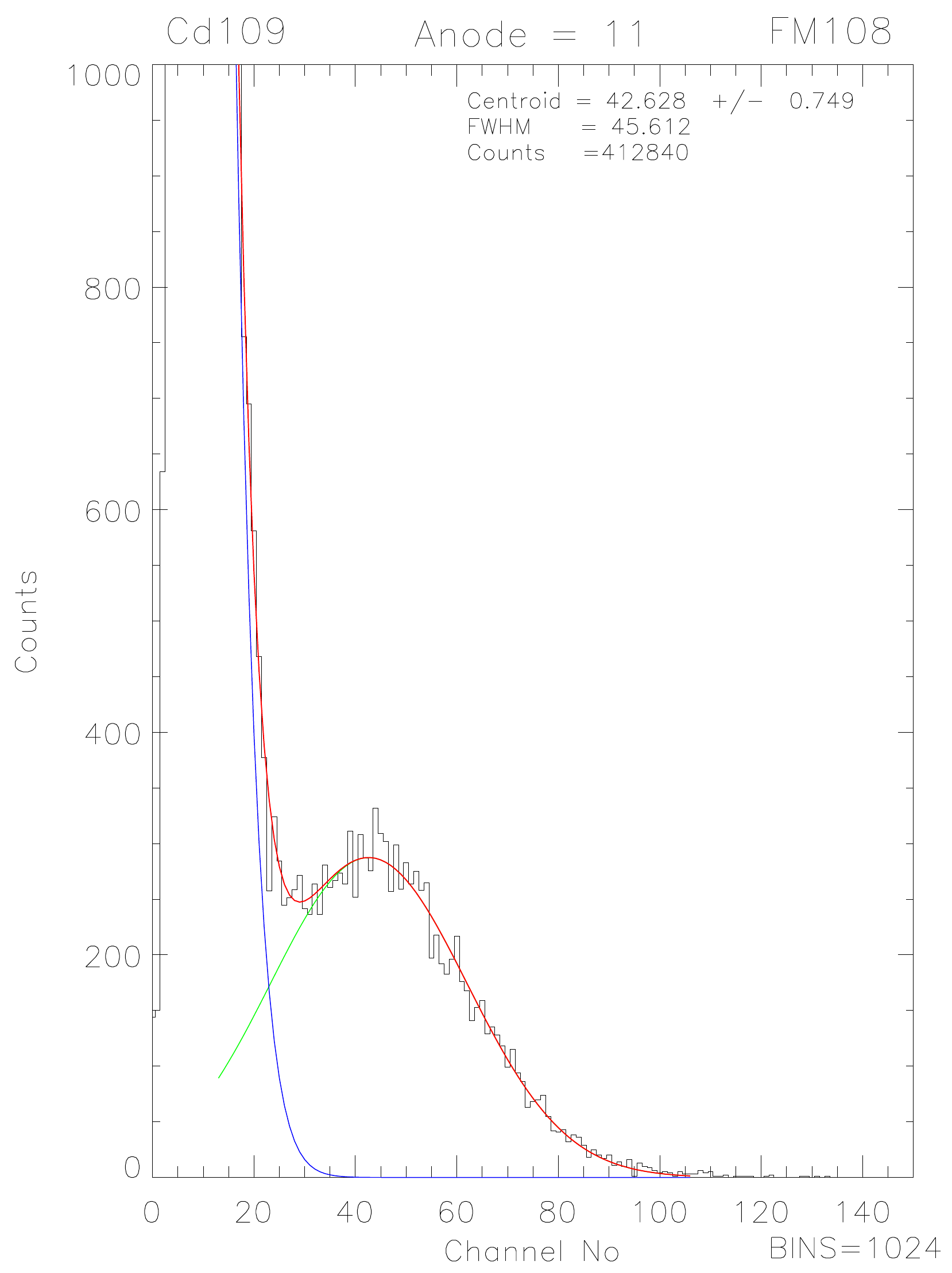}
		 \caption{}
\end{subfigure}  \; \; \; 
 \begin{subfigure}[b]{0.34\textwidth}
 		 \includegraphics[width=1\linewidth]{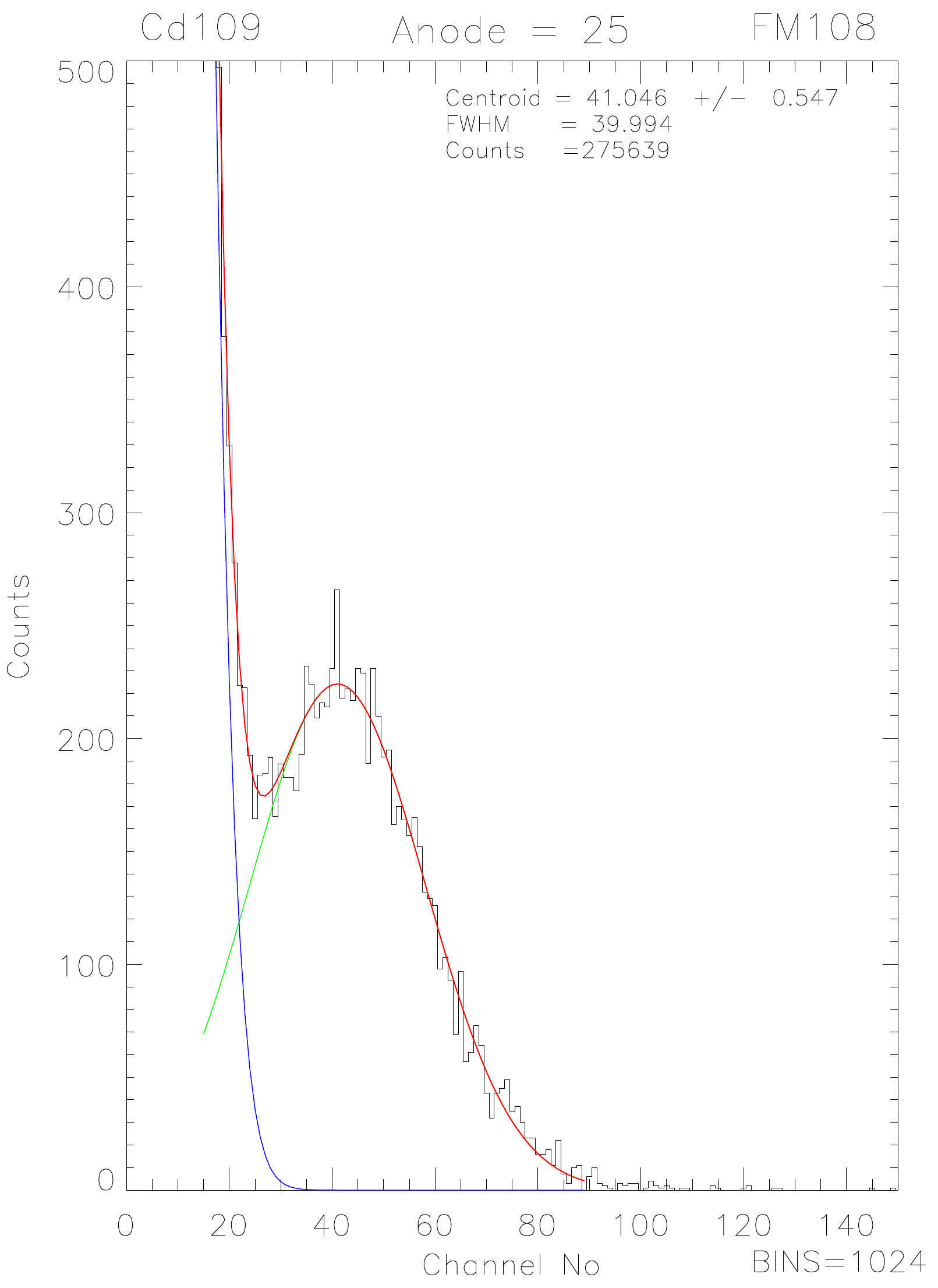}
		 \caption{}
\end{subfigure} 

\begin{subfigure}[b]{0.35\textwidth}
 		 \includegraphics[width=1\linewidth]{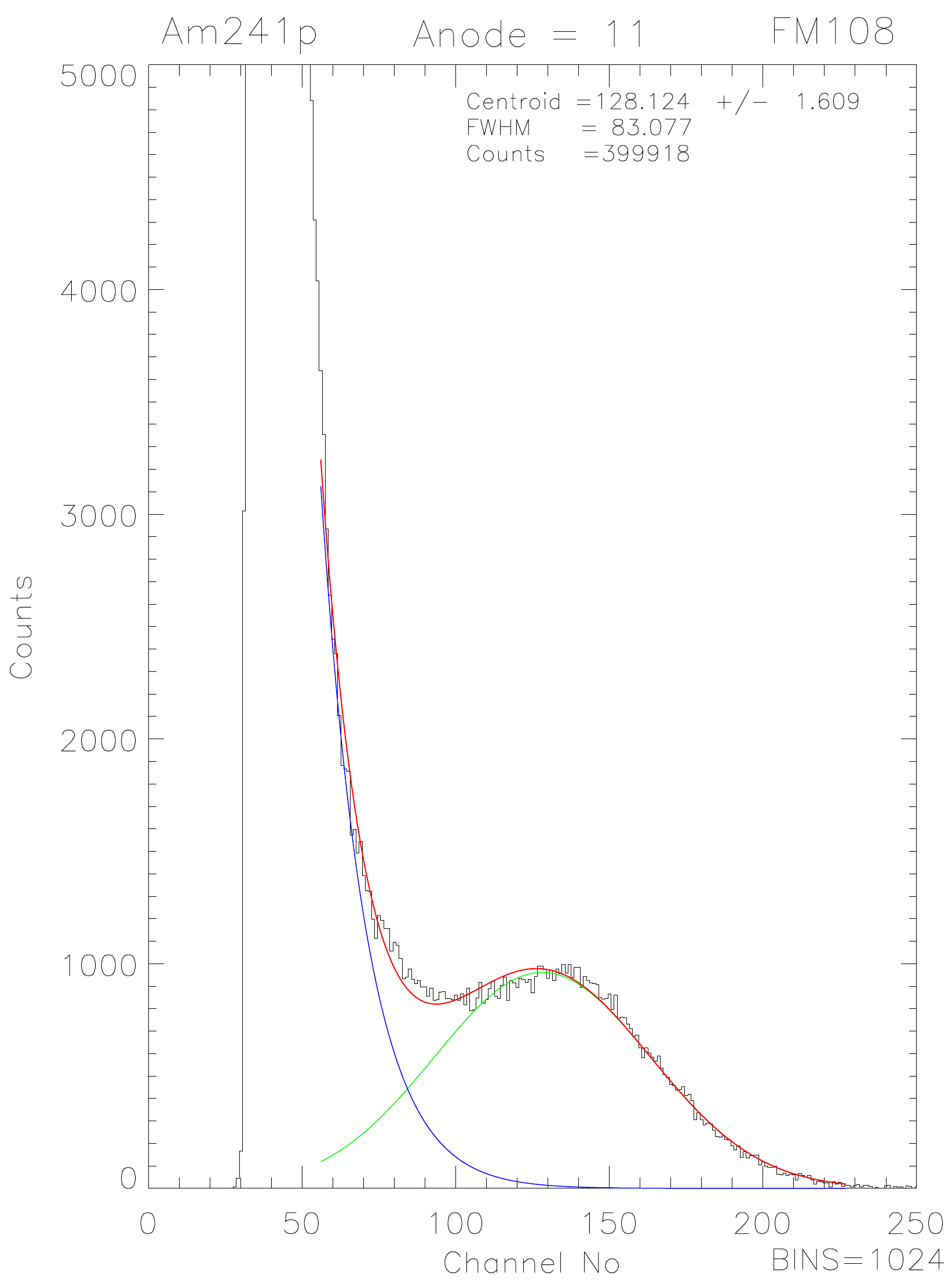}
		 \caption{}
\end{subfigure} \; \; \;
 \begin{subfigure}[b]{0.35\textwidth}
 		 \includegraphics[width=1\linewidth]{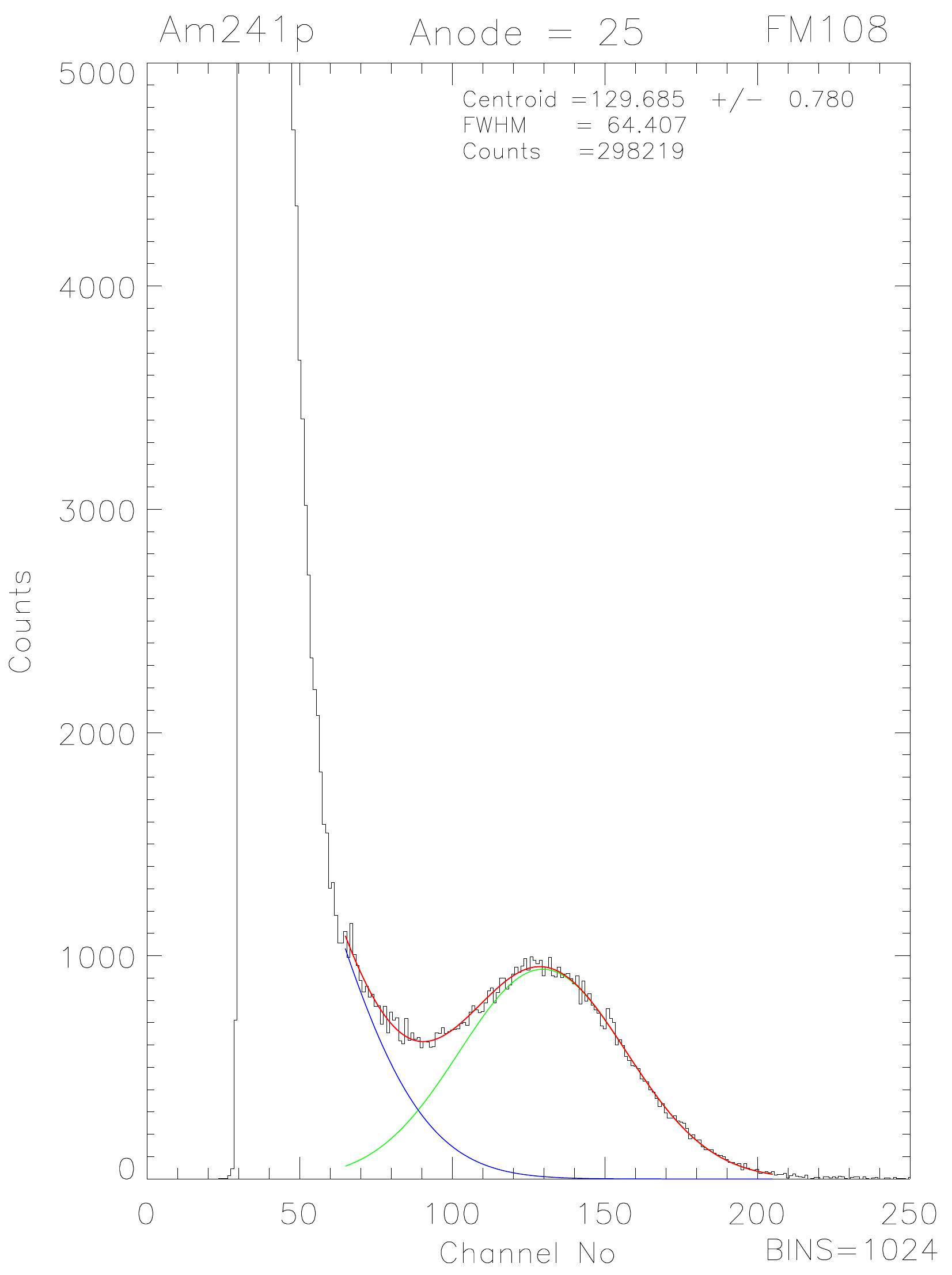}
		 \caption{}
\end{subfigure}  

  \caption{Calibration spectra for two different plastics. (a) and (b) are fits for $^{109}$Cd 22 keV peak for these calorimeters. (c) and (d) are fits for the $^{241}$Am 60 keV peaks.}
\label{fig:sim_cal_fit_pla_1}

\end{figure}

\begin{figure}[hbtp]
 \centering

\begin{subfigure}[b]{0.36\textwidth}
 		 \includegraphics[width=1\linewidth]{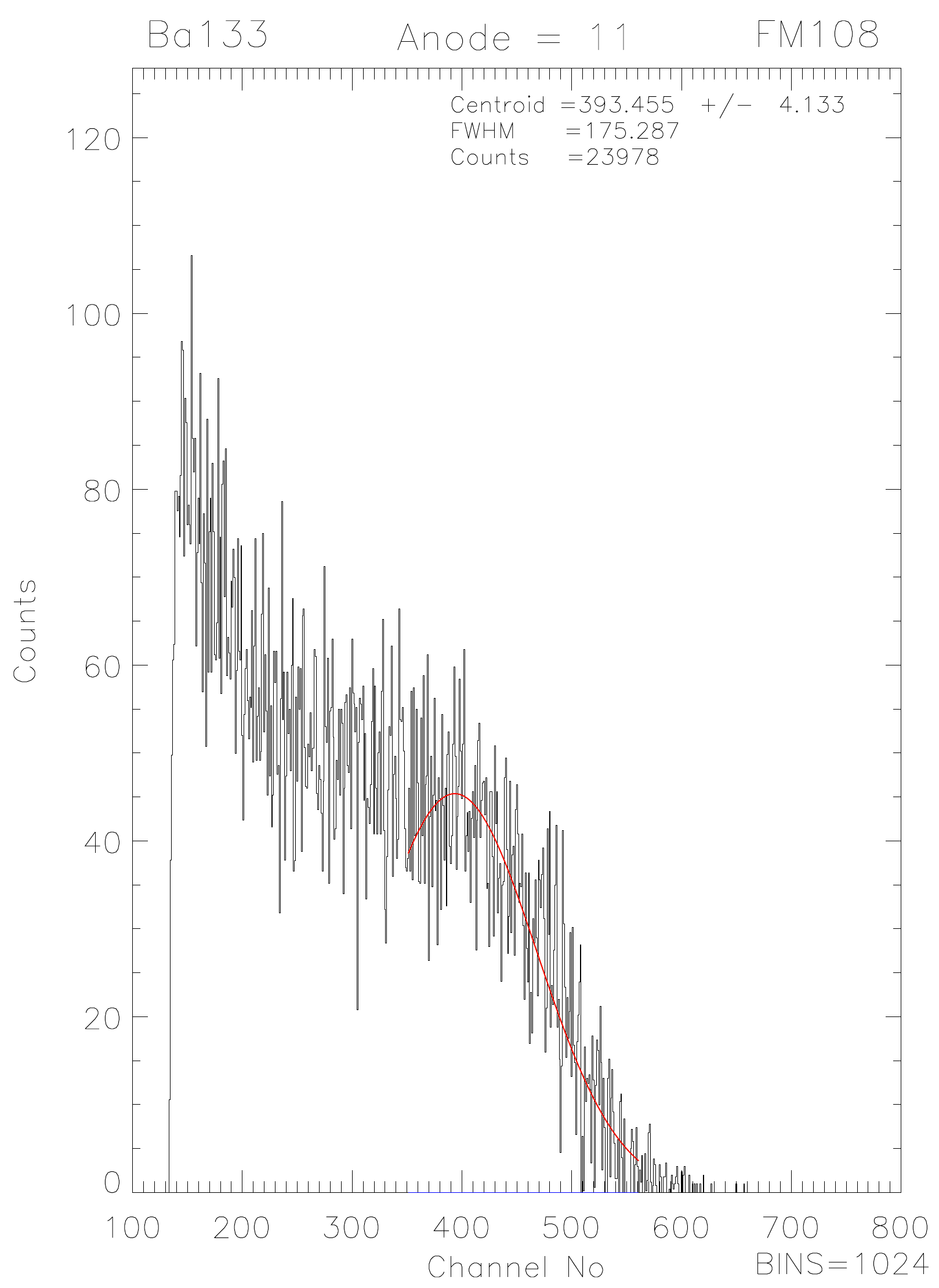}
		 \caption{}
\end{subfigure} \; \; \; 
\begin{subfigure}[b]{0.36\textwidth}
 		 \includegraphics[width=1\linewidth]{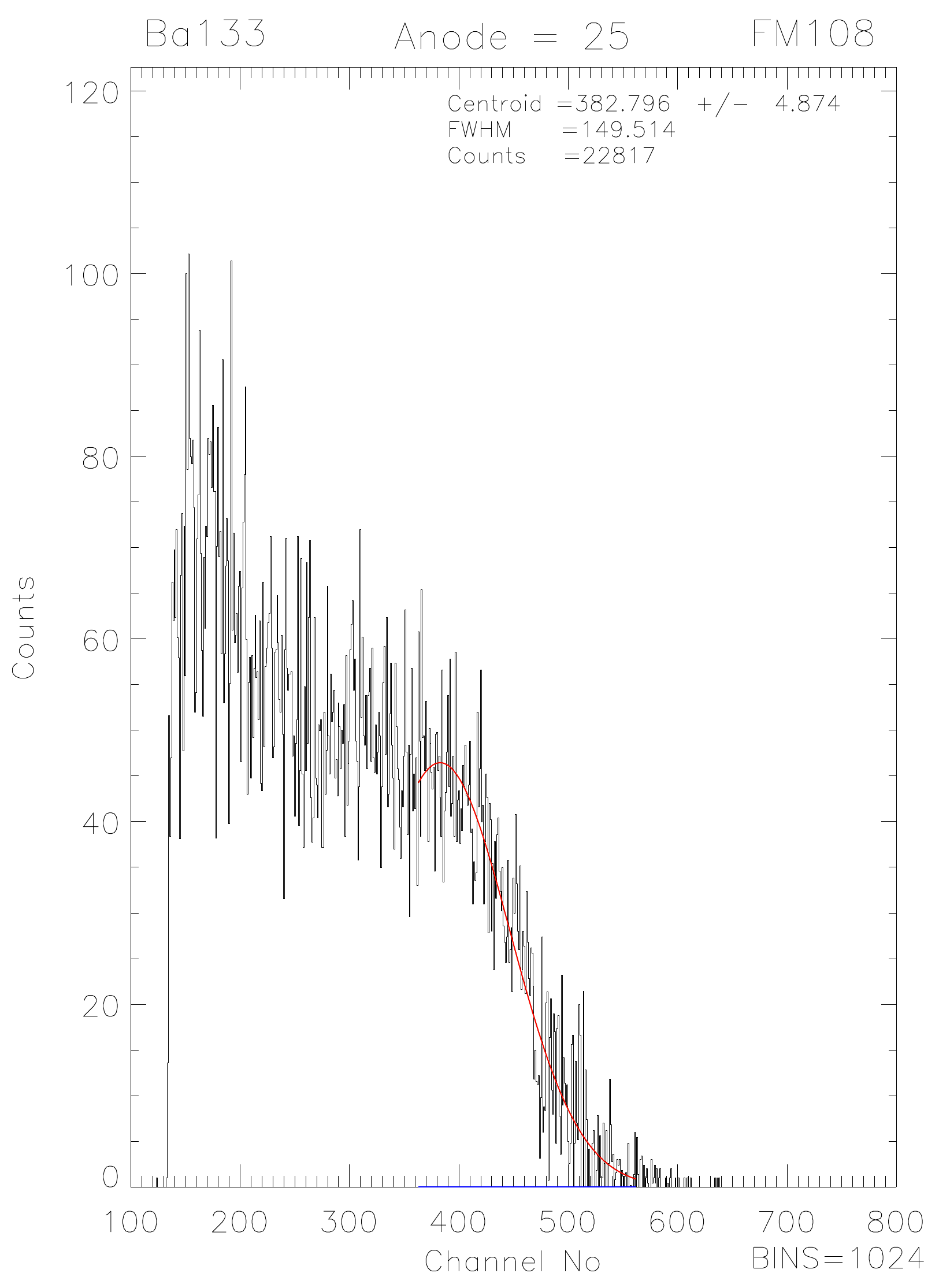}
		 \caption{}
\end{subfigure} 

  \caption{Calibration spectra for two different plastics. The Compton edge of $^{133}$Ba is fitted for both he plastics. }
\label{fig:sim_cal_fit_pla_2}

\end{figure}

\begin{figure}[hbtp]
 \centering
\begin{subfigure}[b]{0.35\textwidth}
 		 \includegraphics[width=1\linewidth]{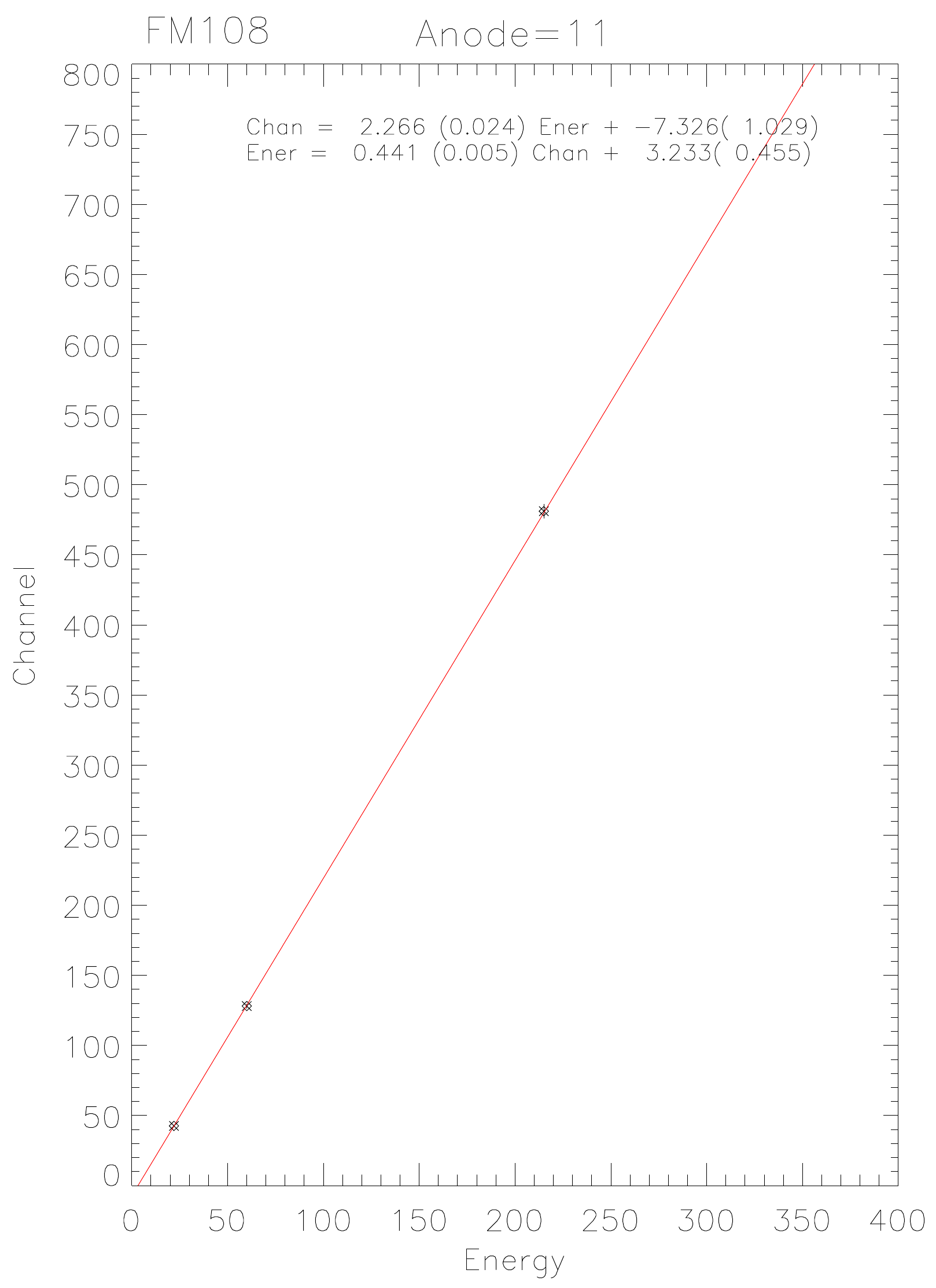}
		 \caption{}
\end{subfigure} \; \; \;  
\begin{subfigure}[b]{0.35\textwidth}
 		 \includegraphics[width=1\linewidth]{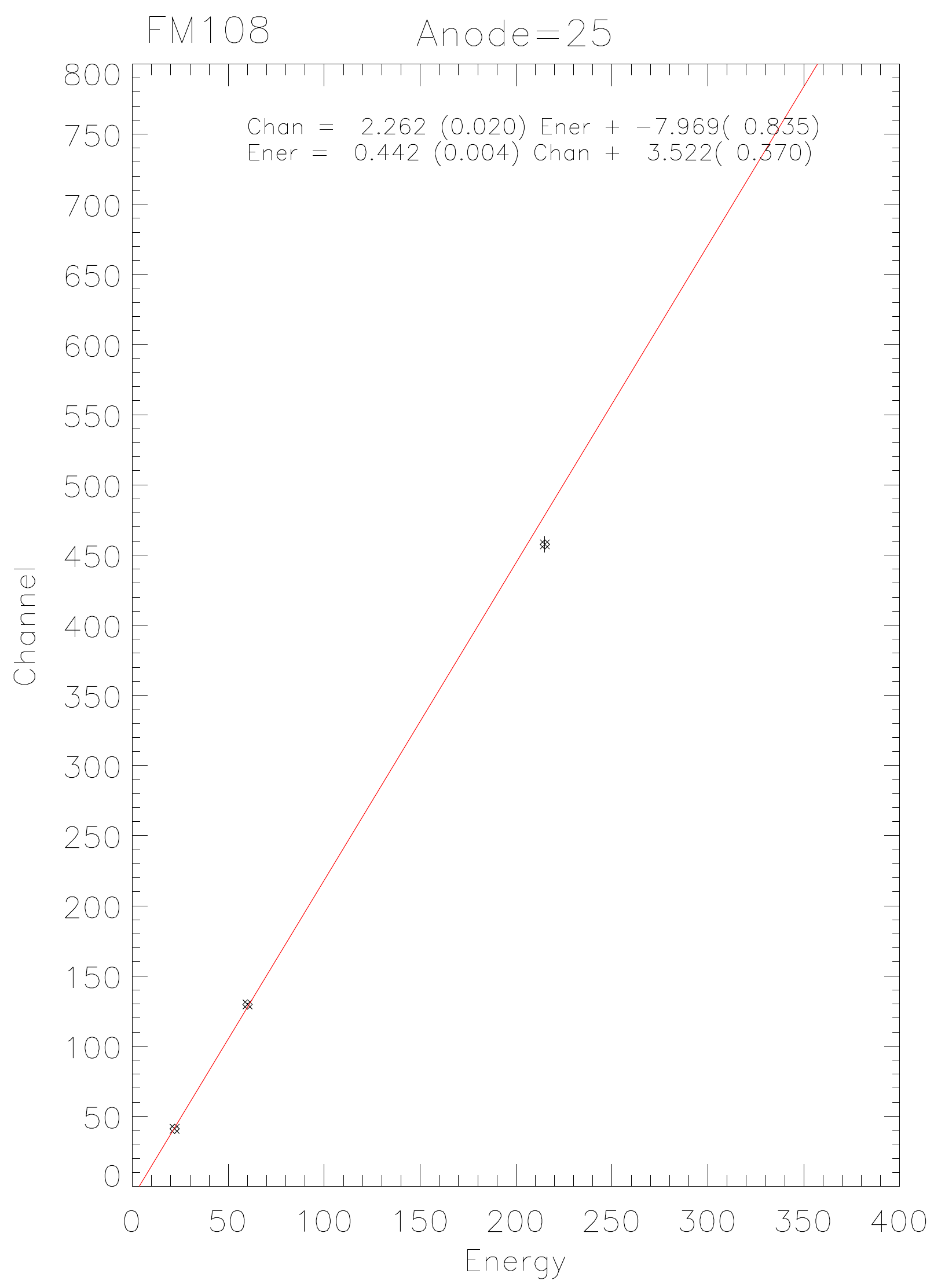}
		 \caption{}
\end{subfigure} 
  \caption{Calibration curves for two different plastic elements.}
\label{fig:sim_cal_fit_pla_nrg}

\end{figure}

					Calibration spectra for two plastic elements are shown  in Figure \ref{fig:sim_cal_fit_pla_1} and Figure \ref{fig:sim_cal_fit_pla_2}. Figure \ref{fig:sim_cal_fit_pla_nrg} shows the energy calibration of the two plastic anodes.
					 Plastic elements with less than two calibration points were  flagged and the events associated with them were excluded from analysis.
					 6 Plastic elements were rejected due to lack of calibration points.

					\subsubsection*{Threshold and Rates}
					\label{sec:ins_perf_rates}
\begin{figure}[tbp]
 \centering

\begin{subfigure}[b]{0.35\textwidth}
 		 \includegraphics[width=1\linewidth]{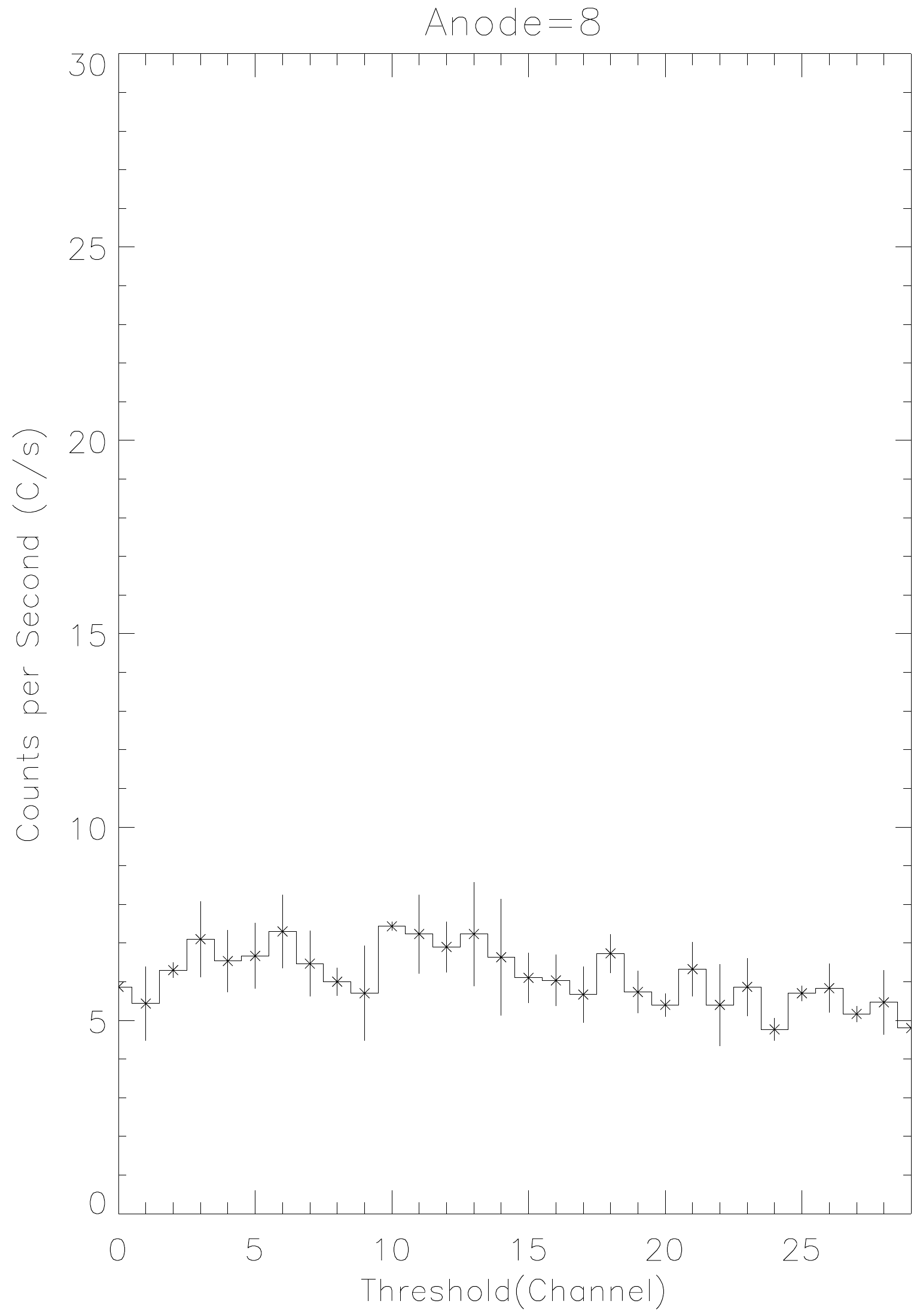}
		 \caption{}
\end{subfigure} \; \; \;   
\begin{subfigure}[b]{0.35\textwidth}
 		 \includegraphics[width=1\linewidth]{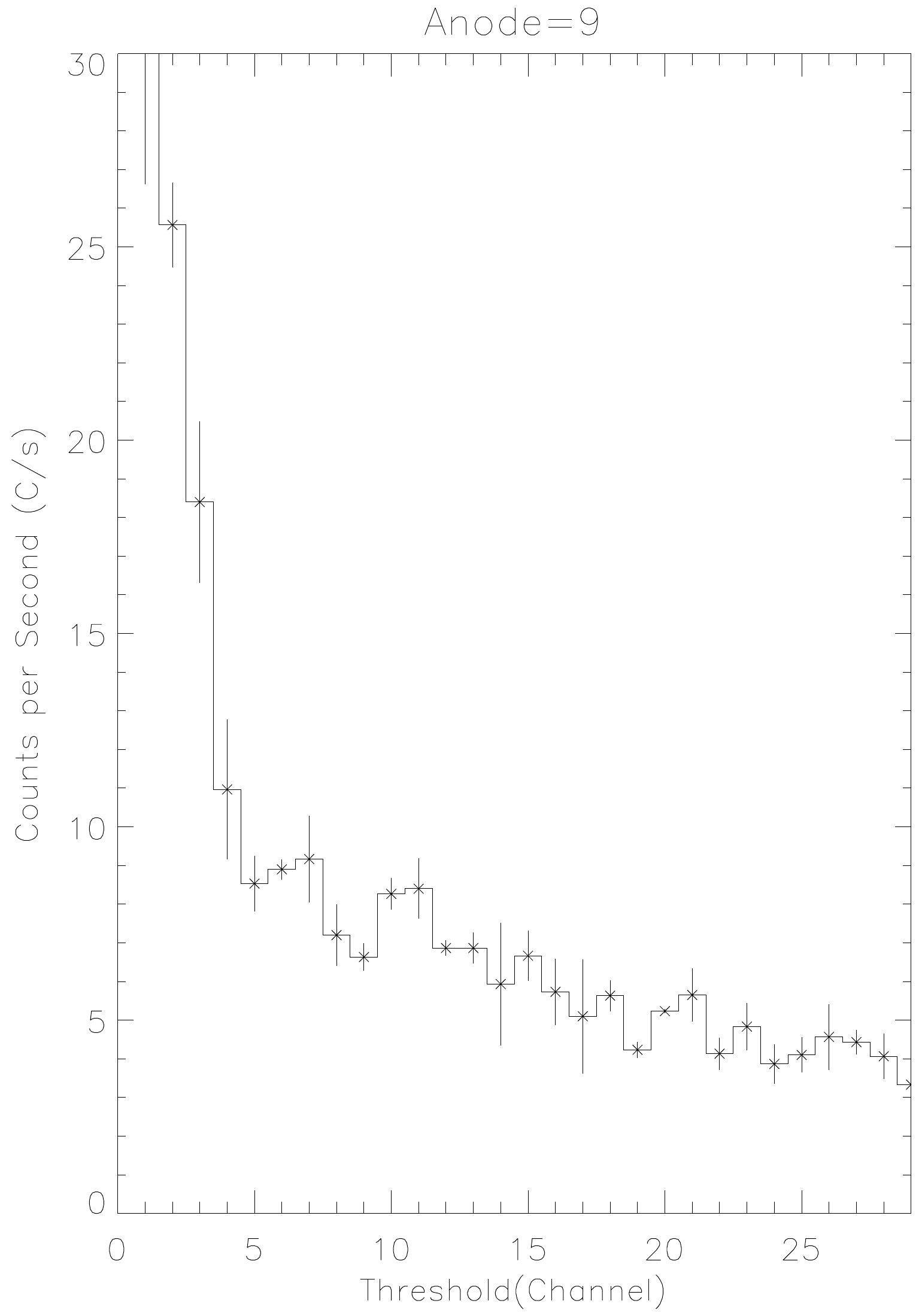}
		 \caption{}
\end{subfigure} 

  \caption{Plot of counts versus threshold rates incrementing threshold values for (a) calorimeter and (b) plastics. These were used to determine a threshold value that would avoid the electrical noise. For plastics, the electrical noise was much more visible as seen in (b).  }
\label{fig:sim_cal_rates}

\end{figure}

					The threshold for each channel was set just above the noise level based on plots of rate vs threshold channel. 
					Goal was to set the calorimeter level at 20 keV and plastic at 6 keV (or as low as possible). 
					Examples of these data can be seen in Figure \ref{fig:sim_cal_rates}, which shows data for both a calorimeter element and a plastic element. 
					For this calorimeters, the electronic noise was not significant so the threshold was set to channel 1. For the plastic, the threshold value of 9 was set. This is the channel number and the corresponding energy value is retrieved from the energy calibration file.  
					Typical values of threshold for calorimeter were around $\sim$20 keV. The plastic varied between 4 keV to 8 keV.

			\subsection{Instrument level calibration at UNH}
			\label{sec:ins_perf_unh}	

\begin{figure}[hbtp]
 \centering
\begin{subfigure}[b]{0.4\textwidth}
 		 \includegraphics[width=1\linewidth]{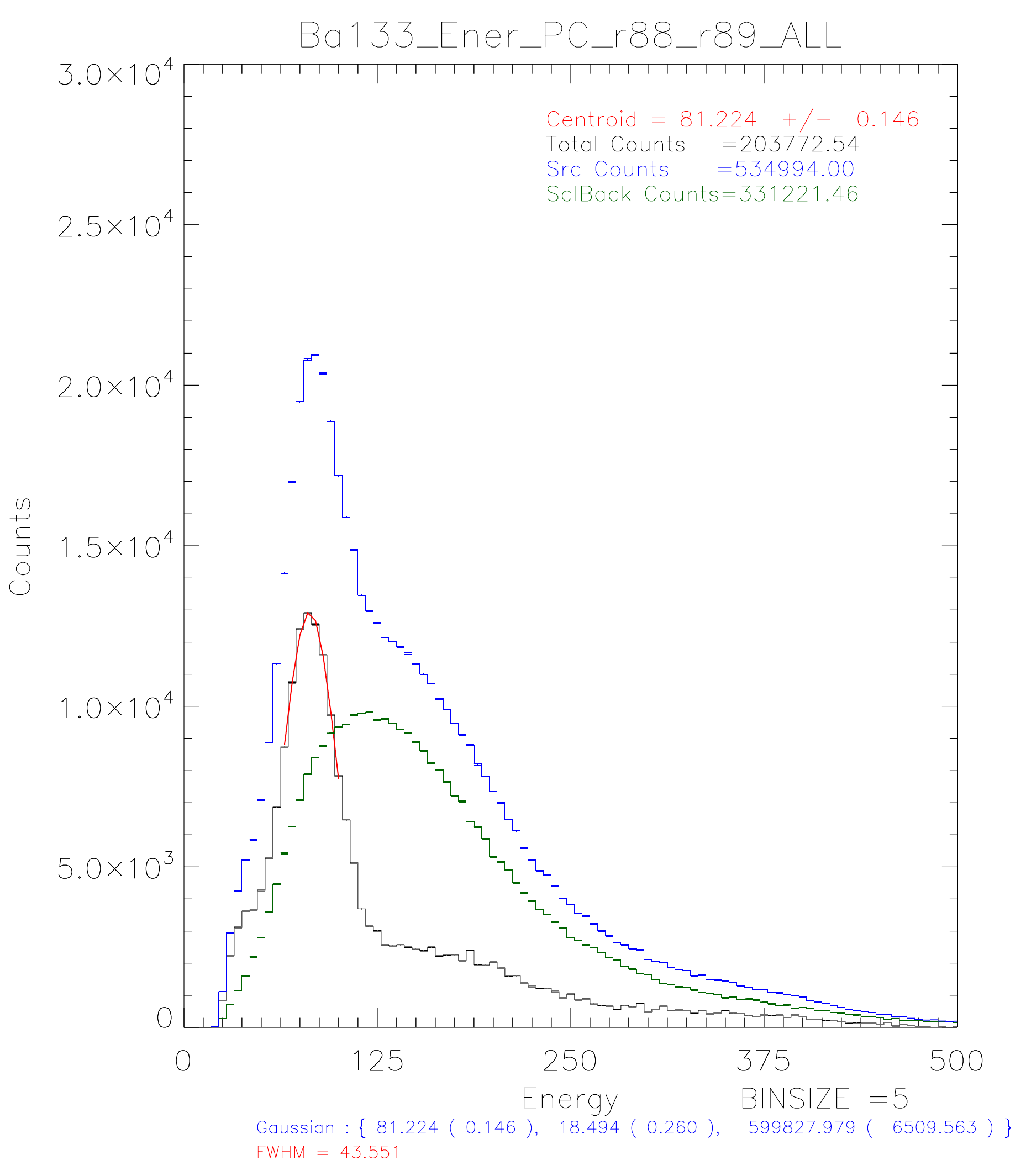}
		 \caption{}
\end{subfigure} 
\begin{subfigure}[b]{0.4\textwidth}
 		 \includegraphics[width=1\linewidth]{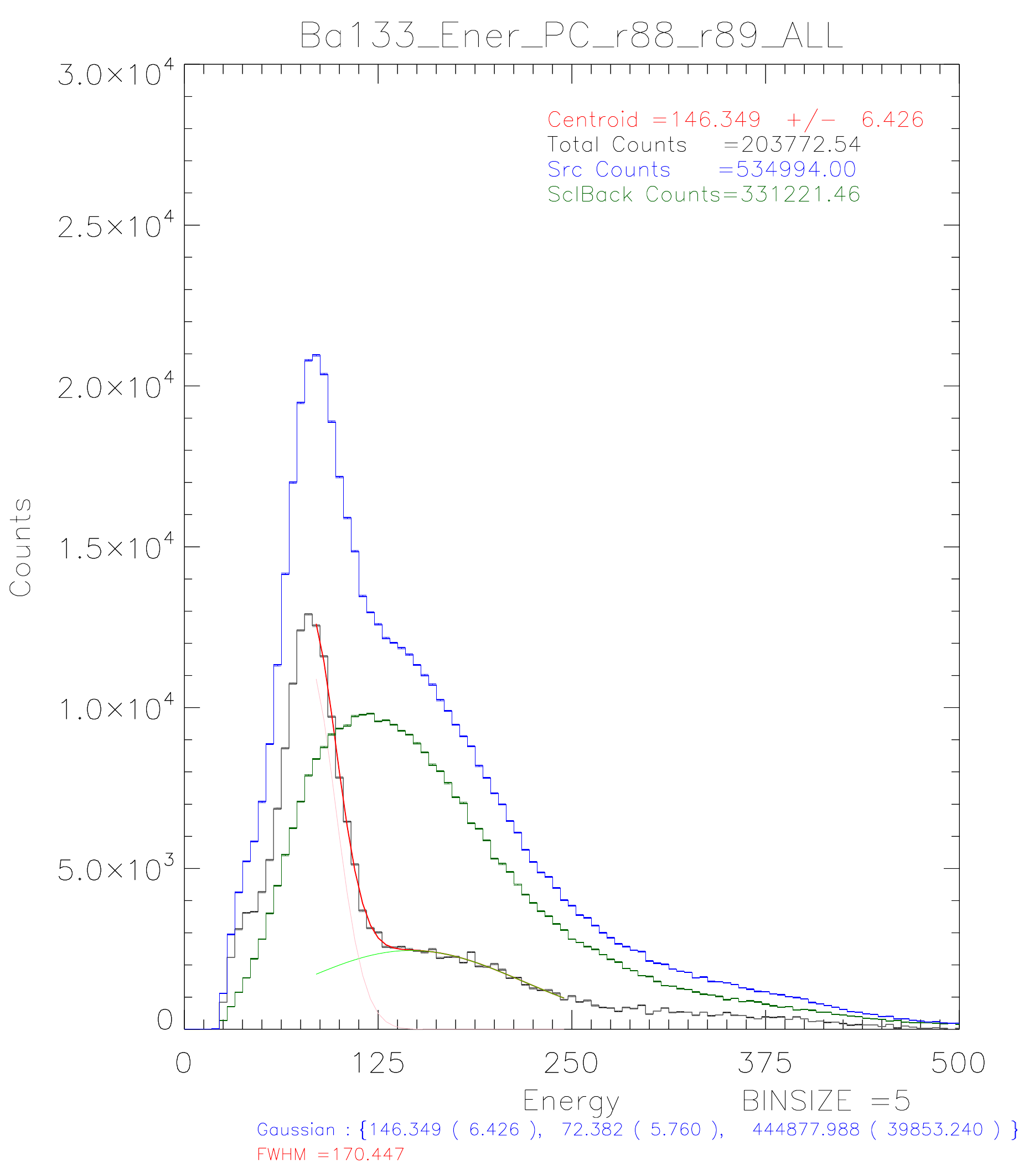}
		 \caption{}
\end{subfigure} 

\begin{subfigure}[b]{0.4\textwidth}
 		 \includegraphics[width=1\linewidth]{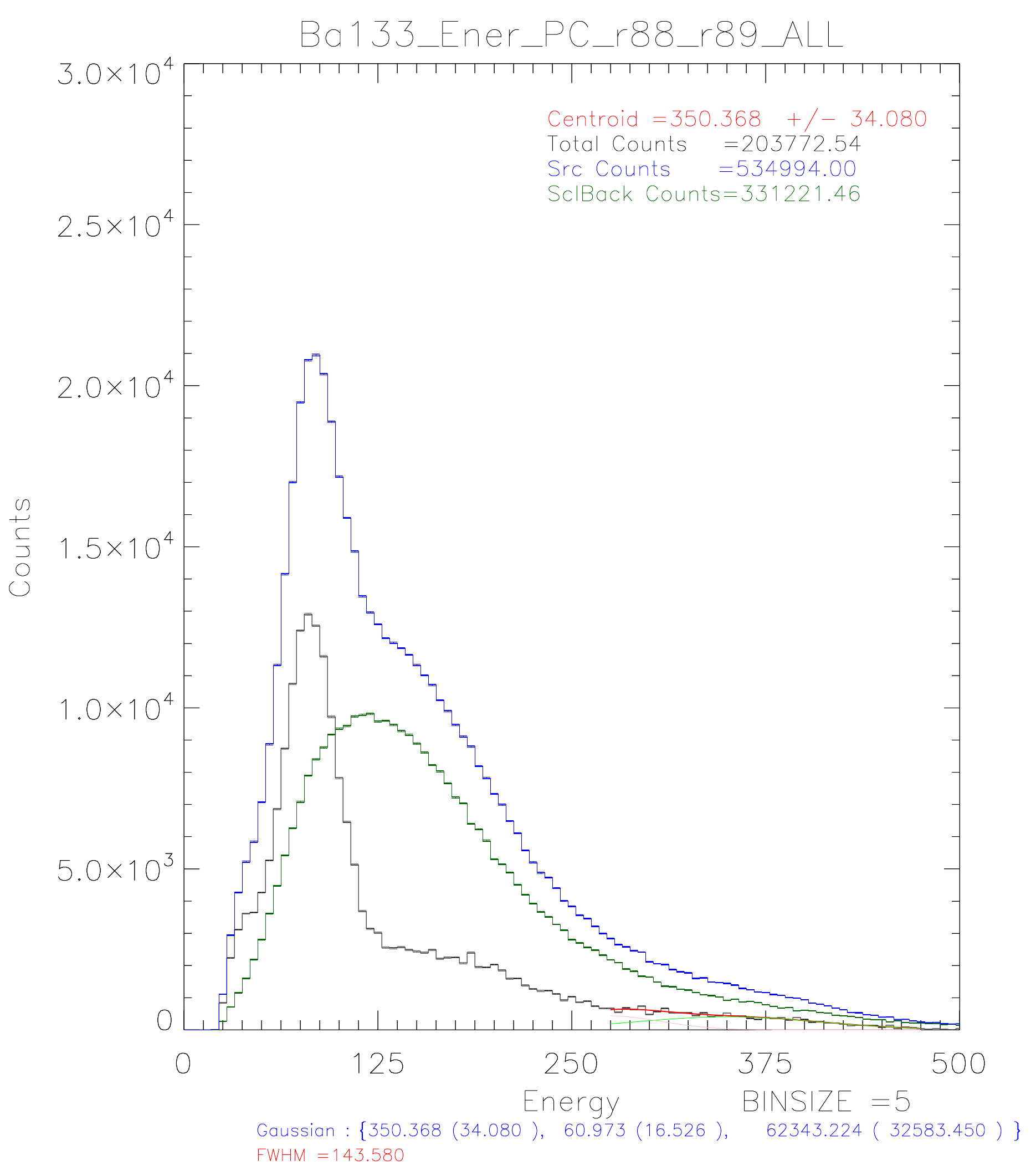}
		 \caption{}
\end{subfigure} 
 \begin{subfigure}[b]{0.4\textwidth}
 		 \includegraphics[width=1\linewidth]{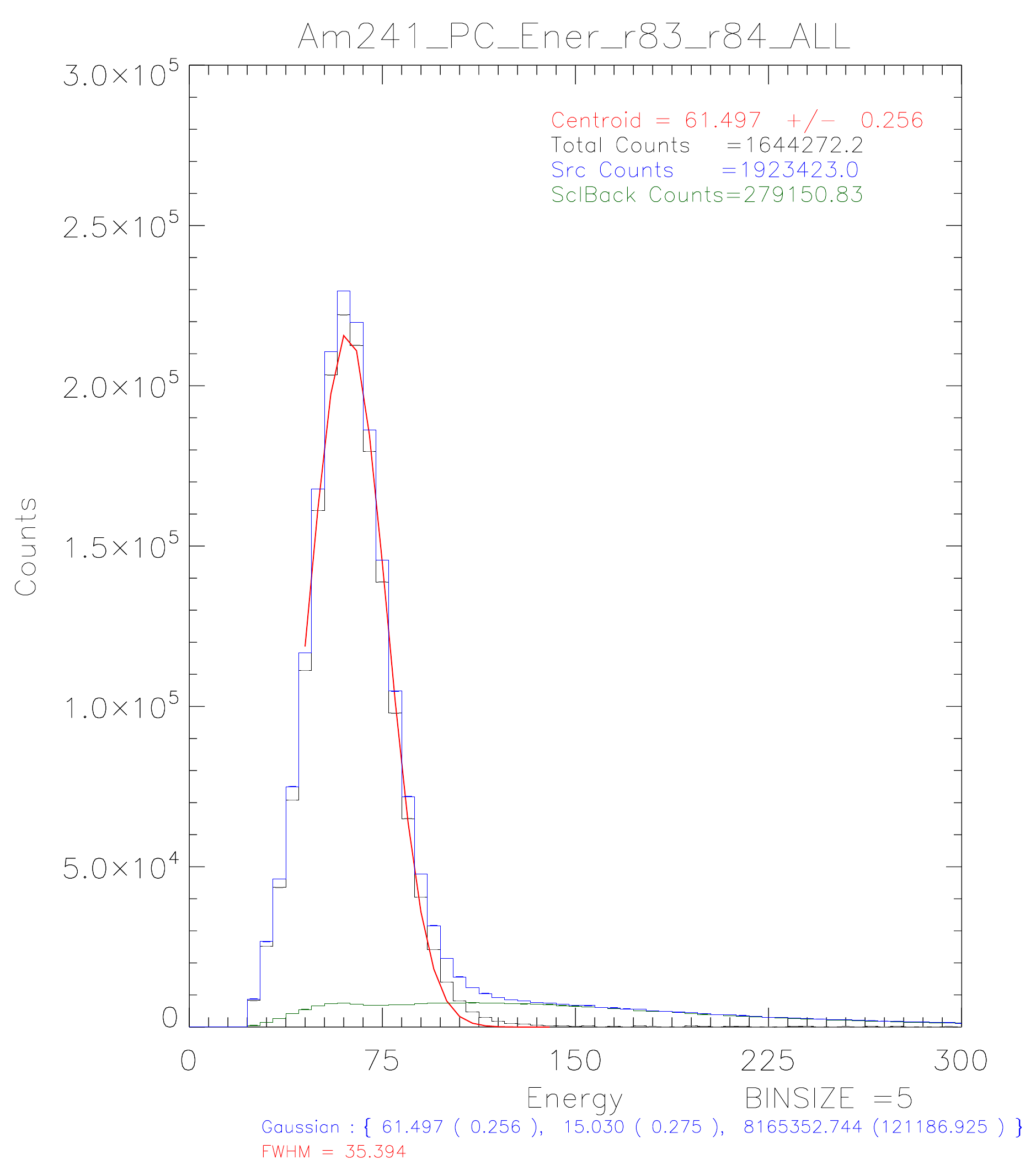}
		 \caption{}
\end{subfigure} 

  \caption{Plot for $^{133}$Ba and $^{241}$Am run for a fully assembled GRAPE instrument at UNH. $^{133}$Ba is fitted for its 3 energy values: 80 keV in (a), 155 keV in (b) and 356 keV in (c). $^{241}$Am is fitted in (d) for 60 keV.}
\label{fig:sim_cal_unh_a}

\end{figure}

			Instrument level calibration refers to the energy calibration of a fully assembled payload. 
			These runs used the calibration files from module level calibration and compared the measured peak energy with the known source energy.
			The module level calibration radiation sources were also used for this purpose.  
			$^{57}$Co,  $^{241}$Am and the $^{133}$Ba were used here. 
			The $^{133}$Ba has the three peaks at 80 keV, 155 keV and the 356 keV. 
			The fitted spectra for each energy are shown in Figure  \ref{fig:sim_cal_unh_a} a, b and c. 
			In each of these plots, the blue data curve represents is the source+background.
			The green data curve represents the data.
			The black curve represents the source data.
			The 80 keV peak intensity dominates the $^{133}$Ba spectrum .
			A fit to the 60 keV peak in $^{241}$Am  is shown in \ref{fig:sim_cal_unh_a}d.

\begin{figure}[!ht]
 \centering
    	\includegraphics[width=.90\textwidth]{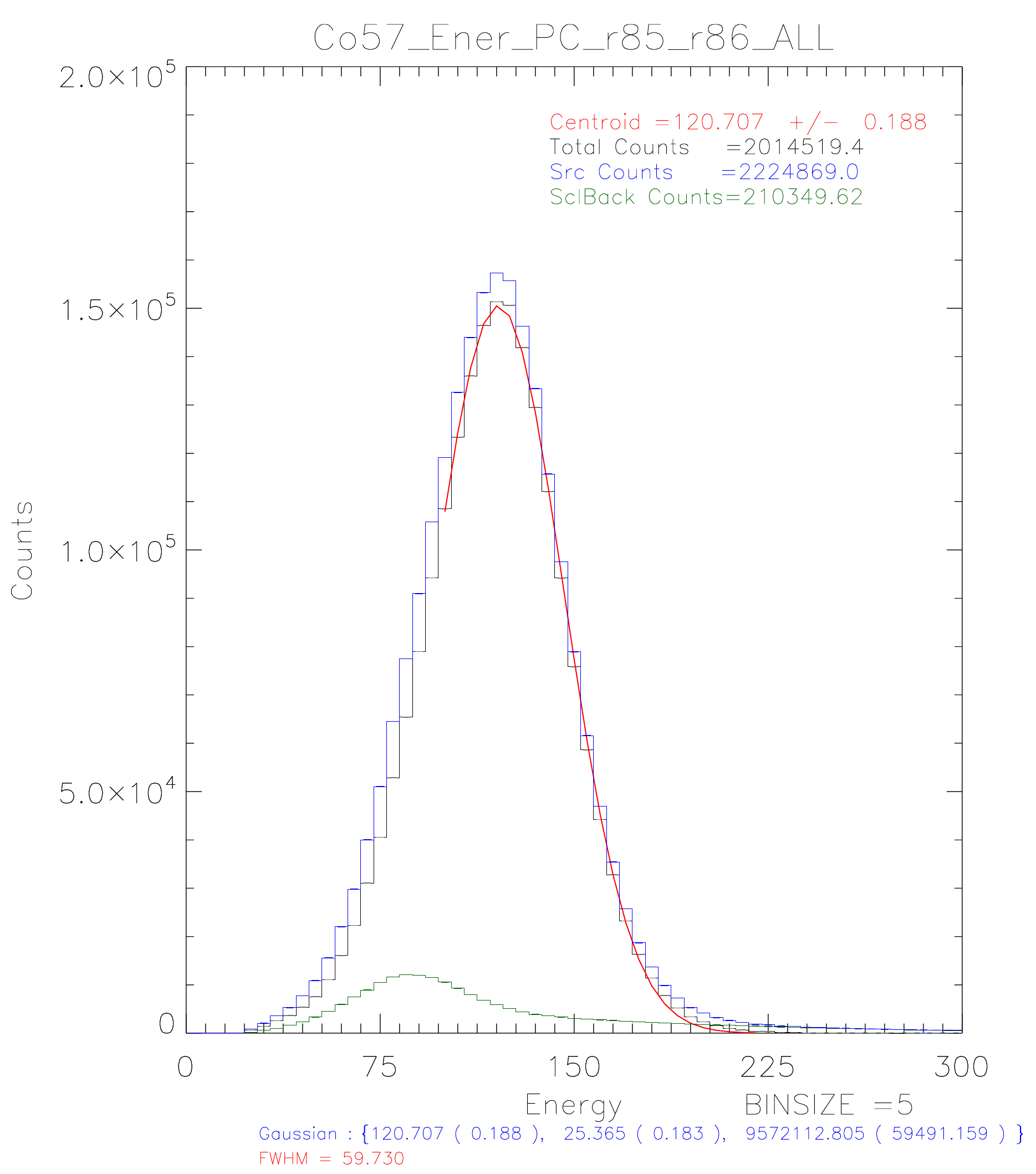}
    	\caption{Energy plot of source (blue) , background (green) and the background subtracted (black) energy histogram of $^{57}$Co. The 122 keV peak is fitted in this figure. }
    	\label{fig:sim_cal_unh_co57_ener_122}
\end{figure}
			The 122 keV peak of $^{57}$Co was used to verify our calibration.
			This is shown in Figure 		\ref{fig:sim_cal_unh_co57_ener_122}.
			These instrument level runs were used to verify our the module level calibration file. 
			Table \ref{table:sim_cal_unh_resolution} shows the source energy, the measured energy and the respective FWHM and resolution.  
			\begin{table}
\begin{center}
\caption{ Table showing the source energy and the measured peak energy with the respective FWHM and the resolution. These values are for fully assembled GRAPE instrument and measurement done at UNH.  }
\label{table:sim_cal_unh_resolution}
 \begin{tabular}{|| c | c | c | c| c | c ||} 
 \hline
Source & Source Energy (keV) &  Measured Peak Energy (keV) & FWHM &Resolution\\ [0.5ex] 
 \hline\hline
$^{133}$Ba & 80 & 81.2 $ \pm$ 0.2& 43.5 & 0.54\\
 \hline
$^{133}$Ba & 155 & 146.0 $ \pm$ 6.4&170.0& 1.16\\
 \hline
$^{133}$Ba & 356 & 350.0 $\pm$ 34.1 & 144.0& 0.41\\
 \hline
 $^{241}$Am & 59.5& 61.4 $\pm$ 0.3& 35.4 & 0.57 \\ 
 \hline
 $^{57}$Co & 122 & 120.7 $\pm$ 0.2& 59.7& 0.49 \\
 \hline
 \end{tabular}
\end{center}
\end{table}

			\subsection{Instrument calibration at FTS}
			\label{sec:ins_perf_fts}	
			
\begin{figure}[hbtp]
 \centering
\begin{subfigure}[b]{0.4\textwidth}
 		 \includegraphics[width=1\linewidth]{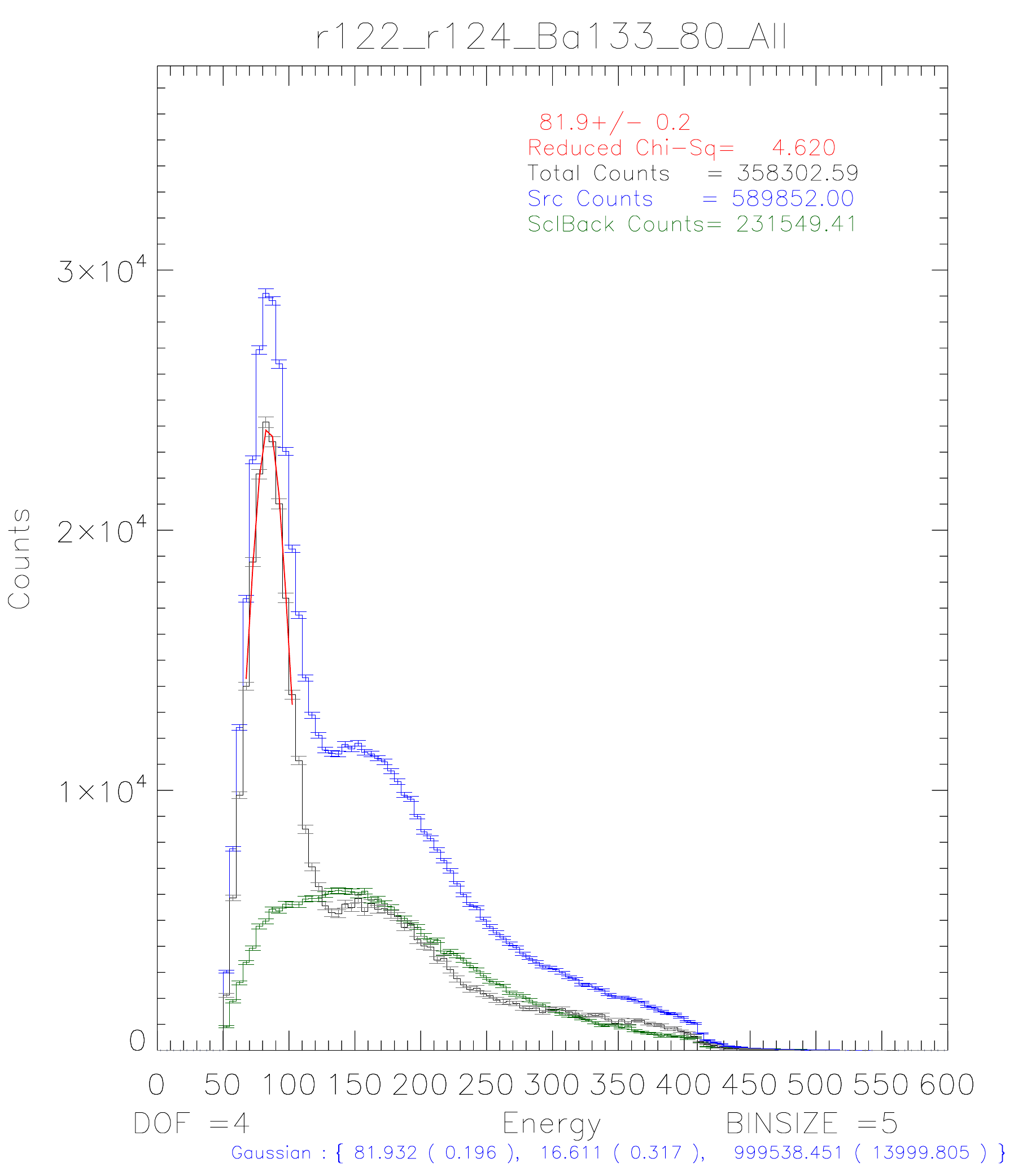}
		 \caption{}
\end{subfigure} \; \; \; 
\begin{subfigure}[b]{0.4\textwidth}
 		 \includegraphics[width=1\linewidth]{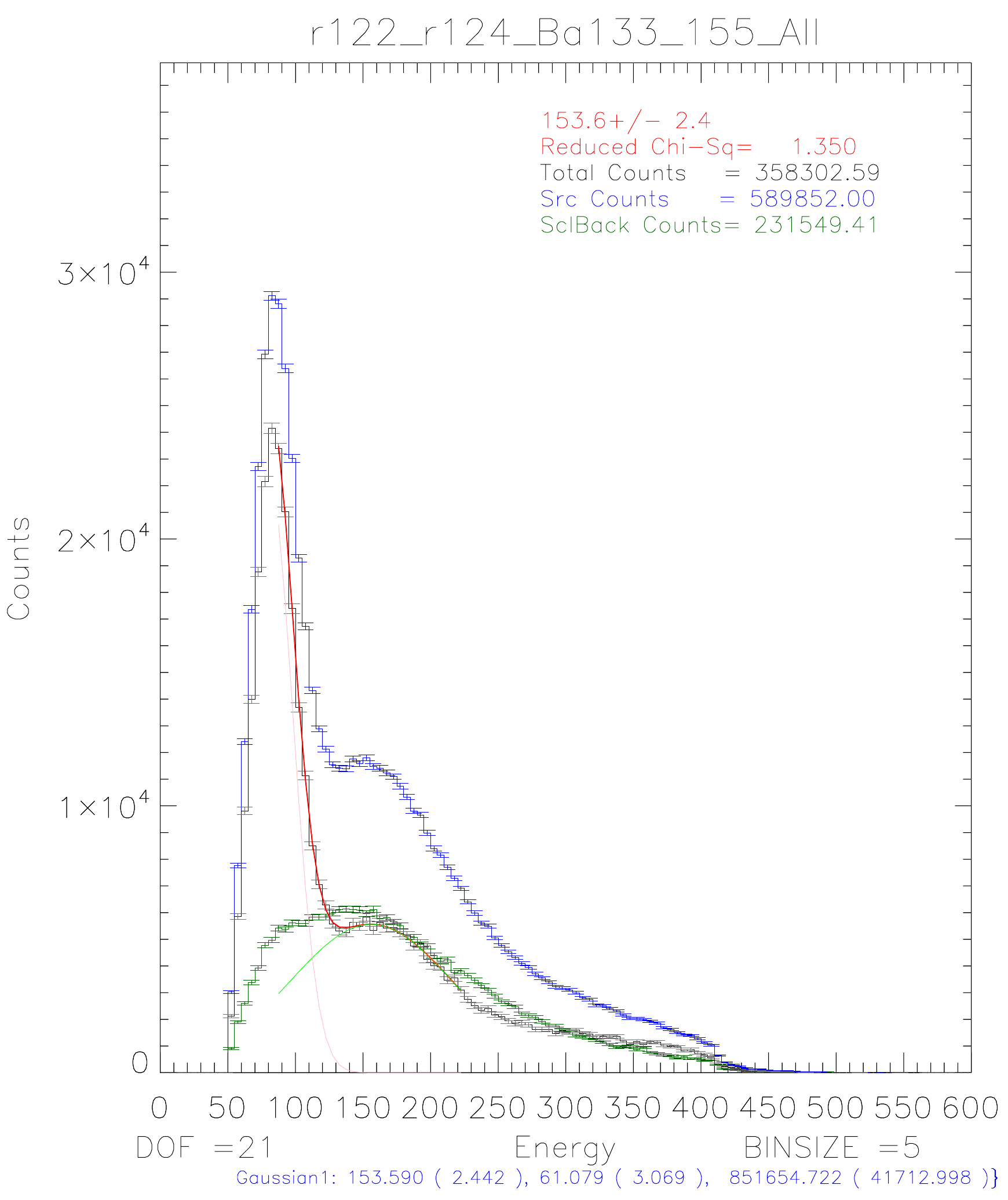}
		 \caption{}
\end{subfigure} 

\begin{subfigure}[b]{0.4\textwidth}
 		 \includegraphics[width=1\linewidth]{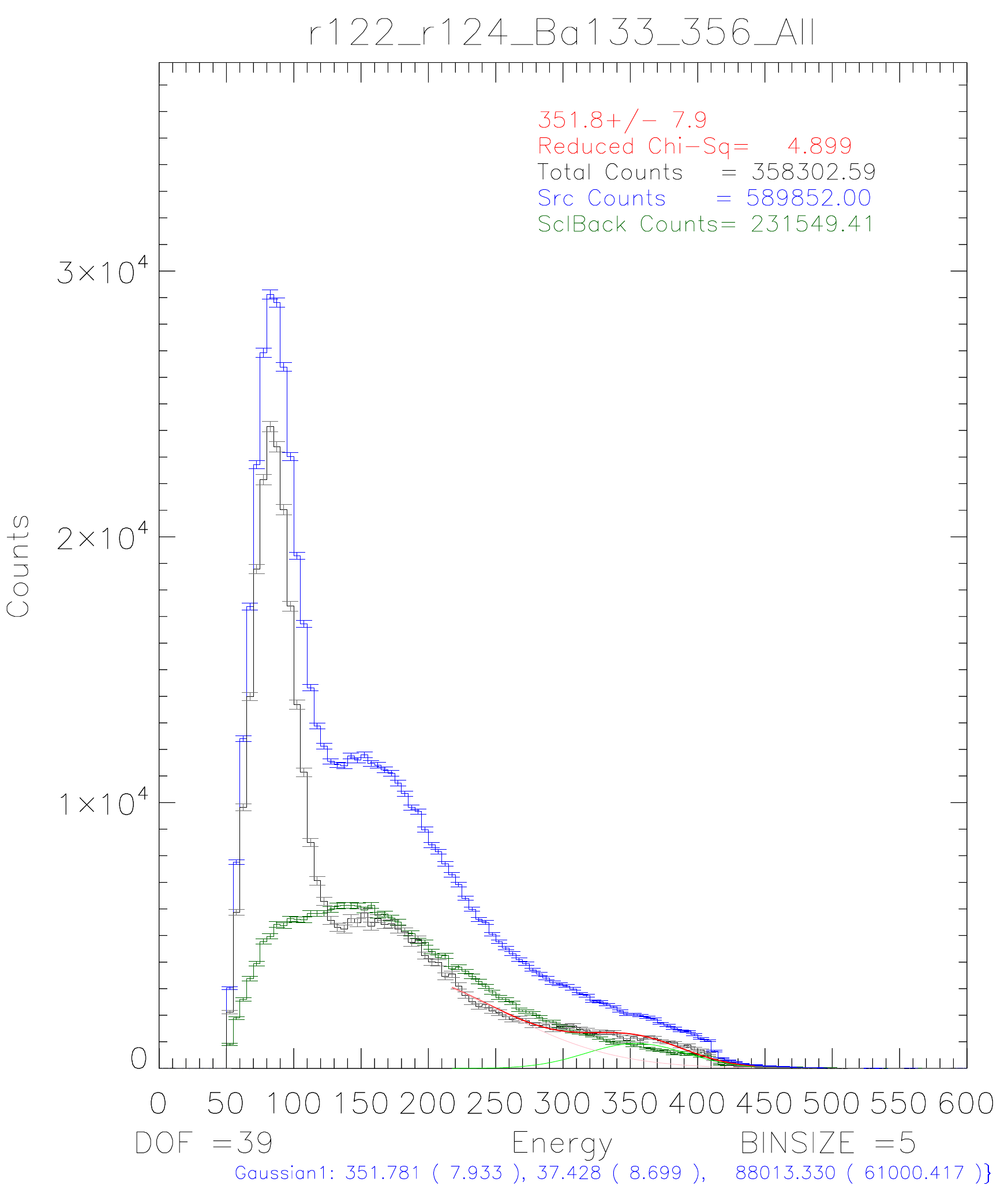}
		 \caption{}
\end{subfigure} \; \; \;
 \begin{subfigure}[b]{0.4\textwidth}
 		 \includegraphics[width=1\linewidth]{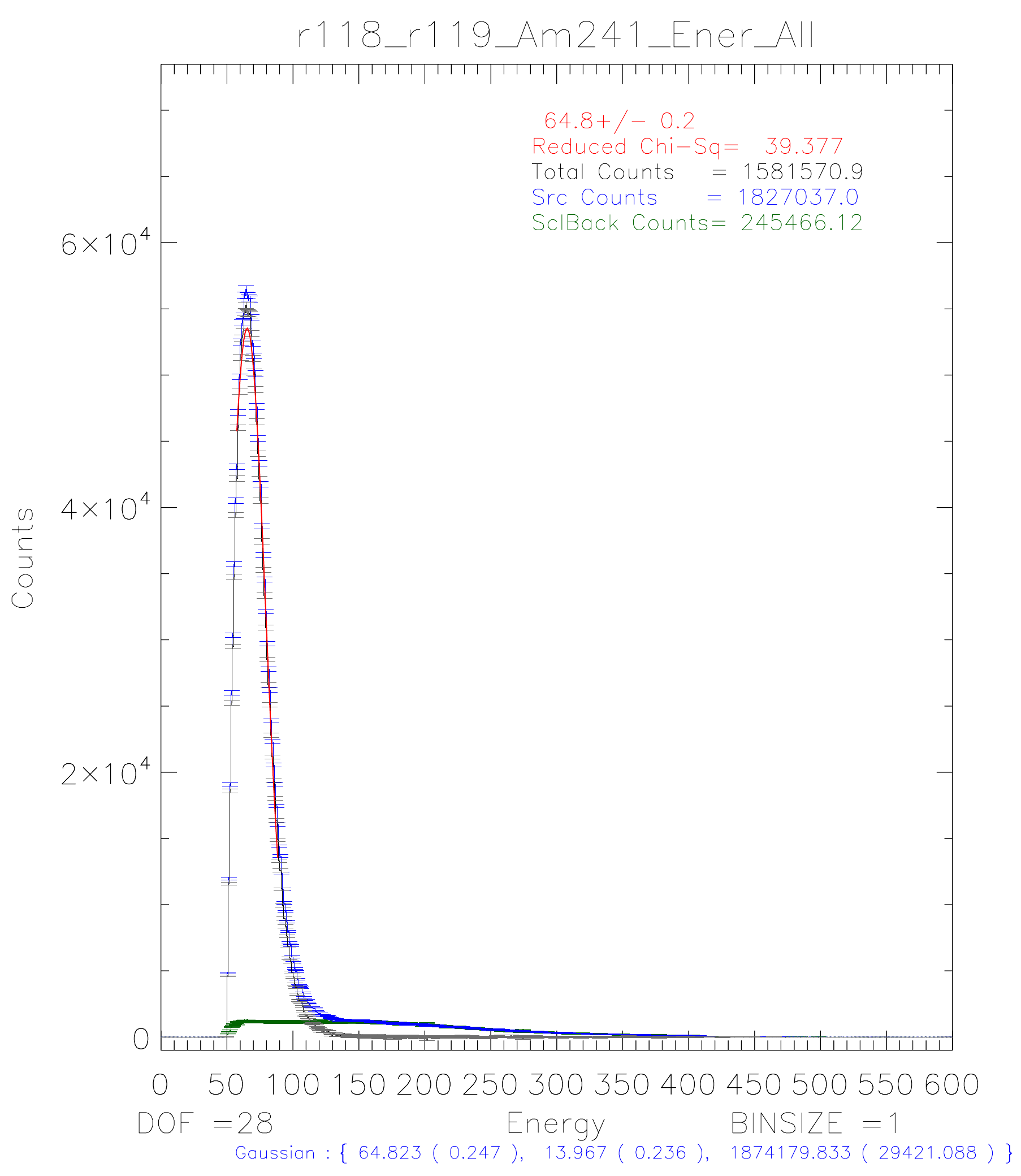}
		 \caption{}
\end{subfigure} 

  \caption{Plot for $^{133}$Ba and $^{241}$Am run for a fully assembled GRAPE instrument at FTS. $^{133}$Ba is fitted for its 3 energy values: 80 keV in (a), 155 keV in (b) and 356 keV in (c). $^{241}$Am is fitted in (d) for 60 keV.}
\label{fig:sim_cal_fts_a}

\end{figure}

\begin{figure}[hbtp]
 \centering
\begin{subfigure}[b]{0.60\textwidth}
 		 \includegraphics[width=1\linewidth]{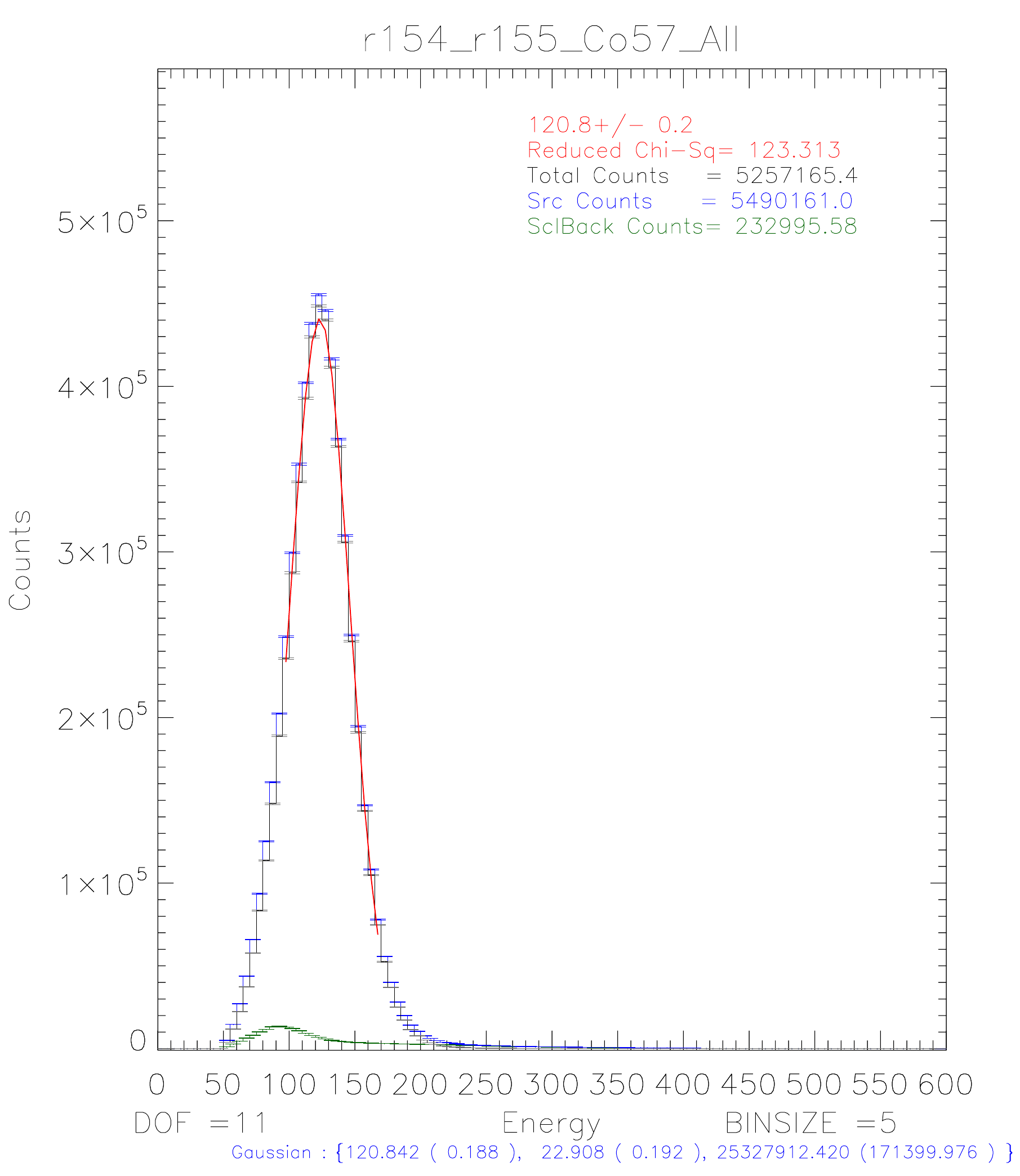}
		 \caption{}
\end{subfigure} \; \; \;     

 \begin{subfigure}[b]{0.60\textwidth}
 		 \includegraphics[width=1\linewidth]{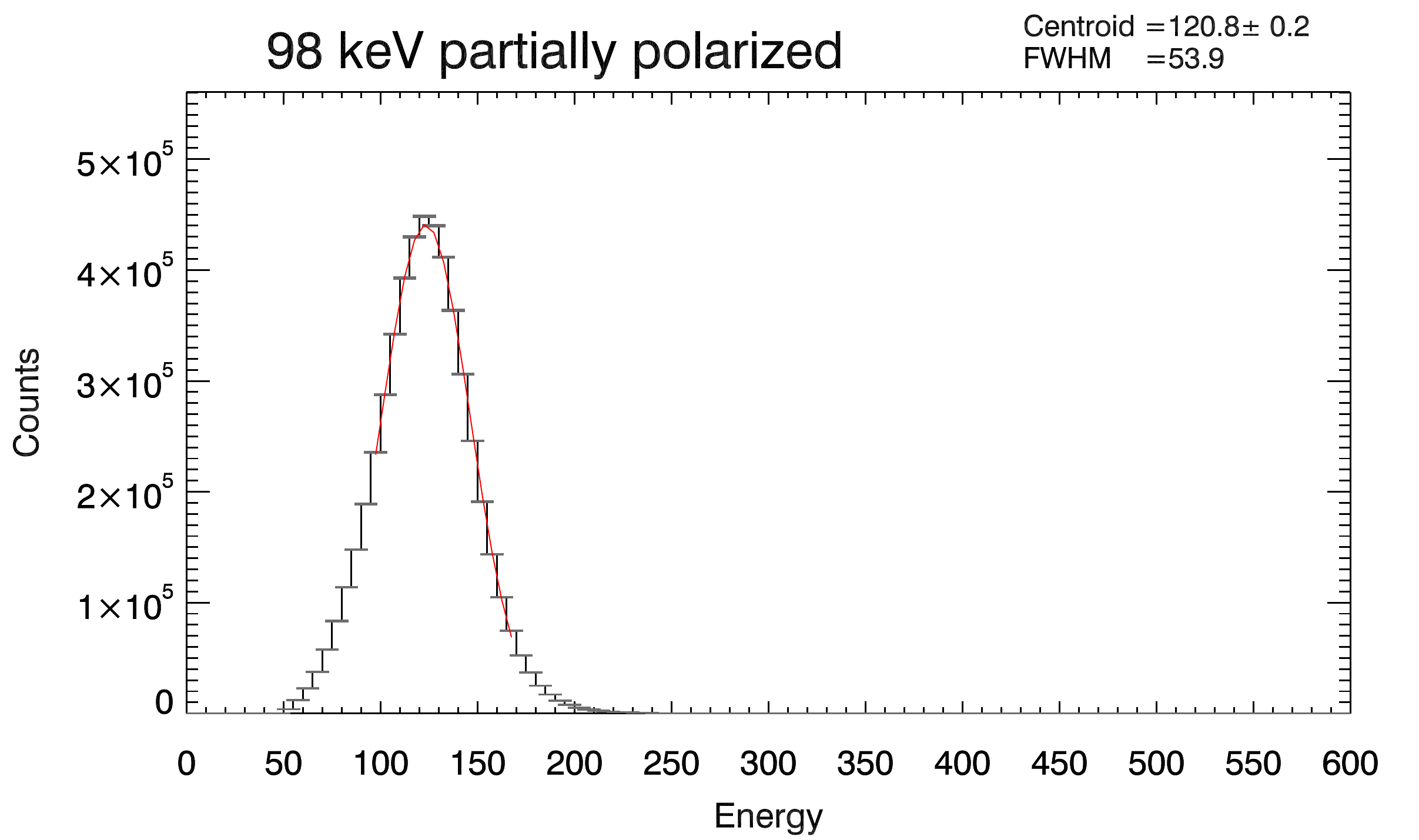}
		 \caption{}
\end{subfigure} 
  \caption{a) Energy plot of source (blue) , background (green) and the background subtracted (black) energy histogram of $^{57}$Co. The 122 keV peak is fitted in this figure. b) This is the same plot with only the background subtracted (black) energy histogram.  }
\label{fig:sim_cal_fts_co57_ener}
\end{figure}

			The whole GRAPE instrument was disassembled after the UNH instrument calibration and then transported to Fort Sumner, NM. 
			Instrument level calibrations were then repeated after final assembly in Fort Sumner, NM. 
			These measurements confirmed that our detectors were working as intended and verified that our calibration were unchanged during the reassembly and transport.
			The runs were performed for $^{133}$Ba, $^{241}$Am and $^{57}$Co.
			We fit the 3 peaks for $^{133}$Ba and one each for $^{241}$Am and $^{57}$Co. 
			The software thresholds were applied before doing these fits. 
			The software threshold applied were 10 keV for plastics, 40 keV for calorimeter and 50 keV for total energy. 
			Therefore the $^{241}$Am 60 keV was very close to the energy threshold. 
			These runs are shown in Figure \ref{fig:sim_cal_fts_a} and \ref{fig:sim_cal_fts_co57_ener}. 
			Table \ref{table:sim_cal_fts_resolution} summarizes these measurements. 
			\begin{table}[h]
\begin{center}
\caption{ Table showing the source energy and the measured peak energy with the respective FWHM and the resolution. These values are for fully assembled GRAPE instrument and measurement done at Fort Sumner (FTS).}
\label{table:sim_cal_fts_resolution}
 \begin{tabular}{|| c | c | c | c| c | c ||} 
 \hline
Source & Source Energy (keV) &  Measured Peak Energy (keV) & FWHM &Resolution\\ [0.5ex] 
 \hline\hline
$^{133}$Ba & 80 & 81.9 $\pm$ 0.2& 39.1 & 0.48\\
 \hline
$^{133}$Ba & 155 & 153.6 $\pm$ 2.4&143.8& 0.93\\
 \hline
$^{133}$Ba & 356 & 350.2 $\pm$ 3.1 & 91.9& 0.26\\
 \hline
 $^{241}$Am & 59.5& 64.2 $\pm$ 0.2& 33.6 & 0.52 \\ 
 \hline
 $^{57}$Co & 122 & 120.8 $\pm$ 0.2& 53.9& 0.44 \\
 \hline
 \end{tabular}
 \end{center}

\end{table}

\clearpage
		\section{Collimator Calibration}
		\label{sec:collimator_calibration}		
\begin{figure}[!t]
 \centering
    	\includegraphics[width=0.8\textwidth]{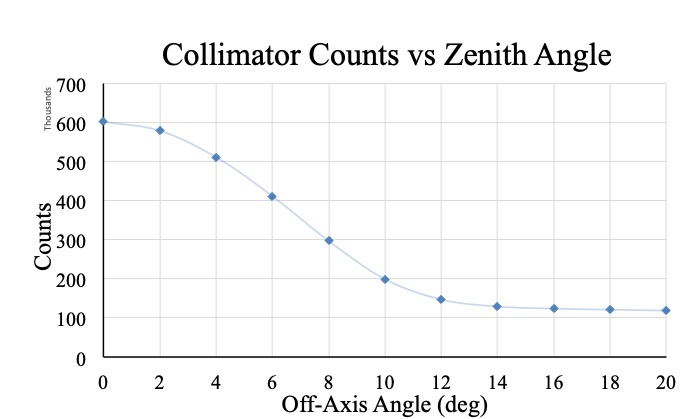}
    	\caption{Plot showing collimator calibration data. It shows counts vs zenith angle incrementing at an angle of 2$^\circ$. A Co-57 source was used on a fully assembled GRAPE instrument.}
    	\label{fig:sim_cal_collimator}
\end{figure}

			GRAPE uses 24 cylindrical collimators, one for each 24 modules.  
			These have been discussed in section  \ref{sec:collimator}.
			Collimator calibration involves  a measurement of the source attenuation as a function of off-axis angle.
			The instrument took  measurements with the source at various angles with respect to the pointing directiono. 
			The result is shown in Figure \ref{fig:sim_cal_collimator}. 
			This Figure represents the total number of recorded events versus off-axis angle. 
			We only scanned the source in one direction.
			We expect the response to be the same for other directions. 
			We can see that at 0$^\circ$ we have the maximum response and that it slowly diminishes until about 12$^\circ$, at which point we are only measuring the background. 
			
			\clearpage

%
%
\chapter{The 2014 Balloon Campaign}
\label{sec:balloon_campaign}

	\section{Flight Payload}
	\label{sec:flt_payload}

\begin{figure}[hbtp]
 \centering
    	\includegraphics[width=0.85\textwidth]{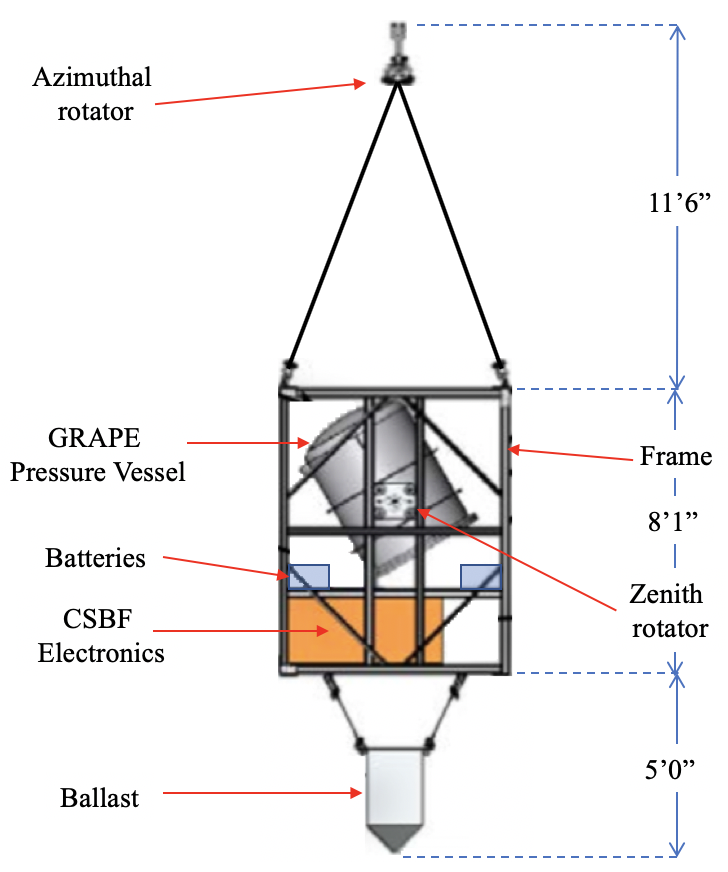}
    	\caption{A drawing of the GRAPE 2014 flight payload with the ballast and the rotators (azimuthal and the zenith). We can also see the CSBF electronics and batteries along with our instrument.}
    	\label{fig:flt_payload_drawing}
\end{figure}

\begin{figure}[tbp]
 \centering
    	\includegraphics[width=0.8\textwidth]{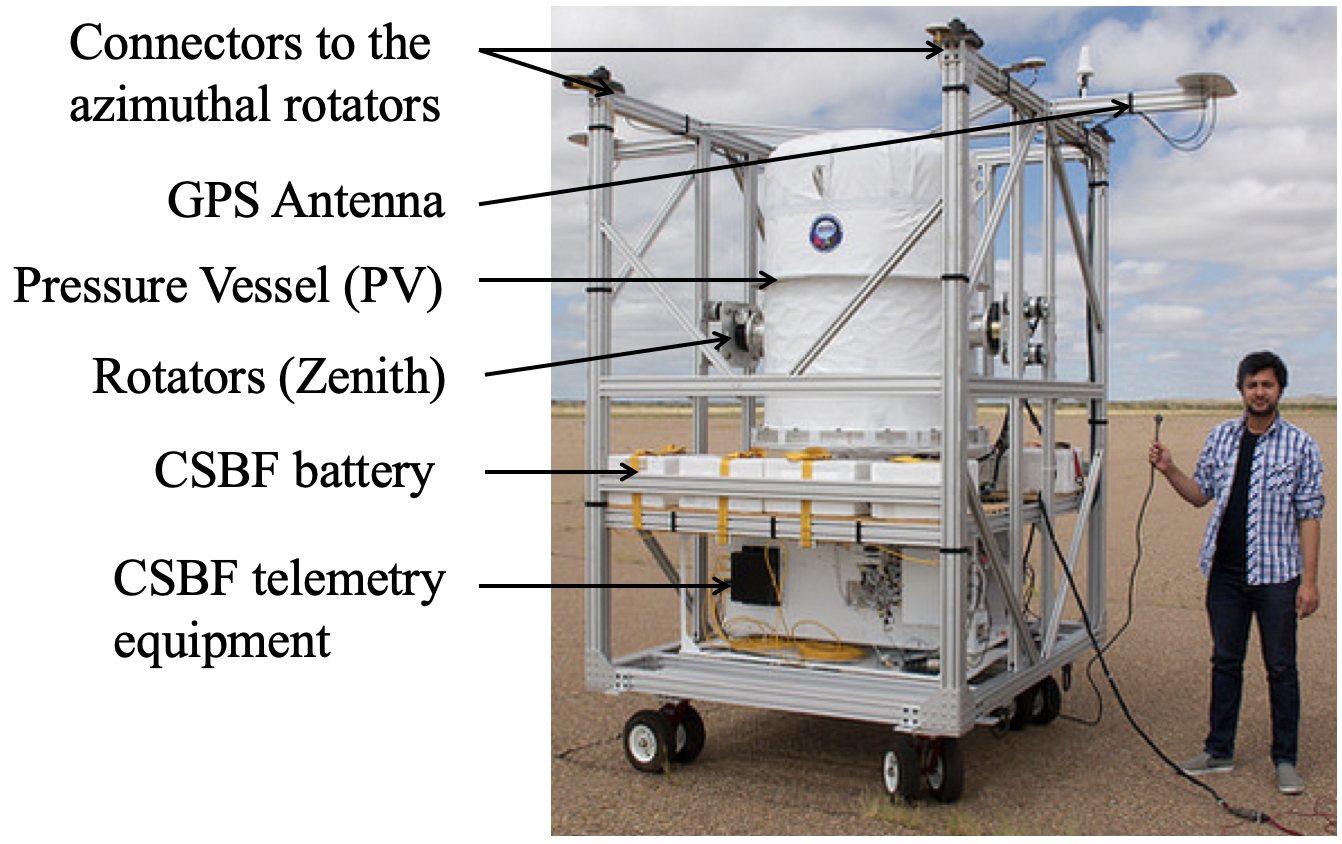}
    	\caption{Flight payload ready to be attached to the balloon. We can see various components of the flight payloads : wrapped Pressure Vessel (PV), Frame, batteries , antenna and CSBF equipments.}
    	\label{fig:flt_payload_frame}
\end{figure}

The instrument payload (enclosed in the the Pressure Vessel (PV)) is described in chapter  \ref{sec:grape_instrumentation}.
The instrument payload is attached to the gondola's aluminum frame on two points. 
This configuration has been described in \ref{sec:gondola}.
Along with the instrument payload, the CSBF telemetry equipments and batteries are also present in the gondola. 
Drawings of these components in the gondola are shown in Figure \ref{fig:flt_payload_drawing} and the actual payload is shown in \ref{fig:flt_payload_frame}.
The ballast is made up of fine steel shots and is used as a dead weight which can be dropped off to  maintain the current or achieve a higher altitude when necessary. 
The ballast is attached to the bottom of the gondola.
The GPS antenna is attached at one of the corners of the frame as seen in Figure \ref{fig:flt_payload_frame}.
The flight payload consists of all these equipments and components present within the gondola. 
There are four connection from the top of the gondola frame to the azimuthal rotator which, along with the elevation motor, provides us the ability to point at the target source.

 	\section{Pointing}
	\label{sec:flt_pointing}

The collimators provided a 10$^\circ$ Field Of View (FOV) (Figure \ref{fig:sim_cal_collimator}). 
The azimuthal orientation of the gondola was set using a NASA supplied rotator.
The rotator is placed between the balloon and the gondola. 
The rotator employed a reaction wheel and momentum transfer unit to control the payload orientation. 
The rotator was developed by NASA for maintaining proper solar panel altitude during Long Duration Balloon (LDB) flights.
The rotation has a precision of $\pm 3^\circ $ and a pointing knowledge uncertainty of $\pm 0.5^\circ $ in azimuth. 
This was adequate for our purpose as our instrument had a FOV of 10$^\circ$. 

Elevation was measured by an inclinometer mounted inside the PV and an elevation drive mechanism controlled the elevation angle (zenith angle). 
A hard limit of 60$^\circ$ was set for the zenith angle as a safety feature. 
The absolute orientation of the gondola was obtained via a commercial differential GPS unit (Ashtech ADU5).
The 3-dimensional orientation of the gondola along with its zenith angle was measured and recorded at regular intervals, time tagged and inserted into the telemetered data and also recorded on board. 

The pointing of the instrument was controlled using the GSE computer.
Updated pointing parameters were calculated  and were sent to the payload at 5 minute interval to update the pointing.
A window in the LabVIEW GUI was dedicated to the calculation of the required pointing direction of the instrument based on the target source position. 
It took the RA, DEC (provided manually through our database of sources) and the current latitude and longitude (feedback from the telemetered data) to give the required azimuth and zenith angle.

	\section{Flight Plan}
	\label{sec:flt_flight_plan}
The 2014 GRAPE flight was labelled flight 655N.
The primary goal of the flight was to make a polarization measurement of the Crab.
The flight was scheduled from Ft. Sumner, New Mexico in the fall of 2014. 
From this location, at that time of the year, the Crab was only visible to us during early morning. 
The Crab set about when we got to float as the balloon launches were only being scheduled in the morning.
Therefore, a 24 hour flight was need to cover one full transit of the Crab. 

	\begin{table}
\begin{center}
\caption{ Right Ascension (RA) and Declination (Dec) of the astrophysical objects (or regions) that are observed during float period of the GRAPE 2014 flight. The Sun's RA and Dec was not constant so it had to be constantly adjusted during the observation hence its not included here.}
\label{table:ra_dec_sources}
 \begin{tabular}{|| c | c | c| c | c ||} 
 \hline
 Source & RA (Deg) & RA (Time)&Dec (Deg) &Dec(Time) \\ [0.5ex] 
 \hline\hline
 Bgd2  & 265.95$^\circ$ & 17h 43m 48s & 47$^\circ$& 3h 8m 0s \\ 
 \hline
 Bgd4 & 0$^\circ$  & 0h 0m 0s & 22.15$^\circ$ & 1h 28m 36s\\
 \hline
 Cygnus X-1& 265.95$^\circ$ & 19h 58m 22s & 35.2$^\circ$& 2h 20m 48s \\
 \hline
Crab& 83.63$^\circ$ & 5h 34m 31s & 22.01$^\circ$& 1h 28m 2s \\
 \hline
\end{tabular}
\end{center}
\end{table}

\begin{figure}[tp]
 \centering
    	\includegraphics[width=.74\textwidth]{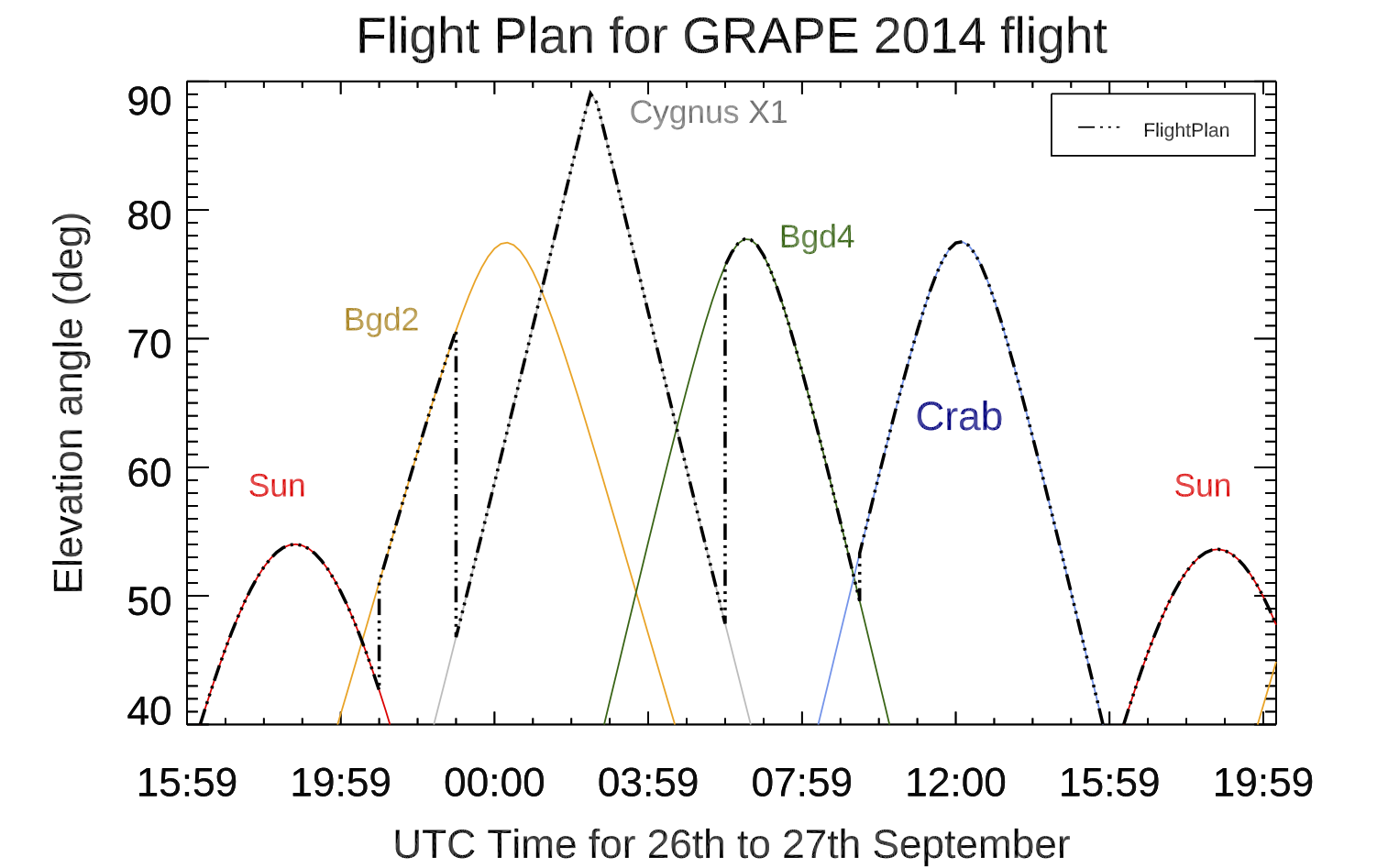}
    	\caption{Flight plan for GRAPE 2014. The elevation angle vs UTC time for our astrophysical sources is plotted here. The solid lines for each sources represents them in our field of view and the dash line is a predicted sources chosen to be observed. The elevation angle is measured from the horizontal plane so the zenith angle is 90$^\circ$ - elevation angle. The elevation angle is set between 40$^\circ$ to 90$^\circ$ (or zenith angle of $< \text{60}^\circ$) as a safety precaution.  }
    	\label{fig:flt_plan}
\end{figure}

We developed a flight flan (Figure \ref{fig:flt_plan}) based on early morning launch and 24 hours at float. 
The Crab was visible to us in the latter part of the flight. 
This provided us with an opportunity to look at various other sources of interest. 
There were 3 primary targets: Crab, Cygnus X-1, and the Sun.
Background regions (Bgd2 and Bgd4) were intended to facilitate background estimates for the source observations.
The two Bgd regions are the area in the sky where we do not have any known sources above our instrument's threshold.
The Right Ascension (RA) and Declination (Dec) are shown in table \ref{table:ra_dec_sources}. 
The Sun's RA and DEC is not constant during our flight so a procedure within the GSE labview tab calculated the RA, DEC and the pointing parameters.
The flight plan is generated by using the RA and DEC of our target source. 
It takes in the time, altitude, latitude and longitude to calculate the source azimuth and zenith. 
During the flight, live values of the altitude, longitude, latitude and time were used to measure and update the pointing parameters for the observation. 

	\section{Improvements from Flight 2011 to 2014}
	\label{sec:flt_improvements}

GRAPE was initially flown in 2011. 
The 2011 flight provided some insights on ways to improve the instrument performance.
These improvements were implemented on 2014 flight.
Improvements included both hardware and changes in operations.

To maximize livetime a decimation of the C and CC (hardware) events was set to 5. 
This meant that only one out of five of these events were recorded. 
The PC (hardware) events were not impacted.
The passive shielding, as discussed in section \ref{sec:instrument_shielding}, was also improved. 
The thickness of the Pb shield was increased from 0.8mm to 4.24mm. 
Eight new modules were added to increase the total number of modules from 16 to 24.

The sweeps were continuous for the 2011 flight even when the target source was switching. 
This meant that the beginning of an observation was often from middle of a sweep.  
For 2018, a command to pause at the end of the sweep was implemented. 
In order to change to a new source, a pause command would be sent. 
The instrument would then complete the sweep and then wait for the instrument to be be repointed, after which a new sweep would be started.
This meant that each full sweep was looking at a particular astrophysical target.
We worked closely with CSBF to insure a flight profile to insure maximum altitude during the Crab observation. 
This was achieved by careful se of ballast. 
 
	\section{Flight Summary}
	\label{sec:flt_flight_summary}
	
\begin{figure}[htbp]
 \centering
    	\includegraphics[width=0.9\textwidth]{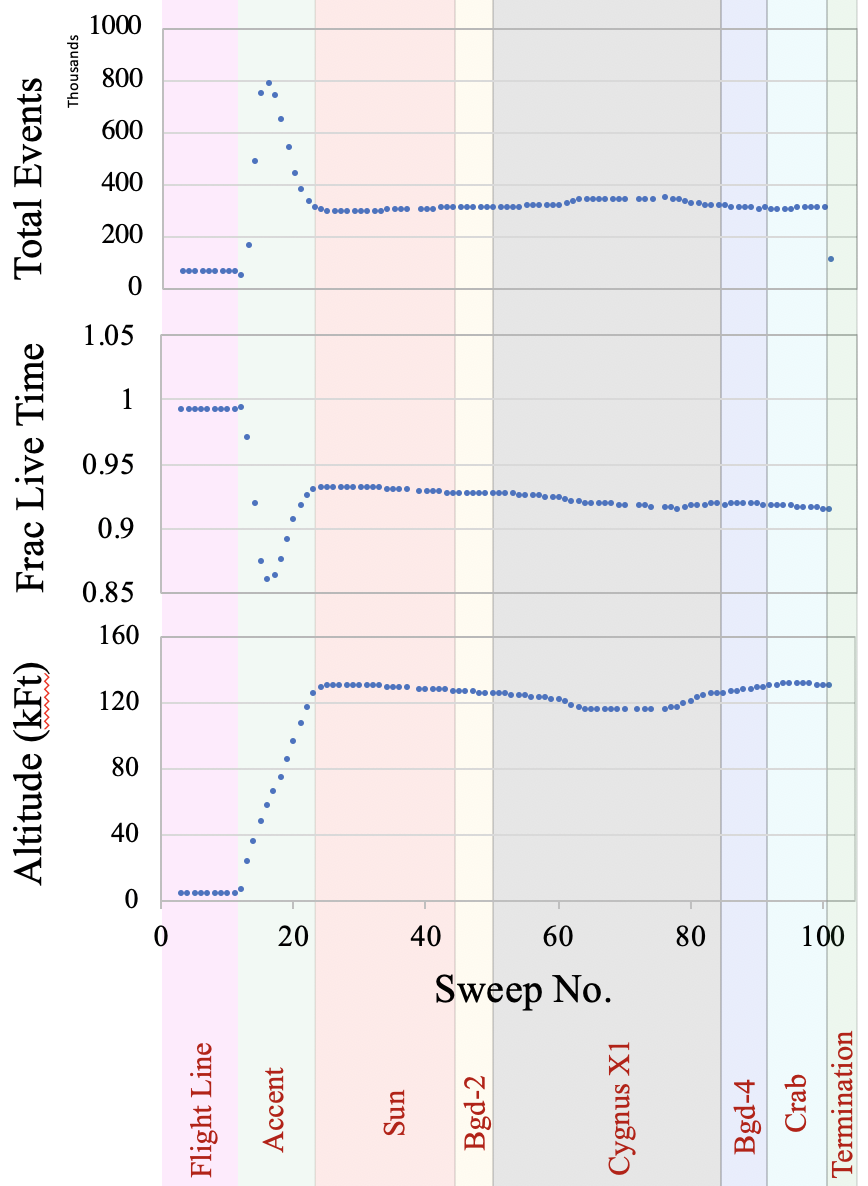}
    	\caption{Flight profile for the 2014 Balloon flight for all stages (from launch to termination). Each data point represents an individual sweep. The shaded regions represents the labelled stage of the flight. The Total Events represents all the events (PC, C, CC).}
    	\label{fig:flt_profile_summary}
\end{figure}

\begin{figure}[ht]
 \centering
    	\includegraphics[width=0.95\textwidth]{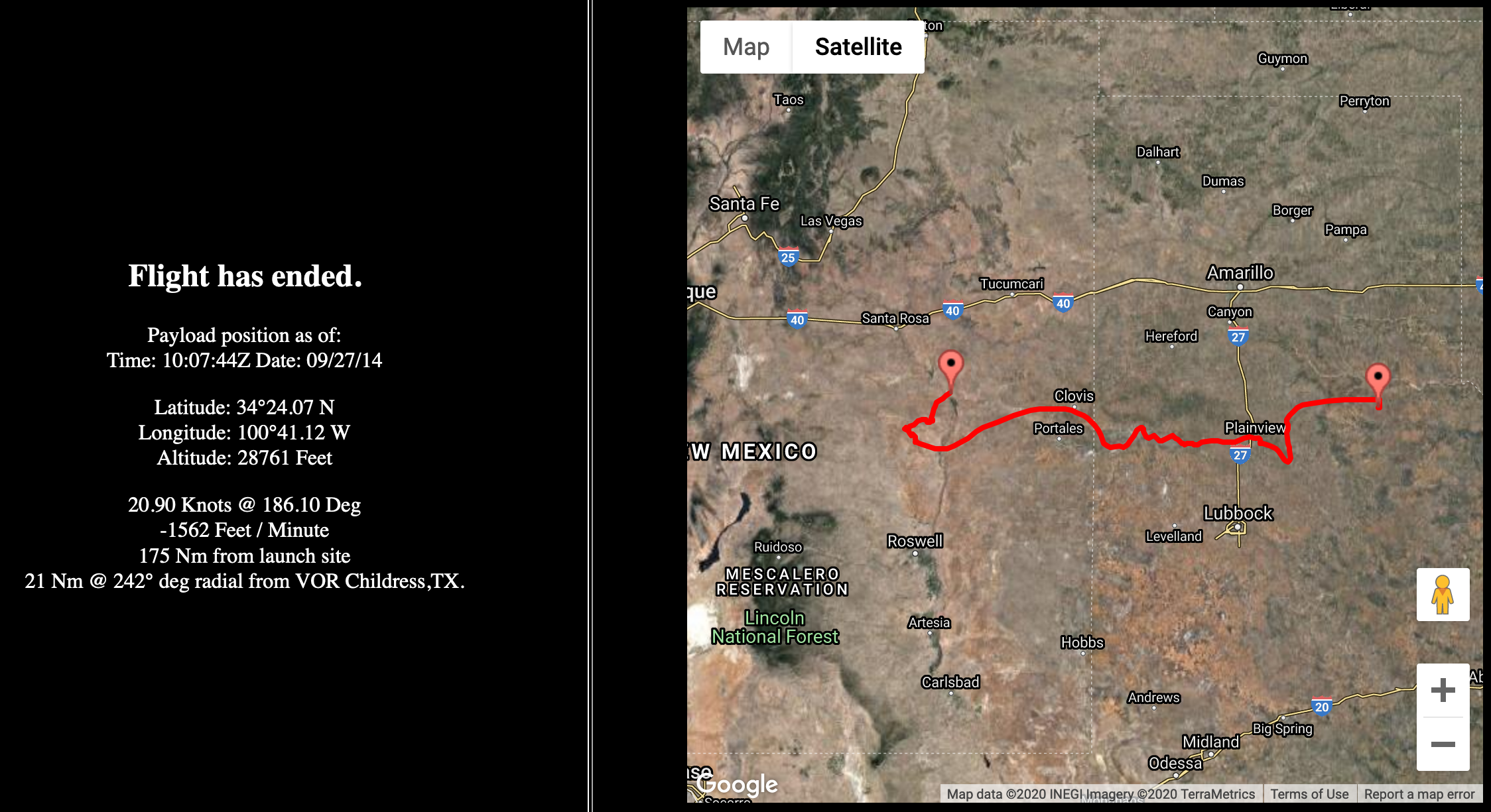}
    	\caption{Figure showing the launch point, flight path and the termination point for GRAPE 2014 flight. The launch was at Ft. Sumner, NM and the termination was near Childress, TX}
    	\label{fig:flt_profile_b}
\end{figure}

GRAPE 2014 (Flight no. 655N) was launched on the morning of September 26, 2014 from Fort Sumner, NM.              
The instrument startup began around 1200 hrs UTC (5:00 AM MST).             
After balloon inflation and flight protocols, the balloon was launched at 1445 hrs UTC (7:00 AM MST).            
The balloon reached float altitude around 1700 hrs UTC (12:00 PM MST) at an altitude of $\sim$130 kft.
The float altitude is where we start to take measurements from our target sources.    
As mentioned previously in the section \ref{sec:flt_flight_plan}, the Crab was only visible to us in the later part of the flight so the Sun, Cygnus X1 and two background regions (Bgd2 and Bgd4) were also observed (Figure \ref{fig:flt_profile_summary}).
The Crab observation started around 0800 hrs UTC (1:00 AM MST). 
The flight was terminated at 0937 hrs UTC (2:37 AM MST),  near Childress, TX as the balloon began to drift towards populated areas.
The flight path of the GRAPE 2014 is shown in Figure \ref{fig:flt_profile_b}.

			\subsection{Flight Profile}
			\label{sec:flt_flight_profile}
The total duration of the 2014 flight was roughly 19.5 hrs (launch to termination).
The float duration was around 15.4 hrs, out of which, the Crab was observed for around 1.8 hrs. 
Table \ref{table:flt_profile_table} shows the observation time and the respective sweep numbers for different stages of the flight.
	\begin{table}
\begin{center}
\caption{ Table showing observational time and sweep information. Each sweep takes 12mins (720s) For few observations, we paused at the end the sweep, pointed our instrument towards the source and resumed from a new sweep. Sweep 38 (Sun observation) and Sweep 71 and 75 (Cygnus X1 observation) was excluded due to instrumental issues. }

\label{table:flt_profile_table}

 \begin{tabular}{|| c | c | c| c | c ||} 
 \hline
 Source Target & Start Swp & End Sweep & No. of Sweep &Total Time\\ [0.5ex] 
 \hline\hline
 Flight Line & 1  & 11& 11 & 2hr 12min\\
 \hline
 Ascent & 12 & 23 & 12 & 2hr 24min\\
 \hline
 Sun & 24 & 44 & 21 & 4hr 12min\\
 \hline
 Bgd2  & 45& 50& 6& 1hr 12min \\ 
 \hline
  Cygnus X-1& 51& 84& 34& 6hr 48min \\
 \hline
 Bgd4 & 85 & 91 & 7& 1hr 24min\\
 \hline
Crab& 92 & 100 & 9& 1h 48min \\
 \hline
 Termination& 101 &101 & 1& 12min\\
 \hline
\end{tabular}

\end{center}
\end{table}

Figure \ref{fig:flt_profile_summary} shows the number of events, average fractional live time and the altitude for each sweep during the full flight.         
The number of events represented on this figure is the sum total of all types of events (PC, C, CC) recorded at hardware level.
The fractional live time is computed for each module and an average of all modules integrated over a sweep and is represented in the plot.
The fractional live time is inversely proportional to counts which is seen in the plot.
We can also see the bump in total events or a dip in fractional live time during accent. 
The peak corresponds to point where our instrument passed the Pfotzer maximum. 
The flight reached float altitude around sweep no. 24 and remained at float until termination at the end of sweep no. 101.

\begin{figure}[!htb]
 \centering
    	\includegraphics[width=0.8\textwidth]{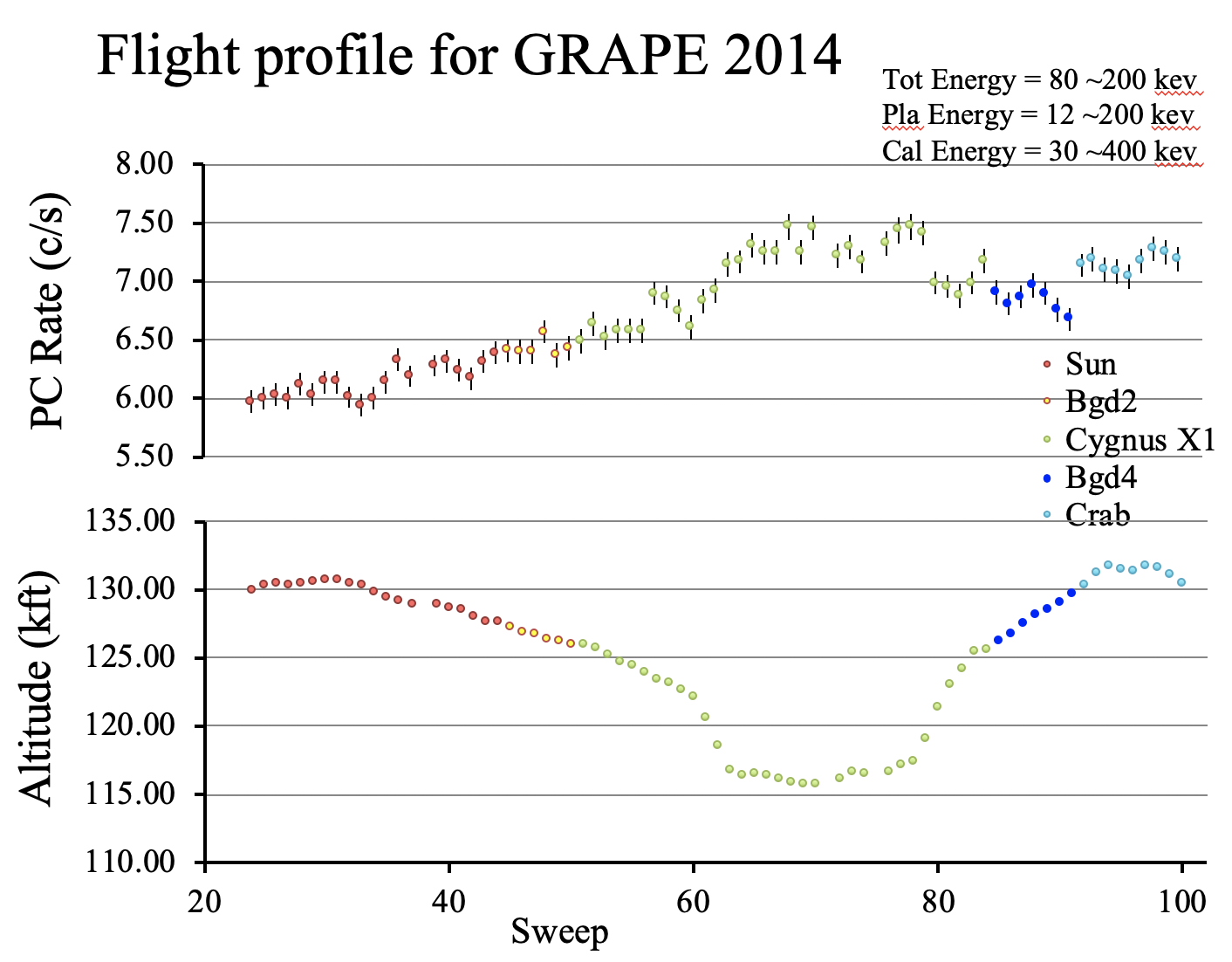}
    	\caption{Flight profile for the 2014 Balloon flight during the float period. Five different targets were observed during this period. The sun, Cygnus X1, Crab and two background regions (Bgd2 and Bgd4). Each data point is for an individual sweep and the PC rate represents the PC events of all types after applying the software threshold. }
    	\label{fig:flt_profile_a}
\end{figure}		

\begin{figure}[t]
 \centering
    	\includegraphics[width=.80\textwidth]{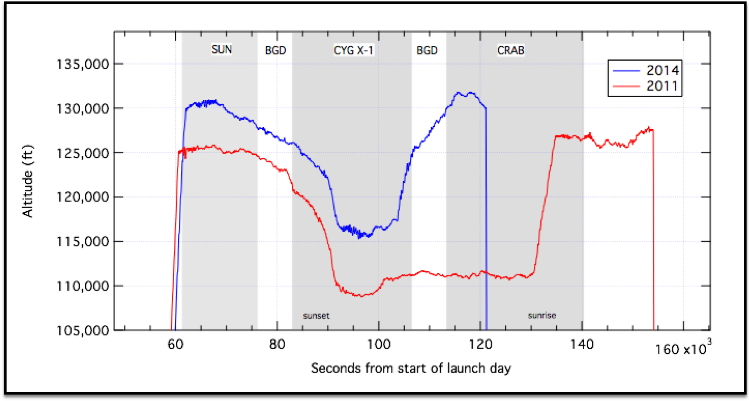}
    	\caption{Comparing the altitude covered by GRAPE 2011 and GRAPE 2014 flight. We can see that the 2014 flight achieved higher altitudes with the implemented improvements.}
    	\label{fig:flt_plan_compare}
\end{figure}

			\subsection{Flight Observations}
			\label{sec:flt_flight_observations}
The flight profile of GRAPE 2014 during the period at float altitude is shown in Figure \ref{fig:flt_profile_a}.
This shows the PC rate (all types) for each sweep at float. 
The PC rate is measured after applying software threshold and defining the energy range.
Plastics were set for 10 - 200 keV, Calorimeters were set from 40 - 400 keV. 
The altitude varied from ~130 kft to ~115 kft (~40 km to ~35 km) for the float.

\begin{figure}[hbt]
 \centering
    	\includegraphics[width=0.93\textwidth]{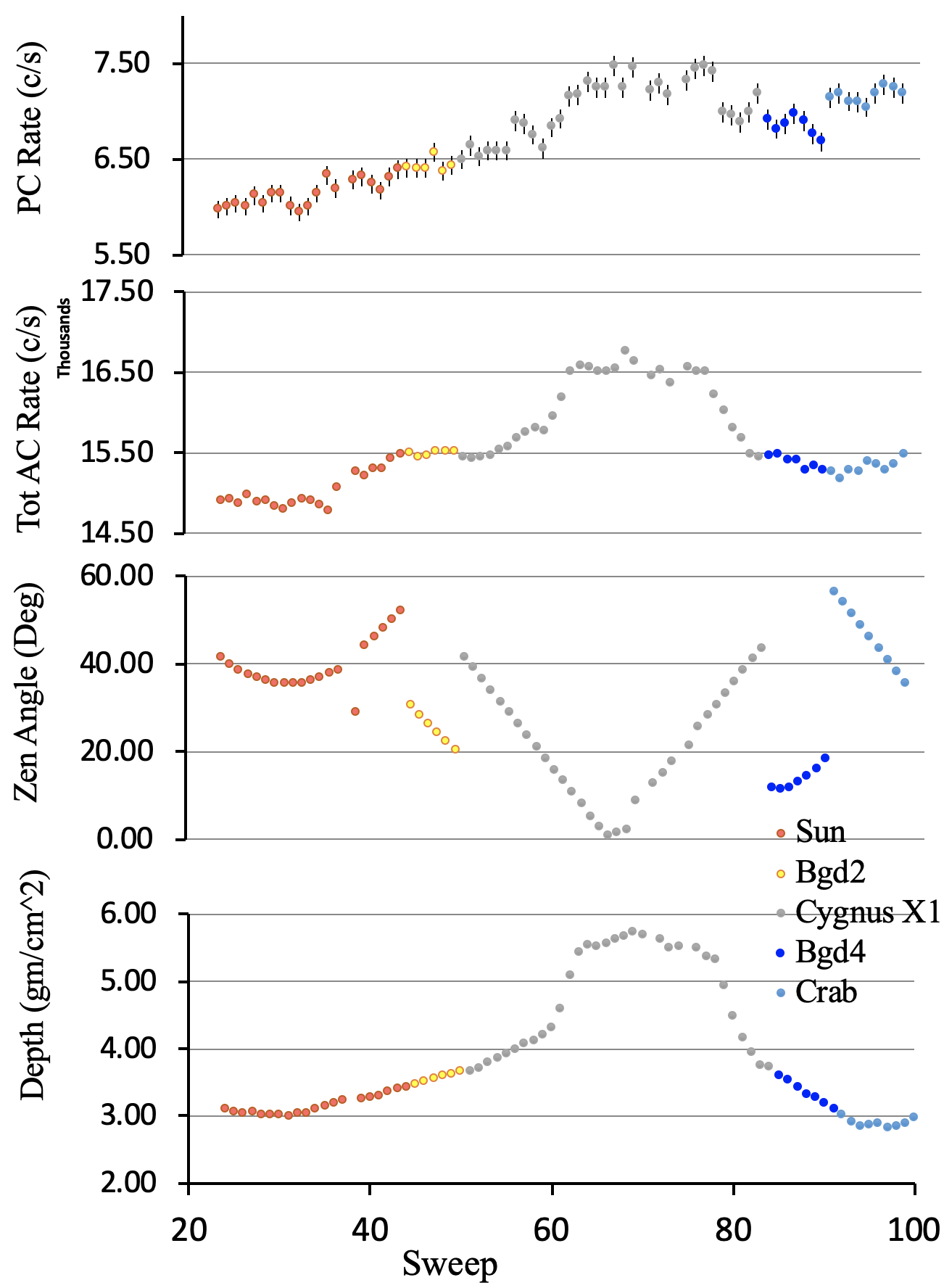}
    	\caption{Plot showing four of the flight parameters that were varying through out the flight (flight period). The total Anti-Coincidence (AC) rate is the summation of rates of all 6 AC panels. }
    	\label{fig:flt_profile_c}
\end{figure}

\begin{figure}[hbt]
 \centering
    	\includegraphics[width=0.93\textwidth]{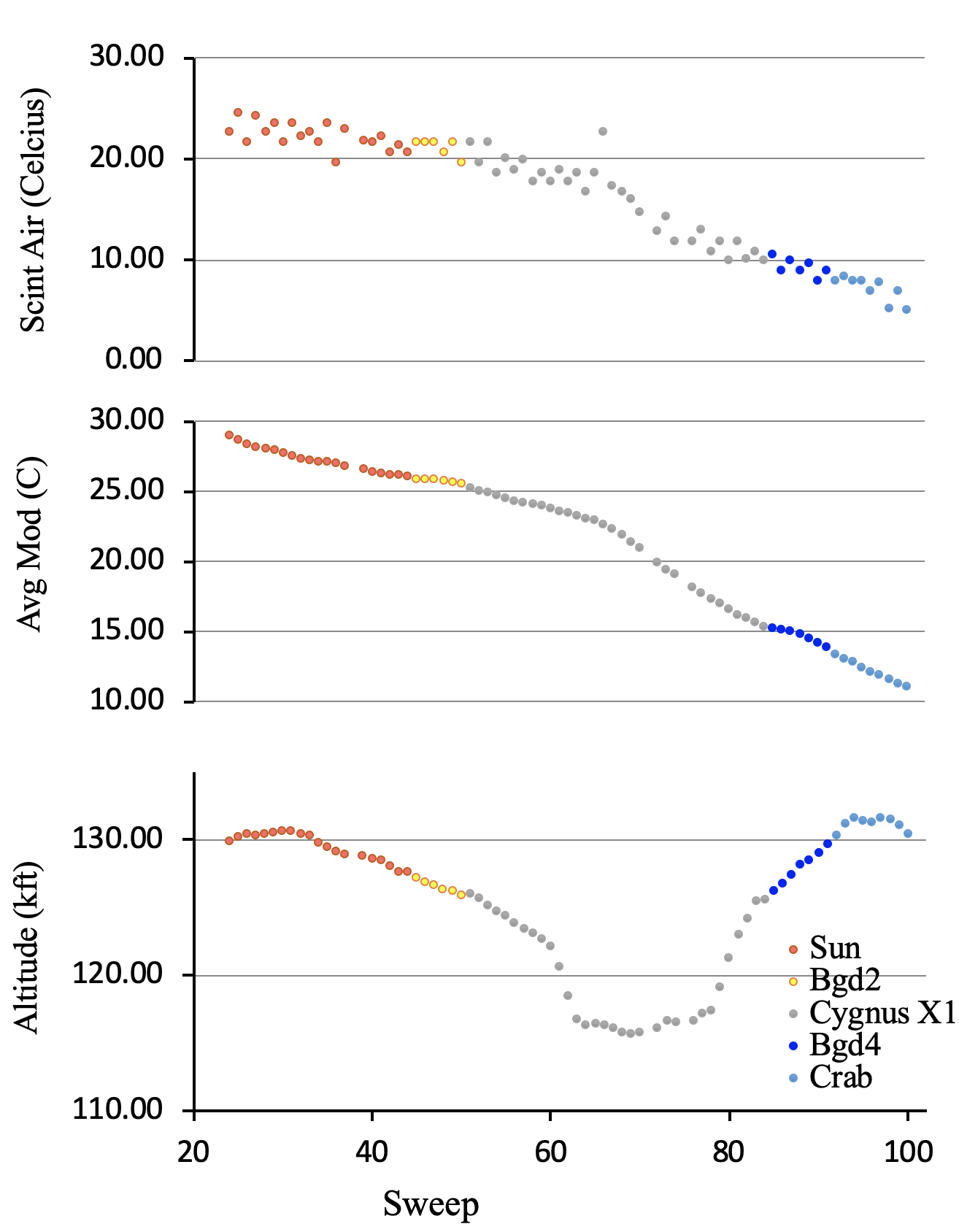}
    	\caption{Three more flight parameters that are varying during the flight. The Scint Air temperature is retrieved from the sensors in the scintillator array and Avg Mod temperature is the average of the temperatures reported by all the modules. }
    	\label{fig:flt_profile_d}
\end{figure}

Carefully planned use of ballast on this flight has allowed us to reach a higher altitude during the crab observation (as compared to the 2011).   
This is more clearly seen in Figure \ref{fig:flt_plan_compare} where the flight profile of 2014 flight is compared with flight profile from 2011 flight.
A number of instrument parameters can be correlated with background rate.
Figure \ref{fig:flt_profile_c} and \ref{fig:flt_profile_d} shows some of these parameters for the float period. 
The total Anti-Coincidence (AC) rate is the summation of all 6 AC panel's rate. 
The zenith angle plot is comparable to the flight plan (Figure \ref{fig:flt_plan}) we had devised.
The depth is the atmospheric depth of the instrument measured in terms of the residual atmosphere (g/cm$^2$). 
The scintillator air temperature is retrieved from the thermistors and the average module temperature is the average of the reported temperature by all the modules. 
These parameters were monitored during the flight.
The elevation motor did not respond at few instances during the flight and it had to be reset. 
The sweeps associated with this issue (1 sweep of Sun and 2 sweeps of Cygnus X-1)  are excluded in the analysis.

\clearpage
	 \chapter{Background Analysis}
	 \label{sec:background_analysis}
	 There are two parts to the background analysis discussed in the chapter. 
	 The first part (discussed in section \ref{sec:bgd_sim_vs_flt_noct}) focuses on a comparison of the simulated  flight background with the measured flight background. 
	This analysis strengthened our understanding of the instrument and led to improvements in the simulations, especially with regards to optical crosstalk. 
	 The second part (discussed in section \ref{sec:bgd_analysis_bgd_estimation}) focuses on the estimation of the flight background  during the Crab observation.

		\section{Simulated versus Measured Flight Background}
		\label{sec:bgd_sim_vs_flt_noct}
\begin{figure}[tbp]
\centering
\includegraphics[width=0.8\textwidth]{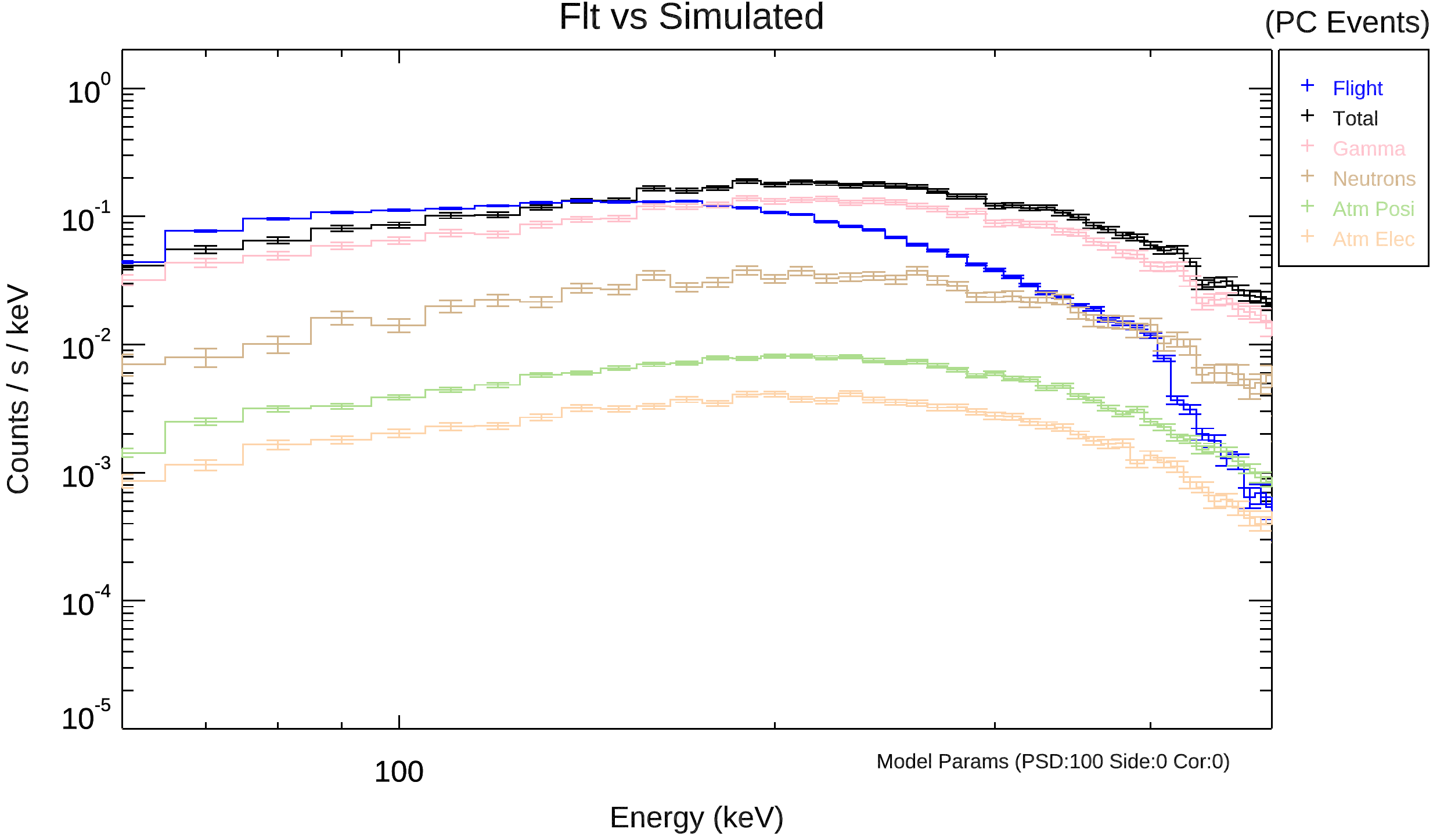}
\caption{The simulated output vs the flight output for PC all types events. The flight background is selected for the Bgd-4 observation. The flight background is shown in blue which is compared with the total simulated (black). We see a clear disagreement in the comparison and we hypothesize that this is due to crosstalk.}
\label{fig:bgd_sim_output_noct_flt}       
\end{figure}

		GRAPE observed two background regions during the 2014 flight, designated Bgd2 and Bgd4. 
		Background observations were designed to cover a full range fo zenith angles, using sky regions with no known sources above the sensitivity threshold.
		Altitude variations during the flight limited the value of these data. 
		The simulations discussed in section \ref{sec:ins_perf_simulation} were for a specific set of flight parameters (zenith angle 0, atmospheric depth of 3.5 g cm$^{-2}$) which approximated flight parameters for the Bgd4 observation. 
		Therefore, the Bgd4 measurement was used as a comparison with the simulated flight background. 
		Figure  \ref{fig:sim_bgd_output_noct} showed all of the simulated background components.
		The sum of all these components is represented by the black histogram labeled `Total'. 
		The comparison of the flight background with the simulated background is shown in Figure \ref{fig:bgd_sim_output_noct_flt}. 
		The flight background is shown in blue and the total simulated background is shown in black.
		The 4 most dominant background components are also shown in the Figure \ref{fig:bgd_sim_output_noct_flt}.

		The same energy and event filters were applied to both the flight and simulated data. 
		The energy range applied for plastics are from 10 keV to 200 keV, calorimeters from 40 keV to 400 keV and total energy from 50 keV to 600 keV. 
		Our analysis focused PC events with total energy from 70 keV to 200 keV. 
		We can notice from the results shown in Figure \ref{fig:bgd_sim_output_noct_flt} that the simulation underestimates the flight data at low energies and overestimates the flight data at higher energies. 
		This transition happens around $\sim$150 keV where the simulation agrees with the flight. 
		The atmospheric depth chosen for the simulation was 3.5 gm cm$^{-2}$, which was about the depth of GRAPE  during the Bgd4 observation. 
		The simulation was also performed for the instrument pointing at the zenith. 
		The zenith angle actually varied between  10-20$^\circ$ during the Bgd4 observation.
		Therefore, some level of disagreements may be expected between the simulated spectrum and the flight spectrum. 
		However, the discrepancies seen in our comparison  are thought to be too large to be attributed to these approximations. 
		We therefore decided to investigate the possibility that optical crosstalk could account for these discrepancies. 
				
		\section{Crosstalk Modeling}	
		\label{sec:bgd_crosstalk_modeling}
		
		Crosstalk was introduced in section \ref{sec:ins_crosstalk}.
		Our instrument has an intrinsic electronic crosstalk (electronic noise) of $\sim$3\%. 
		Additionally, the 1.5 mm glass window interface between the scintillator and the MAPMT anodes allow for optical crosstalk. 
		Ideally, each element of the scintillator grid corresponds to an unique anode in the MAPMT. 
		During crosstalk, the light from the scintillator element can spill over to its neighboring anodes in the MAPMT. 
		If the spilled light corresponds to an energy above the anode's set threshold, it will trigger that anode element.
		This trigger will lead to a misclassification of the event. 
		For example, for an original PC event, crosstalk can occur from the calorimeter to a neighboring plastic  anode where the added signal in the plastic anode is sufficient to trigger it.
		This results in a misclassified PPC event. 
		Because of the additional light output, crosstalk effectis much more clearly seen at higher energies. 
		GRAPE implements a PSD method to detect crosstalk between different type of scintillator elements which is discussed in section \ref{sec:ins_psd}. 
		This mechanism, however, does not deal with crosstalk between same type scintillator elements.
		Additionally, this method might not be 100\% efficient which is taken into consideration in our modeling of the crosstalk effect.
		
		\subsection{Development of the crosstalk model}
\begin{figure}[hbtp]
\centering 
\vspace{0.6cm}
\includegraphics[width=0.9\textwidth]{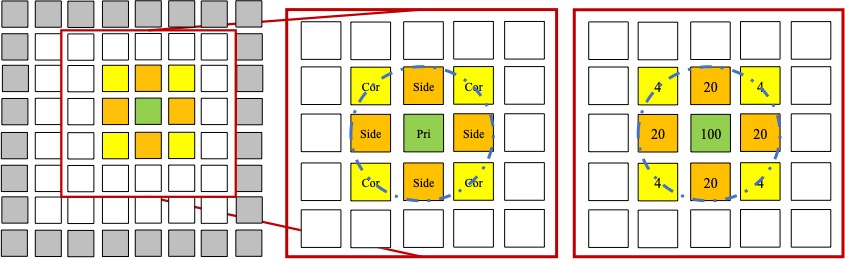}
\caption{Diagram showing the model and the primary anode and its neighboring anodes which are side adjacent and corner adjacent. As the energy (or light output) is distributed symmetrically. An imaginary circle can be drawn to see area covered by corner versus the side adjacent. The corner adjacent anodes received 20\% of the CT that the side adjacent received. As an example, a 100 keV energy deposited in the primary would crosstalk to side adjacent anode for 20 keV (for CT\% of 20\%) and subsequently a 20\% of the 20 keV (4 keV) is deposited in the corner adjacent. }
\label{fig:bgd_ct_model}       
\end{figure}

		The crosstalk model was developed as a part of the simulations to closely replicate the crosstalk process.
		As a first step, the crosstalk model identifies the primary anode along with its neighboring anodes.
		Then it distributes a certain portion of energy detected by the primary anode to its neighboring anodes.
		Energy is distributed to the neighboring anodes instead of light as the energy is proportional to the scintillated light. 
		The algorithm labels these neighboring anodes as "corner adjacent" or "side adjacent" depending on its position relative to the primary anode.
		This is illustrated in Figure \ref{fig:bgd_ct_model}.
		The algorithm also  identifies the type of the neighboring anodes which is needed for crosstalk between different scintillator types as the difference in light yield has to be considered for crosstalk between them.
		The light is distributed symmetrically to the neighboring anodes.
		The lateral area is used to calculate the amount of energy to be distributed to the side adjacent versus the corner adjacent.
		The corner adjacent covers $\sim$20\% of the area covered by the side adjacent.
		Therefore, for the preliminary iteration of the model the corner adjacent anodes received 20\% of the energy crosstalked to side adjacent. 
		In this iteration only the side adjacent crosstalk was a free parameter and not the corner adjacent.
		
\begin{figure}[hbtp]
 \centering
 \vspace{0.6cm}
 	\includegraphics[width=0.65\textwidth]{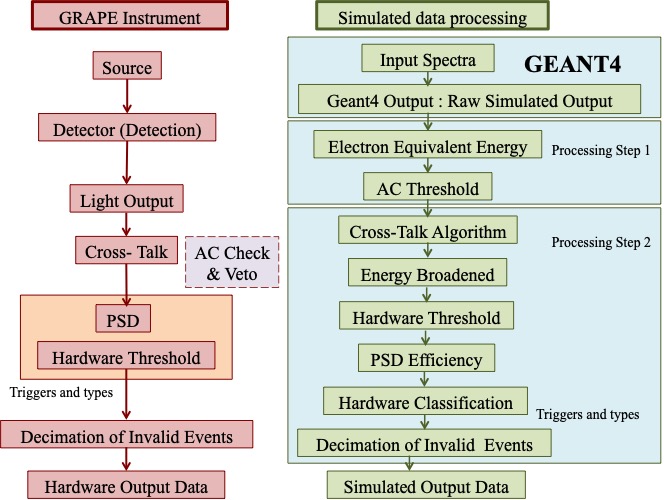}
    	\caption{An updated flow diagram comparing the key steps for the hardware processing (left) and the simulated data processing (right) with crosstalk modeling included in the algorithm. The Cross-Talk Algorithm has been integrated in the processing step to include the crosstalk model.}
    	\label{fig:bgd_sim_flow_process}
\end{figure}

		\subsection*{Updated Simulated Output processing}
		The crosstalk (CT) algorithm is applied to the simulated raw output during the simulated output processing steps discussed in section \ref{sec:ins_perf_sim_out}. 
		The updated flow of this processing is shown in Figure \ref{fig:bgd_sim_flow_process}. 
		To validate the crosstalk model, we compared he percentage of each event class in both the calibration data and the simulations.
		Initial results, comparing the calibration data with the baseline simulations (with no crosstalk) is shown in Table \ref{table:bgd_ct_algo_noct}.
		
		\begin{table}[tbp]
\begin{center}
\caption{Table comparing various event types from Co57 callibration run with the simulated data without crosstalk model.}
\label{table:bgd_ct_algo_noct}
   	\includegraphics[width=0.8\textwidth]{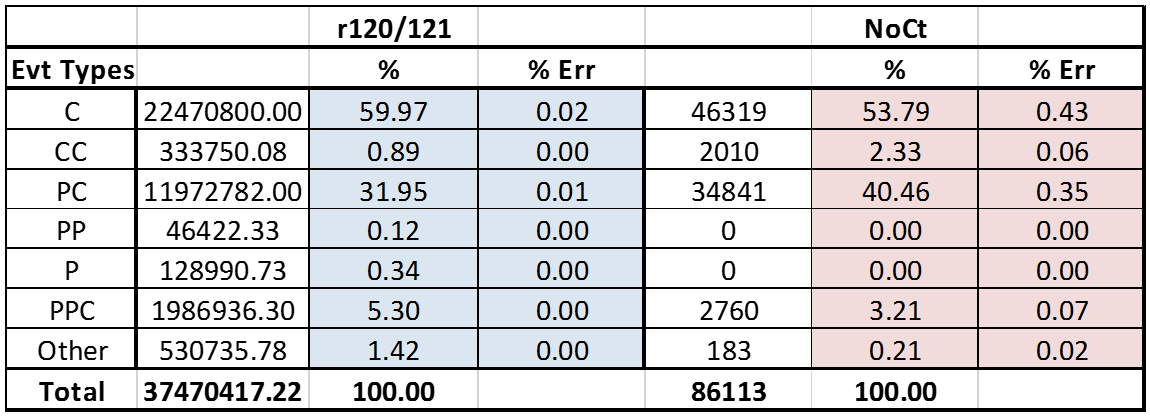}
  \end{center}
\end{table}

		GRAPE also implements a PSD circuit designed to identify crosstalk between two different scintillator types, the plastic  and  the CsI (calorimeters) scintillators. 
		PSD briefly discussed in section \ref{sec:ins_psd}.
		The PSD circuit does not prevent crosstalk between the same scintillator types. 
		PSD efficiency was introduced as a parameter to the crosstalk model assuming that the PSD circuit may not be 100\% effective, allowing some amount of crosstalk between the different scintillator types. 
		A PSD efficiency of 90\% would allow one in ten crosstalk events (between two different scintillator types) to pass through. 
		
		The light yield for plastics and calorimeter element are different so this difference has to be accounted for in the algorithm. 
		The light yield of CsI(Tl) is roughly 5 times higher than plastic scintillator which means that if a 100 keV photon hits a calorimeter, the light generated is five times higher.
		For a defined crosstalk of 10\% , 10\% of the light is transferred to the side adjacent anodes. 
		If the side adjacent anode is the same scintillator type as the primary anode then it would receive light equivalent to 10 keV. 
		However, if the crosstalk was from C to P, the light crosstalk to plastic (which would ideally be representing 10 keV), would be registered as 50 keV by the plastic anode due to the difference in light yield. 
	   A 100 keV photon is in range of GRAPE and the plastics have a threshold of around 10 keV.
	   Therefore it is likely that crosstalk from calorimeter to plastic can occur.
		 Conversely, crosstalk from plastic to calorimeter can be tracked similarly. 
		 The threshold for calorimeters are around 20 keV. 
		 A 10\% crosstalk value representing 20 keV to a side adjacent anode would be from a 200 keV photon in the primary plastic anode. 
		 The algorithm allows for the crosstalk from plastic to calorimeter element, even though it does not contribute significantly to GRAPE.
		 		 
		\begin{table}[tbp]
\begin{center}
\caption{Table comparing the event types for various CT\%. Shaded blue is the calibration data, orange is the one without any crosstalk. The table shows CT\% in 5\% increment and the 15\% was the closest to the calibration. The best fit is then retrieved around this 15\% CT. }
\label{table:bgd_ct_algo4_a}
   	\includegraphics[width=0.8\textwidth]{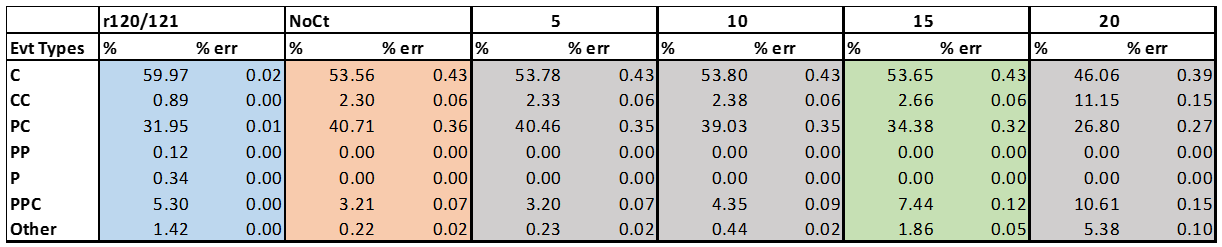}
  \end{center}
\end{table}

		\begin{table}[tbp]
\begin{center}
\caption{Table comparing the event types for various CT\% and PSD efficiency around the 15\% CT value. Each table is for a specific CT\% and each column represents a PSD efficiency (here it is shown from 94-100). The value shown here are the difference from the callibration data. The best fit was defined by sum of this difference squared for all event types and it is shown on the last row. This concluded that for this preliminary model, CT\% of 15 with 100\% efficiency was the best. }
\label{table:bgd_ct_algo4_b}
   	\includegraphics[width=0.68\textwidth]{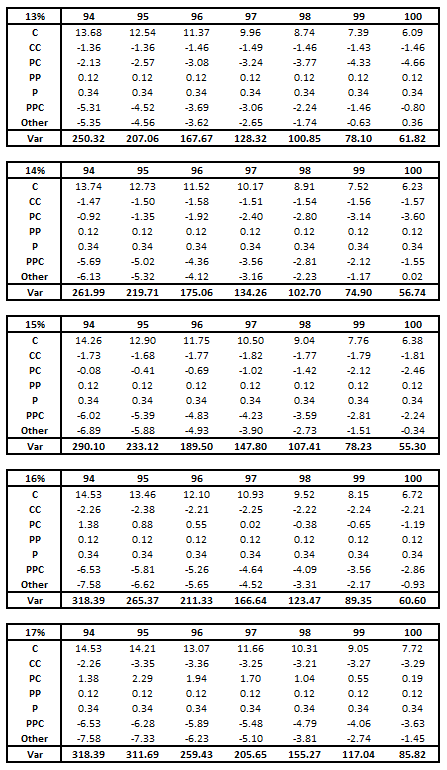}
  \end{center}
\end{table}

		A crosstalk model with two free parameters, the CT\% (side adjacent crosstalk) and PSD\% (PSD efficiency), was used to model the crosstalk.
		The model assumes that the corner adjacent received 20\% of the light received by side adjacent elements. 
		The $^{57}$Co calibration run was used to compare the simulated result to validate the crosstalk model. 
		The percentage of each event class for the callibration run and the initial simulated data (without crosstalk model) is shown in Table \ref{table:bgd_ct_algo_noct}.
		The crosstalk is varied to find the crosstalk percentage that best matches the calibration data.
		The optimization centered around C and PC event as these were the most dominant event types.
		Table \ref{table:bgd_ct_algo4_a}  shows various iterations of crosstalk percentage (at 5\% increments). 
		This table shows the data at 100\% PSD efficiency.
		The C and PC  statistics started to decrease as the crosstalk percentage was increased which is as expected.
		As the crosstalk percentage increases, the amount of light making its way to adjacent elements increases and a larger number of anode elements will be triggered. 
		This results in events with more anodes trigger so a C could become PC, CC, PPC, etc. and PC event could become PPC,  PPPC, PCC, etc. 
		The initial analysis (Table \ref{table:bgd_ct_algo4_a}) suggested a crosstalk level of 15\%. 
		A refined analysis used a finer variations of the crosstalk.
		Additionally, the PSD efficiency was also varied to find the best values of both parameters. 
		Table \ref{table:bgd_ct_algo4_b} shows the results. 
		These data showed that 15\% crosstalk with 100\% PSD efficiency provided the best description of the data.

\begin{figure}[hbtp]
 \centering
\begin{subfigure}[b]{0.95\textwidth}
 		 \includegraphics[width=1\linewidth]{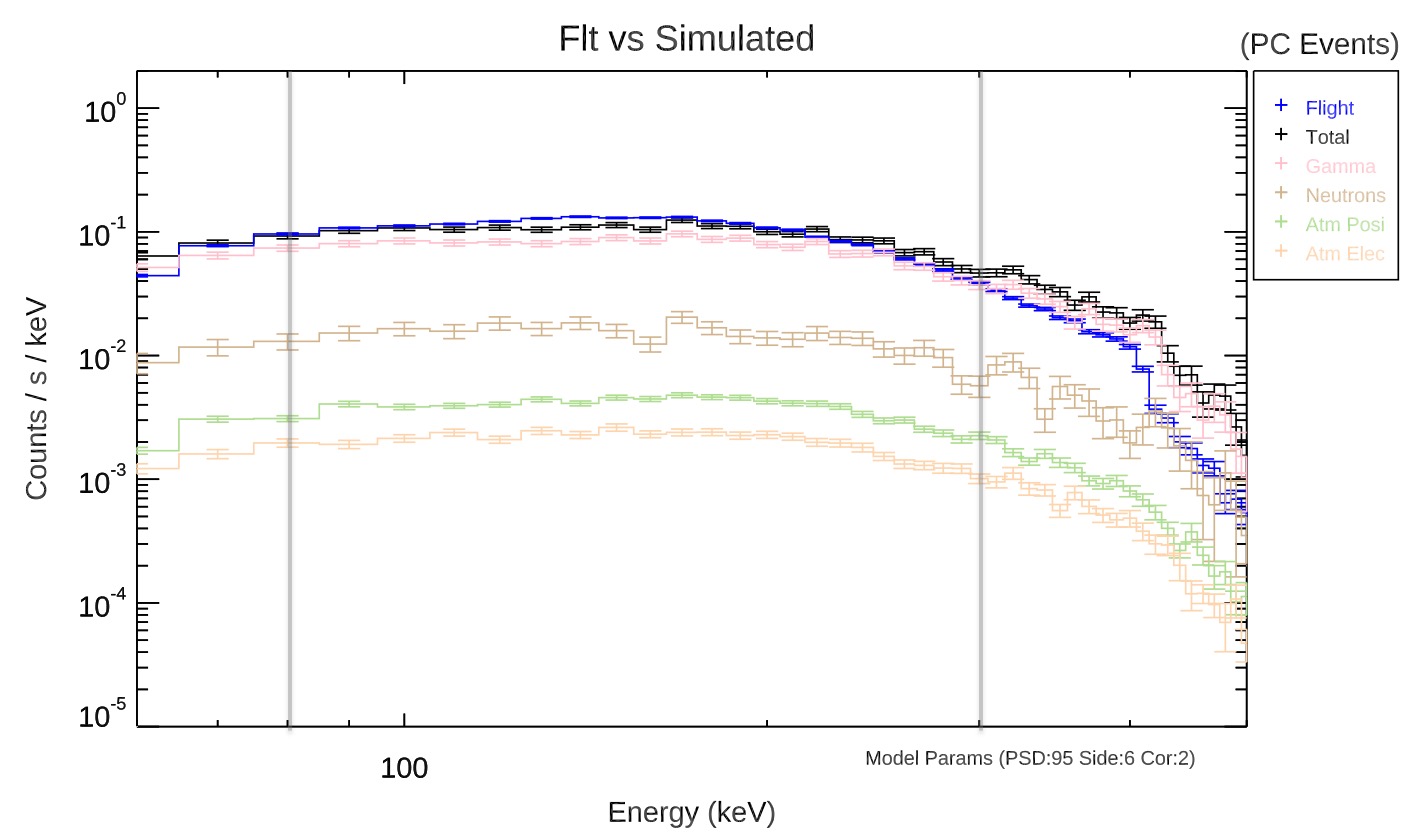}
		 \caption{}
\end{subfigure}

 \begin{subfigure}[b]{0.95\textwidth}
 		 \includegraphics[width=1\linewidth]{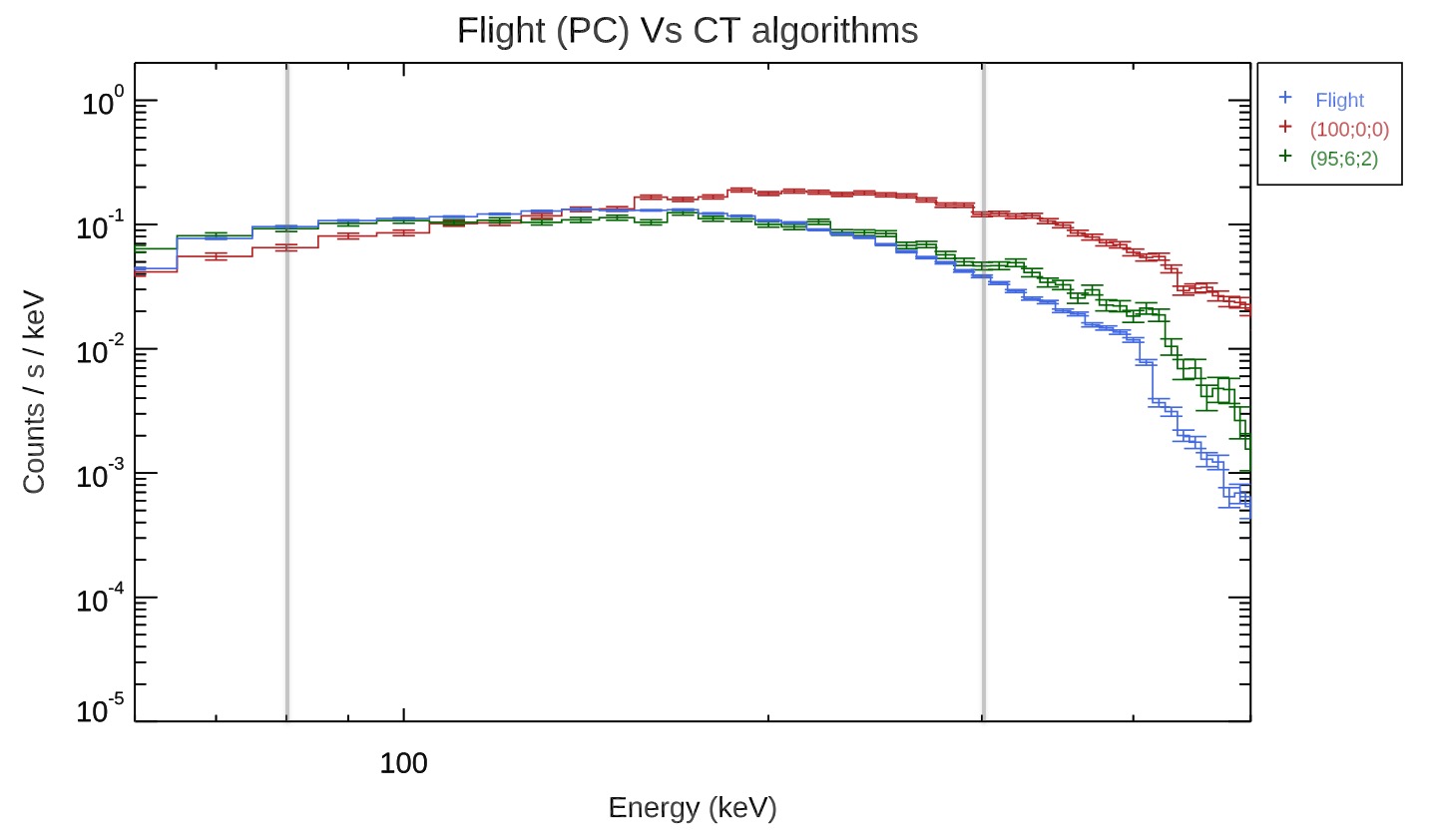}
		 \caption{}
\end{subfigure} 
  \caption{Comparison of flight background versus crosstalk induced simulated energy loss spectra. These are shown for PC events of all type. a) Black is the total simulated data with the crosstalk model of (95;6;2) and blue is the flight data. The optimization is done for 70 to 300 keV. b) This shows the flight data in blue, green for the simulated data with the best fit and red is the simulated data without the crosstalk model. }
\label{fig:bgd_ct_bestfit}
\end{figure}
	
		The comparison with $^{57}$Co data gave us the crosstalk validation at a specific energy.
		A more complete validation of the model could be achieved by considering flight background simulations. 
		PC events were selected for this analysis as those are the events that carry the polarization signature.
		Bgd4 measurements were used to compare with the simulated flight background.
		The model was modified to include crosstalk to the the corner adjacent anodes as a free parameter.
		This described our final crosstalk model with 3 parameters (Side adjacent CT\%, Corner adjacent CT\% and the  PSD efficiency). 
		We expected the PSD efficiency to be around 90\% to 100\%, side adjacent around 10\% and  corner adjacent around 5\%. 
		The model was represented in form of (PSD efficiency; side adjacent ; corner adjacent). 
		A series of simulated data with various combinations of these three parameters was compared with the measured flight data. 
		The analysis was focused  between 70 keV and 300 keV to find the best fit parameters. 
		The best fit parameters were PSD efficiency of 95\%, side crosstalk of 6\% and corner crosstalk of 2\% (95; 6; 2) for PC events. 
		Figure \ref{fig:bgd_ct_bestfit}a showcases this result for various background components simulated.
		Figure \ref{fig:bgd_ct_bestfit}b shows the best modeled fit (in green), along with the flight data (in blue) and the simulated result without the crosstalk model (in red). 
		The simulated data with the crosstalk model (95; 6; 2) is in good agreement with the measured data. 
		Therefore this crosstalk model (95; 6; 2) was applied on all the simulated data used in the analysis.
		
		From Figure \ref{fig:bgd_ct_bestfit}b, we can see that the simulated data without the crosstalk switched from under-estimating the flight data to over-estimating around this energy. 
		Even for the best fit parameters, the data does not change much around the 122 keV.
		Therefore, 122 keV might not have been the best energy to test the crosstalk modeling.
		Additionally the simulation is done for a specific set of flight parameters but the flight parameters are varying for the flight data.
		The flight parameters was also varying during Bgd4 observation (the altitude varied $\sim$5 kft). 
		Therefore, some level of discrepancy is to be expected between the features of the simulated and the measured data.

	\section{Background Estimation}
	\label{sec:bgd_analysis_bgd_estimation}
	The primary objective of GRAPE 2014 was to measure the polarization of the Crab.
	The background dominates the measurements at flight altitude. 
	Therefore a proper estimation of the background is needed to analyze the Crab signal from the flight observation.
	Estimating the background for a balloon borne experiment is a challenging task as the flight parameters (which influence the background) are varying through out the flight. 
	
	Ideally, the background measurements are made during the flight with the flight parameters similar to that of the Crab observation.
	Typically this is achieved by one of two methods. 
	The first method consists of alternating between measuring the background and measuring the source (the Crab) during the source observation.
	This would require the instrument to change the pointing between a background region and the Crab or simply changing azimuth by 180$^\circ$.
	There would some loss of source observation time but this would guarantee that  the flight parameters would be similar between the source and background measurements.
	The second method consists of observing a background regions with a range flight parameters that includes the flight parameters for the Crab observation. 
	The flight plan was carefully designed to provide background measurements over the same range of zenith angles expected for the Crab.
      Unfortunately, altitude differences between the source and background observations rendered the background data unusable. 
	Therefore,  a new approach was needed to estimate the background.
	One option would have been to use the background simulation that we used in the previous section. 
	However, the background simulation were performed for a specific set of flight parameters.
	The flight parameters are varying during the Crab observation.
	The simulation would have to be conducted for each set of the flight parameters for each of the background components separately.
	Furthermore, there might be other flight or instrument parameters that might affect the background measurements.
	Therefore simulation approach was not optimal for this estimation. 
	The flight had a variety of flight parameters  and instrument data (rates, etc) that could be used to define the background. 
	However, not all of these flight parameters affect the background.
	The important flight parameters that affect the background had to be identified.
	Principle Component Analysis (PCA) was used to restructure the data set so that it could be used to define the background.

	\subsection{Principle Component Analysis (PCA)}
		\label{sec:bgd_PCA}
		Principle Component Analysis (PCA) is a multivariate analysis tool used to find correlations amongst parameters that define the dataset.
		PCA can then be used to redefine the dataset with fewer parameters.
		The central step of PCA is to derive an intermediate set of variables (Principle Components), which are mutually-orthogonal and represented by a linear combination of the variables in the data \citep{Whitney1983a}.
		These Principle Components ($\xi$) redefine the data.
		The number of $\xi$ is equivalent to the number of variables.
		Each of the $\xi$ cover some amount of variation present in the data \citep{Francis1999}. 
		The variations covered by the $\xi$ can be used to determine the most important principle components.
		The higher the variation covered, the greater the importance of the $\xi$ \citep{Francis1999}. 
		For a data set with large number of variables, the important principle components is typically less than the total number of variables \citep{Francis1999}. 
		The PCA can be better explained using an example.

	\subsubsection{Example of PCA}
	 Let us assume that we have a data set of variables x, y and z. 
	 This data set is generated by randomizing y about y = x line to show the working of PCA on a correlated data set.
	 This data set is shown in Table \ref{table:bgd_pca_ex_a}a. 
	 Variable \textbf{z} is the variable we are interested in and is defined by some combination of variables \textbf{x} and \textbf{y}. 
	 Let us also assume that there are some new x and y for which the variable z is unknown. 
	 The goal would be to use the x and y values to define the data and further extend it to estimate the new z using the new x and new y values.

\begin{figure}[hbtp]
 \centering
\begin{subfigure}[b]{0.75\textwidth}
 		 \includegraphics[width=1\linewidth]{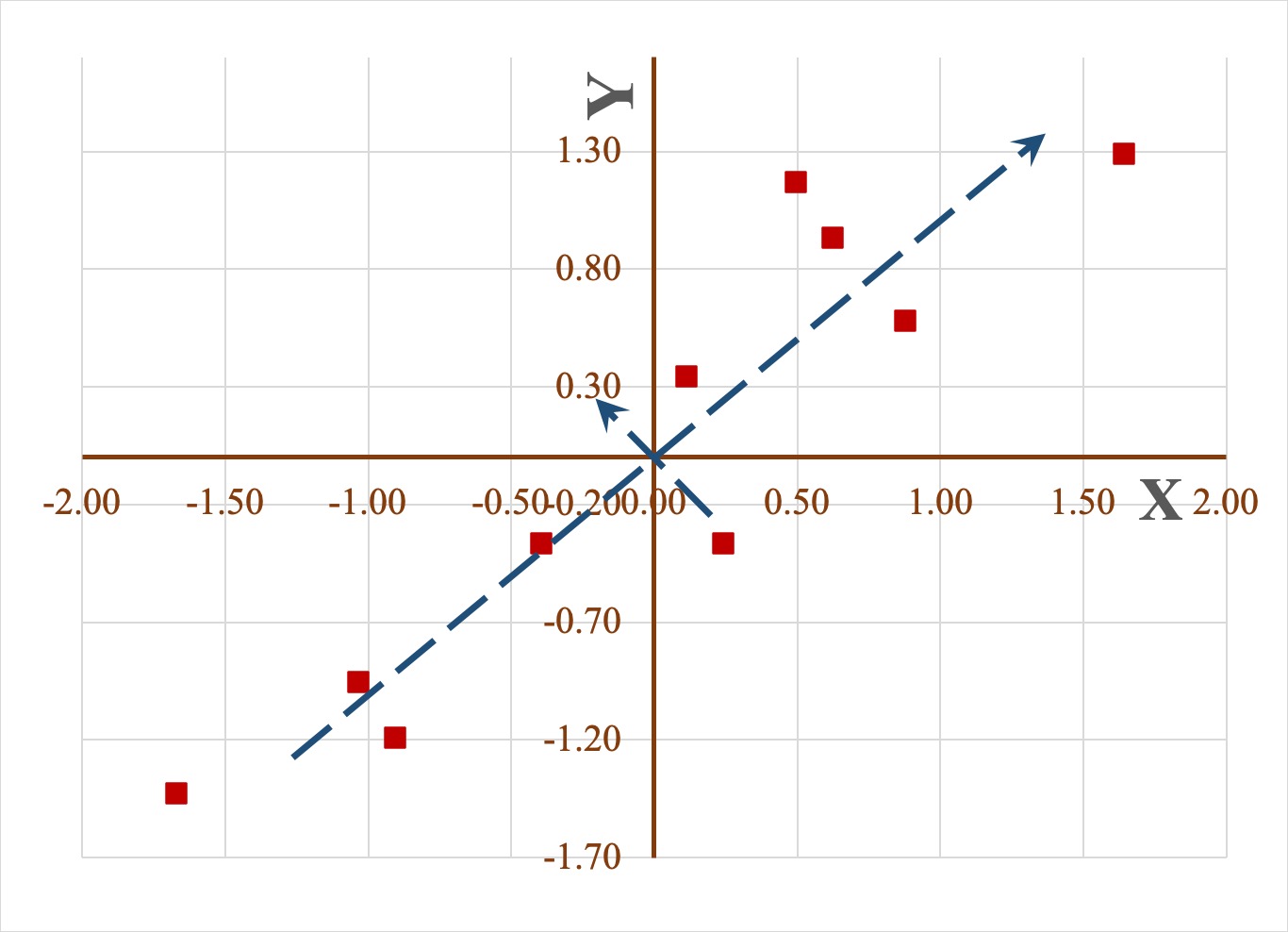}
		 \caption{}
\end{subfigure}     

 \begin{subfigure}[b]{0.75\textwidth}
 		 \includegraphics[width=1\linewidth]{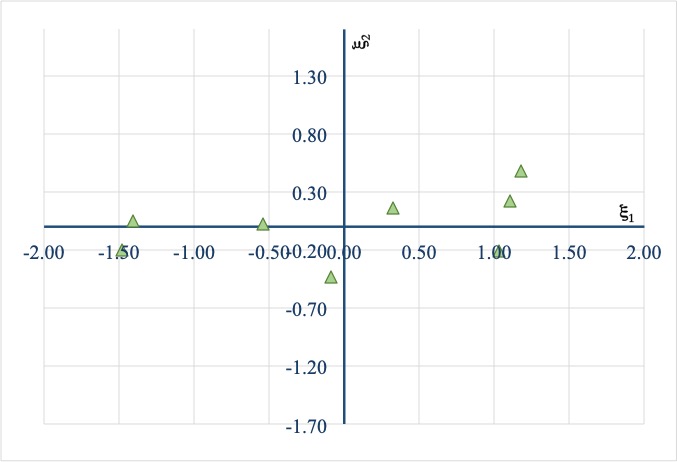}
		 \caption{}
\end{subfigure} 
  \caption{a) Plot of the standardized x and y values. It also illustrates the two eigenvectors that defines the variation (spread) in the data. b) The data being redefined by the two principle components. The principle components redefines the data which is analogous to rotation of the data to be represented by new orthogonal vectors.}
\label{fig:bgd_pca_ex_b}
\end{figure}

	The first step towards this estimation is to standardize the x and y values. 
	This is done by subtracting the mean and dividing by the standard deviation.
	The standardized data is shown under the column standardize x and standardize y in Table \ref{table:bgd_pca_ex_a}a. 
	This standardized data is plotted in in Figure \ref{fig:bgd_pca_ex_b}a where the axis are the standardized X and Y. 
	This process centers the spread of each variable about the origin.
	
	The correlation between the variables is defined using the correlation matrix.
	The correlation matrix and the respective eigenvalues ($\lambda$) and eigenvectors ($\nu$) calculated for this matrix are shown in Table \ref{table:bgd_pca_ex_a}b. 
	The $\nu$ are orthogonal to each other so these are used to generate the PriComps.
	$\xi_1$ is a linear combination of $\nu_1$ and the standardized variables. 
	The number of $\xi$'s is equal to the number of variables, therefore this example will have two $\xi$ ($\xi_1$ and $\xi_2$). 
	The $\xi$'s are tabulated in Table \ref{table:bgd_pca_ex_a}c.
	The redefined data using these principle components is shown in Figure \ref{fig:bgd_pca_ex_b}b where the data is just rotated from the original definitions and the new axis are $\xi_1$ and $\xi_2$.
	\begin{table}[tbp]
\begin{center}
\caption{Table showing various data sets for the PCA example. a) This shows the x, y and z variables with the standardized X and Y.  b) shows the correlation matrix generated from the standardized x and y along with calculated eigenvalues and eigenvectors. c) Two principle components generated using the the two eigenvectors.}
\label{table:bgd_pca_ex_a}
   \begin{subfigure}[b]{0.7\textwidth}
 		 \includegraphics[width=1\linewidth]{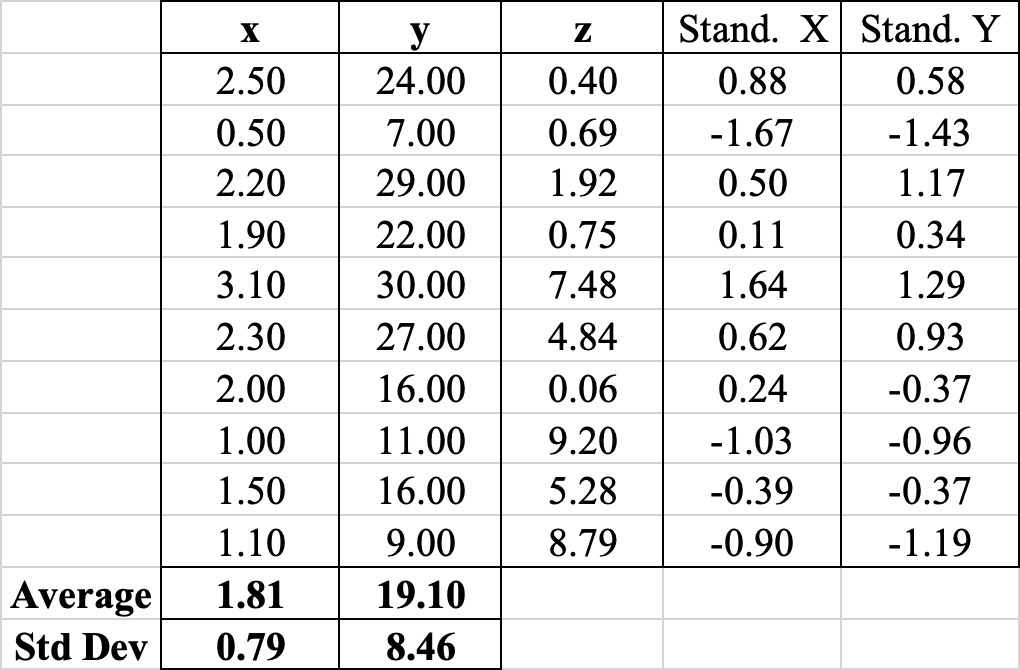}
		 \caption{}
\end{subfigure} %

 \begin{subfigure}[b]{0.6\textwidth}
 		 \includegraphics[width=1\linewidth]{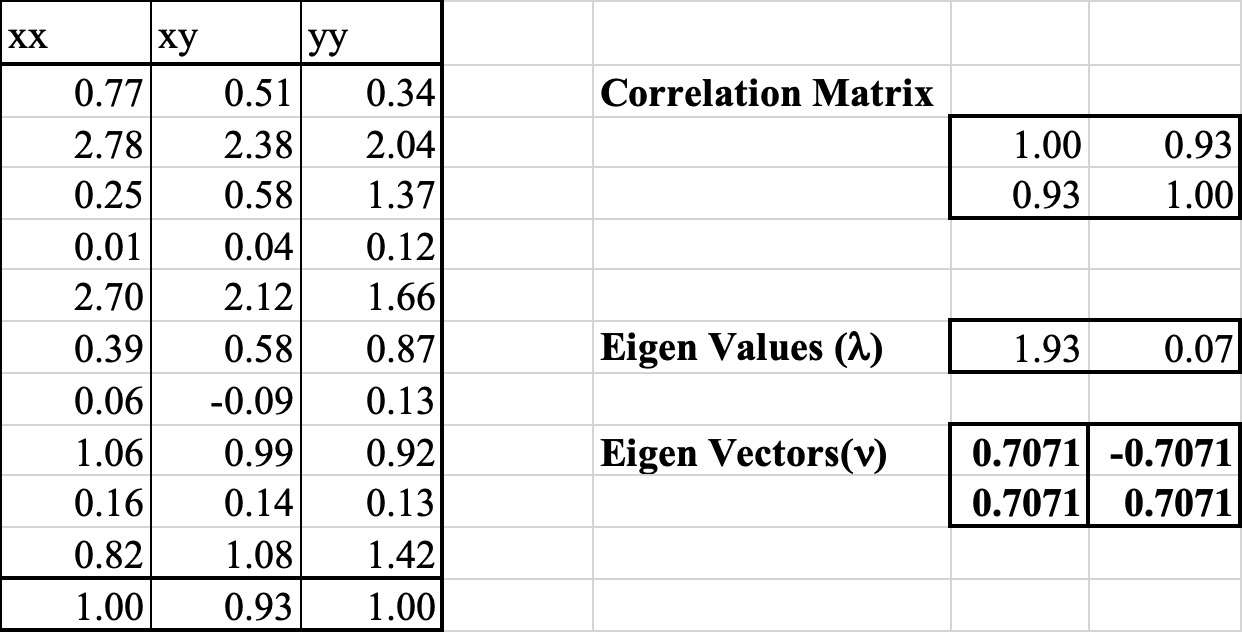}
		 \caption{}
\end{subfigure} \; \; \; 
 \begin{subfigure}[b]{0.2\textwidth}
 		 \includegraphics[width=1\linewidth]{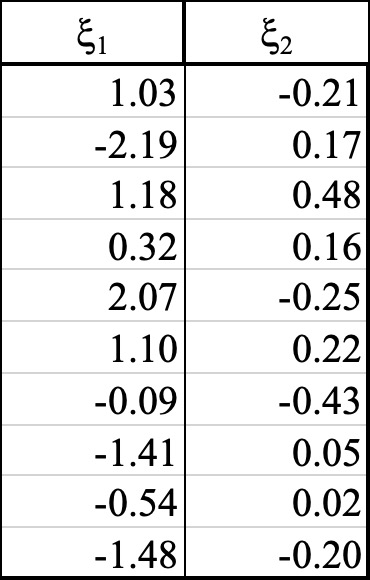}
		 \caption{}
\end{subfigure} 

  \end{center}
\end{table}

\begin{figure}[hbtp]
\centering 
\includegraphics[width=0.70\textwidth]{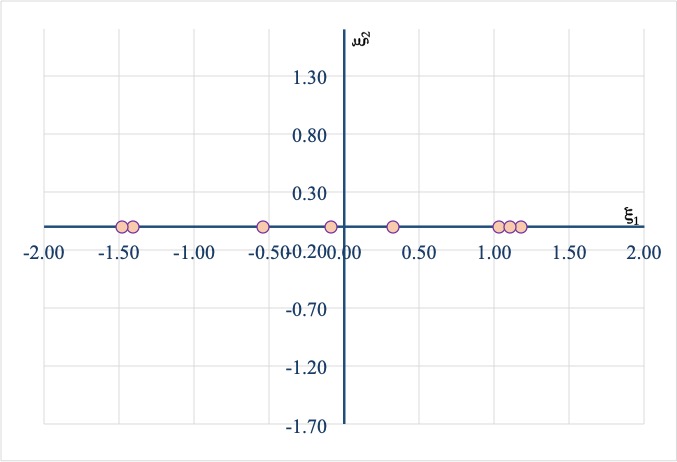}
\caption{Plot representing the data being represented by only the principle component 1 which covers 97\% of the variation (spread) in the data. The data is now defined by only one variable which is reduced from the original two variables.}
\label{fig:bgd_pca_ex_c}       
\end{figure}

	These $\lambda$'s and $\nu$'s define the variation (spread) of data and are plotted in \ref{fig:bgd_pca_ex_b}a with arrows. 
	The $\lambda$s gives a quantitative number for the variation covered by the corresponding $\nu$ and subsequently the corresponding $\xi$. 
	The largest $\lambda$ covers the most variation and this is defined as $\lambda_1$. 
	$\lambda_2$ represents the second highest variation and so on.
	In our example and in the Figure \ref{fig:bgd_pca_ex_b}a, the longer arrow corresponds to $\nu_1$ and its corresponding $\lambda_1$.
	The shorter arrow corresponds to $\nu_2$ and its respective $\lambda_2$. 
	$\lambda_1$ has a value of 1.93 and $\lambda_2$ has a vale of 0.07. 
	The $\lambda$s sum up to 2.00 so the $\nu_1$ (which is related to $\lambda_1$) covers 97\% of the variation and $\nu_2$ (corresponding to $\lambda_2$) covers 3\% of variation in the data. 
	The $\xi$s are generated using these $\nu$s, so $\xi_1$ covers 97\% of the data and $\xi_2$ covers 3\% of the data. 
	$\xi_1$ covers 97\% of the variation so the data can be redefined using only $\xi_1$ (by sacrificing 3\% of the variation). 
	This is represented by the plot in Figure \ref{fig:bgd_pca_ex_c}. 	 

	 In this example, the number of variables that define the data was reduced from 2 to 1. 
	 Therefore the variable z is defined using only $\xi_1$.
	 For a larger data sets with more variables, the reduction of variables is usually more than 1. 
	 The $\xi_1$ is generated from linear combination of both x and y with our $\nu_1$. 
	 Their importance can be evaluated by looking at the elements of the $\nu_1$ which reveals that both of the variables contribute equally to the $\xi_1$.
	 This means that both variables x and y are important but one variable is enough to define the data set.
	 Additionally, in bigger data sets, there could be variables that are analogous to noise which do not define the data.
	 In those cases, the noisy variable's contribution to the most important $\xi$s is significantly low.
	 The amount of variation covered (which is cumulative for more than one $\xi$) defines the number of $\xi$ that are important. 
	 This limit is set manually and depends on the need of the analysis.
	 In this example, $\xi_1$ was enough to cover 97\% of variation. If the limit for variation covered was set to 99\% then both of the $\xi$s ($\xi_1$ and $\xi_2$) would had to be used.
	Once the most important $\xi$s are determined, a regression analysis can be applied to fit the z variable using the $\xi$s. 
	Then it could be further interpolated for the new x and y values. 
	 	 
	\subsection{PCA for GRAPE}
	\label{sec:bgd_anal_pca_grape}
	The PCA for GRAPE follows a similar approach as that of the example provided above. 
	There are a number of flight parameters or measurement of parameter that vary throughout the flight and that might somehow be connected to the instrument background. 
	The goal is to estimate the instrument background using parameters that are measured independent of the background.
	From our example, background counts are equivalent to z variable and the flight parameters are equivalent to the x and y variables in the examples. 
	The Crab was observed with its own flight parameters (equivalent to the new x and y variables in the example).
	The goal here is to estimate background for the Crab observation (equivalent to the new z in example) using the Crab flight parameters. 
	\begin{table}[htbp]
\begin{center}
\caption{Table of the measured parameters from the GRAPE flight used for Principle Component Analysis to estimate the background.}
\label{table:bgd_pca_grp_table}
   	\includegraphics[width=0.72\textwidth]{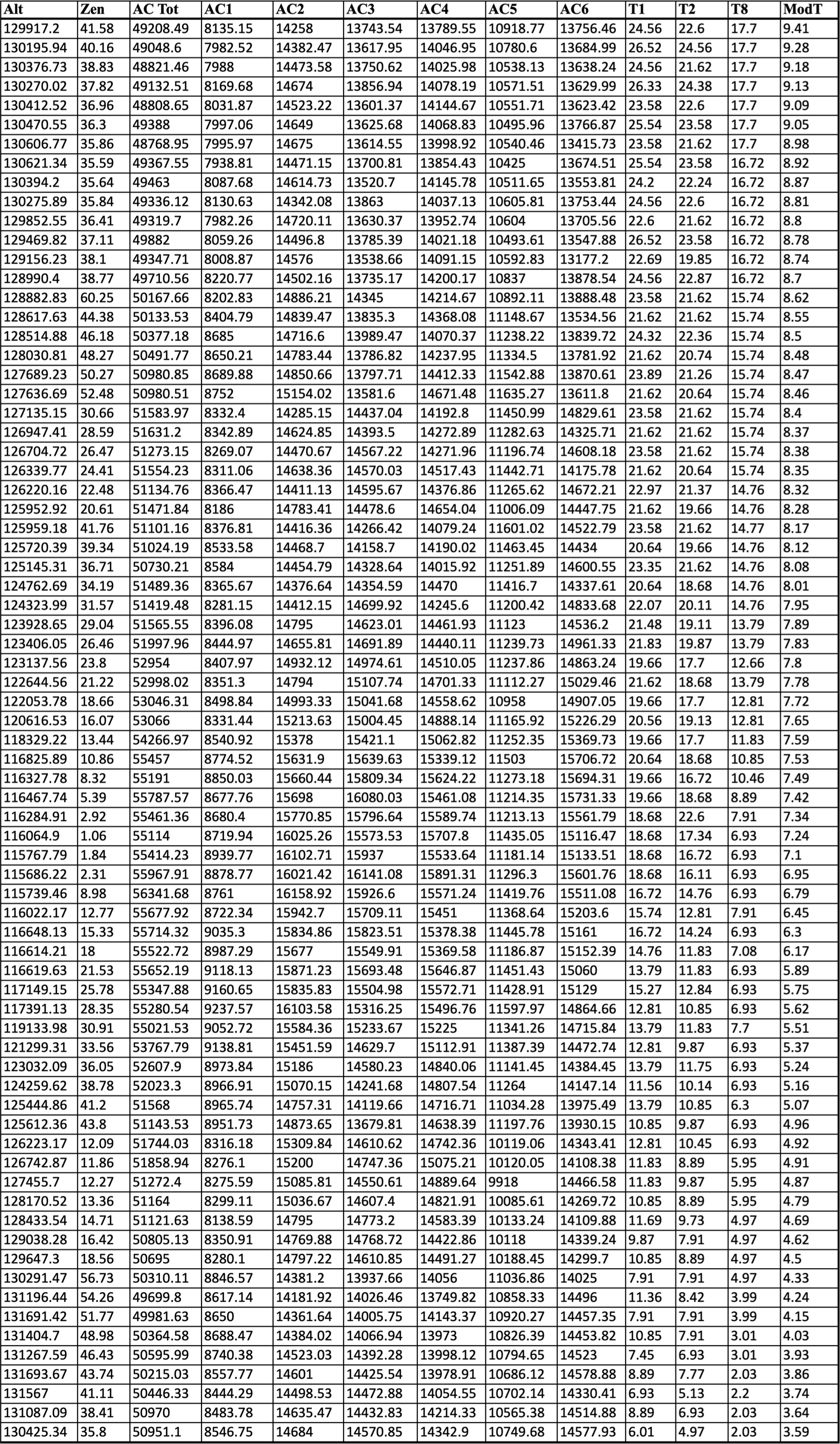}
  \end{center}
\end{table}

	The first step involved collecting the flight parameters that could define or be a surrogate for background during the flight. 
	These parameters were selected because it was thought that they might somehow be correlated with the background.
	There were 13 flight parameters selected, as shown in Table \ref{table:bgd_pca_grp_table}. 
	The 13 parameters  included balloon altitude and zenith pointing angle which are well-known to effect the background rate. 
	Anti-Coincidence (AC) counts were selected because they provide a direct measure of the radiation environment.
	Each of the six AC panel's counts as well as the total AC counts was used for this purpose.   
	The temperatures dropped low at the balloon altitude during the night.
	There were heaters placed at the elevation motor to prevent any malfunctions due to this drop in temperature. 
	The CsI(Tl) scintillator element's light yield varies slightly within the the temperature variation. 
	This variation in light yield is small and we do not expect it to affect the background.
	However, four of the temperatures are used and the PCA is allowed to determine the significance of these temperatures. 
	The insignificant parameters do not contribute the most important principle components. 
	Four temperatures were used which included three different temperatures from the thermistors  and an average temperature from all the modules.
	Some of these varying flight parameters are shown in Figure \ref{fig:flt_profile_c} and \ref{fig:flt_profile_d}.
				\begin{table}[htbp]
\begin{center}
\caption{Table showing the 13 eigenvalues along with individual and cumulative variation covered. The green selection is the biggest seven eigenvalues that cover over the 99\% of the variation  covered in the data.}
\label{table:bgd_pca_grp_eigentable}
   	\includegraphics[width=1\textwidth]{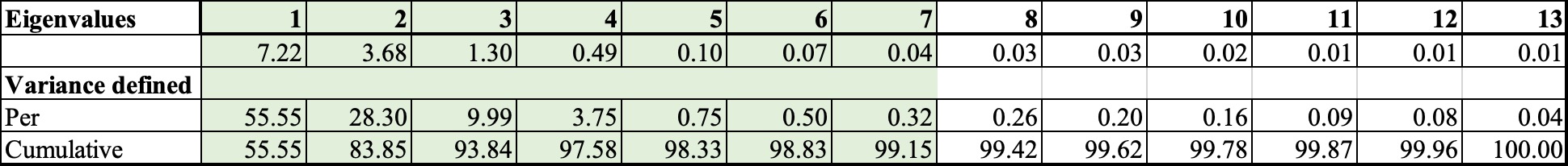}
  \end{center}
\end{table}

		The data set collected was standardized and a correlation matrix was generated along with the eigenvalues ($\lambda$) and eigenvectors ($\nu$). 
		The 13 $\lambda$ calculated are shown in Table \ref{table:bgd_pca_grp_eigentable}.
		It also shows the variation covered by the $\nu$ associated with each of these $\lambda$s along with  the cumulative variation covered by these $\lambda$s.
		The $\lambda_1$ (1st eigenvalue) covers 55\% of the variation in the data and $\lambda_2$ covers 28\% of the variation. 
		Cumulatively, these two covers 83\% of the variation in the data. 
		For our purpose, we set the limit of variation covered to be 99\% so that we would remove only the absolutely unnecessary components. 
		This level was reached by the first 7 $\lambda$s so the seven principle components ($\xi$) that are generated by these $\nu$s (corresponding to these $\lambda$) are used to estimate the background for the Crab.
				
	 	\subsection{Estimating the background for Bgd2 using PCA}
		\label{sec:bgd_est_pca_bgd2}
		One of the background observation from the flight was used to test our background estimation before it could be applied to Crab.
		Part of Bgd2 was selected for this analysis since the flight parameters during Bgd2 observation are spanned by the variables used to generate the principle components. 
		As mentioned previously in section \ref{sec:flt_flight_profile}, measurements done during the Sun and Cygnus X-1 observations can also be used as background measurements (along with Bgd2 and Bgd4) as those flux were estimated to be below our instrument sensitivity.
		Therefore, for this analysis, the flight parameters from Sun, Cygnus X-1 and Bgd4 were used to generate the 7 principle components (using the PCA) that covered 99\% of the variation. 
\begin{sidewaysfigure}[htbp]
\centering 
\includegraphics[width=0.9\columnwidth]{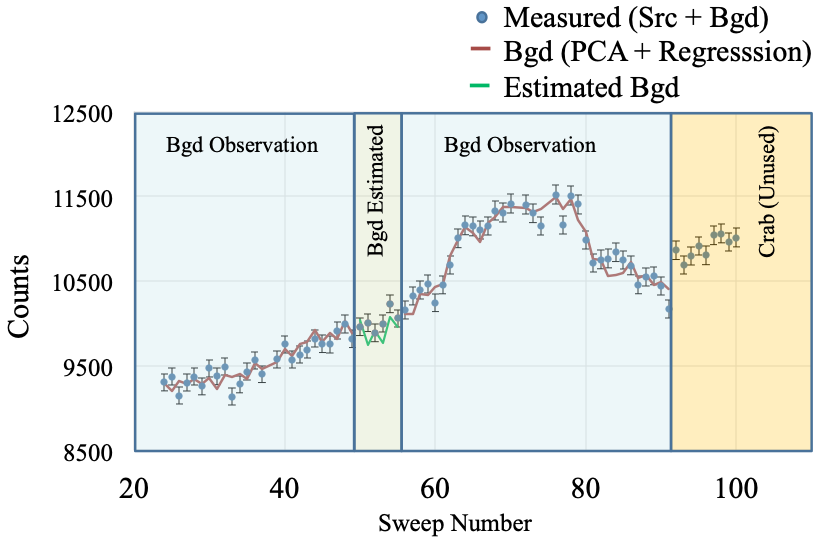}
\caption{Plot showing PCA approach to estimate the background for part of the Bgd region (shown in green). Using the flight variables from rest of the backgrounds (blue shaded region) principle components are generated and a regression fit is done to the background (red line). The fitted parameters are used to interpolate the background for the Bgd2 region (green line). This is done for energy range of 70-200 keV.}
\label{fig:bgd_pca_bgd2_extrap}       
\end{sidewaysfigure}

		Regression analysis was used to fit the background rates using these principle components. 
		The fit was used to estimate the background counts for the Bgd2 observation (shaded green regions) using the Bgd2 flight parameters.
		The background along with the estimated Bgd2 background are shown in Figure \ref{fig:bgd_pca_bgd2_extrap}. 
		Data from the shaded blue regions labeled `Fitted' is used to get the 7 principle components.
		A fit to the background rate during the same period using the principle components as the fit parameters.
		The fit is then used to estimate the background for part of the Bgd2 observation which is labeled as `Estimated'.
		The estimated background counts are shown in green. 
		The estimated background rate is in statistical agreement with the measured background rate as the difference of the total counts is statistically equal to 0. 
		The energy range selected for this analysis is from 70 keV to 200 keV. 
		This approach to estimation of the background is further extended to get an estimation of the background for the Crab observation.

		\subsection{Background estimation for the Crab}
		\label{sec:bgd_anal_pca_crab}
\begin{sidewaysfigure}[htbp]
\centering 
\includegraphics[width=0.9\columnwidth]{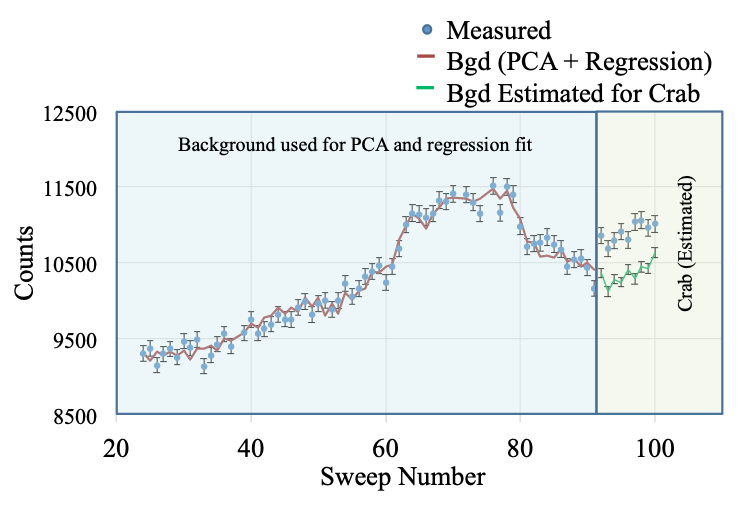}
\caption{Plot showing PCA approach to estimate the background for Crab observation. The background data, Sun, Bgd2, Cygnus X1 and Bgd4 (shaded blue region) is used to retrieve the principle components and a regression fit is done to this data. The fitting parameters are extended to retrieve the background for Crab (shown in green). This plot is for energy range of 70-200 keV. }
\label{fig:bgd_pca_crab_extrap_pol}       
\end{sidewaysfigure}

		The background estimation for the Crab is achieved in similar fashion as that of Bgd2 in section \ref{sec:bgd_est_pca_bgd2}. 
		However, Bgd2 observation is also included as part of the known background.
		Therefore the flight parameters from Sun, Bgd2, Cygnus X-1 and Bgd4 observations were used to generate the principle components for this analysis. 
		This is shown in Figure \ref{fig:bgd_pca_crab_extrap_pol}. 
		In the Figure, the shaded blue region is the background region which includes the Sun, Bgd2, Cygnus X-1 and Bgd4 observations. 
		The flight parameters from these observations were used to generate the seven principle components.
		The seven principle components covered 99\% of the variation in the data.
		The background counts (of the shaded regions) are then fit with these seven principle components using regression analysis. 
		The fit is then used to estimate the background counts for the Crab observation using the flight parameters of the Crab.
		The regression fit is shown in red and the estimated background for the Crab observation is shown in green.
		This analysis is done for PC events of all types and for the energy range 70 keV to 200 keV. 
		This method is used to estimate the background for our polarization and spectral analysis.
		The spectral analysis has several different energy bins (different energy ranges).
		The PCA method is applied to each of these bins separately.	
		The polarization analysis only has one energy range so this approach has to be applied only once.
		
\chapter{Spectral Analysis}
\label{sec:spec_analysis}

\begin{figure}[hbtp]
\centering 
\includegraphics[width=0.8\textwidth]{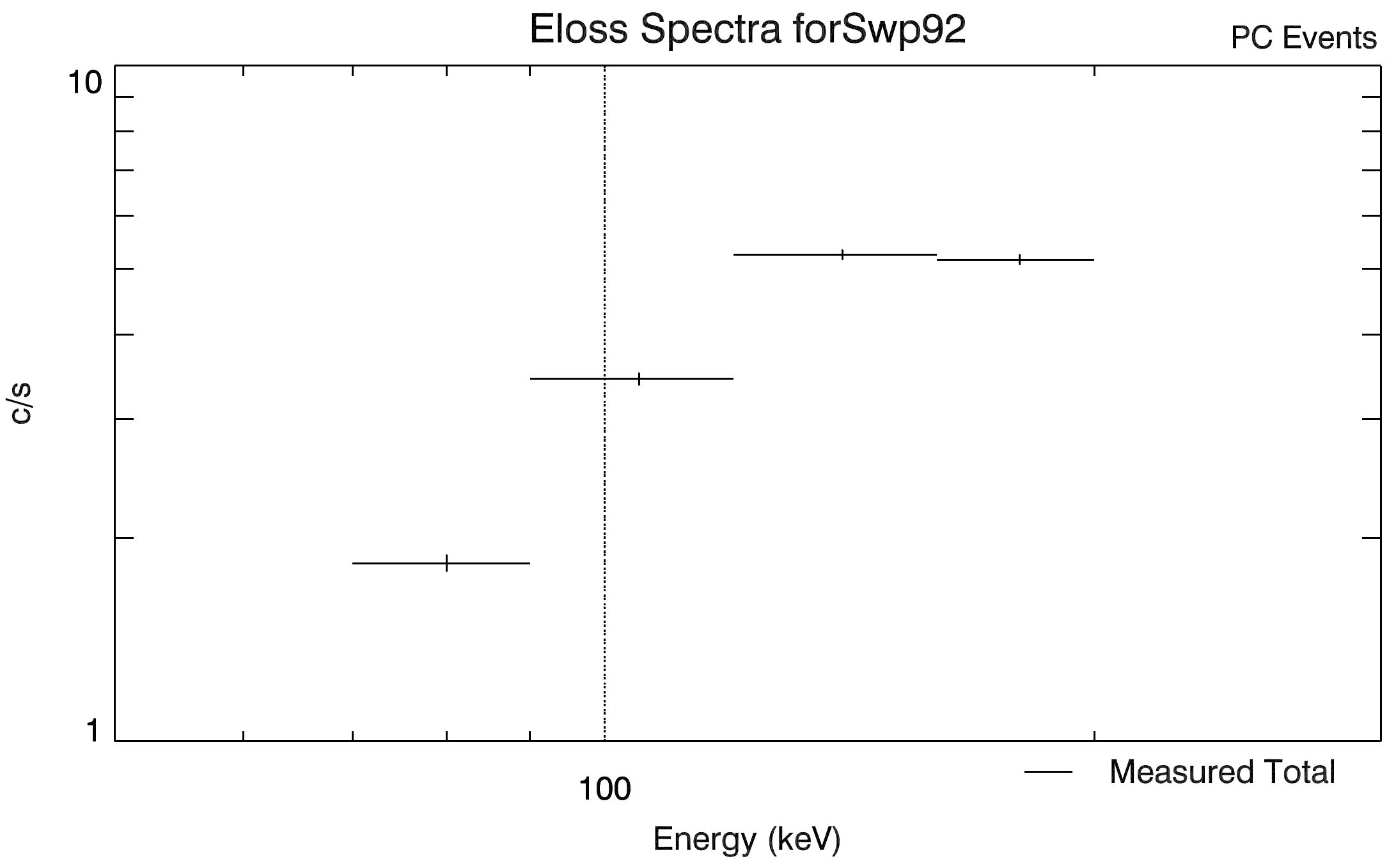}
\caption{Energy loss (Eloss) spectra for sweep number 92. This is one of the 9 sweeps associated with the Crab. This is the total measured which includes background and the source measurements. The analysis is selected for PC events of all types in the energy range of 70-200 keV. This is for PC event.   }
\label{fig:spec_eloss_sweep_total}       
\end{figure}

The Crab was observed towards the end of the 2014 flight for 9 sweeps (1.8 hrs). 
Various flight parameters during the 2014 GRAPE flight are shown in Figures \ref{fig:flt_profile_a}, \ref{fig:flt_profile_c} and \ref{fig:flt_profile_d}. 
The parameter values are displayed as a function of sweep and the sweeps associated with the Crab observation are from sweep 92 through 100. 
Energy loss (Eloss) spectra are generated for each sweep (70 - 200 keV).
The Eloss spectra for one of these sweeps (Sweep 92) is shown in Figure \ref{fig:spec_eloss_sweep_total}. 
The software threshold value of 10 keV was set for the plastics and 40 keV was set for the calorimeters. 
As specified previously, our analysis is focused primarily on PC (1 Plastic and 1 Calorimeter trigger)  event types. 
Similar Eloss spectra for rest of the sweeps associated with the Crab were also generated.

\begin{figure}[hbtp]
\centering 
\includegraphics[width=0.75\textwidth]{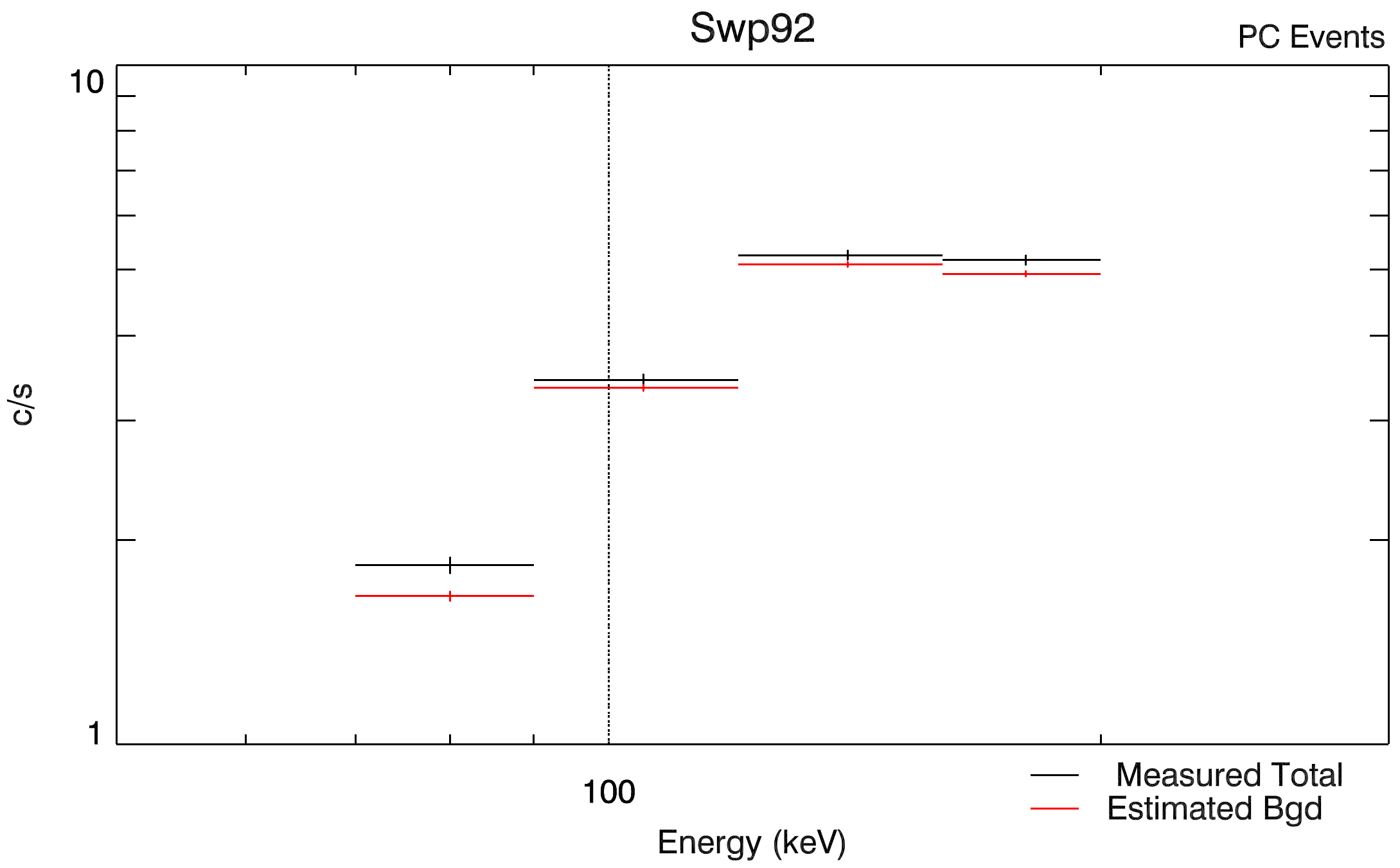}
\caption{Energy loss spectra for sweep number 92. The data in black represents the measured total which is the combination of background and the source. The red is the estimated background from PCA.}
\label{fig:spec_eloss_sweep_a}       
\end{figure}

\begin{figure}[hbtp]
 \centering
\begin{subfigure}[b]{0.9\textwidth}
 		 \includegraphics[width=1\linewidth]{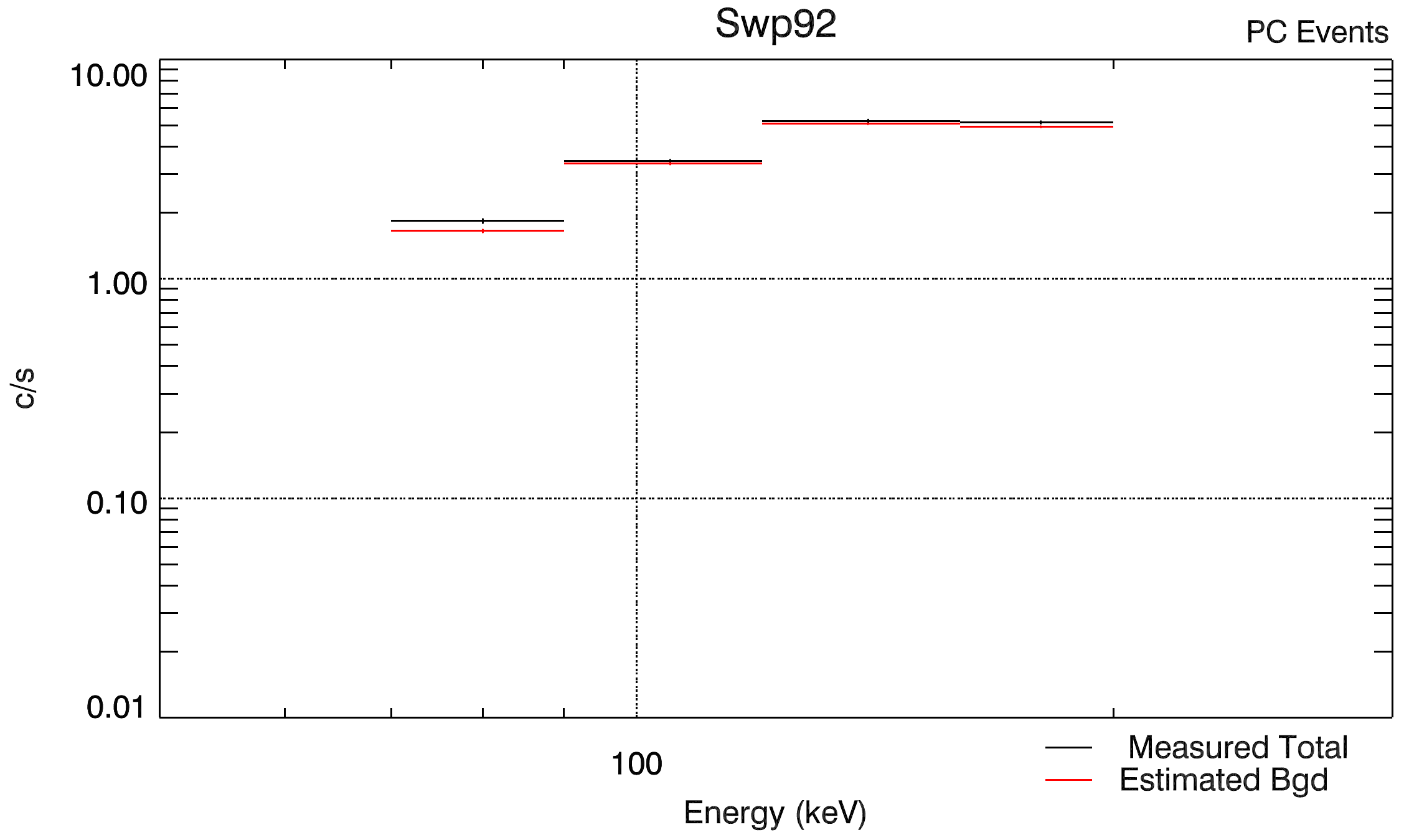}
		 \caption{}
\end{subfigure}  
 \begin{subfigure}[b]{0.9\textwidth}
 		 \includegraphics[width=1\linewidth]{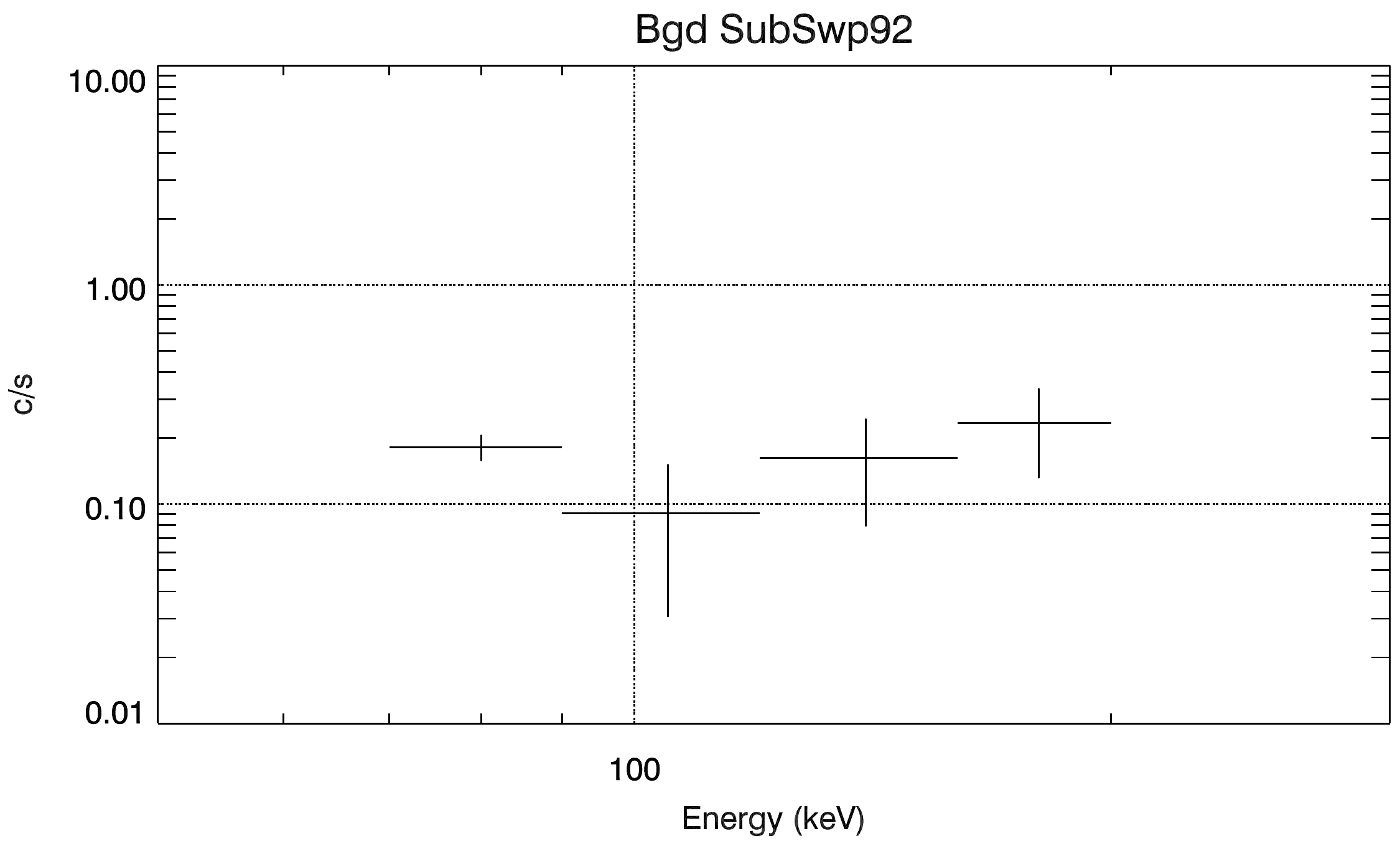}
		 \caption{}
\end{subfigure} 
  \caption{Energy loss spectra for sweep number 92. a) The data in black represents the measured total which is the combination of background and the source. The red is the estimated background from PCA. This is the same plot as shown in figure \ref{fig:spec_eloss_sweep_a} but at different scale which highlights the background dominating the measurements. b) The background subtracted energy loss spectra that represents only the Crab counts. }
\label{fig:spec_eloss_sweep_b}
\end{figure}

The Eloss spectra generated for each sweep is a combination of background and the Crab source signal. 
In order to retrieve the Crab energy loss spectra, we need to subtract the background from this measured total. 
The background estimation is done using Principle Component Analysis (PCA) approach discussed in section \ref{sec:bgd_PCA}. 
The PCA is is used to estimate a background for each sweep. 
The PCA gives a background value for a specified energy range so a separate PCA is conducted to get a background estimation for each individual energy bin.
The estimated background for the Sweep 92 is shown in Figure \ref{fig:spec_eloss_sweep_a} in red. 
The background subtracted energy loss spectra of the Crab observation  is shown in Figure \ref{fig:spec_eloss_sweep_b} b. 
This process is repeated for all 9 sweeps associated with the Crab to end up with 9 background subtracted energy loss spectra. 

\section{Spectral Fitting}
\label{sec:xspec}

In any gamma-ray spectrometer, the detector does not directly measure the  the photon spectrum f(E) but rather counts C(I) for our detector channel (I). The observed energy loss spectrum C(I) is related to the photon spectra f(E) by
\begin{equation}
			C(I) =  \int f(E) R(I,E) dE	
			\label{eqn:spec_xspec_eqn1}		
\end{equation}
 where R(I,E) is the response of the instrument which is briefly discussed in section \ref{sec:ins_perf_instrument_response}. 
The goal is to retrieve the photon spectrum f(E) so one must invert this equation to solve for f(E). 
However, the solutions from this inversion  are non-unique and unstable for small changes in C(I) \citep{Arnaud1996,Arnaud1999}. 
Therefore an alternative method is used where a model of the spectrum is used to fit the data. 
The model parameters are used to generate a predicted counts spectrum (or energy loss). 
This result can then be compared with the data.
The input parameters can be varied and a chi-square analysis is used to determine the parameters that best fits the data. 
This is also known as forward folding deconvolution of a spectrum.  
 
The deconvolution and fitting of the spectrum is done using XSPEC, an X-Ray spectral fitting package included in NASA's HEASARC software \citep{Arnaud1996}. 
XSPEC uses a forward folding method to get a photon spectrum. 
To achieve this, XSPEC requires the count spectrum, the instrument response, and the model for the spectral fitting. 
The count spectra used in our analysis are the background subtracted energy loss spectra for each of the 9 sweeps. 
These spectra are generated in PHA format for use in XSPEC.

The response of the instrument is in redistribution matrix file (RMF) file. 
This matrix has two components, the response matrix R$_\text{D}$ which is the probability density matrix (rsp) and the auxiliary response  A$_\text{D}$ file (arf). 
The R$_\text{D}$ (rsp) defines the probability of the incoming photon of energy E to be detected by our instrument channel I. 
The A$_\text{D}$ (arf file) contains the effective area and has a unit of cm$^2$. Sections \ref{sec:ins_perf_response_matrix} and\ref{sec:ins_perf_effective_area}  provide a detailed descriptions of these files and how they are generated via simulations. 
	
	The photons reaching our instrument suffer through some amount of attenuation as they travel through the atmosphere.  
	The amount of atmosphere, the particle travels through, is defined as air mass ($\rho_a$) which is calculated using the zenith angle and atmospheric depth for each event. 
	The zenith angle and atmospheric depth did not vary a lot during a sweep to change the $\rho_a$ significantly.
	Therefore, an average value of the $\rho_a$ was used for each sweep.
	The attenuation is calculated using the mass attenuation coefficient of air ($\mu/\rho$) and $\rho_a$ by exp$^{-(\mu/\rho)\rho_a}$.
	A table of mass attenuation coefficient (as a function of energy) is taken from the National Institute of Standards and Technology (NIST) database \citep{Hubbell2004}.
	This attenuation was included in the arf file and consequently each sweep has an arf file associated with it. 

A power law model ($AE^{-\alpha}$) was used for fitting the spectrum where $A$ is the normalization constant (ph/keV/s/cm$^2$) and $\alpha$ is the photon index. 
These two parameters are varied to find the best fit for the model.

\begin{figure}[tp]
\centering 
\includegraphics[width=1\textwidth]{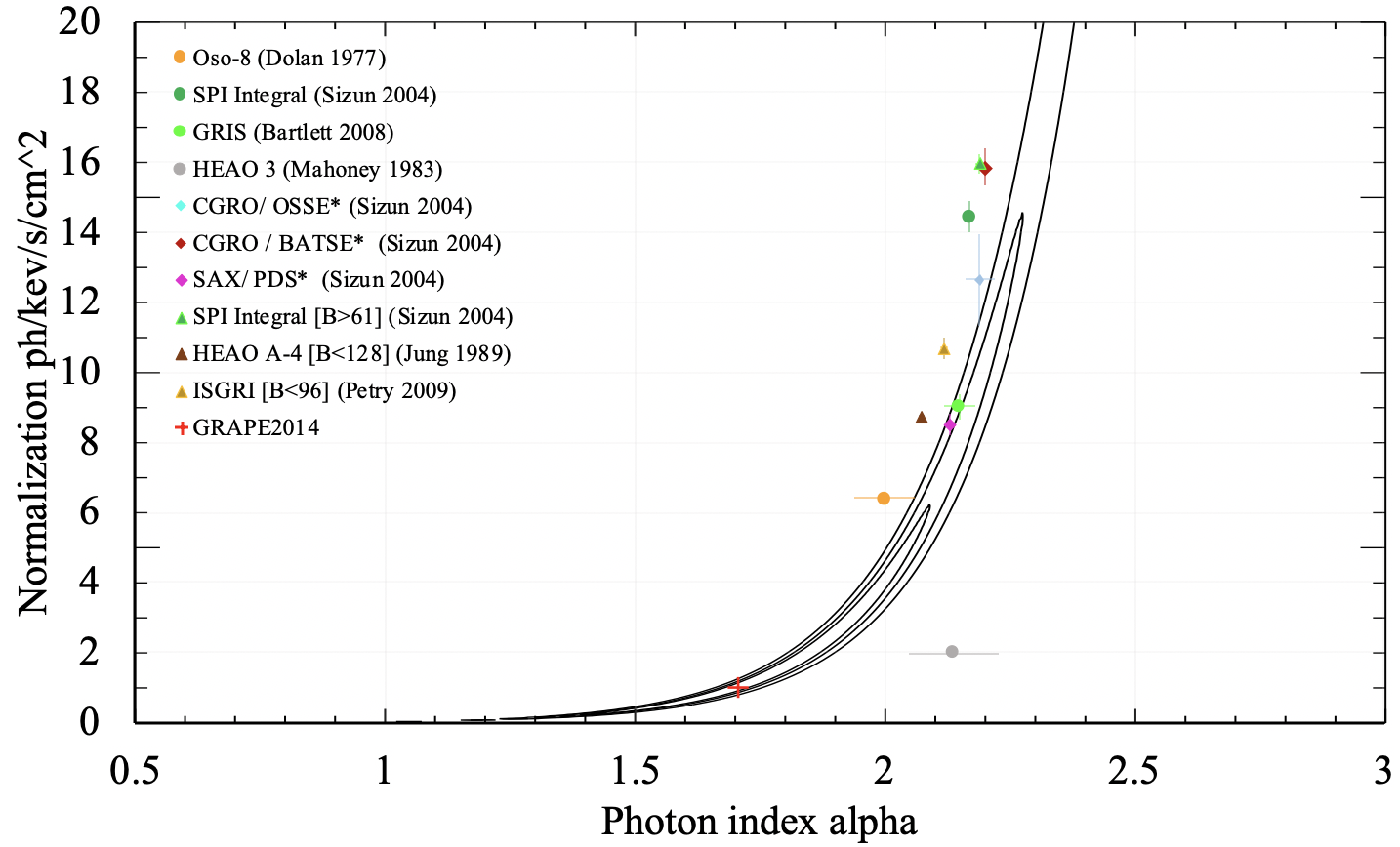}
\caption{Confidence plot of the spectral fit. The red $+$ represents the best fit value for the two parameters (normalization and photon index) for GRAPE. The contours represent 1$\sigma$,  2$\sigma$ and  3$\sigma$ from inner to outwards. The measurements of Crab from other experiments are also shown here. The energy range of these experiments overlap with that of GRAPE (70-200 keV). }
\label{fig:spec_confidence_plot}       
\end{figure}

XSPEC provides us with an ability to do a simultaneous fit to multiple spectra. 
This method was used to fit the power law model to all 9 sweeps. 
Each of the 9 sweeps had a 4 energy bins (4 data points). 
There were some energy bins that had less than 1-sigma statistics which were excluded from this analysis. 
The best power law fit for our spectra was for a photon index of 1.70 $\pm$ 0.24 and a normalization of 1.01 $\pm$ 1.35. 
A confidence plot (including values from other experiments) is shown in Figure \ref{fig:spec_confidence_plot}, where the best fit parameter is represented by a red $+$ sign.  
The contours represents the 68\%(1-sigma), 90\% and 99\% confidence levels. 

None of the experimental measurements shown in Figure \ref{fig:spec_confidence_plot} correspond precisely to that of GRAPE, but all of them overlap with at least some of the 70-200 keV energy range. These discrepancies, plus the limited statistics of the GRAPE data are likely contributors  to the slight disagreement about the spectral fit parameters between GRAPE and the other experiments.

\chapter{Polarization analysis}
	\label{sec:polarization_analysis}
	
	The polarization analysis is performed for the 9 sweeps (sweep 92 through 100) associated with the Crab observation.
	These are the same sweeps that were associated with the spectral analysis in chapter \ref{sec:spec_analysis}.
	The first step in the polarization analysis consist of measuring the Minimum Detectable Polarization (MDP) of the measurement. 
	\subsection{Minimum Detectable Polarization (MDP)}
	\label{sec:pol_anal_mdp}

\begin{figure}[t]
\centering
\includegraphics[width=1\linewidth]{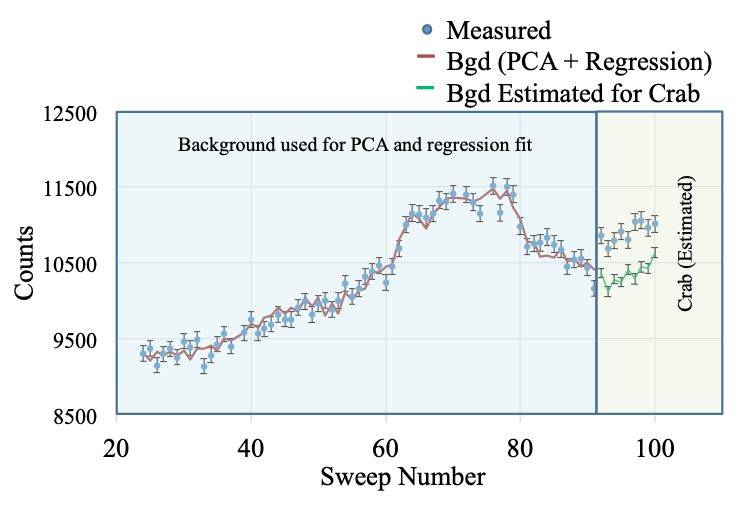}
  \caption{Background estimation for crab using the PCA approach on the various flight parameters during the background observations. The data is shown per sweep. Sweep 92 through 100 are associated with the Crab.}
\label{fig:pol_anal_pca_crab}

\end{figure}
	\begin{table}
\begin{center}
\caption{Table with the total measured, estimated background and the background subtracted Crab counts for each of the 9 Crab sweeps.}
\label{table:pol_anal_pca_table}
   	\includegraphics[width=1\textwidth]{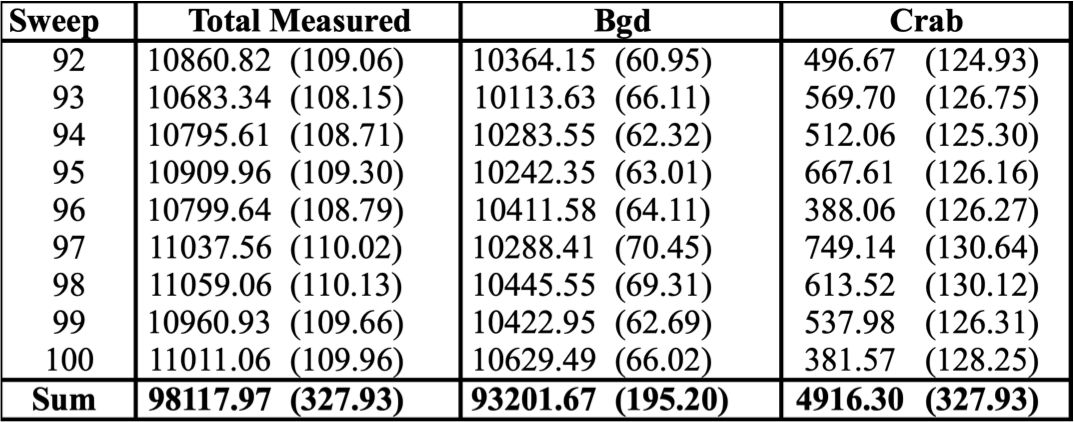}
  \end{center}
\end{table}

	The MDP defines the sensitivity of our instrument for the measurement. 
	It depends both on the characteristics of the instrument and the statistics of the observation.
	MDP is discussed in section \ref{sec:mdp} and is given by the equation 
	 \begin{equation}
		\text{MDP}_{99} = \frac{4.29}{\mu_{100}\ \text{C}_S}{\sqrt{\text{C}_S+\text{C}_B}}
				\label{eqn:pol_anal_mdp}
	\end{equation}
	where C$_S$ is the source counts, C$_B$ is the background counts, and $\mu_{100}$ is the modulation factor for a 100\% polarized source. 
    The background C$_B$ is needed in order to determine the Crab counts (C$_S$) from this measured total. 
    The background is estimated using the Principle Component Analysis (PCA), in the same way as done for the spectral analysis. 
    The energy range for this analysis is from 70-200 keV and the analysis is selected for PC (1 plastic and 1 calorimeter element) event types. 
    The PCA is used to get the background estimation for each of the 9 sweeps which are shown in Figure \ref{fig:pol_anal_pca_crab}. 
    The green data points represent the estimated background for the Crab. 
    The PCA process for the background estimation is described in section \ref{sec:bgd_anal_pca_crab}.
    The Crab counts are retrieved by subtracting the background from the measured total.  
	The total counts for each sweep, the estimated background counts, and the Crab counts retrieved from subtracting the background are tabulated in Table \ref{table:pol_anal_pca_table}. 
	The data represented in the table are for each sweep (720s). 
	The summation of the background and Crab counts gives us a C$_S$ = 4196.30 and C$_B$ = 93201.67 (S/B = 5.3\%). 
\begin{figure}[hbtp]
\centering 
\includegraphics[width=0.8\textwidth]{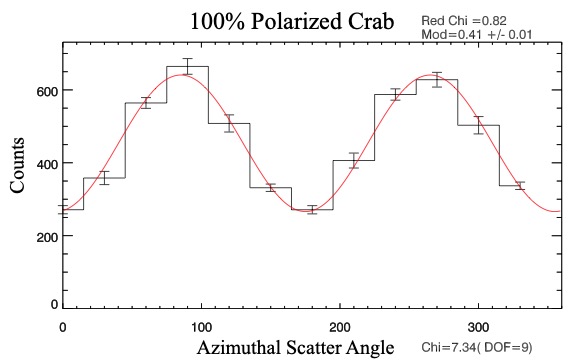}
\caption{Scatter angle histogram for a 100\% polarized Crab retrieved from simulation. The histogram has a modulation factor ($\mu_{100}$) of 0.41 $\pm$ 0.01.}
\label{fig:pol_anal_crab_100}       
\end{figure}

	The final ingredient in calculating the MDP is the $\mu_{100}$ which represents the modulation for the 100\% polarized Crab and is determined via simulation. 
	A simulation was conducted for GRAPE with a 100\% polarized Crab and an azimuthal scattering histogram was generated to calculate the $\mu_{100}$. 
	The simulated azimuthal scattering plot is shown in Figure \ref{fig:pol_anal_crab_100}. 
	The measured modulation factor for this 100\% polarized Crab source is $\mu_{100}$ = 0.41 $\pm$ 0.01. 
	With the aforementioned values, using the equation \ref{eqn:pol_anal_mdp}, the MDP of the measurement is  0.78. 
	This means that for this measurement, our instrument is able to detect a polarization fraction of above 78\% with 99\% confidence. 
	One of the characteristics of the MDP is that the MDP decreases as measurement time increases.
	Although originally planned for 8 hours, the Crab observation lasted only 1.8 hours. 
	We can calculate the MDP for the case where the observation lasted the full 8 hours. 
	The resulting MDP for an 8 hrs of observation period can be represented by 
 \begin{equation}
		\text{MDP}_{99} = \frac{4.29}{\mu_{100}\ \text{4C}_S}{\sqrt{\text{4C}_S+\text{4C}_B}} =\frac{1}{2} \bigg( \frac{4.29}{\mu_{100}\ \text{C}_S}{\sqrt{\text{C}_S+\text{C}_B}}\bigg)
				\label{eqn:pol_anal_mdp_4}
	\end{equation}
	The MDP would have been reduced by a factor $\sim$2.1 (from 78\%  to 37\%) if we were to have observed the Crab for 8 hrs as initially planned. 

	\section{Background Polarization}
	\label{sec:pol_anal_bgd_polarization}
\begin{figure}[!t]
\centering 
\includegraphics[width=0.9\textwidth]{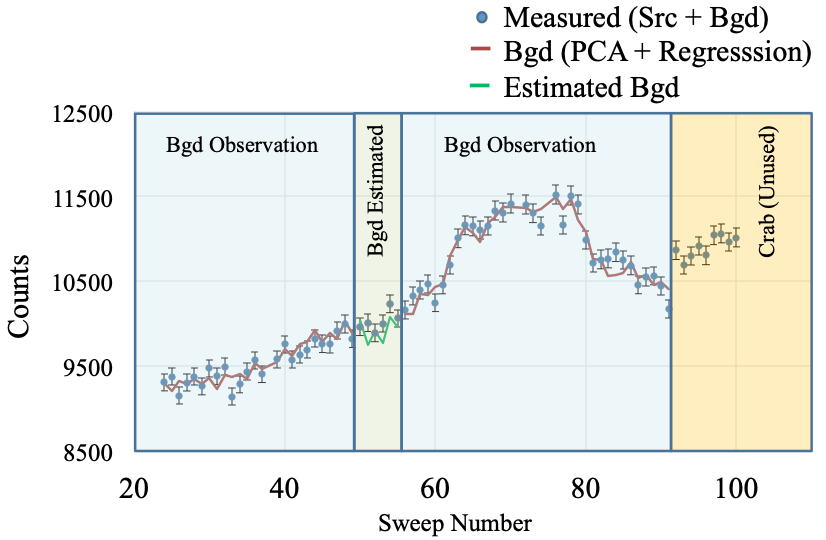}
\caption{Background estimated for sweep 50 through 55 using the PCA. These sweeps represent part of the background observation. }
\label{fig:pol_anal_pca_bgd}       
\end{figure}

	GRAPE rotates through 360$^\circ$ during each sweep. 
	We expect that this rotation eliminates asymmetries of the measurement. 
	To confirm this, we measured the polarization for a set fo background data and compared it with a uniform distribution.
	The background level is estimated using the PCA.
	The PCA gives a total  background count for each sweep. 
	6 background sweeps (sweep 50 through 55) were chosen for the analysis.
	These sweeps were the same sweeps used for PCA verification in section \ref{sec:bgd_est_pca_bgd2}.
	The selected data interval and the estimated background is shown in Figure \ref{fig:pol_anal_pca_bgd}. 
	The PCA process and the background estimation for this  period is discussed in section \ref{sec:bgd_est_pca_bgd2}.	

\begin{figure}[tbp]
 \centering
\begin{subfigure}[b]{0.49\textwidth}
 		 \includegraphics[width=1\linewidth]{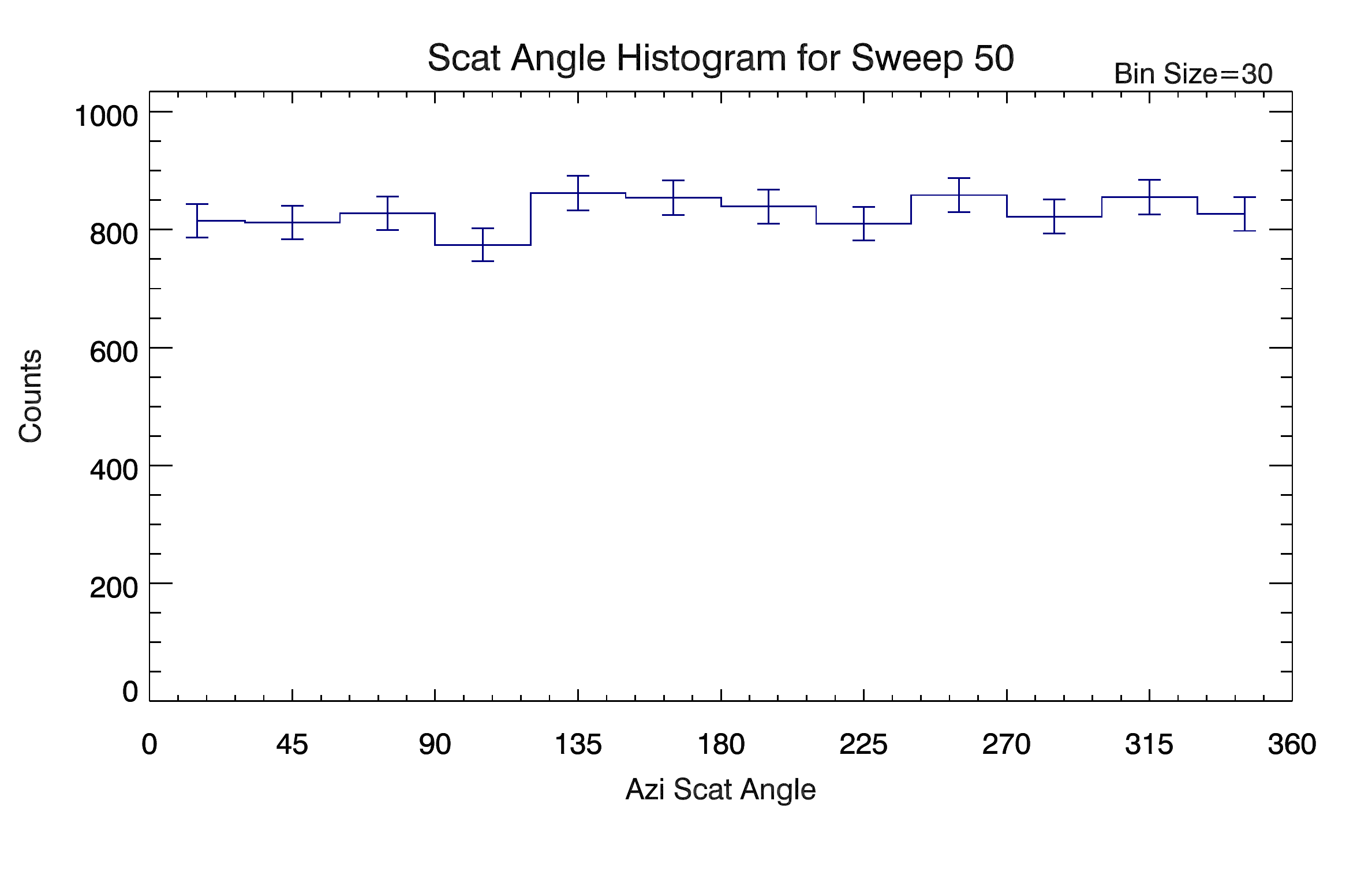}
		 \caption{}
\end{subfigure} 
 \begin{subfigure}[b]{0.49\textwidth}
 		 \includegraphics[width=1\linewidth]{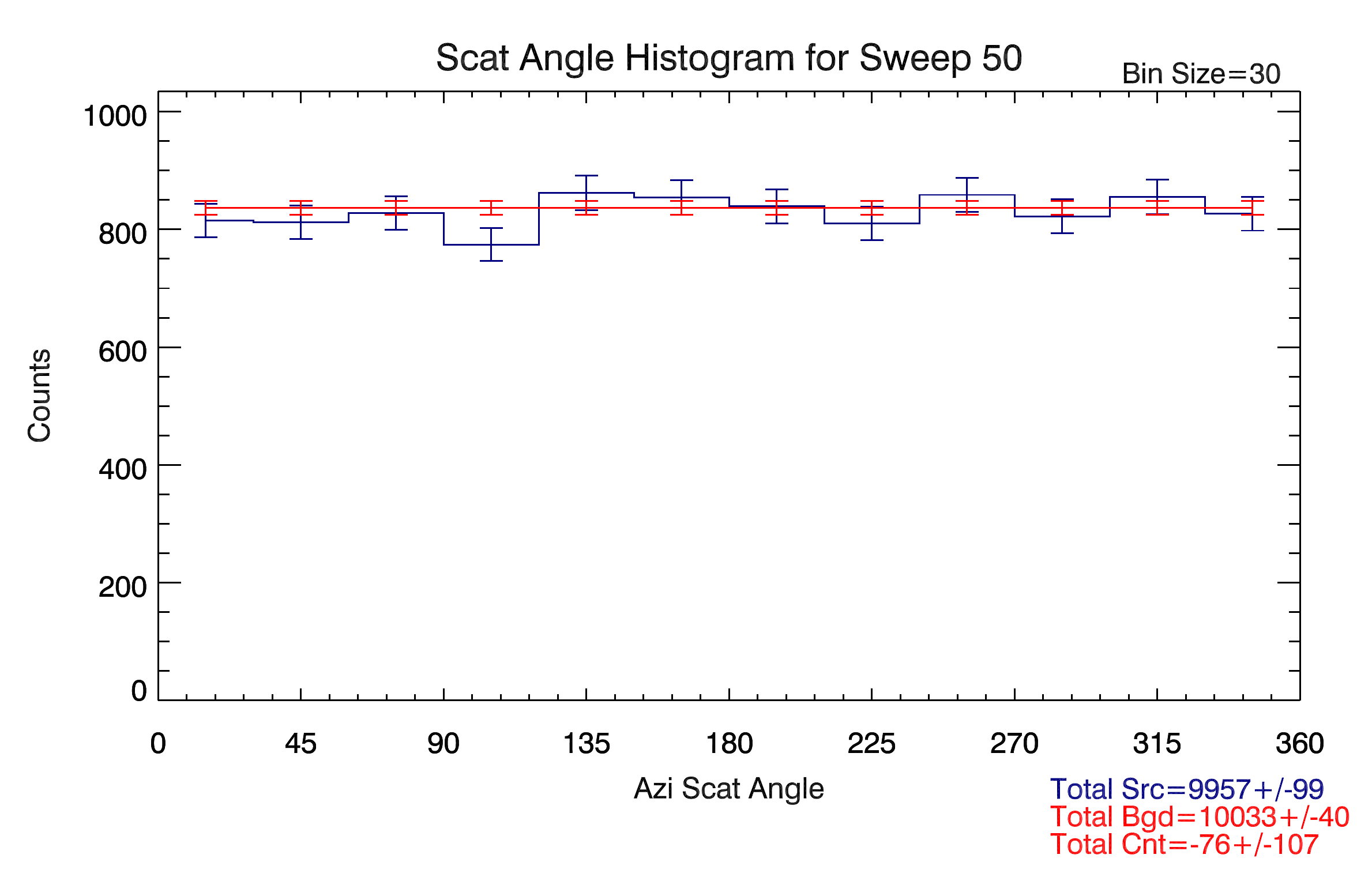}
		 \caption{}
\end{subfigure} 
  \caption{Scatter angle histogram for sweep 50 in blue. The estimated background via PCA is distributed uniformly which is shown in red. }
\label{fig:pol_anal_bgd_pol}
\end{figure}

	An azimuthal scatter angle histogram generated for sweep 50 is shown in Figure \ref{fig:pol_anal_bgd_pol}.
	The background for each of the associated sweep is calculated via PCA and is distributed uniformly to each of the angular bins. 
	The azimuthal scatter angle histogram (blue) for sweep 50 along with the uniformly distributed background (red) is shown in Figure \ref{fig:pol_anal_bgd_pol} b.

\begin{figure}[tbp]
\centering 
\includegraphics[width=0.6\textwidth]{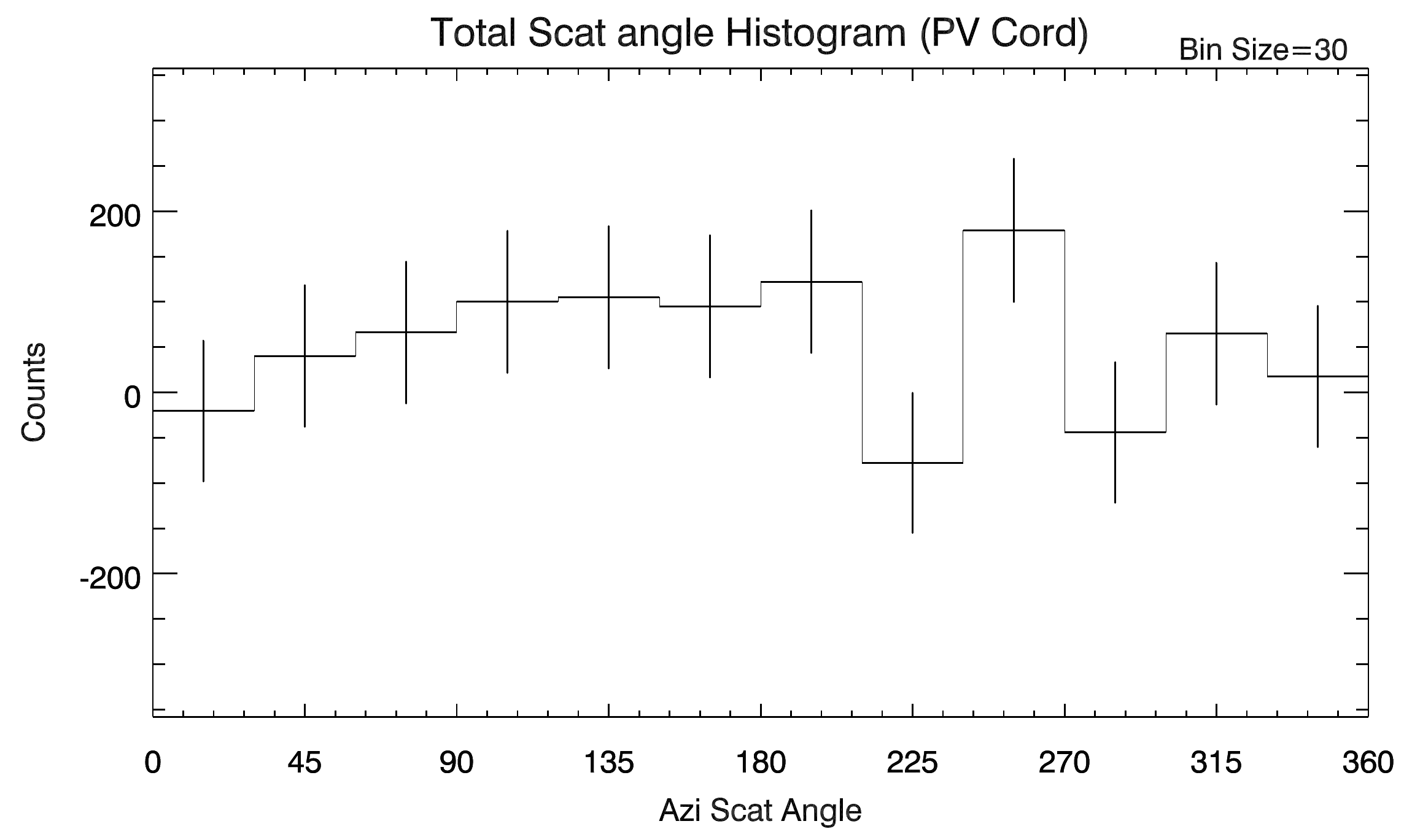}
\caption{The total scatter angle histogram for the background observation generated by summing all the background subtracted histogram for each of the sweeps 50 through 55.}
\label{fig:pol_anal_bgd_pol_sub}       
\end{figure}

\begin{figure}
 \centering
\begin{subfigure}[b]{0.7\textwidth}
 		 \includegraphics[width=1\linewidth]{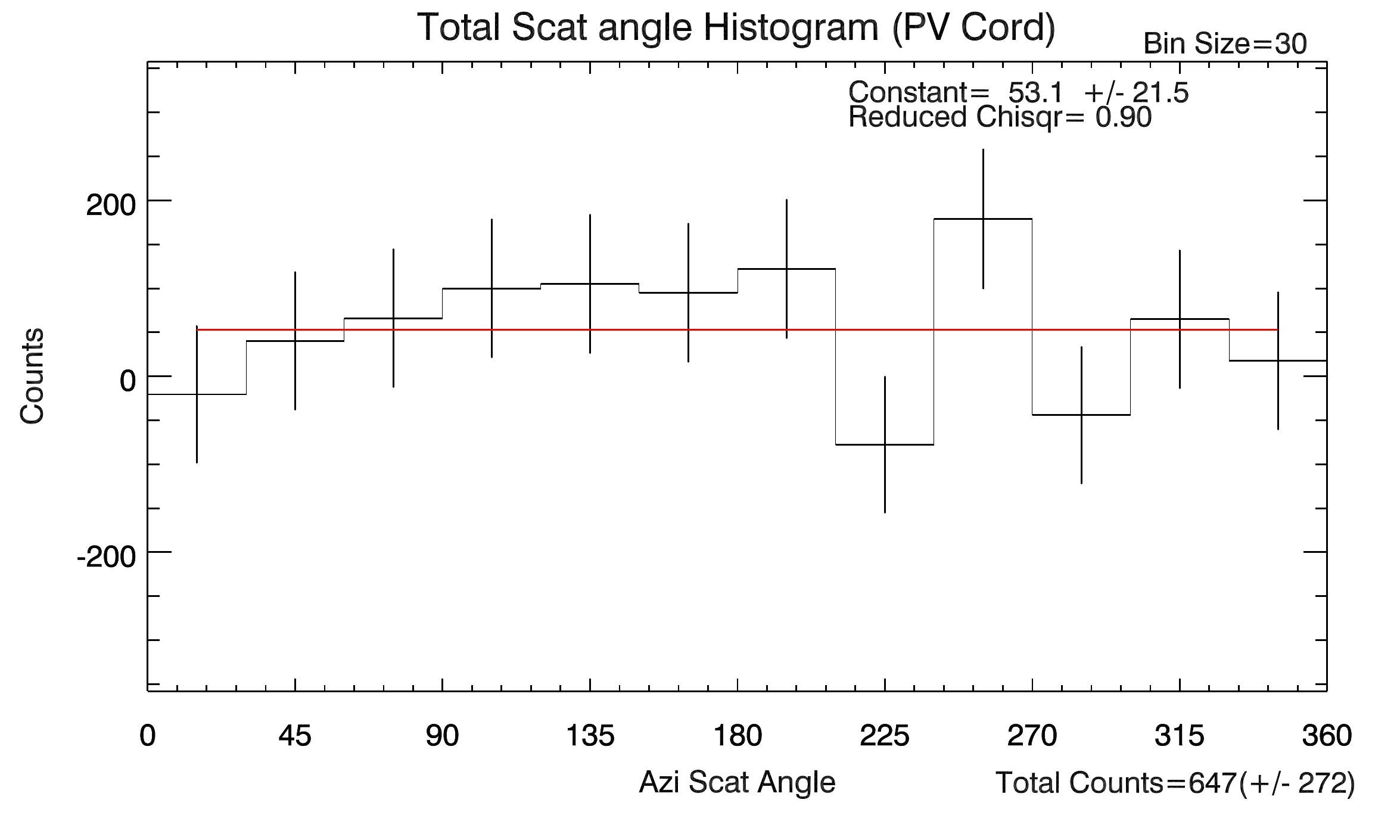}
		 \caption{}
\end{subfigure}    

 \begin{subfigure}[b]{0.7\textwidth}
 		 \includegraphics[width=1\linewidth]{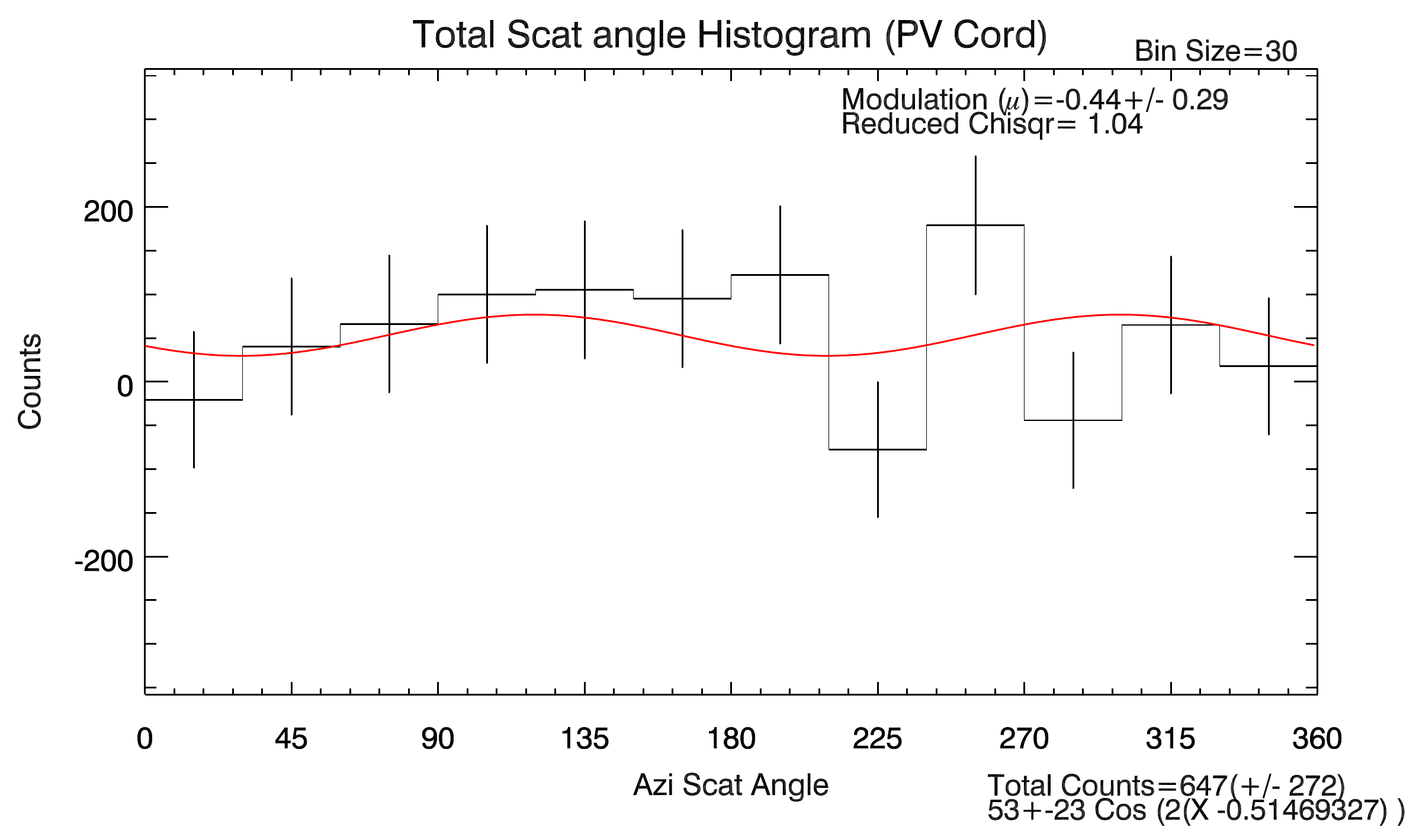}
		 \caption{}
\end{subfigure} 
  \caption{The two fits used for the F-test as a chi-squared test was not enough to identify the better fitting function for the background. a) The histogram is fitted with a flat function. b) A sinusoidal term is added to the flat fitting function. The F-statistics from these plots is used to test the significance of the added term in the fitting. The F-Test revealed that the added term was not significant.}
\label{fig:pol_anal_bgd_pol_ftest}
\end{figure}

	The scatter angle histogram is generated for each sweep and the corresponding background level is determined using the PCA.
	The (uniformly distributed) background is subtracted from the azimuthal scatter angle histogram to get the background subtracted azimuthal scatter angle histogram.
	This process is repeated for each sweep and the resulting distributions are summed together to get the total  azimuthal scatter angle histogram for the full background period. 
	The result is shown in Figure \ref{fig:pol_anal_bgd_pol_sub}.  
	This histogram (Figure \ref{fig:pol_anal_bgd_pol_sub}) is fitted with a flat line (Figure \ref{fig:pol_anal_bgd_pol_ftest}a) and then with an added sinusoidal function (Figure \ref{fig:pol_anal_bgd_pol_ftest}b). 
	A F-test was used to verify that the distribution is uniform and that there is no significant residual polarization in the analysis. 
	It is important to mention that as we have already verified in section \ref{sec:bgd_est_pca_bgd2} that the data represents the background. Therefore, the F-test could be directly applied to the sum of the distributions (without the background subtraction) to check for any significant residual polarization.
	However, this analysis step is going to be applied for Crab analysis in section \ref{sec:pol_anal_crab_polarization}.
	Therefore, in order to be thorough, this analysis step was included.
 
	\subsubsection{F-test}
	\label{sec:pol_anal_bgd_ftest}
	F-test is generally used to determine whether an added term in the fit function is significant or not \citep{Bevington2003}.
	To determine if the histogram  is better represented by a flat line (unpolarized) or a sinusoidal fit (polarized), the F-test was used.
	In our analysis, the histogram is first fit using the function $ f_1 = \text{C}_1$ which represents a flat unpolarized source. 
	Then a sinusoidal term is added to this function as $ f_2 = \text{C}_1 + \text{C}_2 \, \text{cos}(\,2(\text{C}_3 - \text{X}))$ and the fit is repeated.
	A F-statistics (F$_{12}$) is evaluated to determine the significance of the sinusoidal term in function 2 from function 1.
	The F-statistics (F$_{12}$) is defined by 
	\begin{equation}
		\text{F}_{12} = \frac{(\chi_1 ^2 - \chi_2 ^2)/ (\nu_1-\nu_2) }{\chi_2^2 / \nu_2}
		\label{eqn:fstatistic}
	\end{equation}
	where, $\chi^2$ is the chi-squared value from the fit and $\nu$ is the degrees of freedom. 
	The ratio is a measure of how much the additional term improved the fit. 
	If the additional term did not significantly improve the fit then the F-statistic would be small. 
	However, if the F-statistic is large then we can be confident that the added term significantly improved our fit. 
	
	The two functional fits,  f$_1$ and f$_2$ are shown in Figure \ref{fig:pol_anal_bgd_pol_ftest}a and \ref{fig:pol_anal_bgd_pol_ftest}b, respectively. 
	 Fit 1 has $\chi_1^2$= 9.91  for $\nu_1$= 11 and fit 2 has  $\chi_2^2$=9.37  for $\nu_2$= 9 and using these values in equation \ref{eqn:fstatistic}, the F-statistics F$_{12}$ is  0.26. 
	This is lesser than 1 therefore the added term is not significant.
	Therefore the histogram is better represented by a flat line (uniform).

	\section{Crab Polarization}
	\label{sec:pol_anal_crab_polarization}
	
	\begin{figure}[!t]
 \centering
\begin{subfigure}[b]{0.47\textwidth}
 		 \includegraphics[width=1\linewidth]{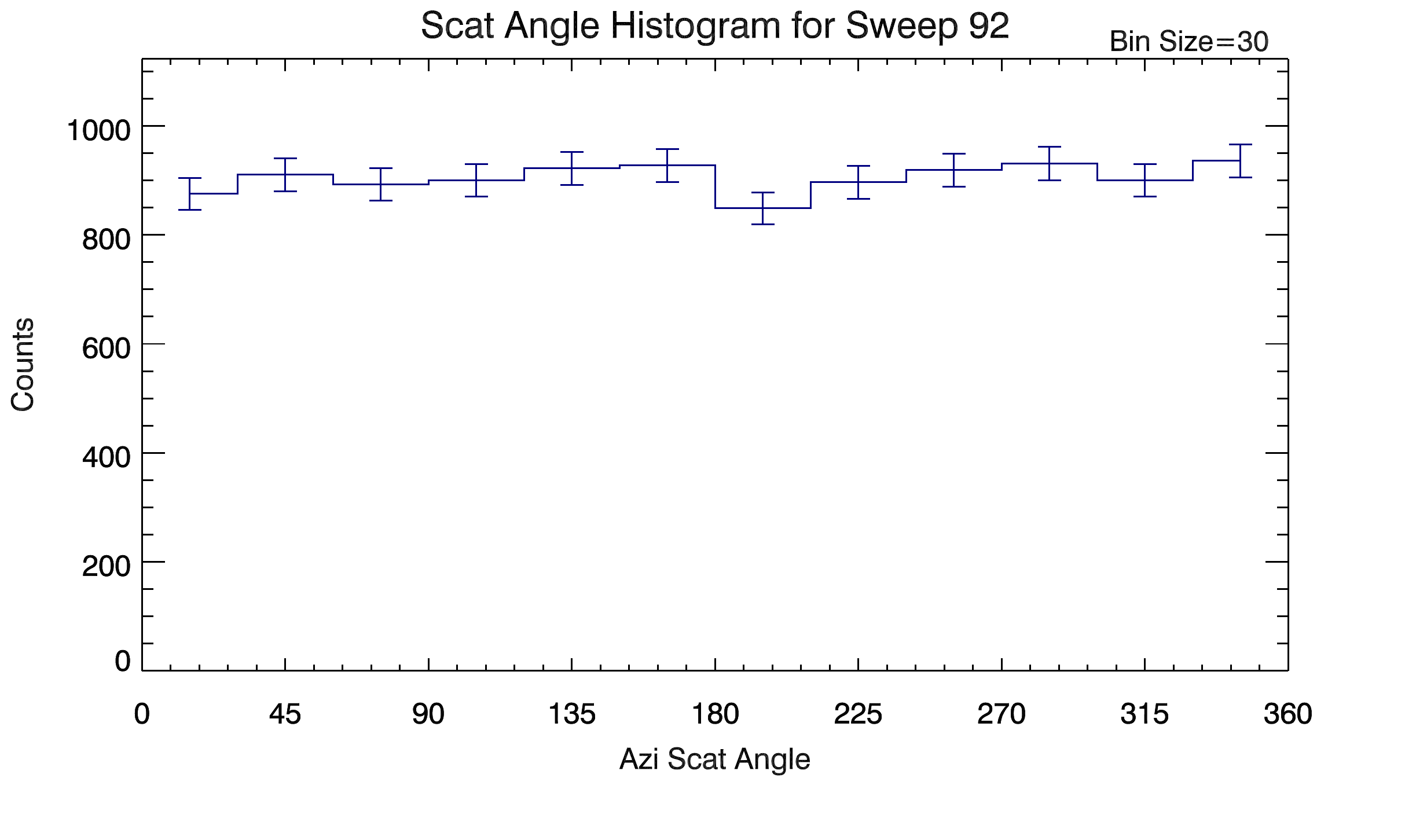}
		 \caption{}
\end{subfigure} 
 \begin{subfigure}[b]{0.49\textwidth}
 		 \includegraphics[width=1\linewidth]{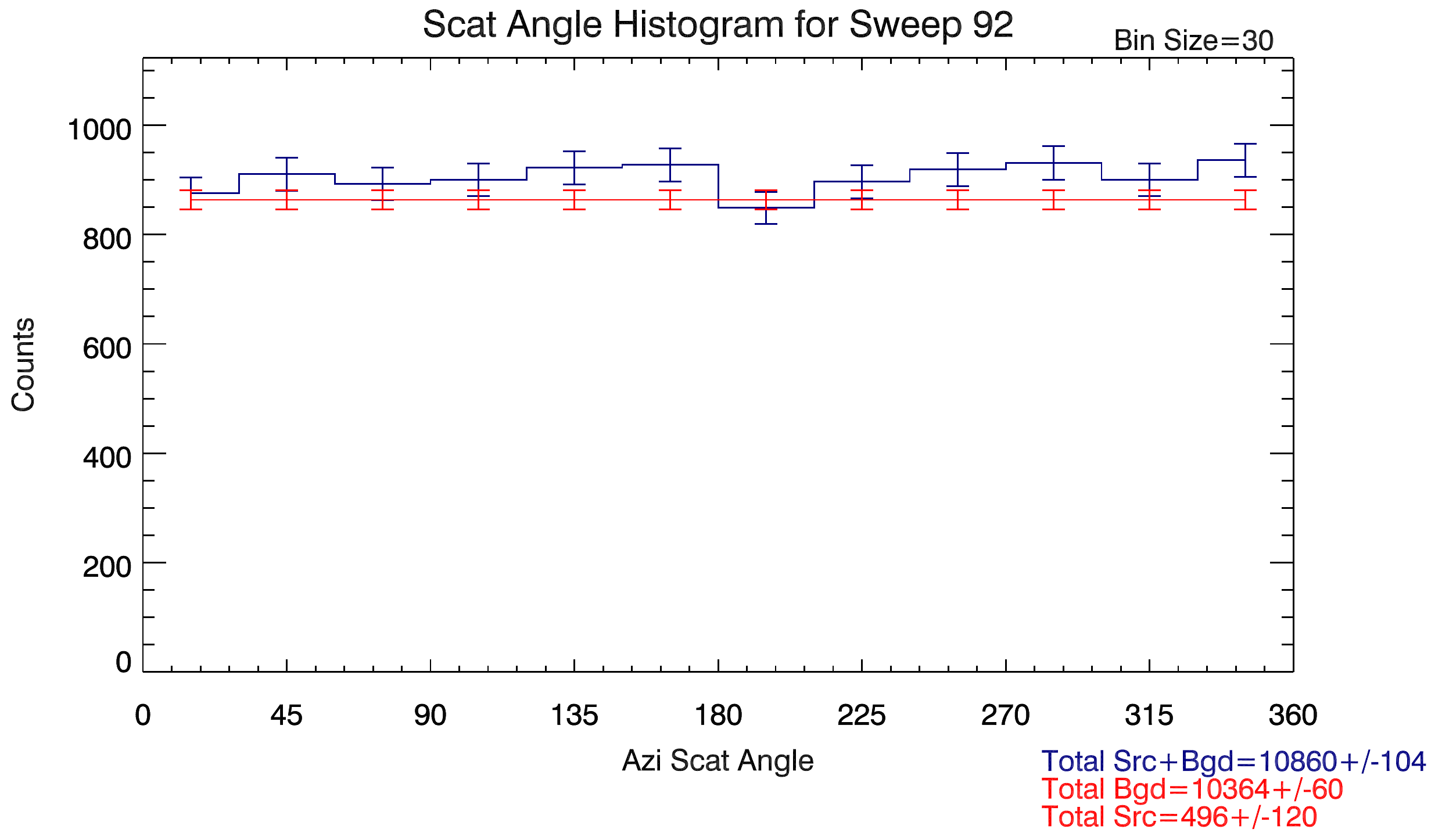}
		 \caption{}
\end{subfigure} 
  \caption{Scatter angle histogram for sweep 92 that is associated with the Crab. The scatter angle for the measured total (Crab + Bgd) is shown in blue and the estimated background via PCA uniformly distributed for each of the angular bin is represented in red}
\label{fig:pol_anal_crab_pol}
\end{figure}
\begin{figure}[tbp]
\centering 
\includegraphics[width=0.6\textwidth]{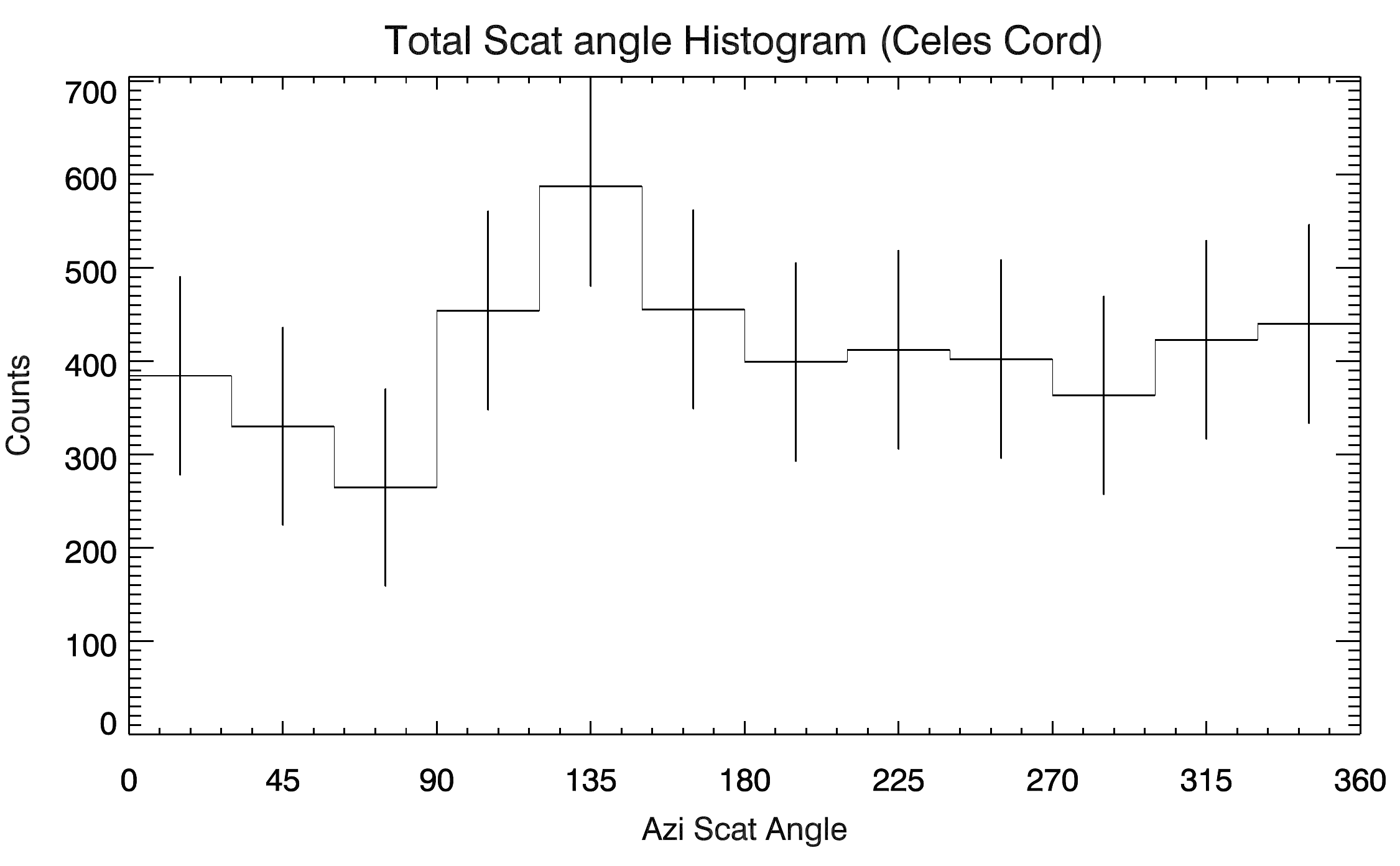}
\caption{Total scatter angle histogram for the whole Crab observation. The plot is the sum of the 9 background subtracted azimuthal scatter angle histogram representing each sweep associated with the Crab.  }
\label{fig:pol_anal_crab_pol_sub}       
\end{figure}

	Analysis for the Crab data follows steps similar to the background analysis in the previous section. 
	During calibration, the azimuthal scatter angle histogram is generated in pressure vessel coordinate system. 
	For the flight data, this has to be transformed to the celestial coordinate system. 
	This is achieved by calculating the parallactic angle. 
	The parallactic angle is the angle between the hour circle of the object and the great circle through a celestial object and the zenith. 
	The parallactic angle for the 9 sweeps of the Crab observations varied between -61$^\circ$ to -63$^\circ$ (the parallactic angle is measured between -180$^\circ$ and 180$^\circ$). 
	An azimuthal scatter angle histogram (in celestial coordinate) is generated for each of the 9 sweeps associated with the Crab observation.
	The scatter angle histogram for one of the sweeps (sweep 92) is shown in Figure \ref{fig:pol_anal_crab_pol} a. 
	The background level for this sweep is determined via PCA and the counts are uniformly distributed among the scatter angle bins.
	The measured data (blue) and the estimated background (red) are shown in Figure \ref{fig:pol_anal_crab_pol} b. 
	The background is subtracted from the measured total  for each of the 9 sweeps and the results are summed to generate the final polarization measurement shown in Figure \ref{fig:pol_anal_crab_pol_sub}. 

	
		\begin{figure}
 \centering
\begin{subfigure}[b]{0.7\textwidth}
 		 \includegraphics[width=1\linewidth]{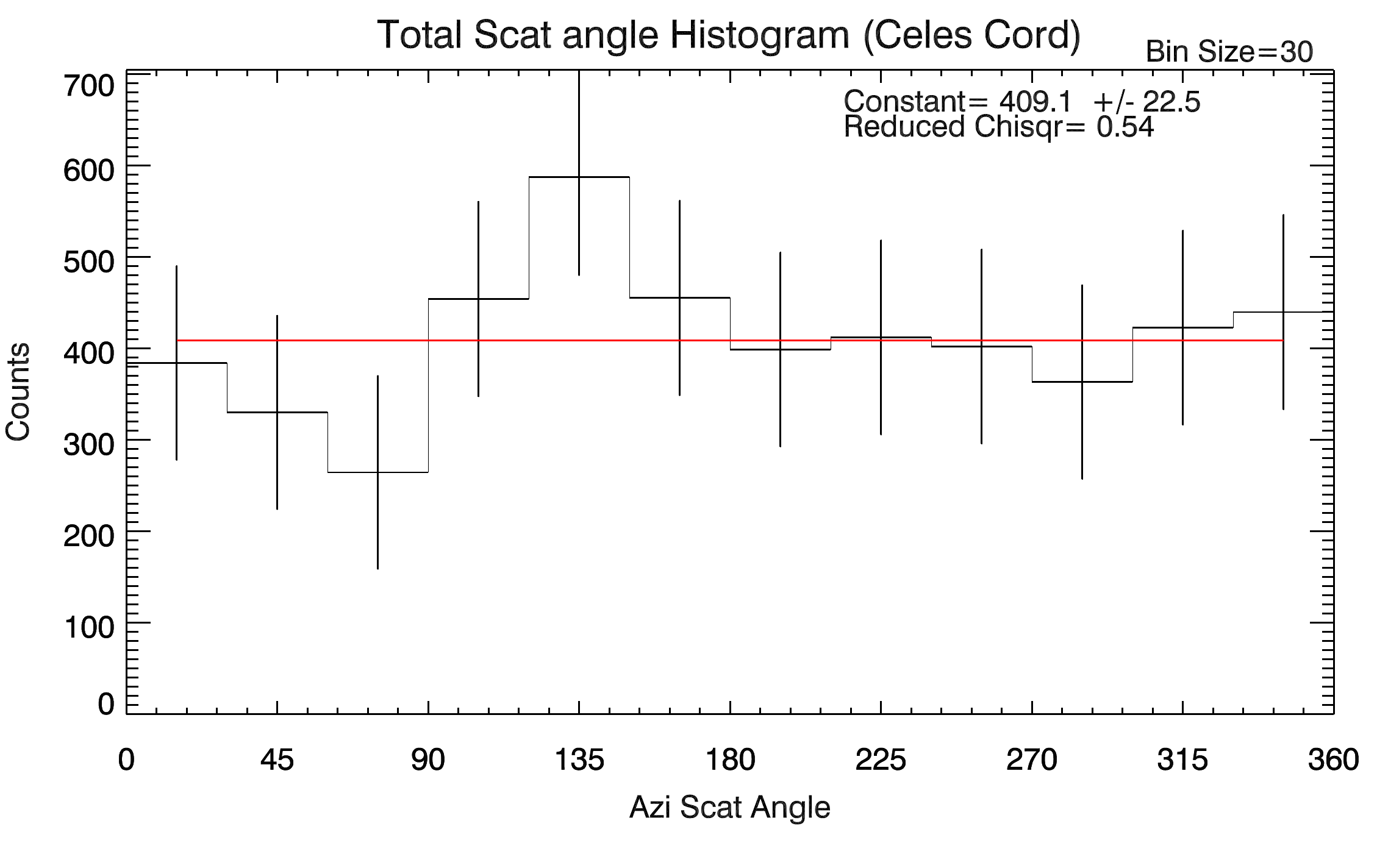}
		 \caption{}
\end{subfigure}    
 \begin{subfigure}[b]{0.7\textwidth}
 		 \includegraphics[width=1\linewidth]{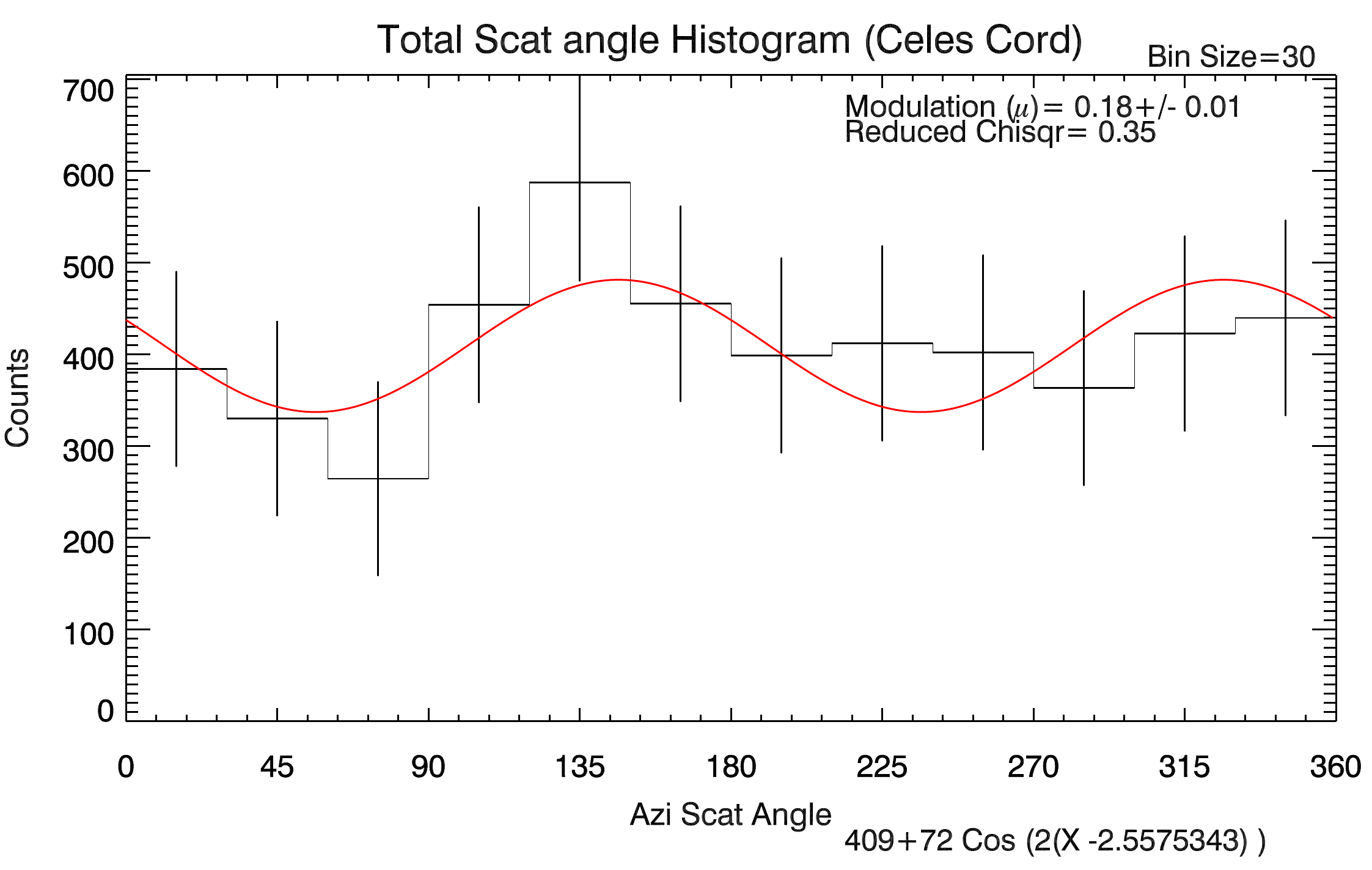}
		 \caption{}
\end{subfigure} 
  \caption{The two fits used for the F-test as a chi-squared test was not enough to identify the better fitting function. a) The histogram is fitted with a flat function. b) A sinusoidal term is added to the flat fitting function. The F-statistics from these plots is used to test the significance of the added term in the fitting.}
\label{fig:pol_anal_crab_pol_ftest}
\end{figure}
	
	The F-Test described in section \ref{sec:pol_anal_bgd_ftest} is also used here to verify the significance of the added term.
    The function $ f_1 = \text{C}_1$, which represents a flat unpolarized source is used to fit the histogram first.
	Then a sinusoidal term is added to this function as $ f_2 = \text{C}_1 + \text{C}_2 \, \text{cos}(\,2(\text{C}_3 - \text{X}))$ and the fit is repeated similar to the background.
	A F-statistics (F$_{12}$ from equation \ref{eqn:fstatistic}) is evaluated to determine the significance of the sinusoidal term in function 2 from function 1.
The two functional fits,  f$_1$ and f$_2$ are shown in Figure \ref{fig:pol_anal_crab_pol_ftest} a and b, respectively. 
	 Fit 1 has $\chi_1^2$= 5.90  for $\nu_1$= 11 and fit 2 has  $\chi_2^2$=3.14  for $\nu_2$= 9 and using these values in equation \ref{eqn:fstatistic}, the F-statistics F$_{12}$ is  3.96. 
	This is greater than 1 therefore the added term is an improvement to the fit.
	
\begin{figure}[hbtp]
\centering 
\includegraphics[width=0.7\textwidth]{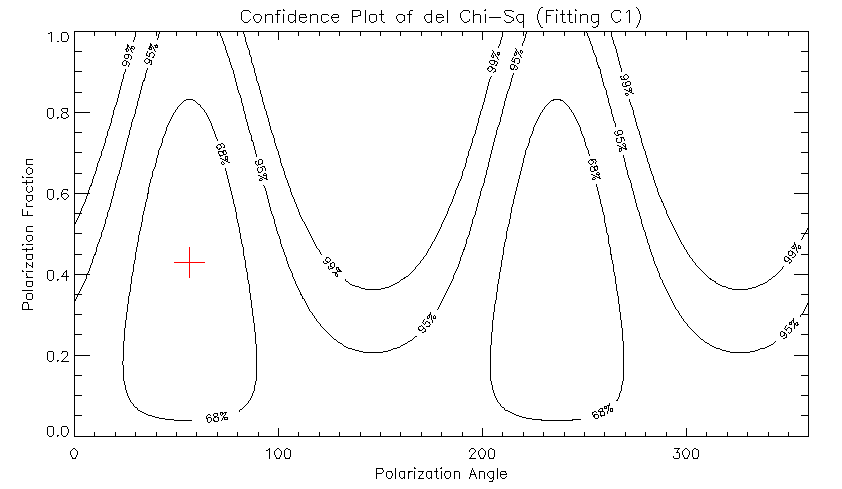}
\caption{A confidence plot to understand the significance of our result. The red cross is our result. The polarization angle has a 180$^\circ$ symmetry. The three contours labelled 68\%, 95\% and 99\% represent the 1-$\sigma$, 2-$\sigma$ and 3-$\sigma$ confidence levels. Our analysis resulted in a polarization fraction of 0.43$^{+0.4}_{-0.4}$  and a polarization angle of 56$^\circ$$^{+30^\circ}_{-30^\circ}$ within 1-$\sigma$.  At the 95\% and 99\% confidence level, these place no constraints on the Crab polarization.  }
\label{fig:pol_anal_crab_confidence}       
\end{figure}

	Therefore the histogram is fit using the second functional form f$_2$. 
	The modulation fraction of this fit is 0.18 $\pm$ 0.01.
	With the simulated modulation fraction (for a 100\% polarized source) of $\mu_{100}$ = 0.41, the resulting polarization fraction is measured to be 0.43 or 43\%. 
	The significance of this measurement can be visualized better using a confidence plot which is shown in Figure \ref{fig:pol_anal_crab_confidence}.  
	The plot is generated by varying the polarization angle and the polarization fraction and fitting the function to fill the parameter space with a $\delta \chi^2$ value.
	The three contours labelled 68\%, 95\% and 99\% represent the 1-$\sigma$, 2-$\sigma$ and 3-$\sigma$ confidence levels. 
	This shows that the polarization fraction is 0.43$^{+0.4}_{-0.4}$ (or is between 0.03 and 0.83) in the 1-$\sigma$ interval. 
	
	\chapter{Results Discussion}
	\label{sec:results_discussion}
	
	\section{Polarization Results}
\begin{figure}[tbp]
\centering 
\includegraphics[width=0.7\textwidth]{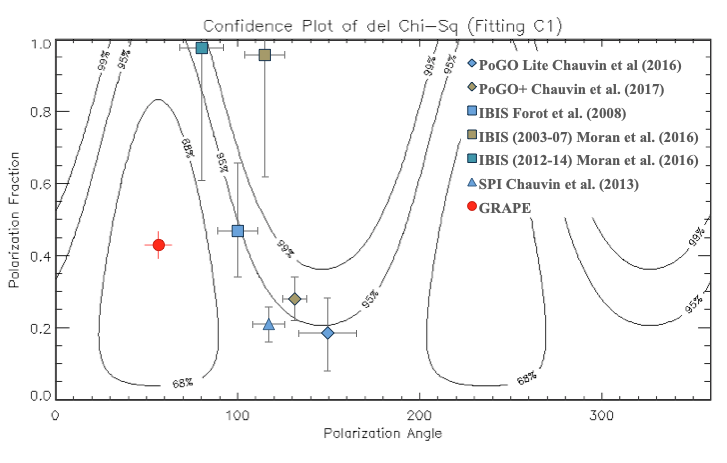}
\caption{Confidence plot for our result of the polarization measurement for the phase-integrated Crab. The red cross represents our result. The measurements from various other experiments are also presented  \citep{Chauvin2017,Chauvin2013,Chauvin2016,Chauvin2016a,Chauvin2016b,Forot2008,Moran2015a}. }
\label{fig:disc_pol_confidence}       
\end{figure}

	A polarization  analysis was done for the Crab using data from 2014 GRAPE flight.
	The flight plan included 8 hours of Crab observation but the flight was terminated before all data could be collected.
	The Crab was observed for only 1.8 hours.
	The collected data was far less than we had hoped for therefore the expectations for this analysis were not very high.
	The initial goal was to do a phase-resolved analysis (analysis for each of the pulsar phases) for the Crab data. 
	The reduced observation time also hindered the phase-resolved analysis. 
	Therefore the analysis was only done for all pulsar phases (phase-integrated analysis). 
		The analysis of the phase-integrated Crab resulted in a polarization fraction of 0.43$^{+0.4}_{-0.4}$ and polarization angle of 56$^\circ$$^{+30^\circ}_{-30^\circ}$. 
		The confidence plot (Figure \ref{fig:pol_anal_crab_confidence}) from our analysis, along with other polarization measurements from other experiments are shown in Figure \ref{fig:disc_pol_confidence} \citep{Chauvin2017,Chauvin2013,Chauvin2016,Chauvin2016a,Chauvin2016b,Forot2008,Moran2015a}.  
		At 1-$\sigma$, the polarization fraction ranges from 5-80\% an the angle varies from 20-90$^\circ$. 
		The data are not constrained at the 2-$\sigma$ level.		
		
\begin{figure}[hbtp]
 \centering
\begin{subfigure}[b]{0.7\textwidth}
 		 \includegraphics[width=1\linewidth]{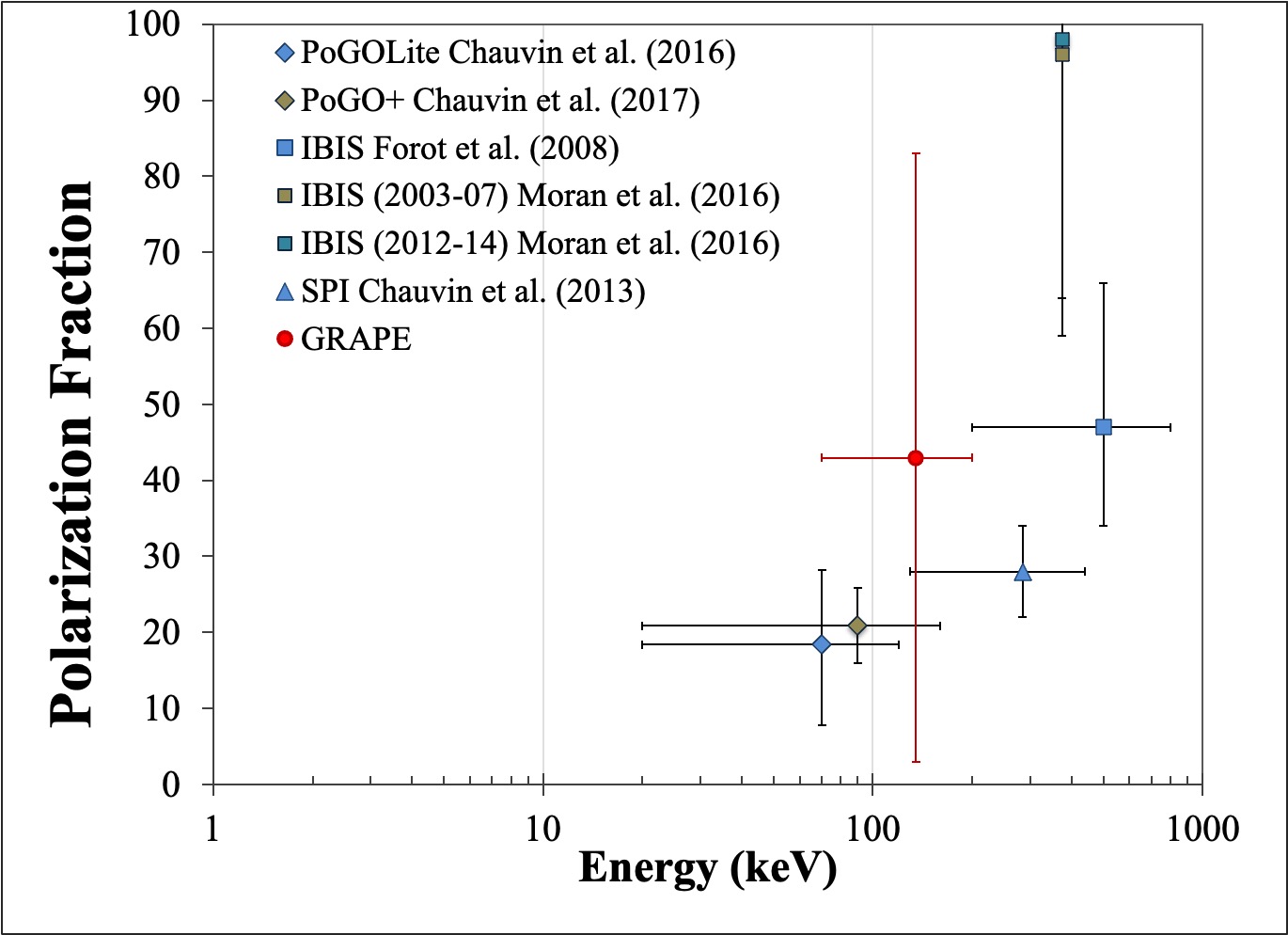}
		 \caption{}
\end{subfigure} 

 \begin{subfigure}[b]{0.70\textwidth}
 		 \includegraphics[width=1\linewidth]{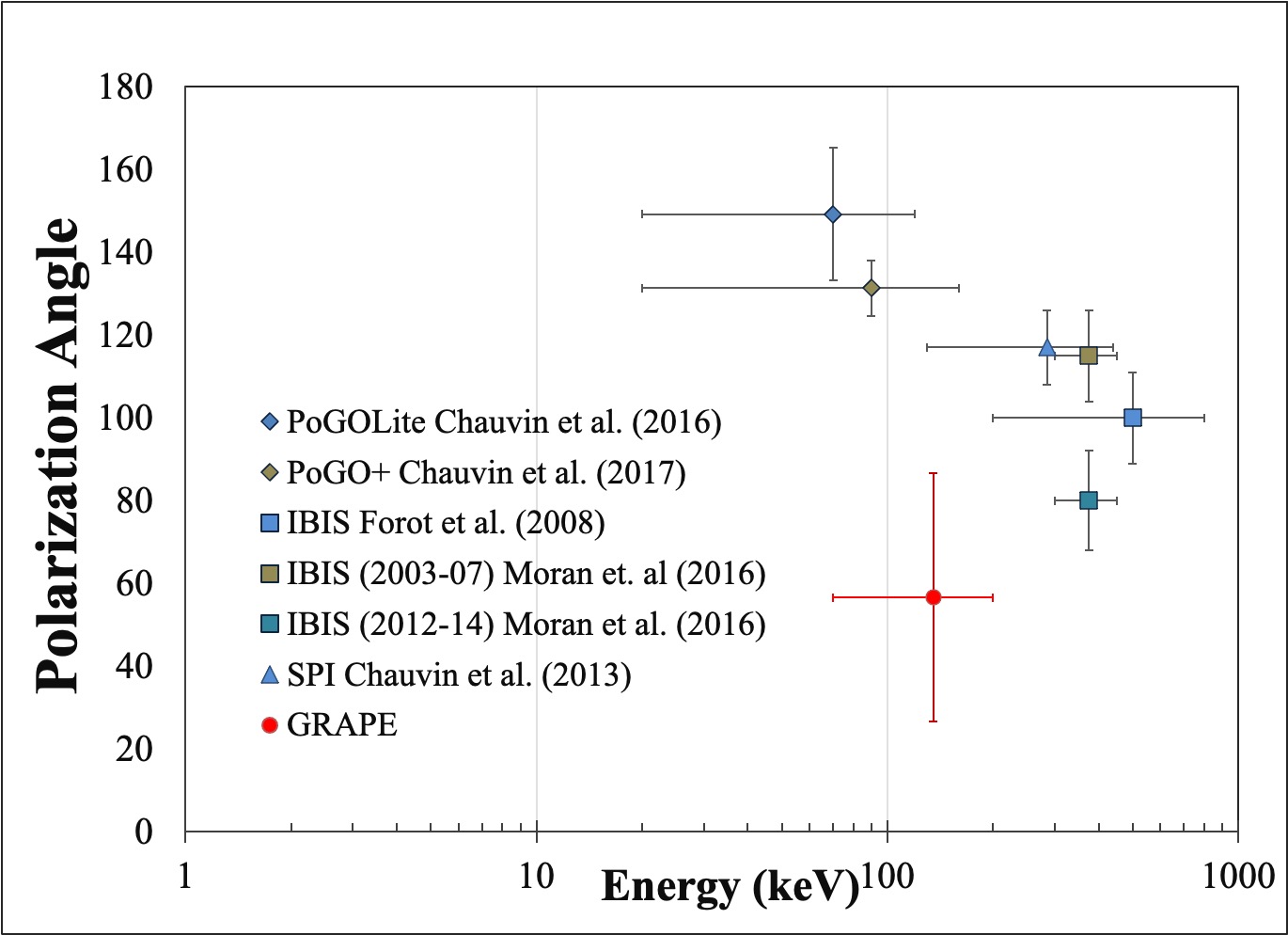}
		 \caption{}
\end{subfigure} 
  \caption{ Plots representing the polarization measurements from various experiments versus the energy range of the instruments. All the data points are for phase-integrated Crab. a) The polarization fraction versus the energy. b) The polarization angle versus the energy \citep{Chauvin2017,Chauvin2013,Chauvin2016,Chauvin2016a,Chauvin2016b,Forot2008,Moran2015a}.  }
\label{fig:disc_pol_ang_frac}
\end{figure}
					
The polarization measurements shown in Figure \ref{fig:disc_pol_confidence} are not all for the same energy range but the energy ranges do overlap. 
The difference in energy range and the measurements from these experiments can be better viewed in Figure \ref{fig:disc_pol_ang_frac}. 
Figure \ref{fig:disc_pol_ang_frac}a shows the polarization fraction (PF) vs energy and Figure \ref{fig:disc_pol_ang_frac}b shows the polarization angle (PA) vs energy.
These results suggest that the polarization parameters have some dependence on energy. 
In particular, the PF appears to increase with energy, and PA appears to decrease with energy.
A more significant polarization measurement from GRAPE may have contributed a data point that might have confirmed these trends. 

The polarization vector is defined by the electric field vector of the photon (which is orthogonal to both the magnetic field vector and the momentum vector). 
Therefore, a measured polarization vector would reveal the orientation of the magnetic fields which could potentially help to identify the origin of the photons.       
\citet{Dean2008} did the polarization measurement for only the Crab Nebula (off-pulse emission) using the SPI. 
This off-pulse data corresponds to the shaded portion of the pulse profile shown in Figure \ref{fig:sci_crab_pulse_profile}b. 
They measured a PA of 123.0$^\circ$ $\pm$ 11$^\circ$ (measured from north, anti-clockwise on the sky).
The Crab's magnetic axis is estimated to be at 124$^\circ$ $\pm$ 1$^\circ$ \citep{Romani1996}.
As the polarization vector was along the magnetic axis, the magnetic field lines responsible for the emission are orthogonal to this magnetic axis. 
\citet{Dean2008} measured the polarization for the off-pulse (non-pulsed) component which meant that the emission region of these photons were outside the pulsar. 
The magnetic field lines along the jets are parallel at the poles but are closer to perpendicular near the termination far away from the pulsar.
They concluded that the emission region of these detected photons were near the termination point of these jets \citep{Dean2008}.

\begin{figure}[!t]
\centering 
\includegraphics[width=0.5\textwidth]{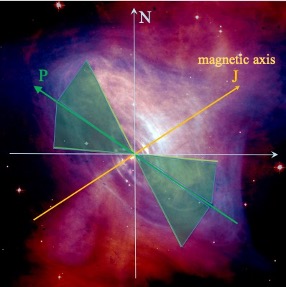}
\caption{A composite picture of the Crab nebula from the Chandra (x-ray in blue) and from Hubble Space Telescope (optical in red). The picture is shown in the celestial coordinate \citep{Hester2008,Dean2008}. The yellow line represents pulsar's magnetic axis and the green arrow is the polarization vector measured by GRAPE. The shaded green region represents the 1-$\sigma$ deviations. }
\label{fig:disc_crab_pic}       
\end{figure}

\citet{Chauvin2017} measured the polarization of the phase-integrated (all pulsar phases) Crab using the PoGO+.
They measured the PF of (20.9$\pm$5.0)\% and the PA of 131$\pm$6.8$^\circ$ which was also close to the spin axis. 
As the measurement was for the phase-integrated Crab, the pulsar structure also had to be considered for locating the emission region. 
Using the high-resolution X-ray images from Chandra, \citet{Chauvin2017} contributed the emission of these photons to two concentric tori at the center position of the pulsars. 
The electrons trapped in these structure would emit synchrotron radiation parallel with the PA (polarization angle) parallel to the spin axis \citep{Nakamura2007} 

\begin{figure}[!t]
\centering 
\includegraphics[width=0.7\textwidth]{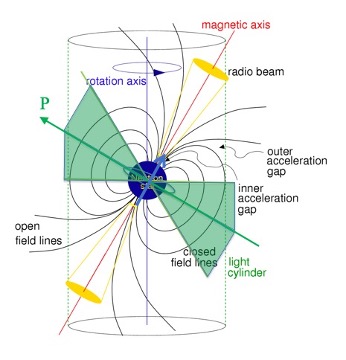}
\caption{A schematic drawing as shown in Figure \ref{fig:sci_pulsar_drawing} from \citet{Lorimer2012}. The green arrow represent the measured polarization vector (P) by GRAPE. The shaded green region is the 1-$\sigma$ deviation. The magnetic field lines responsible for the emission of the detected photons are orthogonal to the polarization vector. This direction is parallel to the magnetic axis and is represented by the blue arrow. The magnetic field lines at the poles are parallel to the magnetic axis which are attributed for the emission of the detected photons.}
\label{fig:disc_crab_pic_b}       
\end{figure}

GRAPE measured the polarization for the phase-integrated Crab data.
The measured PA is 56$^\circ$$^{+30^\circ}_{-30^\circ}$ which is shown in Figure \ref{fig:disc_crab_pic}. 
The yellow arrow labeled J is the magnetic axis and the measured polarization vector from GRAPE is labelled in green as P.
The shaded green represent the 1-$\sigma$ deviations of the measured polarization angle.
Figure \ref{fig:disc_crab_pic_b} shows a pulsar drawing (magnetic dipole) along with the measured polarization vector. 
The direction of the polarization vector (or the polarization angle) suggests that the magnetic field lines responsible for this emission is orthogonal to this vector.
This means that the magnetic field lines responsible for the photon emissions are parallel to the magnetic axis (shown by blue arrow in Figure \ref{fig:disc_crab_pic_b}).  
The possible emission regions, that has the magnetic field lines parallel to the magnetic axis, would be in the equatorial region or in the inner part of the polar region near the poles.

The magnetic field lines at the poles are parallel to the magnetic field axis. 
Additionally, the analysis is done for the phase integrated Crab and the pulsar component 
for the dipole model shown in Figure \ref{fig:disc_crab_pic_b}) are parallel to the magnetic axis. 
The analysis is done for the phase integrated Crab, therefore it is likely that emission region of the detected photons are near the polar caps.
However, the polarization angle measured has a huge upper and lower limit which does not constrain the direction of the magnetic field and subsequently does not help to localize the origin of these photons.

\section{Further studies}
		GRAPE categorizes events into various types depending on the anodes involved.
		The analysis of the 2014 GRAPE data focused mainly on PC (1 plastic and 1 calorimeter) events as the PC events were the most dominant event type that carried a polarization signature.
		Other event types, even though they are not the most dominant, could also be included in this analysis (especially CC and PPC) as these events also carry a polarization signature.
		This will increase the number of counts for the observation. 
		It would also increase the background counts. 
		It is not clear whether the inclusion of these events would increase or decrease the total source to background ratio as this has not been investigated.
		An analysis with the inclusion of these events could potentially improve the statistics of the dataset.
		
		One of the biggest hurdles of the analysis revolved around the optical crosstalk. 
		The optical crosstalk was modeled for the PC event types using the three parameters side-adjacent crosstalk, corner adjacent crosstalk, and the PSD efficiency. 
		The modeled crosstalk was integrated in response simulations used for the analysis.
		Our current model assumed no energy-dependence of the crosstalk. 
		The model could be improved by including energy as an additional parameter which would further improve the instrument response.
	
		Ideally, 	a phase-resolved analysis of the Crab would have provided a better understanding of the origin of the detected photons. 
		Unfortunately, this was not viable with GRAPE due to the limited statistics of the dataset. 
		GRAPE was initially flown in 2011. 
		A full transient ($\sim$8 hrs)  of the Crab was observed during the 2011 flight. 
		This allowed for an attempt to do a phase-resolved analysis of the Crab.
		The off-pulse analysis resulted in a polarization fraction of 0.347 $\pm$ 0.223 and a polarization angle of 87.8$^\circ$ $\pm$ 28.2$^\circ$ \citep{Connor2012}. 
		A phase-integrated analysis resulted in polarization fraction of 0.068 $\pm$ 0.085 and a polarization angle of 118.6$^\circ$ $\pm$ 35.6$^\circ$ \citep{Connor2012}. 
		The polarization measurements were not significant and the poor measurement was attributed to a higher flight background and the crosstalk effect.
		Improvements were implemented to deal with the aforementioned issues for 2014 flight and subsequently the analysis. 
		The flight plan was designed to observe the Crab at higher altitude, thicker shields and additional modules were integrated. 
		The analysis of the 2014 GRAPE data provided us with a better understanding of crosstalk.
		A crosstalk model was included in the response simulations used for data analysis. 
		The background was handled better using the principle component analysis (PCA).
		The 2011 GRAPE flight data could be reanalyzed by integrating the crosstalk model and using the PCA to estimate the background. 
		The data from the 2011 flight and the 2014 flight could be combined for an analysis of the Crab.
		The combined analysis might improve our measurements of the phase integrated results and may also provide  better understanding of the high energy emission using a phase-resolved analysis.

\clearpage
\begin{appendices}
	\chapter{ Coordinate Systems used in GRAPE}
\label{app:appen_coordinate_system}
This appendix lists the coordinate systems and vector transformations used in GRAPE for polarization measurements.
There are two major scatter vector transformations used in polarization measurements. 
The first transformation transforms the scatter vectors to the Pressure Vessel Coordinate (PVC). 
This is used in polarization measurements of non-flight observations. 
For the polarization measurements of the astrophysical objects, the transformed vector in the PVC (from the first part), is further transformed into Celestial Coordinate System (CCS).

\begin{figure}[hbtp]
\centering 
\includegraphics[width=0.9\textwidth]{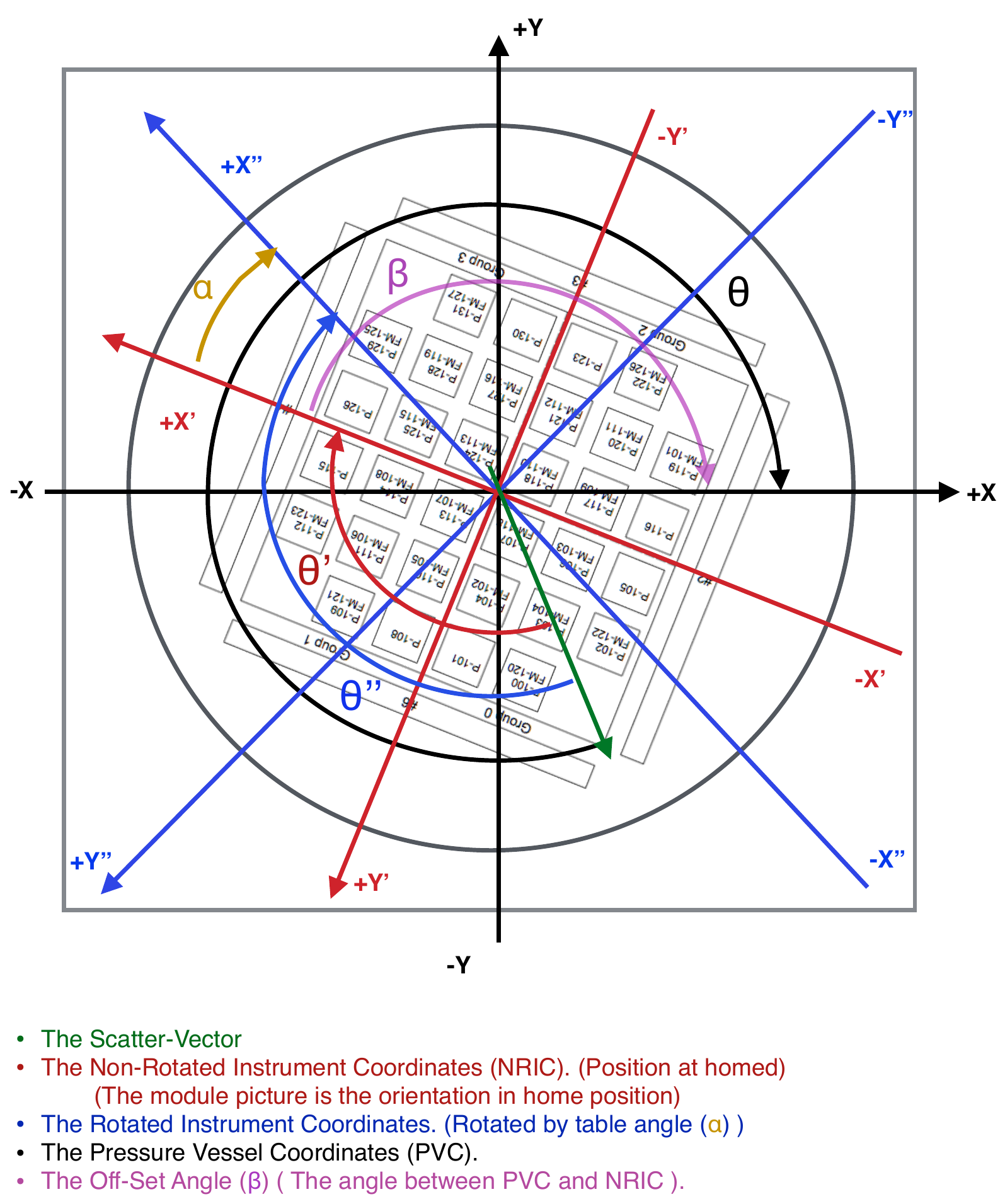}
\caption{Coordinate systems present in the GRAPE instrument. }
\label{fig:appen1_coordinate_system}       
\end{figure}

Figure \ref{fig:appen1_coordinate_system} displays the coordinate systems present in GRAPE instrument which is used for transforming the scatter vector into PVC.
The green arrow is the scatter vector.
The Non-Rotated Instrument Coordinate (NRIC) system is shown in red. 
The instrument assembly rotates about the center and the angle $\alpha$ defines this rotation angle (table angle). 
The Rotated Instrument Coordinate (RIC) is shown in blue. 
At $\alpha$ = 0 (homed), the NRIC equals the RIC. 
The Pressure Vessel Coordinate (PVC) system (shown in black) is defined w.r.t the PV and it is the NRIC with an offset angle $\beta$ which was measured to be 151.6$^\circ$.
The polarization measurements for the pre-flight observations are done in the PVC system.

For the polarization measurements of the astrophysical objects, the scatter vector in the PVC is further transformed to the the Celestial Coordinate (CC) System.
This transformation happens in two steps.
In the first step, the vector in PVC is projected into the sky that is defined by the Altitude/Azimuth coordinate system. 
And in the second step, it is further transformed to the celestial coordinate using the parallactic angle ($\eta$). 
The parallactic angle is the angle between the great circles that pass through the zenith and the celestial north pole \citep{Meadows2007}. 

\begin{figure}[hbtp]
 \centering
\begin{subfigure}[b]{0.35\textwidth}
 		 \includegraphics[width=1\linewidth]{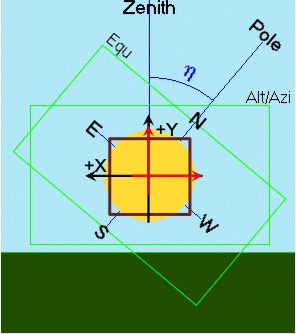}
		 \caption{}
\end{subfigure}   
 \begin{subfigure}[b]{0.5\textwidth}
 		 \includegraphics[width=1\linewidth]{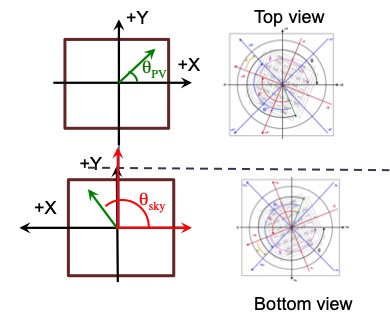}
		 \caption{}
\end{subfigure} 
  \caption{Energy plot of source (blue) , background (green) and the background subtracted (black) energy histogram of Co57. The 122 keV peak is fitted in this figure. }
\label{fig:appen1_parallactic}
\end{figure}

Figure \ref{fig:appen1_parallactic}a illustrates this angle from our instrument's perspective. 
The astrophysical object is the yellow solid circle (Sun in this example) and the green boxes represent the two sky coordinates systems (the Altitude/Azimuth and the Equatorial) \citep{Meadows2007}. 
The parallactic angle $\eta$ is the angle between the zenith and the celestial pole. 
In the Figure \ref{fig:appen1_parallactic}a, we are looking at the astrophysical object through our instrument. 
The PVC in Figure \ref{fig:appen1_coordinate_system} is defined by looking at the instrument from top. 
Therefore the scatter vector is transformed to replicate from looking the bottom (or through) the instrument.
These viewpoints are shown in Figure \ref{fig:appen1_parallactic}a where the "Top view" illustrates the PVC and the "Bottom view" illustrates how the scatter vector transforms when projecting it on the sky. 
The green arrow is the scatter vector, the black arrows represent the axes of PVC and the red arrows represent the axes of the Alt/Azi coordinate system (sky). 
The position angle in sky coordinate system is defined as
\begin{equation}
 \theta_{\text{Sky}} =  180^{\circ} - \theta_{\text{PV}}
 \label{eqn:appen1_sky}
\end{equation}

\begin{figure}[hbtp]
\centering 
\includegraphics[width=0.5\textwidth]{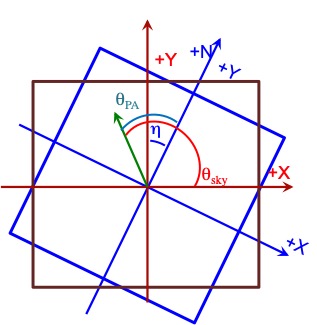}
\caption{Figure illustrating the transformation from Alt/Azi coordinate system  (Red) to celestial coordinate system (Blue) using the parallactic angle ($\eta$). The green is the scatter vector and the $\theta_{\text{PA}}$ is the Position Angle measured from the celestial north.}
\label{fig:appen1_celestial_cord}       
\end{figure}

In the second step the angle $\theta_{\text{Sky}}$ is transformed to the celestial coordinate using the parallactic angle $\eta$. Figure \ref{fig:appen1_celestial_cord} displays the parallactic angle, the celestial coordinate system, the Alt/Azi coordinate system and the scatter vetor. The position angle (PA) is measured from North pole towards east and is represented by
\begin{equation}
 \theta_{\text{PA}} =  \theta_{\text{sky}} - \eta' = \theta_{\text{sky}} -  (90^\circ - \eta) = 180^{\circ} - \theta_{\text{PV}} -  (90^\circ - \eta)  = 90^\circ - \theta_{\text{PV}} + \eta
 \end{equation}
where  $\eta' = 90 - \eta$.

\end{appendices}

\bibliographystyle{mycustombibstyle_3}
	\bibliography{Thesis.bib}

\end{document}